\documentclass[12pt]{article}
\usepackage[a4paper]{geometry}
\usepackage{amsfonts,amssymb,amsmath,epsfig,bbm,enumerate}
\usepackage{cases}
\DeclareMathAlphabet{\mathpzc}{OT1}{pzc}{m}{it} 
\renewcommand{\theequation}{\thesection.\arabic{equation}}
\numberwithin{equation}{section}
\begin{document}
\newcommand{\tr}{\operatorname{tr}}
\newcommand{\vns}{\boldsymbol{\vec{\nabla}'}}
\newcommand{\ds}{\displaystyle}
\newcommand{\scr}{\scriptstyle}
\newcommand{\scs}{\scriptsize}
\newcommand{\e}{\operatorname{e}}
\newcommand{\mustbe}{\stackrel{!}{=}}
\newcommand{\rt}{(r,\vartheta)}
\newcommand{\rts}{(r',\vartheta')}

\newcommand{\sdot}{\,{\scriptscriptstyle{}^{\bullet}}\,}
\newcommand{\A}{\mathcal{A}}
\newcommand{\de}{\delta_{\textrm{e}}}
\newcommand{\ap}{{}^{(a)}\varphi_{\pm}(\vec{r})}
\newcommand{\app}{{}^{(a)}\varphi_{+}(\vec{r})}
\newcommand{\apm}{{}^{(a)}\varphi_{-}(\vec{r})}
\newcommand{\appm}{{}^{(a)}\varphi_{\pm}(\vec{r})}
\newcommand{\sappm}{{}^{(a)}\varphi\,'_{\pm}(\vec{r})}

\newcommand{\apmd}{{}^{(a)}\!\varphi_{\pm}^{\dagger}(\vec{r})}

\newcommand{\amd}{{}^{(a)}\!\varphi_{-}^{\dagger}(\vec{r})}
\newcommand{\apd}{{}^{(a)}\varphi_{+}^{\dagger}(\vec{r})}

\newcommand{\vp}[2]{ {}^{(#1)}\!\varphi_{#2} }
\newcommand{\dvp}[2]{{}^{(#1)}\!\varphi^\dagger_{#2} }

\newcommand{\ivpmp}{{}^{(1)}\varphi_{\mp}(\vec{r})}
\newcommand{\ivppm}{{}^{(1)}\varphi_{\pm}(\vec{r})}
\newcommand{\iivpmp}{{}^{(2)}\varphi_{\mp}(\vec{r})}
\newcommand{\iivppm}{{}^{(2)}\varphi_{\pm}(\vec{r})}
\newcommand{\iiivpmp}{{}^{(3)}\varphi_{\mp}(\vec{r})}
\newcommand{\iiivppm}{{}^{(3)}\varphi_{\pm}(\vec{r})}
\newcommand{\ipp}{{}^{(1)}\varphi_{+}(\vec{r})}
\newcommand{\ipm}{{}^{(1)}\varphi_{-}(\vec{r})}

\newcommand{\iipp}{{}^{(2)}\varphi_{+}(\vec{r})}
\newcommand{\iipm}{{}^{(2)}\varphi_{-}(\vec{r})}

\newcommand{\iiipp}{{}^{(3)}\varphi_{+}(\vec{r})}
\newcommand{\iiipm}{{}^{(3)}\varphi_{-}(\vec{r})}

\newcommand{\sPh}{{\stackrel{*}{\Phi}}(\vec{r}_1,\vec{r}_2)}
\newcommand{\sph}{{\stackrel{*}{\varphi}}}
\newcommand{\sphi}{{\stackrel{*}{\varphi}_1}(\vec{r}_1)}
\newcommand{\sphii}{{\stackrel{*}{\varphi}_2}(\vec{r}_2)}
\newcommand{\sphis}{{\stackrel{*}{\varphi}_1}(\vec{r}\,')}
\newcommand{\sphiis}{{\stackrel{*}{\varphi}_2}(\vec{r}\,')}

\newcommand{\hH}[1]{\hat{\mathcal{H}}_{\textrm{(#1)}}\,}
\newcommand{\shH}[1]{\hat{\mathcal{H}}\,'_{\textrm{(#1)}}\,}
\newcommand{\hh}[1]{\hat{\mathfrak{h}}_{\textrm{(#1)}}\,}
\newcommand{\shh}[1]{\hat{\mathfrak{h}}\,'_{\textrm{(#1)}}\,}
\newcommand{\hR}[1]{\hat{R}_{\textrm{(#1)}}\,}
\newcommand{\shr}[1]{\hat{r}\,'_{\textrm{(#1)}}\,}
\newcommand{\shR}[1]{\hat{R}\,'_{\textrm{(#1)}}\,}
\newcommand{\hr}[1]{\hat{r}_{\textrm{(#1)}}\,}

\newcommand{\aRpm}{{}^{(a)\!}\tilde{R}_\pm}
\newcommand{\aRp} {{}^{(a)\!}\tilde{R}_+{}}
\newcommand{\aRm}{{}^{(a)\!}\tilde{R}_-{}}
\newcommand{\iRp}{{}^{(1)\!}\tilde{R}_+{}}
\newcommand{\iRm}{{}^{(1)\!}\tilde{R}_-{}}
\newcommand{\iiRp}{{}^{(2)\!}\tilde{R}_+{}}
\newcommand{\iiRm}{{}^{(2)\!}\tilde{R}_-{}}
\newcommand{\ntR}{{}^{\{0\}}\!\tilde{R}}
\newcommand{\aSpm}{{}^{(a)\!}\tilde{S}_\pm}
\newcommand{\aSp}{{}^{(a)\!}\tilde{S}_+{}}
\newcommand{\aSm}{{}^{(a)\!}\tilde{S}_-{}}
\newcommand{\iSp}{{}^{(1)\!}\tilde{S}_+{}}
\newcommand{\iSm}{{}^{(1)\!}\tilde{S}_-{}}
\newcommand{\iiSp}{{}^{(2)\!}\tilde{S}_+{}}
\newcommand{\iiSm}{{}^{(2)\!}\tilde{S}_-{}}

\newcommand{\vSm}[1] { {}^{\textrm{(m)}}\!\vec{S}_{\textrm{#1}} }

\newcommand{\ikn}{{}^{(1)}\!k_0(\vec{r})}
\newcommand{\ikp}{{}^{(1)}\!k_\phi(\vec{r})}
\newcommand{\iikp}{{}^{(2)}\!k_\phi(\vec{r})}

\newcommand{\iikn}{{}^{(2)}\!k_0(\vec{r})}
\newcommand{\ikns}{{}^{(1)}\!k_0(\vec{r}\,')}
\newcommand{\iikns}{{}^{(2)}\!k_0(\vec{r}\,')}

\newcommand{\itkn}{{}^{(1)}\!\tilde{k}_0}
\newcommand{\iitkn}{{}^{(2)}\!\tilde{k}_0}

\newcommand{\itkp}{{}^{(1)}\!\tilde{k}_\phi}
\newcommand{\iitkp}{{}^{(2)}\!\tilde{k}_\phi}
\newcommand{\ijn}{{}^{(1)}\!j_0}
\newcommand{\iijn}{{}^{(2)}\!j_0}
\newcommand{\iiijn}{{}^{(3)}\!j_0}

\newcommand{\jj}[2]{ {}^{(#1)} {j^#2} }

\newcommand{\ajn}{{}^{(a)}\!j_0(\vec{r})}
\newcommand{\akn}{{}^{(a)}\!k_0\left(\vec{r}\right)}
\newcommand{\sakn}{{}^{(a)}\!k_0\left(\vec{r}\,'\right)}
\newcommand{\btkn}{{}^{(b)}\!\tilde{k}_0\left(r\right)}
\newcommand{\bkn}{{}^{(b)}\!k_0\left(\vec{r}\right)}
\newcommand{\bknn}{{}^{(b)}\!k_0\left(r\right)}
\newcommand{\btknn}{{}^{(b)}\!\tilde{k}_0\left(r\right)}
\newcommand{\aki}{{}^{(a)}\!k_1(\vec{r})}
\newcommand{\akp}{{}^{(a)}\!k_\phi(\vec{r})}
\newcommand{\akv}{\vec{k}_a(\vec{r})}

\newcommand{\aA}{{}^{(a)}\!A}
\newcommand{\vAa}{\vec{A}_a(\vec{r}}
\newcommand{\aAr}{{}^{(a)}\!A_0}
\newcommand{\aAe}{{}^{[a]}\!A_0}
\newcommand{\aAn}{{}^{(a)}\!A_0(\vec{r})}
\newcommand{\pAn}{{}^{(\textrm{p})}\!A_0(\vec{r})}
\newcommand{\ppAn}{{}^{[\textrm{p}]}\!A_0}
\newcommand{\nF}{{}^{\{0\}}\!F}
\newcommand{\nAn}{{}^{\{0\}}\!A_0}
\newcommand{\ppF}{{}^{[\textrm{p}]}\!F(r)}
\newcommand{\ppFn}{{}^{[\textrm{p}]}\!F(0)}

\newcommand{\iA}{{}^{(1)}\!\!A}
\newcommand{\iiA}{{}^{(2)}\!\!A}
\newcommand{\iiiA}{{}^{(3)}\!\!A}

\newcommand{\iAn}{{}^{(1)}\!\!A_0}
\newcommand{\iiAn}{{}^{(2)}\!\! A_0 }
\newcommand{\iiiAn}{{}^{(3)}\!\! A_0 }
\newcommand{\fAn}{{}^{(1/2)}\!\! A_0 }

\newcommand{\IA}{{}^{\textrm{(I)}}\!\!A}
\newcommand{\IIA}{{}^{\textrm{(II)}}\!\!A}
\newcommand{\IIIA}{{}^{\textrm{(III)}}\!\!A}
\newcommand{\AI}{A_{\textrm{I}}}
\newcommand{\AII}{A_{\textrm{II}}}
\newcommand{\AIII}{A_{\textrm{III}}}
\newcommand{\vAI}{\vec{A}_{\textrm{I}}}
\newcommand{\vAII}{\vec{A}_{\textrm{II}}}
\newcommand{\vAIII}{\vec{A}_{\textrm{III}}}
\newcommand{\AM}{\mathfrak{A}_{\textrm{M}}}
\newcommand{\An}{\mathfrak{A}_0}
\newcommand{\aVHS}{{}^{(a)}V_{\textrm{HS}}(\vec{r})}
\newcommand{\iVHS}{{}^{(1)}V_{\textrm{HS}}(\vec{r})}
\newcommand{\iiVHS}{{}^{(2)}V_{\textrm{HS}}(\vec{r})}

\newcommand{\onp}{\omega_0^\textrm{(+)}}
\newcommand{\onm}{\omega_0^\textrm{(-)}}
\newcommand{\oip}{\omega_1^\textrm{(+)}}
\newcommand{\oim}{\omega_1^\textrm{(-)}}
\newcommand{\onpm}{\omega_0^{(\pm)}}
\newcommand{\oipm}{\omega_1^{(\pm)}}

\renewcommand{\L}{\mathcal{L}}
\newcommand{\B}{\mathcal{B}}
\newcommand{\as}{\alpha_{\textrm{s}}\,}
\newcommand{\aB}{a_{\textrm{B}}\,}
\newcommand{\aM}{a_{\textrm{M}}\,}
\newcommand{\ab}{a_{\textrm{b}}\,}
\newcommand{\Bstar}{\overset{*}{B}{} }
\newcommand{\vXstar}{\overset{*}{\vec{X}}}
\newcommand{\vYstar}{\overset{*}{\vec{Y}}}
\newcommand{\vBstar}{\overset{*}{\vec{B}}}
\newcommand{\Gstar}{\overset{*}{G}}
\newcommand{\gstar}{\overset{*}{g}}
\newcommand{\hstar}{\overset{*}{h}}
\newcommand{\vhstar}{\overset{*}{\vec{h}}}
\newcommand{\CC}{\mathbb{C}}
\newcommand{\GG}{\textnormal I \! \Gamma}
\newcommand{\vfE}{\vec{\mathfrak{E}}}
\newcommand{\vfH}{\vec{\mathfrak{H}}}
\newcommand{\vfA}{\vec{\mathfrak{A}}}
\newcommand{\D}{\mathcal{D}}
\newcommand{\DD}{\mathbb{D}}
\newcommand{\F}{\mathcal{F}}
\newcommand{\FM}{F_{\textrm{M}}}
\renewcommand{\H}{\mathcal{H}}
\newcommand{\I}{\mathcal{I}}
\newcommand{\J}{\mathcal{J}}
\newcommand{\LRST}{\mathcal{L}_{\textrm{RST}}}
\newcommand{\LD}{\mathcal{L}_{\textrm{D}}}
\newcommand{\LG}{\mathcal{L}_{\textrm{G}}}
\newcommand{\U}{\mathcal{U}}
\newcommand{\M}{\mathcal{M}}
\newcommand{\T}{\mathcal{T}}
\newcommand{\Mpe}{M_\textrm{p/e}\,}
\newcommand{\Mp}{M_\textrm{p}\,}
\newcommand{\Mee}{M^\textrm{(e)}}
\newcommand{\Me}{M_\textrm{e}\,}
\newcommand{\MT}{M_\textrm{T}}
\newcommand{\tMT}{\tilde{M}_\textrm{T}}
\newcommand{\MM}[1]{M_{[#1]}}
\newcommand{\Mei}{M_\textrm{I}^\textrm{(e)}}
\newcommand{\Meii}{M_\textrm{II}^\textrm{(e)}}
\newcommand{\Meiii}{M_\textrm{I/II}^\textrm{(e)}}
\newcommand{\Mmi}{M_\textrm{I}^\textrm{(m)}}
\newcommand{\Mmii}{M_\textrm{II}^\textrm{(m)}}
\newcommand{\Mmiii}{M_\textrm{I/II}^\textrm{(m)}}
\newcommand{\TT}{{}^{(\textrm{T})}T}
\newcommand{\DT}{{}^{(\textrm{D})}T}
\newcommand{\GT}{{}^{(\textrm{G})}T}
\newcommand{\hER}{\hat{E}_\textrm{R}}

\newcommand{\eER}{E_\textrm{R}^\textrm{(e)}}
\newcommand{\mER}{E_\textrm{R}^\textrm{(m)}}
\newcommand{\hEC}{E_\textrm{C}^\textrm{(h)}}
\newcommand{\gEC}{E_\textrm{C}^\textrm{(g)}}

\newcommand{\vH}[1]{\vec{H}_{\textrm{#1}}}

\newcommand{\EP}[1]{ E_{\textrm{P}(#1) }   }
\newcommand{\heER}{\hat{E}_\textrm{R}^\textrm{(e)}}
\newcommand{\hmER}{\hat{E}_\textrm{R}^\textrm{(m)}}
\newcommand{\ED}{{E_{\textrm{D}}}}
\newcommand{\EG}{E_{\textrm{G}}}
\newcommand{\Ec}{E_{\textrm{conv}}}
\newcommand{\cEc}{{\cal E}_{\textrm{conv}}}
\newcommand{\ET}{E_{\textrm{T}}}
\newcommand{\ETT}{E_{\textrm{[T]}}}
\newcommand{\tETT}{\tilde{E}_{\textrm{[T]}}}
\newcommand{\tbbETT}{\tilde{\mathbb{E}}_{\textrm{[T]}}}
\newcommand{\bbEIV}{\mathbb{E}^{\textrm{(IV)}}_{\textrm{[T]}}}
\newcommand{\Eiv}{E_\textrm{[T]}^{\textrm{(IV)}}}
\newcommand{\Epot}{ {\cal E}_{\textrm{pot}} }
\newcommand{\epot}{ {\varepsilon}_{\textrm{pot}} }
\newcommand{\tET}{\tilde{E}_\textrm{T}}
\newcommand{\tETi}{\tilde{E}_\textrm{T}^{(|)}}
\newcommand{\tETii}{\tilde{E}_\textrm{T}^{(||)}}
\newcommand{\tEnT}{ {\tilde{E}^{(0)}_\textrm{T}} }
\newcommand{\tEnTmin}{ {\tilde{E}^{(0)}_\textrm{T,min}} }
\newcommand{\ETiii}{E_\textrm{T}^{(|||)}}
\newcommand{\Ekin}{{E_\textrm{kin}}}
\newcommand{\cEk}{{\cal{E}_\textrm{kin}}}
\newcommand{\ES}{{E_{\textrm{S}}}}
\newcommand{\EW}{{E_\textrm{W}}}
\newcommand{\HS}{H_{\textrm{S}}}
\newcommand{\hHin}{\hat{H}_{\textrm{in}}}
\newcommand{\hHS}{\hat{H}_{\textrm{S}}}
\newcommand{\WRST}{W_{\textrm{RST}}}
\newcommand{\WS}{W_{\textrm{S}}}
\newcommand{\lG}{{\lambda_\textrm{G}}}
\newcommand{\lGe}{ {\lambda_\textrm{G}^{(\textrm{e})}}\!}
\newcommand{\lGh}{ {\lambda_\textrm{G}^{(\textrm{h})}}\!}
\newcommand{\lGg}{ {\lambda_\textrm{G}^{(\textrm{g})}}\!}
\newcommand{\lGm}{  {\lambda_\textrm{G}^{(\textrm{m})}}\!\!}
\newcommand{\lGma}{  {\lambda_{\textrm{G}(a)}^{(\textrm{m})}}\!\!}
\newcommand{\lGem}{ {\lambda_\textrm{G}^{(\textrm{e/m})}}}
\newcommand{\lGema}{ {\lambda_{\textrm{G}(a)}^{(\textrm{e/m})}}}
\newcommand{\lD}{{\lambda_\textrm{D}}}
\newcommand{\lS}{{\lambda_\textrm{S}}}
\newcommand{\lpa}{\lambda_{\textrm{p}(a)}}
\newcommand{\NDa}{N_{\textrm{D}(a)}}
\newcommand{\np}{{n_\textrm{p}}}
\newcommand{\NN}{\mathbb{N}}
\newcommand{\NND}{\mathbb{N}_\textrm{D}}
\newcommand{\NNG}{\mathbb{N}_\textrm{G}}
\newcommand{\NNDa}{\mathbb{N}_{\textrm{D}(a)}}
\newcommand{\ND}{{N_\textrm{D}}}
\renewcommand{\NG}{{N_\textrm{G}}}
\newcommand{\NDn}{{N_\textrm{D}^\textrm{(0)}}}
\newcommand{\NGn}{{N_\textrm{G}^\textrm{(0)}}}
\newcommand{\NGe}{{N_\textrm{G}^\textrm{(e)}}}
\newcommand{\NNGe}{{\mathbb{N}_\textrm{G}^\textrm{(e)}}}
\newcommand{\NGh}{{N_\textrm{G}^\textrm{(h)}}}
\newcommand{\NNGh}{{\mathbb{N}_\textrm{G}^\textrm{(h)}}}
\newcommand{\NGg}{{N_\textrm{G}^\textrm{(g)}}}
\newcommand{\NNGg}{{\mathbb{N}_\textrm{G}^\textrm{(g)}}}
\newcommand{\nG}{{n_\textrm{G}}}
\newcommand{\snG}{{\tilde{n}_\textrm{G}}}
\newcommand{\nGe}{{n_\textrm{G}^\textrm{(e)}}}
\newcommand{\snGe}{{\tilde{n}_\textrm{G}^\textrm{(e)}}}
\newcommand{\nGh}{{n_\textrm{G}^\textrm{(h)}}}
\newcommand{\snGh}{{\tilde{n}_\textrm{G}^\textrm{(h)}}}
\newcommand{\nGg}{{n_\textrm{G}^\textrm{(g)}}}
\newcommand{\snGg}{{\tilde{n}_\textrm{G}^\textrm{(g)}}}
\newcommand{\NGm}{{N_\textrm{G}^\textrm{(m)}}}
\newcommand{\NGem}{{N_\textrm{G}^\textrm{(e,m)}}}
\newcommand{\nGm}{{n_\textrm{G}^\textrm{(m)}}}
\newcommand{\nGem}{{n_\textrm{G}^\textrm{(e,m)}}}
\newcommand{\LDk}{{\LD^{(\mathrm{kin})}}}
\newcommand{\LDe}{{\LD^{(\mathrm{e})}}}
\newcommand{\LDm}{{\LD^{(\mathrm{m})}}}
\newcommand{\LDM}{{\LD^{(\mathrm{M})}}}
\newcommand{\Tkin}{{T_\textrm{kin}}}
\newcommand{\oWD}{{\overset{\circ}{W}\!}_\textrm{D}}
\newcommand{\oWG}{{\overset{\circ}{W}\!}_\textrm{G}}
\newcommand{\oWRST}{{\overset{\circ}{W}\!}_\textrm{RST}}
\newcommand{\oWRSTe}{{\overset{\circ}{W}}{}^{\textrm{(e)}}_\textrm{RST}}
\renewcommand{\S}{\mathcal{S}}
\newcommand{\Ph}{\Phi(\vec{r}_1,\vec{r}_2)}
\newcommand{\VN}{V_\NN}
\newcommand{\TN}{T_\NN}
\newcommand{\cTN}{\mathcal{T}_\NN}
\newcommand{\PN}{P_\NN}
\newcommand{\pN}{(p_0,p_1,\ldots p_\NN) }
\newcommand{\cVN}{\mathcal{V}_\NN}
\newcommand{\Z}{\mathcal{Z}}
\newcommand{\tZ}{\tilde{\mathcal{Z}}}
\newcommand{\covnab}{{\boldsymbol{\vec{\nabla}{}'}}}
\newcommand{\covlap}{\boldsymbol{\Delta\!'}}
\newcommand{\dkmat}{\delta^\dagger_\textrm{mat}}
\newcommand{\tEET}{\tilde{\mathbb{E}}_\textrm{[T]}}
\newcommand{\vmSe}{{}^\textrm{(m)}\vec{S}_1}
\newcommand{\vmSz}{{}^\textrm{(m)}\vec{S}_2}
\newcommand{\vmSd}{{}^\textrm{(m)}\vec{S}_3}
\newcommand{\EPa}{E_{{\rm P}(a)}}
\newcommand{\ERe}{{E_\textrm{R}^\textrm{(e)}}}
\newcommand{\ERm}{{E_\textrm{R}^\textrm{(m)}}}
\newcommand{\ECh}{{E_\textrm{C}^\textrm{(h)}}}
\newcommand{\ECg}{{E_\textrm{C}^\textrm{(g)}}}
\newcommand{\lDa}{\lambda_{{\rm D}(a)}}
\newcommand{\lPa}{\lambda_{{\rm P}(a)}}
\newcommand{\lPe}{\lambda_{{\rm P}(1)}}
\newcommand{\lPz}{\lambda_{{\rm P}(2)}}
\newcommand{\lPd}{\lambda_{{\rm P}(3)}}
\newcommand{\MMg}{{\mathbb{M}^\textrm{(g)}}}
\newcommand{\MMgconv}{{\mathbb{M}_\textrm{conv}^\textrm{(g)}}}
\newcommand{\MMgpol}{{\mathbb{M}_\textrm{pol}^\textrm{(g)}}}
\newcommand{\aphip}{{}^{(a)}\varphi_+}
\newcommand{\aphim}{{}^{(a)}\varphi_-}
\newcommand{\epp}{{}^{(1)}\varphi_{+}}
\newcommand{\epm}{{}^{(1)}\varphi_{-}}
\newcommand{\zpp}{{}^{(2)}\varphi_{+}}
\newcommand{\zpm}{{}^{(2)}\varphi_{-}}
\newcommand{\zppm}{{}^{(2)}\varphi_{\pm}}
\newcommand{\dpp}{{}^{(3)}\varphi_{+}}
\newcommand{\dpm}{{}^{(3)}\varphi_{-}}
\newcommand{\bb}{\mathbbm{b}}
\newcommand{\Ih}{\mathbbm{h}}
\newcommand{\zz}{\mathbbm{z}}
\newcommand{\ZZ}{\mathbb{Z}}
\newcommand{\vbb}{\vec{\bb}}
\newcommand{\vIh}{\vec{\Ih}}
\newcommand{\vzz}{\vec{\zz}}
\newcommand{\vZZ}{\vec{\ZZ}}
\newcommand{\vbbstar}{\overset{*}{\vec{\bb}}}
\newcommand{\vIhstar}{\overset{*}{\vec{\Ih}}}
\newcommand{\vzzstar}{\overset{*}{\vec{\zz}}}
\newcommand{\vZZstar}{\overset{*}{\vec{\ZZ}}}
\newcommand{\adn}{{}^{(a)}\!\delta_0}
\newcommand{\idn}{{}^{(1)}\!\delta_0}
\newcommand{\iidn}{{}^{(2)}\!\delta_0}
\newcommand{\iiidn}{{}^{(3)}\!\delta_0}
\newcommand{\Ij}{{}^{\textrm{(I)}}\!j}
\newcommand{\IIj}{{}^{\textrm{(II)}}\!j}
\newcommand{\IIIj}{{}^{\textrm{(III)}}\!j}
\renewcommand{\ij}{{}^{(1)}\!j}
\newcommand{\iij}{{}^{(2)}\!j}
\newcommand{\iiij}{{}^{(3)}\!j}
\newcommand{\adup}{{}^{(a)}\!\delta_\uparrow}
\newcommand{\idup}{{}^{(1)}\!\delta_\uparrow}
\newcommand{\iidup}{{}^{(2)}\!\delta_\uparrow}
\newcommand{\iiidup}{{}^{(3)}\!\delta_\uparrow}
\newcommand{\Bdo}{{}^{(B)}\!\delta_0}
\newcommand{\Bdup}{{}^{(B)}\!\delta_\uparrow}
\newcommand{\Tkina}{{T_{\textrm{kin}(a)}}}
\newcommand{\Tkine}{{T_{\textrm{kin}(1)}}}
\newcommand{\Tkinz}{{T_{\textrm{kin}(2)}}}
\newcommand{\Tkind}{{T_{\textrm{kin}(3)}}}
\newcommand{\tNGm}{{\tilde{N}_\textrm{G}^\textrm{(m)}}}
\newcommand{\hNGm}{{\hat{N}_\textrm{G}^\textrm{(m)}}}
\newcommand{\bbpAo}{{}^{[b/p]}\!A_0}

\newcommand{\bbkphi}{{}^{[b]}k_\phi}
\newcommand{\ppkphi}{{}^{[p]}k_\phi}


\newcommand{\vr}{\vec{r}}
\newcommand{\myrf}[1]{(\ref{eq:#1})}
\newcommand{\eSo}{{\sf {}^1S_0}}
\newcommand{\dSe}{{\sf {}^3S_1}}
\newcommand{\MR}{\mathcal{R}}
\newcommand{\MS}{\mathcal{S}}
\newcommand{\bpppm}{{}^{(b/p)}\varphi_{\pm}(\vec{r})}
\newcommand{\bpphip}{{}^{(b/p)}\varphi_{+}(\vec{r})}
\newcommand{\bpphim}{{}^{(b/p)}\varphi_{-}(\vec{r})}
\newcommand{\bppm}{{}^{(b)}\varphi_{\pm}(\vec{r})}
\newcommand{\bpmp}{{}^{(b)}\varphi_{\mp}(\vec{r})}
\newcommand{\bpp}{{}^{(b)}\varphi_{+}(\vec{r})}
\newcommand{\bpm}{{}^{(b)}\varphi_{-}(\vec{r})}
\newcommand{\bppk}{{}^{(b)}\varphi_{+}^\dagger(\vec{r})}
\newcommand{\bpmk}{{}^{(b)}\varphi_{-}^\dagger(\vec{r})}
\newcommand{\pppm}{{}^{(p)}\varphi_{\pm}(\vec{r})}
\newcommand{\ppmp}{{}^{(p)}\varphi_{\mp}(\vec{r})}
\newcommand{\ppp}{{}^{(p)}\varphi_{+}(\vec{r})}
\newcommand{\ppm}{{}^{(p)}\varphi_{-}(\vec{r})}
\newcommand{\teETT}{{}^{\textrm{(e)}}\!\tilde{E}_{[T]}}
\newcommand{\eETiv}{{{}^{\textrm{(e)}}\!E^\textrm{(IV)}_{\textrm{T}}}}
\newcommand{\bpA}{{}^{(b/p)}\!A}
\newcommand{\bpAo}{{}^{(b/p)}\!A_0}
\newcommand{\bAn}{{}^{(b)}\!A_0}
\newcommand{\bAo}{\bAn}
\newcommand{\bbAo}{{}^{[b]}\!A_0}
\newcommand{\bAphi}{{}^{(b)}\!A_\phi}
\newcommand{\pAo}{{}^{(p)}\!A_0}
\newcommand{\pAphi}{{}^{(p)}\!A_\phi}
\newcommand{\ppAo}{{}^{[p]}\!A_0}
\newcommand{\bpjo}{{}^{(b/p)}j_0}
\newcommand{\bpko}{{}^{(b/p)}k_0}
\newcommand{\bpkphi}{{}^{(b/p)}k_\phi}
\newcommand{\bpk}{{}^{(b/p)}k}
\newcommand{\pk}{{}^{(p)}k}
\newcommand{\bko}{{}^{(b)}k_0}
\newcommand{\bkphi}{{}^{(b)\!}k_\phi}
\newcommand{\pko}{{}^{(p)}k_0}
\newcommand{\pkphi}{{}^{(p)}k_\phi}
\newcommand{\bbko}{{}^{[b]}k_0}
\newcommand{\ppko}{{}^{[p]}k_0}
\newcommand{\ak}{{}^{(a)}k}
\newcommand{\ek}{{}^{(1)}k}
\newcommand{\zk}{{}^{(2)}k}
\renewcommand{\aRpm}{{}^{(a)}R_\pm}
\newcommand{\pRpm}{{}^{(p)}\!R_\pm}
\newcommand{\pRp}{{}^{(p)}R_+}
\newcommand{\pRm}{{}^{(p)}R_-}
\newcommand{\bRpm}{{}^{(b)}R_\pm}
\newcommand{\bRp}{{}^{(b)}R_+}
\newcommand{\bRm}{{}^{(b)}R_-}
\newcommand{\sRpm}{\,'\!R_\pm}
\newcommand{\sRp}{\,'\!R_+}
\newcommand{\sRm}{\,'\!R_-}
\newcommand{\ssRpm}{\,''\!R_\pm}
\newcommand{\ssRp}{\,''\!R_+}
\newcommand{\ssRm}{\,''\!R_-}
\newcommand{\ptRpm}{{}^{(p)}\tilde{R}_\pm}
\newcommand{\ptRp}{{}^{(p)}\tilde{R}_+}
\newcommand{\ptRm}{{}^{(p)}\tilde{R}_-}
\newcommand{\btRpm}{{}^{(b)}\tilde{R}_\pm}
\newcommand{\btRp}{{}^{(b)}\tilde{R}_+}
\newcommand{\btRm}{{}^{(b)}\tilde{R}_-}
\renewcommand{\aSpm}{{}^{(a)}S_\pm}
\newcommand{\pSpm}{{}^{(p)}\!S_\pm}
\newcommand{\pSp}{{}^{(p)}S_+}
\newcommand{\pSm}{{}^{(p)}S_-}
\newcommand{\bSpm}{{}^{(b)}S_\pm}
\newcommand{\bSp}{{}^{(b)}S_+}
\newcommand{\bSm}{{}^{(b)}S_-}
\newcommand{\sSpm}{\,'\!S_\pm}
\newcommand{\sSp}{\,'\!S_+}
\newcommand{\sSm}{\,'\!S_-}
\newcommand{\ssSpm}{\,''\!S_\pm}
\newcommand{\ssSp}{\,''\!S_+}
\newcommand{\ssSm}{\,''\!S_-}
\newcommand{\ptSpm}{{}^{(p)}\tilde{S}_\pm}
\newcommand{\ptSp}{{}^{(p)}\tilde{S}_+}
\newcommand{\ptSm}{{}^{(p)}\tilde{S}_-}
\newcommand{\btSpm}{{}^{(b)}\tilde{S}_\pm}
\newcommand{\btSp}{{}^{(b)}\tilde{S}_+}
\newcommand{\btSm}{{}^{(b)}\tilde{S}_-}
\newcommand{\bpMRpm}{{}^{(b/p)}\!\mathcal{R}_\pm}
\newcommand{\bpMRp}{{}^{(b/p)}\!\mathcal{R}_+}
\newcommand{\bpMRm}{{}^{(b/p)}\!\mathcal{R}_-}
\newcommand{\aMRpm}{{}^{(a)}\!\mathcal{R}_\pm}
\newcommand{\aMRp}{{}^{(a)}\!\mathcal{R}_+}
\newcommand{\aMRm}{{}^{(a)}\!\mathcal{R}_-}
\newcommand{\aMRpS}{{}^{(a)}\!{\overset{*}{\mathcal{R}}{} }_+}
\newcommand{\aMRmS}{{}^{(a)}\!{\overset{*}{\mathcal{R}}{} }_-}
\newcommand{\bMRpS}{{}^{(b)}\!{\overset{*}{\mathcal{R}}{} }_+}
\newcommand{\bMRmS}{{}^{(b)}\!{\overset{*}{\mathcal{R}}{} }_-}
\newcommand{\bMRpm}{{}^{(b)}\!\mathcal{R}_\pm}
\newcommand{\bMRp}{{}^{(b)}\!\mathcal{R}_+}
\newcommand{\bMRm}{{}^{(b)}\!\mathcal{R}_-}
\newcommand{\pMRpm}{{}^{(p)}\!\mathcal{R}_\pm}
\newcommand{\pMRp}{{}^{(p)}\!\mathcal{R}_+}
\newcommand{\pMRm}{{}^{(p)}\!\mathcal{R}_-}
\newcommand{\eMRpm}{{}^{(1)}\!\mathcal{R}_\pm}
\newcommand{\eMRp}{{}^{(1)}\!\mathcal{R}_+}
\newcommand{\eMRm}{{}^{(1)}\!\mathcal{R}_-}
\newcommand{\zMRpm}{{}^{(2)}\!\mathcal{R}_\pm}
\newcommand{\zMRp}{{}^{(2)}\!\mathcal{R}_+}
\newcommand{\zMRm}{{}^{(2)}\!\mathcal{R}_-}
\newcommand{\bpMSpm}{{}^{(b/p)}\!\mathcal{S}_\pm}
\newcommand{\bpMSp}{{}^{(b/p)}\!\mathcal{S}_+}
\newcommand{\bpMSm}{{}^{(b/p)}\!\mathcal{S}_-}
\newcommand{\aMSpm}{{}^{(a)}\!\mathcal{S}_\pm}
\newcommand{\aMSp}{{}^{(a)}\!\mathcal{S}_+}
\newcommand{\aMSm}{{}^{(a)}\!\mathcal{S}_-}
\newcommand{\aMSpS}{{}^{(a)}\!{\overset{*}{\mathcal{S}}{} }_+}
\newcommand{\aMSmS}{{}^{(a)}\!{\overset{*}{\mathcal{S}}{} }_-}
\newcommand{\bMSpS}{{}^{(b)}\!{\overset{*}{\mathcal{S}}{} }_+}
\newcommand{\bMSmS}{{}^{(b)}\!{\overset{*}{\mathcal{S}}{} }_-}
\newcommand{\bMSpm}{{}^{(b)}\!\mathcal{S}_\pm}
\newcommand{\bMSp}{{}^{(b)}\!\mathcal{S}_+}
\newcommand{\bMSm}{{}^{(b)}\!\mathcal{S}_-}
\newcommand{\pMSpm}{{}^{(p)}\!\mathcal{S}_\pm}
\newcommand{\pMSp}{{}^{(p)}\!\mathcal{S}_+}
\newcommand{\pMSm}{{}^{(p)}\!\mathcal{S}_-}
\newcommand{\eMSpm}{{}^{(1)}\!\mathcal{S}_\pm}
\newcommand{\eMSp}{{}^{(1)}\!\mathcal{S}_+}
\newcommand{\eMSm}{{}^{(1)}\!\mathcal{S}_-}
\newcommand{\zMSpm}{{}^{(2)}\!\mathcal{S}_\pm}
\newcommand{\zMSp}{{}^{(2)}\!\mathcal{S}_+}
\newcommand{\zMSm}{{}^{(2)}\!\mathcal{S}_-}
\newcommand{\MCa}{\mathcal{C}_{(a)}}
\newcommand{\MCaS}{{\overset{*}{\mathcal{C}}{} }_{(a)}}
\newcommand{\tZp}{\tilde{\Z}_\mathcal{P}}
\newcommand{\tZO}{\tilde{\Z}_\Omega}
\newcommand{\tNGe}{{\tilde{N}_\textrm{G}^\textrm{(e)}}}
\newcommand{\tTkin}{{\tilde{T}_\textrm{kin}}}
\newcommand{\bTkin}{{{}^{(b)}T_\textrm{kin}}}
\newcommand{\bTr}{{{}^{(b)}T_r}}
\newcommand{\btTr}{{{}^{(b)}\tilde{T}_r}}
\newcommand{\bbtTr}{{{}^{[b]}\tilde{T}_r}}
\newcommand{\bTth}{{{}^{(b)}T_\vartheta}}
\newcommand{\btTth}{{{}^{(b)}\tilde{T}_\vartheta}}
\newcommand{\bbtTth}{{{}^{[b]}\tilde{T}_\vartheta}}
\newcommand{\bTph}{{{}^{(b)}T_\phi}}
\newcommand{\btTph}{{{}^{(b)}\tilde{T}_\phi}}
\newcommand{\bbtTph}{{{}^{[b]}\tilde{T}_\phi}}
\newcommand{\btTkin}{{{}^{(b)}\tilde{T}_\textrm{kin}}}
\newcommand{\bbtTkin}{{{}^{[b]}\tilde{T}_\textrm{kin}}}
\newcommand{\tMe}{\tilde{M}^\textrm{(e)}}
\newcommand{\tND}{\tilde{N}_{\textrm{D}}}
\newcommand{\tNPhi}{\tilde{N}_{\Phi}}
\newcommand{\tNO}{\tilde{N}_{\Omega}}
\newcommand{\lDe}{\lambda_{{\rm D}(1)}}
\newcommand{\lDz}{\lambda_{{\rm D}(2)}}
\newcommand{\NDe}{{N_\textrm{D}^\textrm{(1)}}}
\newcommand{\NDz}{{N_\textrm{D}^\textrm{(2)}}}
\newcommand{\lP}{\ell_\mathcal{P}}
\newcommand{\lO}{\ell_\mathcal{O}}
\newcommand{\dlO}{\dot{\ell}_\mathcal{O}}
\newcommand{\ddlO}{\ddot{\ell}_\mathcal{O}}
\newcommand{\nP}{n_\mathcal{P}}
\newcommand{\tPhi}{\tilde{\Phi}}
\newcommand{\tPhipm}{\tilde{\Phi}_\pm}
\newcommand{\tPhip}{\tilde{\Phi}_+}
\newcommand{\tPhim}{\tilde{\Phi}_-}
\newcommand{\tO}{\tilde{\Omega}}
\newcommand{\tOpm}{\tilde{\Omega}_\pm}
\newcommand{\tOp}{\tilde{\Omega}_+}
\newcommand{\tOm}{\tilde{\Omega}_-}
\newcommand{\tNGee}{{\tilde{N}_\textrm{G}^\textrm{[e]}}}
\newcommand{\tNNGee}{{\tilde{\mathbb{N}}_\textrm{G}^\textrm{[e]}}}
\newcommand{\ERee}{{E_\textrm{R}^\textrm{[e]}}}
\newcommand{\tMee}{\tilde{M}^\textrm{[e]}}
\newcommand{\tMMee}{\tilde{\mathbb{M}}^\textrm{[e]}}
\newcommand{\tEPhi}{\tilde{E}_{[\Phi]}}
\newcommand{\tEsT}{\tilde{E}_{\rm\{T\}}}
\newcommand{\tEO}{\tilde{E}_{[\Omega]}}
\newcommand{\tEEO}{\tilde{\mathbb{E}}_{[\Omega]}}
\newcommand{\EE}{\mathbb{E}}
\newcommand{\EET}{\mathbb{E}_\textrm{T}}
\newcommand{\EEPhi}{\mathbb{E}_\Phi}
\newcommand{\EEivPhi}{\mathbb{E}^\textrm{(IV)}_{[\Phi]}}
\newcommand{\EEivbnk}{\mathbb{E}^\textrm{(IV)}(\beta,\nu_k)}
\newcommand{\EEivbn}{\mathbb{E}^\textrm{(IV)}(\beta,\nu)}
\newcommand{\tEEPhi}{\tilde{\mathbb{E}}_{[\Phi]}}
\newcommand{\tNNPhi}{\tilde{\mathbb{N}}_{\Phi}}
\newcommand{\tNNO}{\tilde{\mathbb{N}}_{\Omega}}
\newcommand{\bbar}{{\mathchoice
{{\vcenter{\offinterlineskip\vskip.1ex\hbox{$\,\tilde{}$}\vskip-1.85ex\hbox{$b$}\vskip.4ex}}}
{{\vcenter{\offinterlineskip\vskip.1ex\hbox{$\,\tilde{}$}\vskip-1.75ex\hbox{$b$}\vskip.4ex}}}
{{\vcenter{\offinterlineskip\vskip.1ex\hbox{$\scriptstyle\,\tilde{}$}\vskip-1.2ex\hbox{$\scriptstyle b$}\vskip.2ex}}}
{{\vcenter{\offinterlineskip\vskip.1ex\hbox{$\scriptscriptstyle\,\tilde{}$}\vskip-.9ex\hbox{$\scriptscriptstyle b$}\vskip.4ex}}}
}}

\newcommand{\woepm}{\omega^{(\pm)}_{0,1}}
\newcommand{\wopm}{\omega^{(\pm)}_0}
\newcommand{\wop}{\omega^{(+)}_0}
\newcommand{\wom}{\omega^{(-)}_0}
\newcommand{\wepm}{\omega^{(\pm)}_1}
\newcommand{\wep}{\omega^{(+)}_1}
\newcommand{\wem}{\omega^{(-)}_1}


\title{\bf Gauge-Invariant Energy Functional\\ in\\ Relativistic Schr\"odinger Theory}
\author{M.\ Mattes and M.\ Sorg} 
\date{ }
\maketitle
\begin{abstract}
  The \emph{non-invariant} energy functional of the preceding paper is improved in order
  to obtain its \emph{gauge-invariant} form by strictly taking into account the
  non-Abelian character of Relativistic Schr\"odinger Theory (RST). As an application of
  the results, the dichotomy of positronium with respect to singlet and triplet states is
  discussed (ortho- and para-positronium). The \emph{degeneracy} of the ortho- and
  para-states occurs in RST if \textbf{(i)} the magnetic interactions are neglected (as in
  the conventional theory) and \textbf{(ii)} the anisotropy of the electric interaction
  potential is disregarded. In view of such a very crude approximation procedure, the
  non-relativistic positronium spectrum in RST agrees amazingly well with the conventional
  predictions. Indeed using here a very simple trial function with only two variational
  parameters yields deviations from the conventional predictions of some 10\% (or less) up
  to large principal quantum numbers~$(n\simeq 100)$. This hints at the possibility that
  the exact RST calculations could yield predictions for the bound systems which come very
  close to (or even coincide with) those of the conventional quantum theory. Such a
  numerical correctness of the RST predictions supports a fluid-dynamic picture of quantum
  matter where \emph{any single particle} appears as the carrier of a wave-like structure
  (such a structure, however, is generally believed to be associated exclusively with
  \emph{statistical ensembles} but not with single particles). Here, more detailed
  predictions of the positronium spectrum would be desirable but this will necessitate to
  first elaborate more exact solutions of the RST eigenvalue problem, especially with
  regard to the anisotropic and magnetic effects.
  \vspace{2.5cm}
 \noindent

 \textsc{PACS Numbers:  03.65.Pm - Relativistic
  Wave Equations; 03.65.Ge - Solutions of Wave Equations: Bound States; 03.65.Sq -
  Semiclassical Theories and Applications; 03.75.b - Matter Waves}

\end{abstract}


\begin{center}
  {\Large\textbf{Contents}}
\end{center}

\begin{itemize}
\item[I] \textbf{Introduction and Survey of Results\dotfill 6}
\item[II] \textbf{Relativistic Schr\"odinger Theory\dotfill 17}
  \begin{itemize}
  \item[1.] \emph{Matter Fields \dotfill 17}
  \item[2.] \emph{Gauge Fields \dotfill 19}
  \end{itemize}
 \item[III] \textbf{Gauge Invariance and Entanglement \dotfill 24}
   \begin{itemize}
   \item[1.] \emph{Gauge Structure \dotfill 25}
   \item[2.] \emph{Mass Eigenvalue Equations \dotfill 29}
   \item[3.] \emph{Gauge Field Equations\dotfill 30}
   \item[4.] \emph{Linearization and Gauge Invariance \dotfill 33}
   \item[5.] \emph{RST Entanglement \dotfill 34}
   \item[6.] \emph{Boundary Conditions and Entanglement Charge \dotfill 38}   
   \end{itemize}
 \item[IV] \textbf{Gauge-Invariant Energy Functional \dotfill 42}
   \begin{itemize}
   \item[1.] \emph{Generalized Poisson Identities \dotfill 43}
   \item[2.] \emph{Exchange Identities \dotfill 45}
   \item[3.] \emph{Relativistic Energy Functional \dotfill 46}
   \item[4.] \emph{Non-Relativistic Variational Principle\dotfill 50}
     \begin{itemize}
     \item[] \emph{Non-Relativistic Eigenvalue Equations \dotfill 51}
     \item[] \emph{Non-Relativistic Gauge Field Equations \dotfill 55}
     \item[] \emph{Non-Relativistic Energie Functional \dotfill 56}
     \end{itemize}
   \end{itemize}
 \newpage
   \item[V] \textbf{Ortho- and Para-Positronium \dotfill 59}
     \begin{itemize}
     \item[1.] \emph{Conventional Multiplet Structure \dotfill 60}
     \item[2.] \emph{Ortho/Para Dichotomy in RST \dotfill 64}
     \item[3.] \emph{Positronium Eigenvalue Problem \dotfill 71}
     \end{itemize}
   \item[VI] \textbf{Energy Difference of Ortho- and Para-States \dotfill 82}
     \begin{itemize}
     \item[1.] \emph{Para-Positronium \dotfill 83}
        \begin{itemize}
        \item[] \emph{Electrostatic Approximation \dotfill 84}
        \item[] \emph{Spherically Symmetric Approximation\dotfill 88}
        \item[] \emph{Non-Relativistic Approximation \dotfill 93}
        \item[] \emph{Non-Relativistic Para-Spectrum \dotfill 97}
        \end{itemize}
     \item[2.] \emph{Ortho-Positronium \dotfill 107}
        \begin{itemize}
        \item[] \emph{Eigenvalue Problem for the Ortho-States \dotfill 108}
        \item[] \emph{Vanishing Angular Momentum $j_z=0$ \dotfill 109}
        \item[] \emph{Ortho-States with $j_z=-1$ \dotfill 112}
        \item[] \emph{Energy Functional for the Ortho-States \dotfill 119}
        \item[] \emph{Anisotropic and Magnetic Effects \dotfill 123}
        \item[] \emph{Non-Relativistic Energy Functional \dotfill 127}
    \end{itemize}
   \end{itemize}
   \end{itemize}
   \vspace{2ex}
   \noindent
   Appendix A:\\ 
   \phantom{Appe}\textbf{Manifest Gauge Invariance of the Energy Functional $\tETT$}\dotfill \textbf{131}\\[1ex]
   Appendix B:\\
     \phantom{Appe}\textbf{Variational Deduction of Non-Relativistic Eigenvalue Equations}
     \dotfill \textbf{134}\\[1ex]
   Appendix C:\\
     \phantom{App}\textbf{Variational Deduction of the Gauge Field Equations} \dotfill \textbf{138}\\[1ex]
   Appendix D\\
     \phantom{Appe}\textbf{Electric Poisson Identity} \dotfill \textbf{143}\\[1ex]
   Appendix E:\\
     \phantom{App}\textbf{Angular Part of the Ortho-Wave Amplitudes $\bMRpm,\bMSpm$} \dotfill \textbf{147}\\[1ex]
   Appendix F:\\
     \phantom{Appe}\textbf{Kinetic Energy ${}^{[b]}\tilde{T}_{\mathrm{kin}}$ of the
       Ortho-States } \dotfill \textbf{153}

\vspace{2ex}    
\noindent
   Figure 1:\\
      \phantom{Figu}\textbf{Azimuthal Para-Current\ \ ${}^{(p)}k_\phi$} \dotfill \textbf{157}\\[1ex]
\noindent
   Figure 2:\\
      \phantom{Figu}\textbf{Non-Relativistic Para-Spectrum} \dotfill \textbf{159}\\[1ex]
\noindent
   Figure 3:\\
      \phantom{Figu}\textbf{Interaction Energy} { \ $\mathbf{\ERee=\tMMee} c^2$ } \dotfill \textbf{161}\\[1ex]
\noindent
   Figure 4a:\\
      \phantom{Figu}\textbf{Energy Curves}\ \ $\mathbf{\EET(\nu)}$ \dotfill \textbf{163}\\[1ex]
\noindent
   Figure 4b:\\
      \phantom{Figu}\textbf{Non-Relativistic Groundstate Energy}\dotfill \textbf{165}\\[1ex]
\noindent
   Figure 5:\\
      \phantom{Figu}\textbf{Angular Factors}\ \ $\mathbf{\bbko(\vartheta)}$ \textbf{for Ortho-Positronium}\dotfill \textbf{167}\\[1ex]
\noindent
   Figure 6:\\
      \phantom{Figu}\textbf{Azimuthal Ortho-Current}\ \ $\mathbf{\bbkphi}$ \dotfill \textbf{169}\\[4ex]

\noindent
\textbf{References  \dotfill 171} \\[4ex]
\textbf{Complete List of RST papers  \dotfill 173}


\section{Introduction and Survey of Results}
\indent

The present paper continues the investigation of the \emph{fluid-dynamic} aspects of
quantum matter which, however, in the literature is mostly described in terms of
\emph{probabilistic} point-particle concepts. The frequently quoted paradigm of
``wave-particle duality'' seems to suggest a certain equivalence of the wave and the
particle picture, but in practice the particle picture appears preferably in combination
with the probabilistic interpretation and thus has overrun the fluid-dynamic approach. But
here one should not overlook the fact that the probabilistic point-particle picture has
its interpretative difficulties, too.

Indeed, even one century after the first tentative steps~\cite{Pl} towards a quantum
theory of microscopic matter there is still considerable uncertainty about the true nature
of the quantum phenomena. Despite the existence of a powerful mathematical apparatus for
the description of the quantum world the controversy persists about the right
understanding of what is really going on during the course of a quantum process. Even
those workers in this field, who are able to handle excellently and successfully the
mathematical apparatus, apparently like to conceive the quantum process as a ``mystery''
from the conceptional viewpoint~\cite{MS,VS}. Thus the proponents of the probabilistic
point-particle picture have to concede (more or less involuntarily) that the core of the
philosophical confusion about the quantum behavior of matter refers to just that notorious
\emph{wave-particle duality} which Feynmann seemingly had in mind when he said:
''...\emph{which has in it the heart of quantum mechanics. In practice, it contains the
  only mystery}~\cite{FLS}''.

Concerning now this mysterious wave-particle duality, there seems to have evolved a
certain wide-spread consensus which may perhaps be best expressed as follows: ''\emph{It
  is frequently said or implied that the wave-particle duality of matter embodies the
  notion that a particle-the electron, for example-propagates like a wave, but registers
  at a detector like a particle. Here one must again exercise care in expression so that
  what is already intrinsically difficult to understand is not made more so by semantic
  confusion. The manifestations of wave-like behavior are statistical in nature and always
  emerge from the collective outcome of many electron events. In the present experiment
  nothing wave-like is discernible in the arrival of single electrons at the observation
  plane. It is only after the arrival of perhaps tens of thousands of electrons that a
  pattern interpretable as wave-like interference emerges.}``~\cite{ref}. According to
this view, the wave-particle duality refers to statistical ensembles but not to individual
``particles''. As a consequence of this general belief, the wave concepts in quantum
theory must not be applied to single particles: ``\emph{Although one can in principal
  measure the mass, charge or energy of a single electron (held, for example, in an
  electromagnetic trap), one cannot measure its de Broglie wavelength except by a
  diffraction or interference experiment employing many such electrons}''.~\cite{ref}.

But in contrast to this generally accepted picture of the wave-particle duality, it seems
to us that those \emph{one-particle self-interference phenomena}~\cite{MS,VS} do distinctly
hint at the fact that already any single particle is equipped with some kind of wave-like
structure which becomes manifest in some special situations (in other situations the
wave-like structure may remain hidden). Naturally, such kind of philosophical basis must
motivate one to look for a \emph{fluid-dynamic} description of quantum particles which is
able to compete in many areas with the conventional probabilistic paradigm. Here the
conventional probabilistic theory of quantum matter and the present fluid-dynamic approach
should not be understood as mutually falsificating logical systems but should rather be
conceived in the sense of Bohr~\cite{NB} as \emph{complementary} ways to look upon the
quantum domain. Indeed, the very successful density functional theory~\cite{PY,DG}
demonstrates that a fluid-dynamic approach to the quantum phenomena is logically possible
and at least as useful in atomic and molecular physics as the conventional probabilistic
approach.

An alternative theoretical framework of the fluid-dynamic type has recently been proposed
in form of the Relativistic Schr\"odinger Theory (RST) (see the list of references at the
end of the paper). This (rigorously relativistic) theory is also based upon the concept of
wave functions, albeit not in the probabilistic sense of the conventional theory, but
rather in the fluid-dynamic sense where the wave functions serve to build up the physical
densities of an N-particle system (e.g.\ the charge and energy-momentum density,
etc.). But clearly, the mathematical formalism of such a fluid-dynamic approach must
necessarily differ considerably from the probabilistic calculus of the conventional theory
(i.e.\ Whitney sums of fibre bundles vs.\ tensor products of Hilbert spaces); however,
despite this mathematical difference the physical phenomena must be describable equally
well in the fluid-dynamic picture. For instance, the outcome for the energy of a bound
N-particle system should be the same, no matter whether it is calculated with reference to
the probabilistic or to the fluid-dynamic terminology. In the first case, one has to look
for the eigenvalues of the N-particle Hamiltonian~$\hat{H}(1\ldots N)$; and in the second
case one has to look for the values of the total energy functional~$\ETT$ upon the
solutions of the RST eigenvalue problem, which itself consists of the coupled mass
eigenvalue equations and the gauge field equations (see below). Obviously, there arises in
both cases a very similar problem, namely \textbf{(i)} either to define the
Hamiltonian~$\hat{H}(1\ldots N)$ for a relativistic N-particle system~\cite{CT,GR} or
\textbf{(ii)} to find the right relativistic energy functional~$\ETT$ for such a system
(which is the main subject of the present paper). The latter object~$\ETT$ does then not
only specify the energy carried by the solutions of the RST eigenvalue problem but
simultaneous it provides also the basis of the \emph{principle of minimal energy}, i.e.\
the variational principle
\begin{equation}
  \label{eq:I.1}
  \delta\ETT = 0\ ,
\end{equation}
whose extremal equations are nothing else than just those coupled field equations of the RST
eigenvalue problem.

The goal of the present paper is now to set up the RST energy functional~$\ETT$ for bound
\emph{two-particle systems}~$(N=2)$; the case of general particle number~$N$ might then be
discussed along quite similar lines of reasoning. Here, an important point refers to the
question whether the considered N-particle system consists of identical or different
particles. From the principal point of view, the interaction of the particles occurs
always via the mechanism of minimal coupling; i.e.\ the N-particle wave function~$\Psi$
enters the RST field equations in form of its gauge-covariant derivative~$\D$
\begin{equation}
  \label{eq:I.2}
  \D_\mu\Psi \doteqdot \partial_\mu \Psi + \A_\mu\Psi\ .
\end{equation}
The wave function~$\Psi$ itself has the status of a section of a spinor bundle; and the
nature of the bundle connection~$\A_\mu$ (\emph{``gauge potential''}) determines the type
of interactions between the particles. For instance, when the particles are subjected to
the electromagnetic interactions one splits up the \emph{structure algebra}~$\U(N)$ (as
the Lie-Algebra valued range of the connection~$\A_\mu$) into its maximal Abelian part
$\mathfrak{a}(N)=\mathfrak{u}(1)\oplus\mathfrak{u}(1)\oplus\ldots\oplus\mathfrak{u}(1)$
and the complement~$\mathfrak{b}(N)$
\begin{equation}
  \label{eq:I.3}
  \mathfrak{u}(N) = \mathfrak{a}(N)\oplus\mathfrak{b}(N)\ .
\end{equation}

The RST philosophy says now that the electromagnetic interactions between the particles
are \emph{always} to be described by the projection of the connection~$\A_\mu$ to the Abelian
subalgebra~$\mathfrak{a}(N)$
(i.e.~$\A_\mu\rightarrow\A_\mu\big|_{\mathfrak{a}}$). Furthermore if all the particles are
\emph{different}, the projection of~$\A_\mu$ to the complement~$\mathfrak{b}$ is required to
vanish: $\A_\mu\big|_{\mathfrak{b}}=0$. However, if the particles are \emph{identical} the
restriction of~$\A_\mu$ to the complement~$\mathfrak{b}$ is non-trivial and then describes
the \emph{additional exchange interactions} between the identical particles. Thus the
gauge interactions of non-identical particles are of Abelian character whereas the
identical particles do feel an additional force (\emph{exchange force}) which requires to
work with a non-Abelian gauge field theory. In the latter case, the theory becomes also
\emph{non-linear} which of course renders more difficult the construction of an adequate
energy functional.

In view of such a logical constellation, it seemed meaningful to first set up an energy
functional~$\ETT$ for a system of \emph{different} particles where the associated gauge
field theory is Abelian~$(\A_\mu\equiv\A_\mu\big|_{\mathfrak{a}})$. The corresponding
results were applied in their non-relativistic form to the positronium spectrum and
yielded physically reasonable results~\cite{MaSo,MaSo2}, albeit the magnetic interactions
have been neglected. Furthermore, in order to have at least an approximative energy
functional for \emph{identical particles}~$(\A_\mu \big|_{\mathfrak{b}}\neq 0)$ the gauge
field theory was linearized~\cite{MaSo2}. But clearly, such a linearization procedure must
necessarily fail to be rigorously gauge-invariant since for this property the non-linear
terms of the non-Abelian gauge theory are indispensable (see below). Therefore the wanted
\emph{gauge-invariant} energy functional~$\tETT$ for a three-particle system with two
identical particles (e.g.\ electrons) and one different particle (e.g.\ positron or
proton) is now constructed in the present paper. As a demonstration for the obtained
results one considers again the case of positronium, but now with emphasis on the
well-known ortho/para degeneracy and its breaking~\cite{CT}. Indeed, this is a crucial test
situation for the constructed RST functional~$\tETT$ because here one can explicitly
display the conditions (or approximations, resp.) by which there first arises the expected
degeneracy which then afterwards is broken by the magnetic (and anisotropic) effects.

Unfortunately, exact solutions of the RST eigenvalue problem are not yet available
(neither analytically nor numerically) so that the subsequent investigation of the
ortho/para degeneracy must necessarily be of a more qualitative character. But despite the
use of very crude approximation techniques, the obtained positronium spectrum is amazingly
close to the conventional predictions and the observations (see fig.~4). This should
provide sufficient motivation for an extensive elaboration of RST in order to test its
predictive power and usefulness to higher accuracy in a wider range of physical
applications.

The scope of the present paper is restricted to the construction of a gauge-invariant
energy functional~($\tETT$) and is organized as follows:
\vspace{4ex}
\begin{center}
\emph{\textbf{A.\ Gauge Structure}}
\end{center}

The intention of setting up a \emph{gauge-invariant} energy functional~$\ETT$ requires to
first define the general gauge structure of the theory. To this end, the basic features of
RST are briefly sketched in \textbf{Sect.~II}. The emphasis lies here on the fact that the
very field equations of RST do automatically imply certain local conservation laws, e.g.\
that of total charge
\begin{equation}
  \label{eq:I.4}
  \nabla^\mu j_\mu \equiv 0\ ,
\end{equation}
or that of total energy-momentum
\begin{equation}
  \label{eq:I.5}
  \nabla^\mu \, \TT_{\mu\nu}\equiv 0\ .
\end{equation}

Indeed, these conservation laws play an important role for the present problems. Namely
they both concern the question whether or not any \emph{entangled} particle can be thought
to carry its own individual properties, e.g.\ charge and energy-momentum, or whether
perhaps only the whole system of entangled particles (quite similarly as one non-entangled
particle) can be considered to be the carrier of well-defined electric charge and
energy-momentum? (The latter question will be answered in the positive). Here, the general
belief says now that those well-defined and observable properties must be gauge-invariant,
whatever the corresponding gauge group should look like. Therefore the gauge structure of
RST in general, and especially also for the bound stationary systems, must first be
clarified in detail. Of course, the wanted energy functional~$\ETT$ must be required to be
invariant under the action of the ultimately established gauge group.

In \textbf{Sect.~III} it is made clear that the proper gauge group of electromagnetically
interacting particles is the magnetic subgroup of $U(1)\times U(1)\times
U(1)\times\ldots\times U(1)$; this means that the \emph{electric} potentials~$\aAn$ are
left invariant and only the \emph{magnetic} potentials~$\vec{A}_a(\vec{r})$ become
transformed, namely in the usual inhomogeneous way (see equations
(\ref{eq:III.20a})-(\ref{eq:III.20c}) below). An important point refers here to the
exchange potential~$B_\mu$ which mediates the exchange forces among the identical
particles; indeed, this potential acts like a tensor object and thus transforms
homogeneously as is the case with the wave functions, too (see equations (\ref{eq:III.6})
and (\ref{eq:III.9}) below).

\begin{center}
\emph{\textbf{B.\ RST Entanglement}}
\end{center}

Quite generally speaking, the phenomenon of entanglement shows up in the occurrence of the
so-called \emph{exchange effects}. From the physical viewpoint, the main difference
between the conventional theory and the present RST refers to the status of these exchange
interactions: The conventional theory incorporates this type of interactions among
identical particles via the postulate that the N-particle wave functions should be
symmetric for bosons and antisymmetric for fermions (or one postulates the anticommutation
relations for fermionic field operators and the commutation relations for bosonic
operators, resp.). In the first line, such commutation postulates are of kinematical
character and their dynamical consequences (e.g.\ exchange effects) are then studied
afterwards, mostly in some approximative way. However, in RST the exchange effects are
treated on the same footing as the electromagnetic interactions! This is briefly sketched
in \textbf{Sect.~III.}

Such an ontological similarity of the electromagnetic and exchange forces implies that the
latter are to be described by some four-potential~$B_\mu$ as the exchange counterpart of the
electromagnetic potentials~$A_\mu$, and both types of four-potentials together combine to
the bundle connection~$\A_\mu$, see equation (\ref{eq:II.2}) below. But clearly, such an
equivalent treatment of the electromagnetic and exchange forces entails the common
emergence of electromagnetic and exchange fields at all the relevant places of theory; for
instance, the source and curl equations of the Maxwellian type for the electric~$(\vec{E})$
and magnetic~$(\vec{H})$ fields must be complemented by a source equation for the electric
exchange field~$\vec{X}$ and a curl equation for the magnetic exchange field~$\vec{Y}$,
see equations (\ref{eq:III.31a})-(\ref{eq:III.31b}) below.

But despite this formal similarity of both kinds of forces there are also important
differences, especially concerning the gauge-fixing conditions. For the inhomogeneously
transforming three-vector potentials~$\vec{A}(\vec{r})$ it is still adequate to resort to
the usual Coulomb gauge~$(\vec{\nabla}\sdot\vec{A}\equiv 0)$; but for the homogeneously
transforming tensor objects~$B_\mu$ the analogous relation must be gauge-invariant, see
equation (\ref{eq:III.33}) below, in order to not spoil the gauge covariance of the RST
eigenvalue problem.

In principle, the latter problem consists of the coupled set of Dirac equations for the
matter fields, see equations (\ref{eq:II.6a})-(\ref{eq:II.6c}) below, and of the
(non-Abelian) Maxwell equations (see (\ref{eq:II.8}) below). But if one splits up these
non-Abelian gauge field equations into its electromagnetic and exchange parts, one observes
the occurrence of an \emph{entanglement current}~$l_\mu$ which is built up by the
complex-valued exchange potential~$B_\mu$ like a Klein-Gordon type of current, see
equation (\ref{eq:III.43}) below.

The phenomenon of entanglement for two identical particles shows up also through the
circumstance that the entanglement current~$l_\mu$ modifies the original electromagnetic
currents~$j_\mu$ of the particles in such a way as if some portion~$(\delta_e)$ of
charge~$(e)$, i.e.\ the \emph{entanglement charge}, would have been exchanged between both
identical particles, see equations (\ref{eq:III.66})-(\ref{eq:III.67b}) below. However
this ``virtual'' charge exchange among the identical particles occurs in such a way that
the total charge~$(e_s)$ of all the entangled particles together is unmodified and equals
the product of the elementary charge times the number of identical particles, see equation
(\ref{eq:III.63}) below. In this way, charge conservation holds for the system of
entangled particles as a whole, albeit not for the individual entangled particles
themselves!

\begin{center}
\emph{\textbf{C.\ Exact Energy Functional}}
\end{center}

It is well known that the non-linear character of the non-Abelian gauge field theories is
an implication of just that non-Abelian structure of the gauge group; and therefore the
transition to the linearized theory must necessarily spoil the desired construction of a
\emph{gauge-invariant} energy functional, see ref.~\cite{MaSo2}. Nevertheless, the preliminary
results of the  linearized theory are very helpful for constructing now in
\textbf{Sect.~IV} the desired gauge-invariant form of the exact energy functional. Namely,
that truly invariant form is essentially the same as for the approximative result of the
linearized theory; however \textbf{(i)} the ordinary derivatives of the RST fields are now
replaced by the gauge-covariant derivatives and \textbf{(ii)} there do emerge additional
exchange terms (which vanish again for non-identical particles because these are unable to
feel the exchange forces).

But apart from this, the general structure of the gauge-invariant functional~$\tETT$ (see
equations (\ref{eq:IV.20}) and (\ref{eq:IV.23}) below) resembles very much its
non-gauge-invariant predecessor~\cite{MaSo2}: there are the contributions with an
immediate physical meaning (denoted by~$\Eiv$, cf.\ (\ref{eq:IV.23})) and these physical
contributions are to be complemented by various constraints, e.g.\ the normalization
constraints for the wave functions. The other constraints do refer to the Poisson and
exchange identities which, roughly speaking, ensure the equality of the gauge field
energies and their ``mass equivalents'', i.e.\ the interaction energies of the coupled
matter and gauge fields.  This is immediately clear for non-identical particles where all
the exchange terms are zero as a consequence of the vanishing of the exchange
potential~$B_\mu$. But the physical meaning of the relationship between the gauge field
energies and their mass equivalents becomes more intricate for identical particles where
some of the constraint terms (i.e.\ those of the type~$\NG$) are subjected to the
variational process while others (i.e.\ those of the type~$\nG$) are to be treated as true
constants, see equations (\ref{eq:IV.1}), (\ref{eq:IV.4}), (\ref{eq:IV.11}) and
(\ref{eq:IV.14}) below. Thus it seems that the true constants (of the type~$\nG$) are to
be conceived as a kind of quantum numbers for the exchange interactions.

However, under strict regard of all these constraints, the variational process~$(\delta\tETT=0)$
yields again \emph{both} the gauge-covariant mass eigenvalue equations for the matter
fields \emph{and} the gauge field equations. This pleasant result does refer to both the
relativistic situation and its non-relativistic approximation, where fortunately the gauge
covariance survives the passage to the non-relativistic limit. A \emph{manifestly}
gauge-invariant form of the energy functional~$\tETT$ is presented in
Appendix~\textbf{A}. But observe here again, that the notion of ``gauge-invariance'' does
refer here exclusively to the gauge transformations of the \emph{magnetic} type!

\begin{center}
\emph{\textbf{D.\ Ortho- and Para-Positronium}}
\end{center}

As a plausibility test for the constructed energy functional~$\tETT$ one will select some
simple but sufficiently non-trivial system (here: positronium). As is well known, this
simple two-particle system does occur in two forms: ortho- and para-positronium; see, e.g.,
ref.s~\cite{CT,GR}. Since these two forms have slightly different binding energies, the
corresponding ortho/para level splitting appears to provide an ideal test situation for
the present energy functional~$\tETT$ which, if physically relevant, should be able to
account for just this kind of splitting.

However, a closer inspection of this ortho/para dichotomy in \textbf{Sect.~V} reveals its
complicated appearance within the framework of RST. Here, it turns out that the energy
difference between the ortho- and para-configurations is not only a consequence of the
magnetic interactions between both positronium constituents but is also caused by their
different anisotropic charge distributions. This difference of the charge distributions is
itself an implication of the different angular momenta of the ortho- and para-states:
whereas the para-states must always have vanishing total angular momentum, cf.\
(\ref{eq:V.46}), the ortho-states may carry an integer number of angular momentum quanta,
cf.\ (\ref{eq:VI.119}). Here it seems plausible that the charge distribution of the
para-states is concentrated in the immediate vicinity of the symmetry axis, which then
implies lower angular momentum; on the other hand, the charge distribution of the
ortho-states splits into two subforms (fig.~5) which are both zero on the symmetry axis
and adopt their maximal value aside of this axis. Clearly such a larger distance of the
rotating matter from the axis of rotation will entail also a larger angular momentum.

But the crucial point is now that the difference of the anisotropic charge distributions
entails a corresponding difference of the \emph{electrostatic} interaction energies and
these are normally thousand times greater than the experimentally verified difference of
the magnetic interaction energies (see the discussion of this point in connection with the
ortho/para degeneracy (\ref{eq:V.13}) below). This hierarchy problem, concerning the
strength of the electric and magnetic interaction, forces one to reconsider the ortho/para
degeneracy and to think about some approach which at least approximately attributes the
same electrostatic interaction energy to the different charge distributions of the ortho-
and para-states: this is the spherically symmetric approximation where the electric
interaction potential~$\bpAo(r,\vartheta)$ is replaced by a spherically symmetric
potential~$\bbpAo(r):\ \bpAo(r,\vartheta)\Rightarrow \bbpAo(r)$ (the wave
functions~$\psi_a(\vec{r})$ are still admitted to be non-spherically symmetric). And
indeed, this assumption of an SO(3) symmetric interaction potential~$\bbpAo(r)$ lets
coincide the eigenvalue problems for the ortho- and para-states to one and the same form,
cf.\ the para-system (\ref{eq:VI.23a})-(\ref{eq:VI.23b}),(\ref{eq:VI.38})-(\ref{eq:VI.40})
and its ortho-counterpart (\ref{eq:VI.115a})-(\ref{eq:VI.115b}), (\ref{eq:VI.139}).

The common non-relativistic spectrum of both eigenvalue problems is sketched in
fig.~2. For a rough quantitative estimate of the energy levels a two-parameter variational
solution is chosen (see equation (\ref{eq:VI.55})). This choice predicts the RST energy
levels with deviations not more than some 10\% in comparison to the conventional
predictions, see fig.s 4a-4b and the table on p.~\pageref{tabE}.

A difference between both level systems refers to the quantum numbers~$\lP$ and~$\lO$ for
orbital angular momentum: while for para-positronium the associated quantum number
is~$\lP=0,1,2,3,4\ldots$, the range of the corresponding ortho-number~$\lO$
is~$\lO=2,4,6,8,\ldots$ (see appendix~E). It is true, the range of the ortho-number~$\lO$
is half the range of the para-number~$\lP$, but for any possible value of the
ortho-number~$\lO$ there exist two different charge distributions, see fig.~5, which are
supposed to cause different binding energies~$\ET$. However, a discussion of the
corresponding breaking of the ortho/para degeneracy must be deferred until more exact
solutions of the RST eigenvalue problems can be elaborated, especially with regard to the
anisotropic and magnetic effects (for a first estimate of the magnitude of the latter
effects see ref.~\cite{MaSo3}). Concerning the relative magnitude of both effects, the
anisotropic electric correction is found to be of order 1~[eV] (i.e.\ roughly 10\% of the
binding energy~\cite{MaSo}), whereas the magnetic corrections amount to
roughly~$10^{-3}$~[eV]~\cite{MaSo3}.


\section{Relativistic Schr\"odinger Theory}
\indent

All those important symmetry principles such as Lorentz and gauge invariance, being
considered as indispensable for any successful field theory of quantum matter, are encoded
manifestly in RST through the general form of the basic field equations. The subsequent
brief account of those field equations will be taken as the point of departure for the
construction of the desired gauge-invariant energy functional~$\ET$. (For a more detailed
discussion of the RST field equations, see the preceding papers~\cite{MaSo,MaSo2}.

\begin{center}
  \emph{\textbf{1.\ Matter Fields}}
\end{center}

First, the key equation for the N-particle wave function~$\Psi$ is chosen to be the Dirac
equation
\begin{equation}
  \label{eq:II.1}
  i\hbar c\GG^\mu\D_\mu\Psi = \M c^2\Psi\ .
\end{equation}
Here, the derivative~$(\D)$ of the wave function~$\Psi$ has already been specified by
equation (\ref{eq:I.2}) where the gauge potential (\emph{bundle connection})~$\A_\mu$
takes its values in the Lie algebra~$\U(N)$ of the unitary group~$U(N)$, provided all the
particles are identical. When some ($N-N'$, say) of the~$N$ particles are non-identical
(e.g.\ differing by their masses or charges), then the \emph{structure group}~$U(N)$ of
the fibre bundles becomes reduced to some subgroup~$U(N,N')$. In order to present a
concrete example, consider a three-particle system~$(N=3)$ where two particles are
identical~($N'=2$) and one particle is different (think, e.g., of a positron or proton and
two electrons). For such a situation, the original nine-dimensional \emph{structure
  group}~$U(3)$ for three identical particles is reduced to the product group~$U(1)\times
U(2)$ (for the reduction process, see ref.~\cite{BeSo}). Accordingly, the gauge
potential~$\A_\mu$ for such a three-particle system can be decomposed into a
five-dimensional Lie algebra basis~$\{\tau_a,\chi,\bar{\chi}\}$
\begin{equation}
  \label{eq:II.2}
  \A_\mu = \sum_{\alpha=1}^5 {A^\alpha}_\mu\tau_\alpha = \sum_{a=1}^3 {A^a}_\mu\tau_a + B_\mu\chi - \Bstar_\mu\bar{\chi}
\end{equation}
where the \emph{electromagnetic generators}~$\tau_a (a=1,2,3)$ do commute and are
anti-Hermitian $(\bar{\tau}_a=-\tau_a)$; and the \emph{exchange
  generators}~$\chi,\bar{\chi}$ are taken to be (pseudo-) Hermitian conjugates. The
anti-Hermiticy of the gauge potential~$\A_\mu\,(=-\bar{\A}_\mu)$ requires then that the
\emph{electromagnetic potentials}~${A^a}_\mu\,(a=1,2,3)$ are real-valued whereas the
\emph{exchange potential}~$B_\mu$ is admitted to be complex-valued.

Returning to the $N$-particle Dirac equation (\ref{eq:II.1}) there remain two elements to
be explained, i.e.\ the \emph{total velocity operator}~$\GG_\mu$ and the \emph{mass
  operator}~$\M$. The first one of these objects is the direct sum of the ordinary Dirac
matrices~$\gamma_\mu$, e.g.\ for the mentioned three-particle system
\begin{equation}
  \label{eq:II.3}
  \GG_\mu = \left(-\gamma_\mu\right) \oplus \gamma_\mu\oplus\gamma_\mu\ .
\end{equation}
Furthermore, the mass operator~$\M$ can be adopted to be diagonal in a suitably
chosen~$U(3)$ gauge
\begin{equation}
  \label{eq:II.4}
  \M =
  \begin{pmatrix}
    \Mp & 0 & 0\\*
    0 &\Me & 0\\*
    0 & 0 & \Me\\*
  \end{pmatrix}
\end{equation}
where~$\Mp\,(\Me)$ is the rest mass of the positively (negatively) charged particles. But
contrary to the conventional (\emph{probabilistic}) theory, the~$N$-particle wave
function~$\Psi$ in RST (as a \emph{fluid-dynamic} theory) is not built up by suitable
tensor products of single-particle wave functions but rather appears as the Whitney sum of
single-particle bundle sections~$\psi_a$, e.g.\ for the considered three-particle system
\begin{equation}
  \label{eq:II.5}
  \Psi(x) = \psi_1(x)\oplus\psi_2(x)\oplus\psi_3(x)\ .
\end{equation}
As a consequence, the matter equation (\ref{eq:II.1}) may be resolved in component form
and then yields the following coupled system of ordinary Dirac equations (for the
considered three-particle system)
\begin{subequations}
  \begin{align}
    \label{eq:II.6a}
      i\hbar c\gamma^\mu D_\mu\psi_1 &= -\Mp c^2 \psi_1\\*
    \label{eq:II.6b}
      i\hbar c \gamma^\mu D_\mu\psi_2 &= \Me c^2 \psi_2\\*
    \label{eq:II.6c}
      i\hbar c \gamma^\mu D_\mu\psi_3 &= \Me c^2 \psi_3
  \end{align}
\end{subequations}
where the gauge-covariant derivatives of the wave function components~$\psi_a\,(a=1,2,3)$
are given by
\begin{subequations}
  \begin{align}
    \label{eq:II.7a}
    D_\mu\psi_1 &= \partial_\mu\psi_1 - i\left({A^2}_\mu+{A^3}_\mu \right)\psi_1\\*
    \label{eq:II.7b}
    D_\mu\psi_2 &= \partial_\mu\psi_2 - i\left({A^1}_\mu+{A^3}_\mu \right)\psi_2-iB_\mu\psi_3\\*
    \label{eq:II.7c}
    D_\mu\psi_3 &= \partial_\mu\psi_3 - i\left({A^1}_\mu+{A^2}_\mu\right)\psi_3-i\Bstar_\mu\psi_2\ .
  \end{align}
\end{subequations}

Observe here that the exchange potential~$B_\mu$ establishes a \emph{direct} coupling of
the wave functions~$\psi_2$ and~$\psi_3$ of the \emph{identical} particles
(\ref{eq:II.7b})-(\ref{eq:II.7c}), whereas the non-identical particle (\ref{eq:II.7a})
does couple only \emph{indirectly} (i.e.\ via the electromagnetic potentials~${A^2}_\mu$
and~${A^3}_\mu$) to both identical particles. This effect of \emph{RST entanglement} of
the identical particles must of course be discussed in more detail below.

\begin{center}
  \emph{\textbf{2.\ Gauge Fields}}
\end{center}

The role of the Dirac equation (\ref{eq:II.1}) for the matter field~$\Psi$ is played for
the gauge field~$\A_\mu$ by the (non-Abelian) \emph{Maxwell equations}
\begin{gather}
  \label{eq:II.8}
  \D^\mu\F_{\mu\nu} = -4\pi i\as\J_\nu\\*
  \left(\as \doteqdot \frac{e^2}{\hbar c} \right)\ .\notag
\end{gather}
Here the bundle curvature (\emph{field strength})~$\F_{\mu\nu}$ is defined as usual in terms of
the connection~$\A_\mu$ as
\begin{equation}
  \label{eq:II.9}
  \F_{\mu\nu}\doteqdot \nabla_\mu\A_\nu - \nabla_\nu\A_\mu + \left[\A_\mu,\A_\nu \right]\ .
\end{equation}
The curvature may be decomposed for the presently considered three-particle system quite
similarly to the connection~$\A_\mu$~(\ref{eq:II.2}) as
\begin{equation}
  \label{eq:II.10}
  \F_{\mu\nu} = \sum_{\alpha=1}^5 {F^\alpha}_{\mu\nu}\tau_\alpha = \sum_{a=1}^3 {F^a}_{\mu\nu}\tau_a + G_{\mu\nu}\chi - \Gstar_{\mu\nu}\bar{\chi}
\end{equation}
so that the curvature components~${F^a}_{\mu\nu},G_{\mu\nu}$ appear in terms of the
connection components~${A^a}_\mu,B_\mu$ in the following way
\begin{subequations}
  \begin{align}
    \label{eq:II.11a}
    {F^1}_{\mu\nu} &= \nabla_\mu{A^1}_\nu-\nabla_\nu{A^1}_\mu\\*
    \label{eq:II.11b}
    {F^2}_{\mu\nu} &= \nabla_\mu{A^2}_\nu-\nabla_\nu{A^2}_\mu+i\left(B_\mu\Bstar_\nu-B_\nu\Bstar_\mu\right)\\*
    \label{eq:II.11c}
    {F^3}_{\mu\nu} &= \nabla_\mu{A^3}_\nu-\nabla_\nu{A^3}_\mu-i\left(B_\mu\Bstar_\nu-B_\nu\Bstar_\mu\right)\\*
    \label{eq:II.11d}
    G_{\mu\nu} &=  \nabla_\mu B_\nu-\nabla_\nu B_\mu + i\left({A^2}_\mu-{A^3}_\mu\right)B_\nu 
    - i\left({A^2}_\nu-{A^3}_\nu\right) B_\mu\ .
  \end{align}
\end{subequations}
Similarly, the Maxwell equations~(\ref{eq:II.8}) read in component form
\begin{subequations}
  \begin{align}
    \label{eq:II.12a}
    \nabla^\mu {F^1}_{\mu\nu} &= 4\pi\as{j^1}_\nu \\*
    \label{eq:II.12b}
    \nabla^\mu {F^2}_{\mu\nu} + i\left(B^\mu\Gstar_{\mu\nu}-\Bstar^\mu G_{\mu\nu}\right) &= 4\pi\as{j^2}_\nu \\*
    \label{eq:II.12c}
    \nabla^\mu {F^3}_{\mu\nu} -i\left(B^\mu\Gstar_{\mu\nu}-\Bstar^\mu G_{\mu\nu}\right)  &= 4\pi\as{j^3}_\nu \\*
    \label{eq:II.12d}
    \nabla^\mu G_{\mu\nu} - i\left({A^2}_\mu-{A^3}_\mu\right){G^\mu}_\nu -
    i\left({F^2}_{\mu\nu}-{F^3}_{\mu\nu}\right)B^\mu &= 4\pi\as \hstar_\nu\ .
  \end{align}
\end{subequations}
Here the current operator~$\J_\mu$~(\ref{eq:II.8}) has been decomposed according to
\begin{equation}
  \label{eq:II.13}
  \J_\mu = i\sum_{\alpha=1}^5 {j^\alpha}_\mu\tau_\alpha = i\left(\sum_{a=1}^3 {j^a}_\mu\tau_a + \hstar_\mu\chi - h_\mu\bar{\chi} \right)\ ,
\end{equation}
where the required charge conservation law~(\ref{eq:I.4}) emerges in operator form as
\begin{equation}
  \label{eq:II.14}
  \D^\mu\J_\mu \equiv 0\ .
\end{equation}
Indeed this is an automatique implication of the Maxwell equations~(\ref{eq:II.8}) since
the curvature~$\F_{\mu\nu}$ must obey the bundle identity
\begin{equation}
  \label{eq:II.15}
  \D^\nu\D^\mu\F_{\mu\nu}\equiv 0\ .
\end{equation}

Surely, the emergence of the charge conservation~(\ref{eq:II.14}) as an immediate
consequence of the field equations represents a pleasant feature of RST and therefore
deserves some closer inspection. First observe here that the current components (\emph{Maxwell
currents}) ${j^\alpha}_\mu=\{{j^a}_\mu,h_\mu,\hstar_\mu\},\alpha=1\ldots 5,a=1\ldots 3$,
play the role of the sources for the gauge fields~${A^\alpha}_\mu$ but cannot be immediately
identified with the well-known \emph{Dirac currents}~$k_{a\mu}$
\begin{equation}
  \label{eq:II.16}
  k_{a\mu} \doteqdot \bar{\psi}_a\gamma_\mu\psi_a \ .
\end{equation}
The precise relationship is~\cite{BeSo}
\begin{subequations}
  \begin{align}
    \label{eq:II.17a}
    {j^1}_\mu &\equiv k_{1\mu} = \bar{\psi}_1\gamma_\mu\psi_1\\*
    \label{eq:II.17b}
    {j^2}_\mu &\equiv -k_{2\mu} = -\bar{\psi}_2\gamma_\mu\psi_2\\*
    \label{eq:II.17c}
    {j^3}_\mu &\equiv -k_{3\mu} = -\bar{\psi}_3\gamma_\mu\psi_3\\*
    \label{eq:II.17d}
    {j^4}_\mu &\equiv \hstar_\mu \doteqdot \bar{\psi}_3\gamma_\mu\psi_2\\*
    \label{eq:II.17e}
    {j^5}_\mu &\equiv -h_\mu \doteqdot -\bar{\psi}_2\gamma_\mu\psi_3\ .
  \end{align}
\end{subequations}
With respect to these identifications, the operator form of the charge conservation
law~(\ref{eq:II.14}) transcribes to the corresponding component form in the following way:
\begin{subequations}
  \begin{align}
    \label{eq:II.18a}
    \nabla^\mu k_{1\mu} &\equiv 0\\*
    \label{eq:II.18b}
    \nabla^\mu k_{2\mu} - i\left(B^\mu h_\mu - \Bstar^\mu\hstar_\mu \right) &\equiv 0\\*
    \label{eq:II.18c}
    \nabla^\mu k_{3\mu} + i\left(B^\mu h_\mu - \Bstar^\mu\hstar_\mu \right) &\equiv 0\\*
    \label{eq:II.18d}
    \nabla^\mu h_\mu -  i\left({A^2}_\mu-{A^3}_\mu\right)h^\mu + i\Bstar^\mu
    \left(k_{2\mu}-k_{3\mu}\right) &\equiv 0\ . 
  \end{align}
\end{subequations}

A further pleasant feature of the RST dynamics concerns the automatic conservation of
energy-momentum~(\ref{eq:I.5}). Indeed, this works as follows: defining the
energy-momentum density~$\DT_{\mu\nu}$ carried by the Dirac matter field~$\Psi$  through
\begin{equation}
  \label{eq:II.19}
  \DT_{\mu\nu} \doteqdot \frac{i\hbar c}{4}\left[ 
    \bar{\Psi}\GG_\mu \left(\D_\nu\Psi\right) -  \left(\D_\nu\bar{\Psi}\right) \GG_\mu\Psi +
    \bar{\Psi}\GG_\nu \left(\D_\mu\Psi\right) -  \left(\D_\mu\bar{\Psi}\right) \GG_\nu\Psi  
\right] 
\end{equation}
and, analogously, the energy-momentum density~$\GT_{\mu\nu}$ concentrated in the gauge
field~$\A_\mu$ through~\cite{MaSo3}
\begin{equation}
  \label{eq:II.20}
  \GT_{\mu\nu} = \frac{\hbar c}{4\pi\as}\,K_{\alpha\beta}\left(
    {F^\alpha}_{\mu\lambda}{F^\beta}_\nu{}^\lambda - \frac{1}{4}{F^\alpha}_{\sigma\lambda}F^{\beta\sigma\lambda} 
  \right)
\end{equation}
then the total density~$\TT_{\mu\nu}$
\begin{equation}
  \label{eq:II.21}
  \TT_{\mu\nu} \doteqdot \DT_{\mu\nu}+\GT_{\mu\nu}
\end{equation}
actually obeys the required (local) law~(\ref{eq:I.5}) of energy momentum conservation,
provided that only both sub-dynamic equations for the matter subsystem~(\ref{eq:II.1}) and
for the gauge field subsystem~(\ref{eq:II.8}) are satisfied! Clearly with this
splitting~(\ref{eq:II.21}) of the total density~$\TT_{\mu\nu}$, the total
energy~$\ET$~\cite{MaSo3} also splits up into two constituents, i.e.
\begin{equation}
  \label{eq:II.22}
  \ET = \ED + \EG\ ,
\end{equation}
with the obvious definitions of the partial energies (for stationary bound systems)
\begin{subequations}
  \begin{align}
    \label{eq:II.23a}
    \ED &= \int d^3\vec{r}\;\DT_{00}(\vec{r})\\*
    \label{eq:II.23b}
    \EG &= \int d^3\vec{r}\;\GT_{00}(\vec{r})\ .
  \end{align}
\end{subequations}

This aims now just at the central point of the present paper: namely the question, to what
extent can \emph{both} the stationary matter equations~(\ref{eq:II.6a})-(\ref{eq:II.6c})
\emph{and} the (non-Abelian) Maxwell equations~(\ref{eq:II.12a})-(\ref{eq:II.12d}) be
conceived as the extremal equations due to the total field energy~$\ET$ (\ref{eq:II.22}),
to be understood as an \emph{energy functional}~$\ETT$? Or in other words, do the
stationary solutions of the coupled matter and gauge field
systems~(\ref{eq:II.6a})-(\ref{eq:II.6c}) \emph{plus}~(\ref{eq:II.12a})-(\ref{eq:II.12d})
actually extremalize the total energy functional~$\ET$~(\ref{eq:II.22})?  For
\emph{non-identical} particles~$(\leadsto B_\mu\equiv 0)$, this question has been answered
in the positive so that it is not necessary to modify the form of that functional~$\ETT$;
but for \emph{identical} particles~$(\leadsto B_\mu\neq 0)$ it could be shown (albeit only
in the \emph{linear approximation}) that the original
definition~(\ref{eq:II.22})-(\ref{eq:II.23b}) of the energy functional~$\ETT$ must be
slightly modified~\cite{MaSo2}. But of course, one wishes to have the question clarified for
the original non-Abelian (and therefore \emph{non-linear}) theory and this is just what we
intend to do by means of the subsequent elaboration. Observe here the fact that the
complication, being caused by the presence of some identical particles, is due to the
non-vanishing exchange potential~$B_\mu$ which renders the theory non-Abelian and
therefore non-linear! However, this specific feature of the theory is indispensable for
the \emph{gauge-covariant} description of the exchange interactions which are to be
conceived as the RST equivalent of the conventional phenomenon of entanglement.


\section{Gauge Invariance and Entanglement}
\indent

It should be obvious that there must exist an intimate relationship between the structure
group of the theory and the exchange effects which can occur in such a theory of the RST
type. The point here is that the interaction potential~$\A_\mu$ takes its values in the
Lie algebra of the structure group~$U(N,N')$, cf.~(\ref{eq:II.2}) for the presently
considered three-particle system, where this structure group results by reduction of the
original structure group~$U(N)$ for~$N$ fictively identical particles. This process of
reduction my be attributed to the fact that only the permutation of \emph{identical}
particles leaves invariant the physical properties of the~$N$ particle system where (in
the true spirit of the gauge theories) the discrete set of permutations or relabelings,
resp., within the subset of identical particles is generalized to a continuous
group~$U(N')$. Therefore, avoiding the permutations of non-identical particles, the
reduced structure group of the bundle arrangement must contain~$U(N')$ as a factor
group. If we further assume that all the residual~$N-N'$ non-identical particles do feel
exclusively the ordinary (i.e.\ Abelian) electromagnetic interactions, we come to the
conclusion that the reduced structure group should be the product~$U(N,N')=U(1)\times
U(1)\times\ldots\times U(1)\times U(N')$, i.e.\ for our presently considered
three-particle system~$U(1)\times U(2)$. Indeed, the permutative decoupling of the first
(positively charged) particle~$a=1$ from the other two (negatively charged)
particles~$a=2,3$ become manifest in \emph{both}
relationships~(\ref{eq:II.11a})-(\ref{eq:II.11d}) between the curvature and the connection
components \emph{and} becomes manifest also in the component form of the Maxwell
equations~(\ref{eq:II.12a})-(\ref{eq:II.12d}); moreover an analogous observation does
apply also to the source equations~(\ref{eq:II.18a})-(\ref{eq:II.18d}). Notice however,
that the total fields of the identical particles obey an ordinary Abelian structure!

In order to realize this more clearly, add up both equations~(\ref{eq:II.11b})
and~(\ref{eq:II.11c}) in order to find
\begin{equation}
  \label{eq:III.1}
  {F^s}_{\mu\nu} = \nabla_\mu {A^s}_\nu - \nabla_\nu {A^s}_\mu
\end{equation}
where the total fields due to both identical particles are defined through
\begin{subequations}
  \begin{align}
    \label{eq:III.2a}
    {A^s}_\mu &\doteqdot {A^2}_\mu + {A^3}_\mu\\*
    \label{eq:III.2b}
    {F^s}_{\mu\nu} &\doteqdot {F^2}_{\mu\nu} + {F^3}_{\mu\nu}\ .
  \end{align}
\end{subequations}
Indeed, these relationships for the total fields obviously avoid any reference to
the exchange potential~$B_\mu$. A similar conclusion does hold also for both the Maxwell
equations~(\ref{eq:II.12b})-(\ref{eq:II.12c}) whose sum appears as
\begin{equation}
  \label{eq:III.3}
  \nabla^\mu {F^s}_{\mu\nu} = 4\pi\as\left({j^2}_\nu + {j^3}_\nu \right) \doteqdot 4\pi\as{j^s}_\nu\ ,
\end{equation}
or similarly for the sum of equations~(\ref{eq:II.18b}) plus~(\ref{eq:II.18c})
\begin{equation}
  \label{eq:III.4}
  \nabla^\mu \left(k_{2\mu}+k_{3\mu}\right) \doteqdot \nabla^\mu k_{s\mu}\equiv 0
\end{equation}
which is now a proper conservation equation for the total Dirac current~$k_{s\mu}$ of
identical particles. Thus the subsystem of identical particles acts in an Abelian way like
\emph{one} (albeit compound) particle relative to the outside world.

\begin{center}
  \emph{\textbf{1.\ Gauge Structure}}
\end{center}

The reduction of the original structure group~$U(N)$ to~$U(N,N')$ does not mean that we
adopt the latter structure group as our proper gauge group because the general RST
philosophy says that any particle of the system must occupy a well-defined one-particle
state. This requirement excludes the state mixing of (both the identical and
non-identical) particles through the gauge action of a general~$U(N,N')$ or~$U(N')$
element. This becomes especially clear by considering the stationary bound systems which
are to be described by a set of~$N$ one-particle wave functions~$\psi_a(\vec{r},t)$
appearing as products of a time-independent factor~$\psi_a(\vec{r})\,(a=1,\ldots N)$ times
the usual exponential time factor, e.g.\ for the presently considered three-particle
system~$(a=1,2,3)$
\begin{equation}
  \label{eq:III.5}
  \psi_a(\vec{r},t) = \exp\left(-\frac{iM_a c^2}{\hbar}t \right)\cdot\psi_a(\vec{r})\ .
\end{equation}
Indeed it is easy to see that an~$U(2)$ rotation of the wave functions~$\psi_a(\vec{r},t)$
of both identical particles~$(a=2,3)$ with different mass eigenvalues~$M_a$ would in
general generate a mixed time dependence violating the \emph{manifest} stationarity.

From this reason, we dismiss the \emph{structure group}~$U(N,N')$ as the \emph{gauge
  group} and in place of it take the simple product~$U(1)\times U(1)\times U(1)\ldots
U(1)$ as our proper gauge group. Thus a change of gauge acts upon the one-particle wave
functions in the following form:
\begin{equation}
  \label{eq:III.6}
  \psi_a(x)\Rightarrow \psi'_a(x) = e^{-i\alpha_a(x)}\cdot\psi_a(x)
\end{equation}
where the group parameters~$\alpha_a\,(a=1,2,\ldots N)$ are real-valued functions over
space-time. As is well-known, the gauge potentials~${A^\alpha}_\mu$ do not transform
homogeneously as do the wave functions~$\psi_a$~(\ref{eq:III.6}) but rather obey some
inhomogeneous transformation law, e.g.\ for our three-particle system:
\begin{subequations}
  \begin{align}
    \label{eq:III.7a}
    \IA_\mu &\Rightarrow \IA'_\mu = \IA_\mu - \partial_\mu\alpha_1\\*
    \label{eq:III.7b}
    \IIA_\mu &\Rightarrow \IIA'_\mu = \IIA_\mu - \partial_\mu\alpha_2\\*
    \label{eq:III.7c}
    \IIIA_\mu &\Rightarrow \IIIA'_\mu = \IIIA_\mu - \partial_\mu\alpha_3\ ,
  \end{align}
\end{subequations}
where the \emph{modified} electromagnetic potentials are defined through
\begin{subequations}
  \begin{align}
    \label{eq:III.8a}
    \IA_\mu &\doteqdot {A^2}_\mu + {A^3}_\mu\\*
    \label{eq:III.8b}
    \IIA_\mu &\doteqdot {A^1}_\mu + {A^3}_\mu\\*
    \label{eq:III.8c}
    \IIIA_\mu &\doteqdot {A^1}_\mu + {A^2}_\mu\ .
  \end{align}
\end{subequations}
However in contrast to these electromagnetic potentials, the exchange potential~$B_\mu$
undergoes a homogeneous transformation law, i.e.
\begin{equation}
  \label{eq:III.9}
  B_\mu \Rightarrow B'_\mu = e^{-i\left(\alpha_2-\alpha_3 \right)} \cdot B_\mu\ .
\end{equation}

For instance, by means of this gauge structure it becomes possible to gauge off the time
dependence of the stationary wave functions~$\psi_a(\vec{r},t)$ (\ref{eq:III.5}), namely
by simply putting for the gauge parameters~$\alpha_a(t)$
\begin{equation}
  \label{eq:III.10}
  \alpha_a = -\frac{M_a c^2}{\hbar}\,t
\end{equation}
which then leaves the spatial components of the modified four potentials
(\ref{eq:III.7a})-(\ref{eq:III.7c}) unchanged but recasts the time components to
\begin{subequations}
  \begin{align}
    \label{eq:III.11a}
    \IA_0 &\Rightarrow \IA'_0 = \IA_0 + \frac{M_1 c}{\hbar}\\*
    \label{eq:III.11b}
    \IIA_0 &\Rightarrow \IIA'_0 = \IIA_0 + \frac{M_2 c}{\hbar}\\*
    \label{eq:III.11c}
    \IIIA_0 &\Rightarrow \IIIA'_0 = \IIIA_0 + \frac{M_3 c}{\hbar}\ .
  \end{align}
\end{subequations}
We will not make use of such a purely temporal gauge transformation for the following
discussions but one point deserves here some attention. Subtracting the equations
(\ref{eq:III.11c}) minus (\ref{eq:III.11b}) yields a special combination~$\AM$ of
potentials
\begin{gather}
  \label{eq:III.12}
  \AM \doteqdot \IIIA'_0 - \IIA'_0 = \An - \frac{1}{\aM}\\*
  \left(\An \doteqdot \iiAn - \iiiAn \right)\notag
\end{gather}
where the \emph{exchange radius}~$\aM$ is evidently defined by
\begin{equation}
  \label{eq:III.13}
  \aM = \frac{\hbar}{(M_2-M_3)c}\ .
\end{equation}
Indeed, this potential~$\AM$ (\ref{eq:III.12}) will frequently occur in connection with
the wanted energy functional~$\ETT$. The reason here is that, for the stationary bound
systems, the RST field equations must be time-independent, and this time independence may
be conceived to be the result of a temporal gauge transformation
(\ref{eq:III.10})-(\ref{eq:III.11c}).

Finally, observe also that the time dependence of the exchange potential~$B_\mu$ can
easily be determined from its gauge behavior: assuming that the new exchange
potential~$B'_\mu$ (\ref{eq:III.9}) is time-independent, i.e.~$B'_\mu\equiv
B_\mu(\vec{r})$, it follows immediately from the equations
(\ref{eq:III.9})-(\ref{eq:III.10}) that the original time-dependent~$B_\mu(\vec{r},t)$
must look as follows
\begin{equation}
  \label{eq:III.14}
  B_\mu(\vec{r},t) = \exp(-i\frac{ct}{\aM})\cdot B_\mu(\vec{r})\ .
\end{equation}
This is the temporal countercurrent of the exchange current~$h_\mu(\vec{r},t)$
(\ref{eq:II.17d})-(\ref{eq:II.17e}) whose time behavior for the stationary states is
given by
\begin{equation}
  \label{eq:III.15}
  h_\mu(\vec{r},t) = \exp\left(i\frac{ct}{\aM}\right)\cdot h_\mu(\vec{r})\ .
\end{equation}
On the other hand, the Dirac currents~$k_{a\mu}$ (\ref{eq:II.16}) will be found to be
time-independent for the stationary states, i.e.
\begin{equation}
  \label{eq:III.16}
  k_{a\mu} = \left\{\akn;-\akv \right\}\ ,
\end{equation}
and since these are the sources of the electromagnetic potentials~${A^a}_\mu$, cf.\
(\ref{eq:II.12a})-(\ref{eq:II.12c}), one will find the latter objects also being
time-independent for the stationary states
\begin{equation}
  \label{eq:III.17}
  {A^a}_\mu = \left\{\aAn;-\vAa \right\}\ .
\end{equation}
Clearly, the Dirac currents~$\akv$ (\ref{eq:III.16}) must have vanishing divergence for
such a situation
\begin{equation}
  \label{eq:III.18}
  \nabla^\mu k_{a\mu} = \vec{\nabla}\sdot\akv \equiv 0\ ,
\end{equation}
and if this is combined with the former source equations
(\ref{eq:II.18b})-(\ref{eq:II.18c}) one concludes that the Lorentz scalar products~$B_\mu
h^\mu$ must be real numbers:
\begin{equation}
  \label{eq:III.19}
  B_\mu h^\mu = \Bstar_\mu \hstar^\mu\ .
\end{equation}

However, for the subsequent discussion we do not wish to modify the standard time behavior
of the wave functions~$\psi_a(\vec{r},t)$ (\ref{eq:III.5}) and of the exchange
potentials~$B_\mu(\vec{r},t)$ (\ref{eq:III.14}). Therefore we restrict ourselves to the
\emph{purely magnetic} gauge transformations where all the gauge parameters~$\alpha_a$ are
time-independent~$(\alpha_a=\alpha_a(\vec{r}))$. In this case, the magnetic three-vector
potentials~$\vAI,\vAII,\vAIII$ (\ref{eq:III.7a})-(\ref{eq:III.7c}) transform as follows
\begin{subequations}
  \begin{align}
    \label{eq:III.20a}
    \vAI &\Rightarrow \vAI' = \vAI + \vec{\nabla}\alpha_1\\*
    \label{eq:III.20b}
    \vAII &\Rightarrow \vAII' = \vAII + \vec{\nabla}\alpha_2\\*
    \label{eq:III.20c}
    \vAIII &\Rightarrow \vAIII' = \vAIII + \vec{\nabla}\alpha_3\ ,
  \end{align}
\end{subequations}
whereas the electric potentials~$\IA_0,\IIA_0,\IIIA_0$ remain invariant. \emph{Thus the gauge
invariance of the wanted energy functional will refer exclusively to those gauge
transformations of the magnetic kind!}

\begin{center}
  \emph{\textbf{2.\ Mass Eigenvalue Equations}}
\end{center}

This restriction to the magnetic gauge transformations is also important for the gauge
covariance of the \emph{mass eigenvalue equations} which are nothing else than the
stationary form of the original matter field equations
(\ref{eq:II.6a})-(\ref{eq:II.6c}). In order to set up this stationary  form, one first
splits up the Lorentz covariant derivatives (\ref{eq:II.7a})-(\ref{eq:II.7c}) into their
time and space components and combines this  with the fact that the Dirac four-spinor
fields~$\psi_a(\vec{r})$ can be conceived as the Whitney sum of two-component Pauli
spinors~$\ap$, i.e.
\begin{gather}
  \label{eq:III.21}
  \psi_a(\vec{r}) = \app\oplus\apm\\*
  \Big(a=1,2,3\Big)\ .\notag
\end{gather}
This then lets appear the stationary version of the coupled system of Dirac equations
(\ref{eq:II.6a})-(\ref{eq:II.6c}) in the following gauge-covariant Pauli form
\begin{subequations}
    \begin{align}
    \label{eq:III.22a}
    &i\vec{\sigma}\sdot\!\!\left(\vec{\nabla} + i\vAI\right)\ivppm + \IA_0\cdot\ivpmp =
    \frac{\pm\Mp-M_1}{\hbar}\,c\cdot\ivpmp\\*
    \label{eq:III.22b}
    &i\vec{\sigma}\sdot\!\!\left(\vec{\nabla} + i\vAII\right)\iivppm + \IIA_0\cdot\iivpmp 
    +B_0\cdot\iiivpmp - \vec{B}\sdot\vec{\sigma}\,\iiivppm\notag\\* 
    & \hspace{8cm}= -\frac{M_2 \pm \Me}{\hbar}\,c \cdot \iivpmp\\*
    \label{eq:III.22c} 
    &i\vec{\sigma}\sdot\!\!\left(\vec{\nabla} + i\vAIII\right)\iiivppm + \IIIA_0\cdot \iiivpmp
    +\Bstar_0\cdot\iivpmp - \vBstar\sdot\vec{\sigma}\,\iivppm\notag\\*
    &\hspace{8cm}= -\frac{M_3\pm \Me}{\hbar}\,c\cdot \iiivpmp\ .
 \end{align}
\end{subequations}

Observe here again that the gauge covariance of this coupled mass-eigenvalue system refers
exclusively to the magnetic gauge transformations (\ref{eq:III.20a})-(\ref{eq:III.20c})
whereas the electric potentials~$\IA_0,\IIA_0,\IIIA_0$ remain invariant, in contrast to
the case of the temporal gauge transformations
(\ref{eq:III.11a})-(\ref{eq:III.11c}). Notice also that for the mass eigenvalue equations
(\ref{eq:III.22a})-(\ref{eq:III.22c}) we preferred the use of the original electric
potentials~$\IA_0,\IIA_0,\IIIA_0$ in order that the mass eigenvalues~$M_a\, (a=1,2,3)$
become explicitly displayed. But observe that the latter do appear in such a specific way
that they in principle could be absorbed also into the transformed potentials
(\ref{eq:III.11a})-(\ref{eq:III.11c}) after performing the gauge transformation
(\ref{eq:III.7a})-(\ref{eq:III.7c}) in combination with (\ref{eq:III.10}).

\begin{center}
  \emph{\textbf{3.\ Gauge Field Equations}}
\end{center}

Concerning the magnetic gauge covariance, the situation with the gauge field equations is
somewhat different from the matter field equations. Namely, building up the Lorentz
covariant objects~${F^a}_{\mu\nu}$ (or~$G_{\mu\nu}$, resp.) by the usual electric and
magnetic three-vector fields~$\vec{E_a},\vec{H}_a$ (or~$\vec{X}$ and~$\vec{Y}$, resp.) one
finds the three-vector versions of the former four-vector relations
(\ref{eq:II.11a})-(\ref{eq:II.11d}) to look as follows, namely for the electric fields
\begin{subequations}
  \begin{align}
    \label{eq:III.23a}
    \vec{E}_1 &= -\vec{\nabla}\,\iAn\\*
    \label{eq:III.23b}
    \vec{E}_2 &= -\vec{\nabla}\,\iiAn - i\left[B_0\vBstar -\Bstar_0\vec{B}\right]\\*
    \label{eq:III.23c}
    \vec{E}_3 &= -\vec{\nabla}\,\iiiAn +i \left[B_0\vBstar -\Bstar_0\vec{B}\right]
  \end{align}
\end{subequations}
and similarly for the magnetic fields
\begin{subequations}
  \begin{align}
    \label{eq:III.24a}
    \vec{H}_1 &= \vec{\nabla}\times\vec{A}_1\\*
    \label{eq:III.24b}
    \vec{H}_2 &= \vec{\nabla}\times\vec{A}_2 - i\vec{B}\times\vBstar\\*
    \label{eq:III.24c}
    \vec{H}_3 &= \vec{\nabla}\times\vec{A}_3 + i\vec{B}\times\vBstar\ .
  \end{align}
\end{subequations}
Obviously these electromagnetic field strengths~$\vec{E}_a(\vec{r})$
and~$\vec{H}_a(\vec{r})$ are invariant under the magnetic gauge group~$U(1)\times
U(1)\times U(1)$ and consequently one expects this invariance to hold also for the
corresponding field equations (see below). In contrast to this, the exchange field
strengths~$\vec{X}$ and~$\vec{Y}$ of both the electric and magnetic type are sensitive
against a magnetic gauge transformation
\begin{subequations}
  \begin{align}
    \label{eq:III.25a}
    \vec{X}(\vec{r}) &= -\vns B_0(\vec{r}) - i\,\AM \vec{B}(\vec{r})\\*
    \label{eq:III.25b}
    \vec{Y}(\vec{r}) &= \vns \times\vec{B}(\vec{r})\ .
  \end{align}
\end{subequations}
Here~$\AM$ is the electric potential difference introduced by equation
(\ref{eq:III.12}) and~$\vns$ is the gauge-covariant generalization of the
ordinary gradient operator~$\vec{\nabla}$, i.e.
\begin{gather}
  \label{eq:III.26}
  \vns \doteqdot \vec{\nabla} -i \vec{\mathfrak{A}}\\*
  \Big(\vec{\mathfrak{A}} \doteqdot \vec{A}_2 - \vec{A}_3\Big)\ .\notag
\end{gather}

Evidently, both exchange field strengths~$\vec{X}$ and~$\vec{Y}$ obey the homogeneous
(tensorial) gauge transformation law
\begin{subequations}
  \begin{align}
    \label{eq:III.27a}
    \vec{X} &\Rightarrow \vec{X}' = e^{-i\left(\alpha_2-\alpha_3\right)} \cdot\vec{X}\\*
    \label{eq:III.27b}
    \vec{Y} &\Rightarrow \vec{Y}' = e^{-i\left(\alpha_2-\alpha_3\right)} \cdot\vec{Y}\ ,
  \end{align}
\end{subequations}
and consequently one expects \emph{gauge-covariant} field equations for both
objects~$\vec{X},\vec{Y}$ in contrast to the situation with the \emph{gauge-invariant}
electromagnetic field strengths~$\vec{E}_a,\vec{H}_a$. Let us remark here that the check
of the claimed transformation laws (\ref{eq:III.27a})-(\ref{eq:III.27b}) is a nice
exercise where the vector potential difference~$\vec{\mathfrak{A}}$ can be shown to change
like
\begin{equation}
  \label{eq:III.28}
  \vec{\mathfrak{A}}\Rightarrow \vec{\mathfrak{A}}' = \vec{\mathfrak{A}}
  - \vec{\nabla}\left(\alpha_2-\alpha_3 \right)\ .
\end{equation}

Of course, the field equations of those three-vector objects must again be deduced from
their Lorentz-covariant predecessors (\ref{eq:II.12a})-(\ref{eq:II.12d}). Here, the gauge
\emph{invariance} of the electromagnetic part is again manifest, i.e.\ one has for the
electric part
\begin{subequations}
  \begin{align}
    \label{eq:III.29a}
    \vec{\nabla}\sdot\vec{E}_1 &= 4\pi\as\cdot\ijn\\*
    \label{eq:III.29b}
    \vec{\nabla}\sdot\vec{E}_2 + i\left(\vBstar\sdot\vec{X}-\vec{B}\sdot \vXstar\right) &= 4\pi\as\cdot\iijn\\*
    \label{eq:III.29c}
    \vec{\nabla}\sdot\vec{E}_3 - i\left(\vBstar\sdot\vec{X}-\vec{B}\sdot\vXstar\right) &=
    4\pi\as\cdot\iiijn\ ,
  \end{align}
\end{subequations}
and similarly for the magnetic part
\begin{subequations}
  \begin{align}
    \label{eq:III.30a}
    \vec{\nabla}\times\vec{H}_1 &= 4\pi\as\vec{j}_1\\*
    \label{eq:III.30b}
    \vec{\nabla}\times\vec{H}_2 -i\left(B_0\vXstar-\Bstar_0\vec{X} + \vec{B}\times\vYstar
      - \vBstar\times\vec{Y} \right) &= 4\pi\as\vec{j}_2\\*
    \label{eq:III.30c}
    \vec{\nabla}\times\vec{H}_3 +i\left(B_0\vXstar-\Bstar_0\vec{X} + \vec{B}\times\vYstar
      - \vBstar\times\vec{Y} \right) &= 4\pi\as\vec{j}_3\ .
  \end{align}
\end{subequations}
(For the link of the Maxwell currents~$\{{j^a}_\mu\}=\{\ajn;-\vec{j}_a(\vec{r})\}$ to the
Dirac currents $\{k_{a\mu}\}=\{\akn;-\vec{k}_a(\vec{r})\}$ see equations
(\ref{eq:II.17a})-(\ref{eq:II.17e})). However the corresponding equations for the exchange
field strengths display the expected gauge \emph{covariance} manifestly through the
emergence of that gauge-covariant derivative~$\vns$ (\ref{eq:III.26})
\begin{subequations}
  \begin{align}
    \label{eq:III.31a}
    \vns\sdot\vec{X}+i\vfE\sdot\vec{B} &= 4\pi\as\hstar_0\\*
    \label{eq:III.31b}
    \vns\times\vec{Y} - i\AM\vec{X}+iB_0\vfE-i\vfH\times\vec{B}&=4\pi\as\vhstar\ .
  \end{align}
\end{subequations}
Here the three-vectors~$\vfE$ and~$\vfH$ are defined through the differences of the
electric and magnetic fields due to the identical particles~$(a=2,3)$, i.e.
\begin{subequations}
  \begin{align}
    \label{eq:III.32a}
    \vfE &\doteqdot \vec{E}_2 - \vec{E}_3\\*
    \label{eq:III.32b}
    \vfH &\doteqdot \vec{H}_2 - \vec{H}_3\ .
  \end{align}
\end{subequations}

Summarizing, the matter equations (\ref{eq:III.22a})-(\ref{eq:III.22c}) and the gauge field
equations (\ref{eq:III.29a})-(\ref{eq:III.31b}) represent the essential RST field
equations for the stationary bound systems; and the task to be tackled now is to construct
an energy functional~$\ETT$ in such a way that its extremal equations turn out to be
identical to just this coupled set of matter and gauge field equations! However before this can
be done there remains one point to be settled which refers to the different
transformation behavior of the electromagnetic potentials~${A^a}_\mu$ and the exchange
potential~$B_\mu$, cf.~(\ref{eq:III.7a})-(\ref{eq:III.7c}) vs.\  (\ref{eq:III.9}).
\newpage
\begin{center}
  \emph{\textbf{4.\ Linearization and Gauge Invariance}}
\end{center}

Both the electromagnetic potentials~${A^a}_\mu$ and the exchange potential~$B_\mu$ appear
as independent parts of the bundle connection~$\A_\mu$ (\ref{eq:II.2}), but there arises
an important difference when the structure group~$U(1)\times U(2)$ becomes reduced to the
proper gauge group~$U(1)\times U(1)\times U(1)$. Namely, through this reduction process
both bundle constituents~${A^a}_\mu$ and~$B_\mu$ do acquire a rather different geometric
meaning: On the one hand, the electromagnetic potentials~${A^a}_\mu$ preserve their
property as connection components for the reduced bundle and therefore do transform
\emph{inhomogeneously}, cf.~(\ref{eq:III.7a})-(\ref{eq:III.7c}), in the true spirit of the
generalized equivalence principle for the gauge theories. But on the other hand, the
exchange potential~$B_\mu$ acquires tensorial properties under the reduction of the
structure group and therefore transforms \emph{homogeneously}, cf.~(\ref{eq:III.9}). Thus
it is no longer possible to locally gauge off the exchange potential~$B_ \mu$ as it is the
case with the electromagnetic potentials~${A^a}_\mu$. Clearly, such a change of the
geometric meaning of~$B_\mu$ must have its consequences for the field equations to be
obeyed by this object.

Recall here that the field equations (\ref{eq:II.12a})-(\ref{eq:II.12c}) for the
potentials~${A^a}_\mu$ are based upon the ``curls'' (\ref{eq:II.11a})-(\ref{eq:II.11c})
and such a type of equation is just sufficient for the connection components. Indeed, it
is not viable to fix also the divergences of~${A^a}_\mu$ by means of some source equation
since this would unduly restrict the set of possible gauge transformations of the reduced
bundle connection. However, since~$B_\mu$ has lost its inhomogeneous transformation
properties through the reduction of the structure group to the gauge group, this object
becomes thereby fixed to a greater extent than is the case with the
potentials~${A^a}_\mu$. As a consequence there must exist, besides the ``curl type'' of
field equation (\ref{eq:II.12d}), also a further equation for~$B_\mu$ which is of the
``source type''. We will satisfy now this requirement by selecting such an additional
``source type'' equation for~$B_\mu$, namely
\begin{equation}
  \label{eq:III.33}
  \DD^\mu B_\mu \equiv 0\ ,
\end{equation}
which is nothing else than the vanishing of the invariant divergence (i.e.\ a kind of
invariant Lorentz gauge condition). The covariant derivative~$(\DD)$ is defined here as
\begin{gather}
  \label{eq:III.34}
  \DD_\mu B_\nu = \nabla_\mu B_\nu + i \mathfrak{A}_\mu B_\nu\\*
  \label{eq:III.35}
  \Big(\mathfrak{A}_\mu \doteqdot {A^2}_\mu - {A^3}_\mu \Big)
\end{gather}
so that the source relation (\ref{eq:III.33}) may be also rewritten as
\begin{equation}
  \label{eq:III.36}
  \vns\sdot\vec{B} = -i\AM B_0
\end{equation}
with the covariant gradient operator~$\vns$ being defined by equation (\ref{eq:III.26})
and the potential difference~$\AM$ by (\ref{eq:III.12}). This is to be conceived not as an
additional dynamical equation but rather as an auxilary constraint (in the sense of a gauge
constraint) in order to ensure the existence and uniqueness of the wanted energy
functional~$\ETT$ (see below).

The latter relation (\ref{eq:III.36}) is well-suited in order to explicitly demonstrate
how the gauge invariance of the theory may become spoiled through the process of
linearization. Namely, the explicit form of (\ref{eq:III.36}) reads
\begin{equation}
  \label{eq:III.37}
  \vec{\nabla}\sdot \vec{B} -i\vec{\mathfrak{A}}\sdot\vec{B} = -i\left( \An -
    \frac{1}{\aM}\right)B_0\ .
\end{equation}
But linearizing the theory in the literally sense means that we omit the products of gauge
potentials~$\vec{\mathfrak{A}}\sdot\vec{B}$ and~$A_0 B_0$ so that the residual divergence
relation reads
\begin{equation}
  \label{eq:III.38}
  \vec{\nabla}\sdot\vec{B} = \frac{i}{\aM}B_0\ .
\end{equation}
Indeed this truncated non-invariant relation was used for the linearized theory (see equation
(\ref{eq:III.18}) of ref.~\cite{MaSo2}), but here it is revealed now as one of those places
where the strict gauge invariance is spoiled through the linearization process.

\begin{center}
  \emph{\textbf{5.\ RST Entanglement}}
\end{center}

Attention to the strange phenomenon of ``entanglement'' has been drawn for the first time
by Schr\"odinger. He discussed this effect within the conceptual framework of the
conventional (i.e.\ probabilistic) quantum theory, where its observable consequences refer
to the well-known \emph{exchange effects} (e.g.\ shift of atomic energy levels by the
``exchange energy''~\cite{CT}). However, this can not mean that other (non-conventional)
quantum approaches could not take account of such exchange effects, too. Indeed, it has
already been demonstrated in some preceding papers that those exchange effects can on
principle also be described within the framework of RST. But whereas these demonstrations
referred mainly to concrete numerical examples, it seems now time to discuss the exchange
phenomenon on a more general level.

Here, the point of departure is the component form (\ref{eq:II.12a})-(\ref{eq:II.12d}) of
the non-Abelian Maxwell equations (\ref{eq:II.8}). It seems rather obvious that these
equations suggest to introduce an \emph{entanglement vector}~$G_\mu$, namely
through
\begin{equation}
  \label{eq:III.39}
  G_\mu \doteqdot \frac{i}{4\pi\as}\left(B^\nu\Gstar_{\nu\mu}-\Bstar^\nu G_{\nu\mu}
  \right)\ .
\end{equation}
By this arrangement, the Maxwell equations (\ref{eq:II.12b})-(\ref{eq:II.12c}) for both
``entangled'' particles~$(a=2,3)$ adopt the following form
\begin{subequations}
  \begin{align}
    \label{eq:III.40a}
    \nabla^\mu {F^2}_{\mu\nu} + 4\pi\as G_\nu &= 4\pi\as {j^2}_\nu\\*
    \label{eq:III.40b}
    \nabla^\mu {F^3}_{\mu\nu} - 4\pi\as G_\nu &= 4\pi\as {j^3}_\nu\ .
  \end{align}
\end{subequations}
This could be interpreted in such a way that the Maxwell currents~${j^a}_\mu\,(a=2,3)$ as
the sources of the field strengths~${F^a}_{\mu\nu}$, become modified through some kind of
``entangling current''~$(\sim G_\mu)$ and thus generate a modified electromagnetic
potential~${A^a}_\mu\,(a=2,3)$ as the solution of just those equations
(\ref{eq:III.40a})-(\ref{eq:III.40b}). This RST entanglement effect would then consist
partly in the modification of the electromagnetic interactions~$(\sim {A^a}_\mu)$ of the
entangled particles and partly in the emergence of a new additional kind of interaction
being mediated directly through the exchange potential~$B_\mu$, see the covariant
derivatives (\ref{eq:II.7b})-(\ref{eq:II.7c}). However observe here that the entanglement
vector~$G_\mu$ formally drops out for generating the total field strength~${F^s}_{\mu\nu}$
(\ref{eq:III.3}) of the entangled particles so that the influence of the entangled
subsystem on the outside world (here the first particle,~$a=1$) works exclusively via the
electromagnetic potential~${A^s}_\mu$ (\ref{eq:III.2a}) which coincides with the former
potential~$\IA_\mu$ (\ref{eq:III.8a}).

There arises now a certain ambiguity concerning the normalization of the wave function for
an entangled subsystem; but this problem can be enlightened by a closer look at the properties
of the entanglement vector~$G_\mu$ (\ref{eq:III.39}). For this purpose, introduce the
skew-symmetric tensor~$B_{\mu\nu}\,(=-B_{\nu\mu})$ through
\begin{equation}
  \label{eq:III.41}
  B_{\mu\nu}\doteqdot \frac{i}{4\pi\as}\left(\Bstar_\mu B_\nu - B_\mu \Bstar_\nu\right)\ ,
\end{equation}
and then find by straightforward differentiation and observation of the vanishing
divergence (\ref{eq:III.33})
\begin{equation}
  \label{eq:III.42}
  G_\nu = -\nabla^\mu B_{\mu\nu} + \frac{i}{4\pi\as}\left[\Bstar^\mu \Big(\DD_\nu B_\mu\Big) -
    B^\mu\Big(\DD_\nu\Bstar_\mu\Big) \right]\ .
\end{equation}
From here it becomes obvious that the entanglement vector~$G_\mu$ can be written as the
sum of the divergence of a (four-dimensional) \emph{entanglement field
  strength}~$B_{\mu\nu}$ and an \emph{entanglement current}~$l_\nu$ which is defined through
\begin{equation}
  \label{eq:III.43}
  l_\nu \doteqdot -\frac{i}{4\pi\as} \left[\Bstar^\mu\left(\DD_\nu B_\mu \right)
  -B^\mu \left(\DD_\nu\Bstar_\mu \right)\right]\ .
\end{equation}
Thus the splitting of~$G_\nu$ (\ref{eq:III.42}) reads
\begin{equation}
  \label{eq:III.44}
  G_\nu = -\left(\nabla^\mu B_{\mu\nu} + l_\nu \right)\ ,
\end{equation}
and if this is inserted in both equations (\ref{eq:III.40a})-(\ref{eq:III.40b}) they do
reappear in the form
\begin{subequations}
  \begin{align}
    \label{eq:III.45a}
    \nabla^\mu{F^2}_{\mu\nu} &= 4\pi\as\left({j^2}_\nu + l_\nu +\nabla^\mu B_{\mu\nu}
    \right) \doteqdot 4\pi\as\cdot \jj{+}{2}_\nu\\*
    \label{eq:III.45b}
    \nabla^\mu{F^3}_{\mu\nu} &= 4\pi\as\left({j^3}_\nu - l_\nu -\nabla^\mu B_{\mu\nu}
    \right) \doteqdot 4\pi\as\cdot \jj{-}{3}_\nu
  \end{align}
\end{subequations}
But here the \emph{entangled currents}~$\jj{+}{2}_\mu$ and~$\jj{-}{3}_\mu$ have now vanishing
divergence
\begin{equation}
  \label{eq:III.46}
  \nabla^\mu \jj{+}{2}_\mu = \nabla^\mu \jj{+}{3}_\mu \equiv 0\ ,
\end{equation}
for by straightforward differentiation one finds
\begin{equation}
  \label{eq:III.47}
  \nabla^\mu G_\mu = -\nabla^\mu l_\mu = i\left(\Bstar^\mu\hstar_\mu - B^\mu h_\mu\right)
\end{equation}
which then just annihilates the sources of~${j^2}_\mu\,(\equiv -k_{2\mu})$
and~${j^3}_\mu\,(\equiv -k_{3\mu})$, cf.~(\ref{eq:II.18b})-(\ref{eq:II.18c}), so that the
source equations (\ref{eq:III.46}) can actually be true. The ordinary Maxwell equations
(\ref{eq:III.45a})-(\ref{eq:III.45b}) then say that the field strengths~${F^2}_{\mu\nu}$
and~${F^3}_{\mu\nu}$ can be thought to be generated by those entangled and source-free
currents~$\jj{+}{2}_\mu$ and~$\jj{-}{3}_\mu$. Such a situation would provide us with the
possibility to postulate normalization conditions for the entangled wave
functions~$\psi_a(\vec{r})\,(a=2,3)$ of the following kind
\begin{equation}
  \label{eq:III.48}
  \int\limits_{(S)}\jj{+}{2}_\mu\,dS^\mu = \int\limits_{(S)}\jj{-}{3}_\mu\,dS^\mu = -1\ .
\end{equation}
Here, the integrals over some three-dimensional hypersurface~$(S)$ would then be
independent of the choice of that hypersurface~$(S)$. However, we will not resort to this
construction but rather will prefer the following normalization conditions:
\begin{gather}
  \label{eq:III.49}
  \ND(a) \doteqdot \int\limits_{t=\mathrm{const.}}d^3\vec{r}\,\akn - 1\equiv 0\\*
  \big(a=1,2,3\big)\ .\notag
\end{gather}
The reason for this preference, where the independence of the 3-surface of integration is
lost, will become clearer below in connection with the discussion of the gauge-invariant
energy functional.  

Finally, it is also interesting to remark that the ``external'' field
strength~${F^s}_{\mu\nu}$ (\ref{eq:III.2b}) of the entangled two-particle system may be
complemented by an ``internal'' counterpart~${f^s}_{\mu\nu}$
\begin{equation}
  \label{eq:III.50}
  {f^s}_{\mu\nu} \doteqdot {F^2}_{\mu\nu} - {F^3}_{\mu\nu}\ .
\end{equation}
The combination of both non-Abelian Maxwell equations(\ref{eq:III.45a})-(\ref{eq:III.45b}) 
then yields
\begin{equation}
  \label{eq:III.51}
  \nabla^\mu\left({f^s}_{\mu\nu} - 2 B_{\mu\nu} \right) = 4\pi\as\left({j^2}_\nu -
    {j^3}_\nu + 2l_\nu \right)\ .
\end{equation}
The comparison of this ``internal'' Maxwell equation to its ``external'' counterpart
(\ref{eq:III.3}) says that the internal gauge interactions do not obey the ordinary
(Abelian) Maxwell scheme but take account of the non-Abelian character whose imprint on
the internal gauge dynamics results in the additional objects~$B_{\mu\nu}$ and~$l_\nu$
occurring in equation (\ref{eq:III.51}).

\begin{center}
  \emph{\textbf{6.\ Boundary Conditions and Entanglement Charge}}
\end{center}

The preceding arguments were based upon a Lorentz-covariant presentation of the RST
entanglement, but the wanted energy functional~$\ETT$ can naturally not fit into this
scheme because the concept of energy is quite generally not a Lorentz-invariant object,
even in the relativistic case. Therefore, in search of a gauge-invariant energy functional,
it is more convenient to represent the preceding discussion also in three-vector
notation. This will be helpful for building up the RST energy~$\ET$ from its various
constituents.

First, decompose the tensor~$B_{\mu\nu}$ (\ref{eq:III.41}) into its ``electric'' and
``magnetic'' part according to
\begin{subequations}
  \begin{align}
    \label{eq:III.52a}
    \vec{W} &= i\left(\Bstar_0\vec{B}-B_0\vBstar \right)\\*
    \label{eq:III.52b}
    \vec{V} &= i\left(\vec{B}\times\vBstar\right)\ ,
  \end{align}
\end{subequations}
so that the electric fields~$\vec{E}_a$ (\ref{eq:III.23b})-(\ref{eq:III.23c}) for the
entangled particles appear now as
\begin{subequations}
  \begin{align}
    \label{eq:III.53a}
    \vec{E}_2 &= -\vec{\nabla}\,\iiAn + \vec{W}\\*
    \label{eq:III.53b}
    \vec{E}_3 &= -\vec{\nabla}\,\iiiAn - \vec{W}\ ,
  \end{align}
\end{subequations}
and similarly for the magnetic fields~$\vec{H}_a$ (\ref{eq:III.24b})-(\ref{eq:III.24c})
\begin{subequations}
  \begin{align}
    \label{eq:III.54a}
    \vec{H}_2 &= \vec{\nabla}\times\vec{A}_2 - \vec{V}\\*
    \label{eq:III.54b}
    \vec{H}_3 &= \vec{\nabla}\times\vec{A}_3 + \vec{V}\ .
  \end{align}
\end{subequations}

Furthermore, it becomes obvious from equation (\ref{eq:III.34}) that, after bundle
reduction, the homogeneously transforming exchange potential~$B_\mu$ acts like a charged
\emph{matter} field with current~$l_\mu$ (\ref{eq:III.43}) and thus feels the induced
connection~$\mathfrak{A}_\mu\,(\doteqdot {A^2}_\mu - {A^3}_\mu)$. The corresponding
induced curvature follows from equations (\ref{eq:III.53a})-(\ref{eq:III.54b}) in terms of
the original curvature components~$\vfE,\vfH$ (\ref{eq:III.32a})-(\ref{eq:III.32b}) and
the entanglement fields~$\vec{W},\vec{V}$ (\ref{eq:III.52a})-(\ref{eq:III.52b}) as
\begin{equation}
  \label{eq:III.55}
  -\vec{\nabla}\An = \vfE - 2\vec{W}
\end{equation}
and 
\begin{equation}
  \label{eq:III.56}
  \vec{\nabla}\times\vec{\mathfrak{A}} = \vfH + 2\vec{V}\ .
\end{equation}
Conversely, the time component of the entanglement vector~$G_\mu$ (\ref{eq:III.44}) reads
in terms of the electric entanglement field strength~$\vec{W}$ and density~$l_0$
\begin{equation}
  \label{eq:III.57}
  G_0 = \frac{\vec{\nabla}\cdot\vec{W}}{4\pi\as} - l_0\ ,
\end{equation}
and similarly for the space component
\begin{equation}
  \label{eq:III.58}
  \vec{G}= -\left(\frac{\vec{\nabla}\times\vec{V}}{4\pi\as} + \vec{l}\right)
\end{equation}
where evidently~$\vec{V}$ plays the part of a magnetic entanglement field strength
and~$\vec{l}$ is the entanglement three-current. Obviously, the latter two equations
represent the three-vector reformulation of the Lorentz-covariant equation
(\ref{eq:III.44}). In a similar way, the ``internal'' Maxwell equation (\ref{eq:III.51})
may be understood as the field equation for the induced
connection~$\{\An,\vfA\}=\{\mathfrak{A}^\mu\}$ and appears in three-vector notation as
\begin{subequations}
  \begin{align}
    \label{eq:III.59a}
    -\Delta\An+ 4\left(\vec{\nabla}\cdot\vec{W}\right) &= \vec{\nabla}\cdot\left(\vfE+2\vec{W} \right) =
    4\pi\as\left({}^{(2)}\!j_0-{}^{(3)}\!j_0+2l_0 \right)\\*
    \label{eq:III.59b}
    -\Delta\vfA-4\left(\vec{\nabla}\times\vec{V}\right) &= \vec{\nabla}\times\left(\vfH-2\vec{V} \right) =
    4\pi\as\left(\vec{j}_2-\vec{j}_3+2\vec{l}\, \right)\ .
  \end{align}
\end{subequations}

This three-vector formalism for the induced ``exchange bundle'' is also well-suited in
order to consider (albeit somewhat superficially) the question of boundary
conditions. Surely, one will require that the electrostatic potentials~$\aAr$
(\ref{eq:III.17}) will decay at infinity~$(r\to\infty)$ like an ordinary Coulomb
potential~$(a=1,2,3)$
\begin{equation}
  \label{eq:III.60}
  \lim\limits_{r\to\infty} \aAr(\vr) = e_a\cdot\frac{\as}{r}
\end{equation}
where~$e_a$ is a \emph{``pseudo-charge number''}. Of course for the non-identical
particle~$(a=1)$ one will put~$e_1=1$ which is the actual charge number, because the
non-identical particle is not subjected to the exchange forces; but for the two entangled
particles one naturally expects~$e_a\neq -1\,(a=2,3)$. Recall here the fact that the
total potential~${A^s}_\mu$ (\ref{eq:III.2a}) of both entangled particles feels
exclusively the total current~${j^s}_\mu$, cf.~(\ref{eq:III.3})
\begin{equation}
  \label{eq:III.61}
  \nabla^\mu{F^s}_{\mu\nu} = \square {A^s}_\nu = - \Delta{A^s}_\nu = 4\pi\as{j^s}_\nu
\end{equation}
so that we can require at spatial infinity~$(r\to\infty)$
\begin{equation}
  \label{eq:III.62}
  {}^{(s)}\!A_0 \doteqdot\ \iiAn + \iiiAn = -2\cdot\frac{\as}{r}\ .
\end{equation}
This means that one always has
\begin{equation}
  \label{eq:III.63}
  e_s \doteqdot e_2+e_3 = -2
\end{equation}
but not necessarily~$e_2=e_3=-1$ since the exchange interaction can transfer
\emph{entanglement charge}~$\delta_e$ from one identical particle to the other.

In order to discuss this charge exchange a little bit more thoroughly, integrate over the
equation (\ref{eq:III.59a}) and apply also Gau\ss' integral theorem in order to find under
observation of the normalization conditions (\ref{eq:III.49})
\begin{equation}
  \label{eq:III.64}
  \oint\limits_{(S_\infty)} \left(\vfE+2\vec{W} \right)\cdot d\vec{S} =
  8\pi\as\int d^3\vec{r}\;l_0(\vec{r})\ ,
\end{equation}
where~$S_\infty$ is a closed two-surface at infinity~$(r\to\infty)$ with surface
element~$d\vec{S}$. Here the two-dimensional surface integral of the vector
field~$\vec{W}$ (\ref{eq:III.52a}) can now be assumed to be zero because it is a nearby
hypothesis that~$B_0$ should decay at spatial infinity~$(r\to\infty)$ at least
like~$r^{-1}$ (monopole) and~$\vec{B}$ at least like~$r^{-2}$ (dipole), which
yields~$\vec{W}\sim r^{-3}$. But for such a situation the equation (\ref{eq:III.63}) then
links the entanglement charge~$\de=\frac{(e_2-e_3)}{2}$ to the integrated
entanglement charge density~$l_0(\vec{r})$ through
\begin{equation}
  \label{eq:III.65}
  \oint\limits_{(S_\infty)}\vfE\cdot d\vec{S} = 8\pi\as\de=8\pi\as\int
  d^3\vec{r}\;l_0(\vec{r})\ ,
\end{equation}
i.e.\ the entanglement charge
\begin{equation}
  \label{eq:III.66}
  \de = \int d^3\vec{r}\; l_0(\vec{r})
\end{equation}
is shifted from one identical particle to the other.

Applying quite similar arguments as for the present equation (\ref{eq:III.59a}) also to
the former equations (\ref{eq:III.29b})-(\ref{eq:III.29c}) for the individual particles
one concludes that the pseudo-charges~$e_a$ (\ref{eq:III.60}) for the identical particles
are given in terms of the entanglement charge~$\de$ (\ref{eq:III.66}) through
\begin{subequations}
  \begin{align}
    \label{eq:III.67a}
    e_2 &= -1 + \de\\*
    \label{eq:III.67b}
    e_3 &= -1 - \de\ .
  \end{align}
\end{subequations}
Thus summarizing the results, one arrives at the conclusion that it is only the total
charge number~$e_s$ (\ref{eq:III.63}) (of a subsystem of identical particles) which is
\emph{generally} fixed and leads to a well-defined electrostatic potential~${}^{(s)}A_0$
(\ref{eq:III.62}) at spatial infinity, whereas the individual particles do carry charge
numbers (here~$e_2$ and~$e_3$) which are neither integer nor are generally fixed but
depend upon the considered situation (i.e.\ the concrete quantum state). Clearly, the
tacitly made presumption is here that the charge number~$e_a$ of any particle~(a)
is defined by that asymptotic form~$\aAn$ of its emitted potential
(\ref{eq:III.60}). Here, it should become also obvious that future work has to be spent
upon a more thorough inspection of the boundary conditions of all the concerned gauge
fields.


\section{Gauge-Invariant Energy Functional}
\indent

It seems unlikely that the wave functions (or even densities) in an atom or molecule can
be observed directly in order to test the physical truth of the RST solutions; and
therefore one has to look for some other physical observable which is essentially
determined by the RST solutions but on the other hand provides also a handle for the
observations. Surely, such a quantity will be the total field energy~$\ET$ being
concentrated in any RST solution since this quantity determines uniquely the optical
spectrum of the bound system under consideration. A nearby definition of the wanted RST
energy~$\ET$ refers of course to the total energy momentum density~$\TT_{\mu\nu}$, cf.\
equation (\ref{eq:II.21}). This proposal has been studied extensively in the preceding
paper~\cite{MaSo2}, however only for the \emph{linearized} theory which has been shown (by the
preceding arguments) to miss the exact gauge invariance. Therefore we will readily set up
once more the wanted energy functional~$\tETT$, but this time with properly taking into
account the non-Abelian (and therefore non-linear) character of RST.

The non-relativistic limit~$\tbbETT$ of the relativistic~$\tETT$ is also worked out; and
here it is important to preserve the gauge covariance just in this limit. It should also
be a matter of course that the subsequent exact result for~$\tETT$ must be closely related
to the previous approximative result of the linearized theory. Indeed, this relationship
shows up in the \emph{formal identity} of the present exact result (\ref{eq:IV.20}) with
the analogous expression of the linearized theory, cf. equation (\ref{eq:IV.14}) of
ref.~\cite{MaSo2}. Thus the present progress refers here mainly to the individual
contributions to the energy functional \emph{which are now made rigorously
  gauge-invariant!}  This invariance becomes manifest through the rigorous use of the
gauge-covariant derivative~$\DD$ and~$\vns$, see equations (\ref{eq:III.26}) and
(\ref{eq:III.34}).  Clearly, the expectation is now that such an energy functional, which
meets with the symmetry requirements of all the modern gauge field theories, should also have
something to do with physical reality.

\begin{center}
  \emph{\textbf{1.\ Generalized Poisson Identities}}
\end{center}

The Poisson and exchange identities are important constraints for the \emph{principle of
  minimal energy} and therefore must be considered before the corresponding energy
functional can be set up. The electric Poisson identity is obtained by multiplying through
any of the three electric source equations (\ref{eq:III.29a})-(\ref{eq:III.29c}) with the
corresponding electrostatic potentials~$\IA_0(\vec{r}),\IIA_0(\vec{r}),\IIIA_0(\vec{r})$
and integrating over whole three-space with finally adding up all three results. This
procedure yields the \emph{electric Poisson identity} in the following form
\begin{equation}
  \label{eq:IV.1}
  \NGe \equiv \nGe
\end{equation}
where the left-hand side is defined in terms of the electric potentials and charge
densities~$\ajn$ through
\begin{equation}
  \label{eq:IV.2}
  \begin{split}
  \NGe \doteqdot\, &\frac{\hbar c}{4\pi\as}\int d^3\vec{r}\left[ \left(\vec{\nabla}\,\iAn \right)
    \sdot\left(\vec{\nabla}\,\iiAn\right) + \left(\vec{\nabla}\,\iiAn\right)\sdot
    \left(\vec{\nabla}\,\iiiAn \right) +  \left(\vec{\nabla}\,\iiiAn \right) \sdot \left
      (\vec{\nabla}\,\iAn\right)\right]\\*
  & -\frac{\hbar c}{2}\int d^3\vec{r}\left[ \IA_0\cdot {}^{(1)}j_0 +  \IIA_0\cdot
    {}^{(2)}j_0 +  \IIIA_0\cdot {}^{(3)}j_0 \right]
  -\frac{\hbar c}{4\pi\as}\int d^3\vec{r}\;\;\An\left(\vec{\nabla}\sdot\vec{W}\right)\ .
  \end{split}
\end{equation}
And furthermore, the right-hand side of the identity (\ref{eq:IV.1}) is given in terms of
the exchange objects through
\begin{equation}
  \label{eq:IV.3}
  \nGe \doteqdot -\frac{\hbar c}{2}\int d^3\vec{r}\;\;\An\cdot l_0\ .
\end{equation}

It should be obvious that here must exist also a magnetic analogy to that electric
identity (\ref{eq:IV.1}). Indeed this \emph{magnetic Poisson identity} may be deduced from
the magnetic part (\ref{eq:III.30a})-(\ref{eq:III.30c}) of the (non-Abelian) Maxwell
equations in just the same manner and appears then as follows:
\begin{equation}
  \label{eq:IV.4}
  \NGm \equiv \nGm\ ,
\end{equation}
where the left-hand side of this identity is defined quite analogously to its electric
counterpart~$\NGe$ (\ref{eq:IV.2}) through
\begin{equation}
  \label{eq:IV.5}
  \begin{split}
    \NGm &\doteqdot \frac{\hbar c}{4\pi\as}\int d^3\vec{r}\,
     \left[
      \left( \vec{\nabla}\times\vec{A}_1\right) \sdot \left(\vec{\nabla}\times\vec{A}_2\right) +
      \left( \vec{\nabla}\times\vec{A}_2\right) \sdot
      \left(\vec{\nabla}\times\vec{A}_3\right)\right. \\*
      &+\left. \left( \vec{\nabla}\times\vec{A}_3\right) \sdot \left(\vec{\nabla}\times\vec{A}_1\right) 
    \right]
    - \frac{\hbar c}{2}\int d^3\vec{r}\,
    \left( \vec{j}_1\sdot\vAI + \vec{j}_2\sdot\vAII +\vec{j}_3\sdot\vAIII\right)\ ,
  \end{split}
\end{equation}
and the right-hand side appears again as the magnetic counterpart of~$\nGe$ (\ref{eq:IV.3}),
i.e.
\begin{equation}
  \label{eq:IV.6}
  \nGm\doteqdot -\frac{\hbar c}{4\pi\as}\int d^3\vec{r}\,\vfA\sdot\left(\vec{\nabla}\times\vec{V}\right)
  -\frac{\hbar c}{2}\int d^3\vec{r}\;\vec{\mathfrak{A}}\sdot \vec{l}\ .
\end{equation}
Now with a little bit intuition concerning the relativistic formalism, one easily realizes that the
\emph{electromagnetic Poisson identity}
\begin{equation}
  \label{eq:IV.7}
  \NGe - \NGm - \left(\nGe - \nGm\right) \equiv 0
\end{equation}
is built up by certain Lorentz invariants; namely this identity reads explicitly
\begin{equation}
  \label{eq:IV.8}
  \begin{split}
    &\int d^3\vec{r}\,\left( \iA_{\mu\nu}\cdot\iiA^{\mu\nu}+\iiA_{\mu\nu}\cdot\iiiA^{\mu\nu} 
      +\iiiA_{\mu\nu}\cdot\iA^{\mu\nu} + 8\pi\as \mathfrak{A}^{\mu\nu} B_{\mu\nu}\right)\\*
      +4\pi\as &\int d^3\vec{r}\,\left(\IA_\mu\cdot\jj{1}{\mu} + \IIA_\mu\cdot\jj{2}{\mu} +
      \IIIA_\mu\cdot\jj{3}{\mu}-\mathfrak{A}_\mu l^\mu\right)\equiv 0\ .    
  \end{split}
\end{equation}
Here it is made use of the source equation for the object~$B_{\mu\nu}$ (\ref{eq:III.44})
\begin{equation}
  \label{eq:IV.9}
  \nabla^\nu B_{\nu\mu} = -\left( G_\mu + l_\mu\right)
\end{equation}
whose space-time splitting is given by equations (\ref{eq:III.57})-(\ref{eq:III.58});
and furthermore the skew-symmetric tensors~$\aA_{\mu\nu}\,(=-\aA_{\nu\mu})$ are defined
through~$(a=1,2,3)$
\begin{equation}
  \label{eq:IV.10}
  \aA_{\mu\nu}\doteqdot \nabla_\mu {A^a}_\nu - \nabla_\nu {A^a}_\mu\ .
\end{equation}
But let us stress here the somewhat strange fact that it is only the left-hand side~$\NGe$
(\ref{eq:IV.2}) and~$\NGm$ (\ref{eq:IV.5}) of the electric and magnetic Poisson identities
(\ref{eq:IV.1}) and (\ref{eq:IV.4}) which play the role of a constraint for the subsequent
\emph{RST principle of minimal energy}. Clearly one expects that a similar
constraint must be active also for the exchange subsystem.

\begin{center}
  \emph{\textbf{2.\ Exchange Identities}}
\end{center}

The procedure for deducing the exchange identities is quite similar to that for the
preceding electric and magnetic Poisson identities. More concretely, one multiplies
through the electric exchange equation (\ref{eq:III.31a}) with~$\Bstar_0$ (and its complex
conjugate with~$B_0$) and then one arrives after some simple mathematical manipulations at
the following \emph{electric exchange identity}
\begin{equation}
  \label{eq:IV.11}
  \NGh \equiv \nGh\ .
\end{equation}
Here the left-hand side is given by
\begin{equation}
  \label{eq:IV.12}
  \begin{split}
    \NGh &\doteqdot \frac{\hbar c}{4\pi\as}\int d^3\vec{r}
    \left[
      \left(\vns\Bstar_0\right)\sdot \left(\vns B_0\right) -
      \AM^2\Bstar_0\cdot B_0
    \right]\\*
    &-\frac{\hbar c}{2}\int d^3\vec{r} \left(B_0 h_0 + \Bstar_0\hstar_0 \right)
    + \frac{\hbar c}{8\pi\as}\int d^3\vec{r}\;\mathfrak{A}_0\left(\vec{\nabla}\sdot\vec{W}\right)
  \end{split}
\end{equation}
and the right-hand side by
\begin{equation}
  \label{eq:IV.13}
  \nGh \doteqdot -\frac{\hbar c}{8\pi\as}\int d^3\vec{r}\;\mathfrak{A}_0
  \left(\vec{\nabla}\sdot\vec{W}\right) - \frac{\hbar c}{4\pi\as}\int
  d^3\vec{r}\;\vec{W}^2\ .
\end{equation}

Quite analogously, the \emph{magnetic exchange identity} is obtained from the magnetic
exchange equation (\ref{eq:III.31b}) by multiplying through with~$\vBstar$ (or~$\vec{B}$,
resp.); and this then yields the following identity
\begin{equation}
  \label{eq:IV.14}
  \NGg \equiv \nGg
\end{equation}
with the left-hand side~$\NGg$ being defined by
\begin{equation}
  \label{eq:IV.15}
  \begin{split}
    \NGg &\doteqdot \frac{\hbar c}{4\pi\as}\int d^3\vec{r}
    \left[
      \left(\vns\times\vBstar\right)\sdot\left(\vns\times\vec{B}\right) +
      \left(\vns\sdot\vBstar\right)\cdot\left(\vns\sdot\vec{B}\right) -
      \AM^2\left(\vBstar\sdot\vec{B}\right)
    \right]\\*
      &-\frac{\hbar c}{8\pi\as}\int d^3\vec{r}\;\An\left(\vec{\nabla}\sdot\vec{W}\right)
       -\frac{\hbar c}{4\pi\as}\int d^3\vec{r}\;\vec{\mathfrak{A}}\sdot\left(\vec{\nabla}\times\vec{V}\right)\\*
       &+\frac{\hbar c}{4\pi\as}\int d^3\vec{r}\,\vec{V}^2
       -\frac{\hbar c}{4\pi\as}\int d^3\vec{r}\,\vec{W}^2 
       -\frac{\hbar c}{2}\int d^3\vec{r}\left(\vec{B}\sdot\vec{h}+\vBstar\sdot\vhstar \right)
  \end{split}
\end{equation}
and similarly the right-hand side~$\nGg$ by
\begin{equation}
  \label{eq:IV.16}
  \nGg \doteqdot -\frac{\hbar c}{4\pi\as}\int d^3\vec{r}\;\vec{V}^2 +
  \frac{\hbar c}{8\pi\as}\int d^3\vec{r}\;\An\left(\vec{\nabla}\sdot\vec{W}\right)\ .
\end{equation}
The exchange identity (\ref{eq:IV.11}) of the electric type can again be combined with the
exchange identity of the magnetic type (\ref{eq:IV.14}) in order to yield the \emph{total
exchange identity}
\begin{equation}
  \label{eq:IV.17}
  \NGh - \nGh -\left(\NGg - \nGg\right)\equiv 0\ .
\end{equation}
This may be written again in terms of Lorentz invariants as
\begin{equation}
  \label{eq:IV.18}
  \int d^3\vec{r}\,\left(\Gstar_{\mu\nu}G^{\mu\nu}-4\pi\as\mathfrak{A}^{\mu\nu}B_{\mu\nu}
  \right)
  +4\pi\as\int d^3\vec{r}\left(B^\mu h_\mu + \Bstar^\mu\hstar_\mu + 8\pi\as
    B^{\mu\nu}B_{\mu\nu}\right)
  \equiv 0
\end{equation}
and is of course the exchange counterpart of the electromagnetic case
(\ref{eq:IV.7})-(\ref{eq:IV.8}). Here we have made use also of the fact that the exchange
field strength~$G_{\mu\nu}$ (\ref{eq:II.11d}) can be expressed in terms of the
gauge-covariant derivative (\ref{eq:III.34}) as
\begin{equation}
  \label{eq:IV.19}
  G_{\mu\nu} = \DD_\mu B_\nu - \DD_\nu B_\mu\ .
\end{equation}
However observe again that for the subsequent energy functional~$\tETT$ one cannot use the
Lorentz invariants (\ref{eq:IV.18}) immediately but rather one has to resort to their
non-invariant parts~$\NGh$ and~$\NGg$ separately; the complementary parts~$\nGh$
and~$\nGg$ do work as simple constants which must be held fixed for the variational
process!

\begin{center}
  \emph{\textbf{3.\ Relativistic Energy Functional}}
\end{center}

The essential structure of the RST energy functional $\tETT$ has already been discussed in
the preceding paper~\cite{MaSo2}, albeit only for the linearized theory. But this principal
form of~$\tETT$ does also apply to the original non-Abelian (and therefore non-linear)
theory, i.e.\ we put
\begin{equation}
  \label{eq:IV.20}
  \begin{split}
  \tETT &= \Eiv + \sum\limits_{a=1}^3 \lD(a)\cdot\ND(a) +
  \lGe\cdot\left(\NGe - \nGe\right)\\*
  &+ \lGm \cdot\left(\NGm - \nGm\right) + \lGh \cdot\left(\NGh - \nGh\right) +
  \lGg \cdot\left(\NGg - \nGg\right)\ .
  \end{split}
\end{equation}

The leading feature of this functional consists in its composition of a physical part
(i.e.~$\Eiv$) and of the constraints of electromagnetic and exchange type, besides the
normalization conditions (\ref{eq:III.49}) for the wave functions. The four gauge field
constraints are based upon the demand that the objects~$\NGe$ (\ref{eq:IV.2}),~$\NGm$
(\ref{eq:IV.5}),~$\NGh$ (\ref{eq:IV.12}) and~$\NGg$ (\ref{eq:IV.15}) must be kept fixed
for the variational process, i.e.\ one interpretes the functionals~$\nGe$
(\ref{eq:IV.3}),~$\nGm$ (\ref{eq:IV.6}),~$\nGh$ (\ref{eq:IV.13}) and~$\nGg$
(\ref{eq:IV.16}) as constants \emph{not} undergoing the variational manipulations
($\leadsto \delta\nGe=\delta\nGm=\delta\nGh=\delta\nGg\mustbe 0 $). Especially, for a
system of non-identical particles (such as the previously considered positronium~\cite{MaSo2})
all exchange fields do vanish and thus the real numbers $\nGe,\nGm,\nGh,\nGg$ are all zero
per se. For such a situation the variational procedure does reproduce correctly the RST
system of mass eigenvalue and gauge field equations, see ref.~\cite{MaSo2}. But the progress
with the present generalized case (\ref{eq:IV.20}) is now that both the RST mass
eigenvalue system (\ref{eq:III.22a})-(\ref{eq:III.22c}) \emph{plus} the gauge field system
(\ref{eq:III.29a})-(\ref{eq:III.31b}) does reappear in a gauge-invariant form from the
extremalization procedure with \emph{fixed numbers} $\nGe,\nGm,\nGh,\nGg$ (not to be
interpreted as functionals of the gauge fields):
\begin{gather}
  \label{eq:IV.21}
  \delta\tETT\mustbe 0\\*
  \big( \delta\nGe=\delta\nGm=\delta\nGh=\delta\nGg = 0 \big)\ .\notag
\end{gather}
Thereby the Lagrangean multipliers turn out as
\begin{subequations}
  \begin{align}
    \label{eq:IV.22a}
    \lD(1) = M_1 c^2,\qquad \lD(2) &= - M_2c^2,\qquad \lD(3)=-M_3 c^2\\*
    \label{eq:IV.22b}
    \lGm &= -\lGe = 2\\*
    \label{eq:IV.22c}
    \lGh &= -\lGg = 2\ .
  \end{align}
\end{subequations}
Recall here also the former discussion below equation (\ref{eq:III.47}) concerning the
right choice of the normalization condition for the wave functions~$\psi_a(\vec{r})$. The
reason for that preference of (\ref{eq:III.49}) over (\ref{eq:III.48}) becomes now clear
because it is just this specific condition (\ref{eq:III.49}) which (by virtue of equations
(\ref{eq:IV.22a})) lets emerge the mass eigenvalues~$M_a$ as desired in the mass
eigenvalue equations (\ref{eq:III.22a})-(\ref{eq:III.22c}).

If the system of extremal equations due to (\ref{eq:IV.21}) could be solved exactly, one
could substitute the exact solutions back into the energy functional~$\tETT$
(\ref{eq:IV.20}) and would then find that the total RST energy~$\tET$, concentrated in that
hypothetical exact solution, is determined by the physical part~$\Eiv$ alone. This physical
part itself is the sum of all the partial energies of the coupled RST field system, i.e.
\begin{equation}
  \label{eq:IV.23}
  \Eiv = \Z_{(1)}^2\cdot \Mp c^2 + \Z_{(2)}^2\cdot\Me c^2 + \Z_{(3)}^2\cdot\Me c^2 +
  2\Tkin + \eER -\mER - \hEC + \gEC\ .
\end{equation}
Here, the mass renormalization factors~$\Z_{(a)}^2\,(a=1,2,3)$ are defined in terms of the Dirac
spinors~$\psi_a(\vec{r})$ as follows
\begin{equation}
  \label{eq:IV.24}
  \Z_{(a)}^2 = \int d^3\vec{r}\,\bar{\psi}_a(\vec{r})\psi_a(\vec{r})\ .
\end{equation}
Furthermore, the total kinetic energy~$\Tkin$ is the sum of the three single-particle
energies~$\Tkin(a)$
\begin{equation}
  \label{eq:IV.25}
  \Tkin = \Tkin(1) + \Tkin(2) + \Tkin(3)\ ,
\end{equation}
with the kinetic energy of the first particle~$(a=1)$ being given by~\cite{MaSo2}
\begin{equation}
  \label{eq:IV.26}
  \Tkin(1) = i\frac{\hbar c}{2}\int d^3\vec{r}\,\bar{\psi}_1(\vec{r})\vec{\gamma}\sdot\vec{\nabla}\psi_1(\vec{r})
\end{equation}
and similarly for both identical particles~$(a=2,3)$
\begin{equation}
  \label{eq:IV.27}
  \Tkin(a) = -i\frac{\hbar c}{2}\int d^3\vec{r}\,\bar{\psi}_a(\vec{r})\vec{\gamma}
  \sdot\vec{\nabla}\psi_a(\vec{r})\ .
\end{equation}

The last contribution to the physical energy~$\Eiv$ (\ref{eq:IV.23}) comes from the gauge
field and subdivides into the electrostatic energy~$\eER$
\begin{equation}
  \label{eq:IV.28}
  \eER = \frac{\hbar c}{4\pi\as}\int d^3\vec{r}\,\left(\vec{E}_1\sdot\vec{E}_2 +
    \vec{E}_2\sdot\vec{E}_3 + \vec{E}_3\sdot\vec{E}_1\right)\ ,    
\end{equation}
magnetostatic energy~$\mER$
\begin{equation}
  \label{eq:IV.29}
  \mER = \frac{\hbar c}{4\pi\as}\int d^3\vec{r}\,\left(\vec{H}_1\sdot\vec{H}_2 +
    \vec{H}_2\sdot\vec{H}_3 + \vec{H}_3\sdot\vec{H}_1\right)\ ,    
\end{equation}
exchange energy~$\hEC$ of the electric type
\begin{equation}
  \label{eq:IV.30}
  \hEC = \frac{\hbar c}{4\pi\as}\int d^3\vec{r}\,\left(\vXstar\sdot\vec{X}\right)\ ,
\end{equation}
and finally the exchange energy~$\gEC$ of the magnetic type
\begin{equation}
  \label{eq:IV.31}
  \gEC = \frac{\hbar c}{4\pi\as}\int d^3\vec{r}\,\left(\vYstar\sdot\vec{Y}\right)\ .
\end{equation}

An interesting point of this \emph{RST principle of minimal energy} (\ref{eq:IV.21})
refers now to the fact that only the Lagrangean multipliers~$\lD(a)$ (\ref{eq:IV.22a}),
being linked to the wave function normalization (\ref{eq:III.49}), depend upon the special
RST solution; but the four multipliers~$\lGe,\lGm,\lGh,\lGg$ referring to the Poisson and
exchange constraints (\ref{eq:IV.1}), (\ref{eq:IV.4}), (\ref{eq:IV.11}) and
(\ref{eq:IV.14}) are universal constants, cf.~(\ref{eq:IV.22b})-(\ref{eq:IV.22c}).
Therefore it is possible to collect all four gauge field constraints into a single one,
i.e.\ we may put for the RST functional~$\tETT$ (\ref{eq:IV.20})
\begin{equation}
  \label{eq:IV.32}
  \tETT = \Eiv + \sum\limits_{a=1}^3\lD(a)\cdot\ND(a) + \lG\cdot\left(\NG-\nG\right)
\end{equation}
where for the single residual multiplier one has~$\lG=-2$. The compact gauge field
constraint is then given by the identity
\begin{equation}
  \label{eq:IV.33}
  \NG\equiv\nG\ ,
\end{equation}
with~$\NG$ working as the proper constraint condition whereas~$\nG$ has the status of a
fixed number \mbox{$(\delta\nG=0)$}
\begin{subequations}
  \begin{align}
    \label{eq:IV.34a}
    \NG&\doteqdot\NGe-\NGm-\NGh+\NGg\\*
    \label{eq:IV.34b}
    \nG &\doteqdot \nGe - \nGm - \nGh + \nGg\ .
  \end{align}
\end{subequations}

Finally it is also worth while to point out to the circumstance that the mass
eigenvalues~$M_a$ do not only emerge in the mass eigenvalue equations
(\ref{eq:III.22a})-(\ref{eq:III.22c}) but also in the gauge field equations
(\ref{eq:III.29b})-(\ref{eq:III.31b}), namely implicitly in form of the exchange
length~$\aM$ (\ref{eq:III.13}) occurring in the transformed potential~$\AM$
(\ref{eq:III.12}) and also in the exchange field strength~$\vec{X}$
(\ref{eq:III.25a}). This entails certain complications when one wishes to exploit the
\emph{principle of minimal energy} (\ref{eq:IV.21}) for developing variational
techniques in order to obtain approximate solutions for the RST eigenvalue problem
(consisting of the matter equations (\ref{eq:III.22a})-(\ref{eq:III.22c}) \emph{plus} the
gauge field equations (\ref{eq:III.29a})-(\ref{eq:III.31b})). It is true, the exchange
length~$\aM$ is to be kept fixed during the variational procedure (\ref{eq:IV.21}), but
for solving the extremal equations due to the functional~$\tETT$ one has to express $\aM$
in terms of the \emph{mass functionals}, i.e.\ one has to render more precise the former
definition of~$\aM$ (\ref{eq:III.13}) to
\begin{equation}
  \label{eq:IV.35}
  \frac{1}{\aM} = \frac{\left(\MM{2}-\MM{3}\right)c}{\hbar}
\end{equation}
(for a discussion of the mass functionals~$\MM{a}$ see ref.~\cite{MaSo2}). Clearly by this
arrangement (\ref{eq:IV.35}) the RST eigenvalue problem is revealed to be essentially an
integro-differential system. However this complication is not present for systems of
\emph{different} particles where the exchange subsystem becomes trivial (i.e.~$B_\mu\equiv
0$), see below for the treatment of positronium. See also Appendix~A for the manifestly
gauge-invariant form of the energy functional~$\tETT$.

\begin{center}
  \emph{\textbf{4.\ Non-Relativistic Variational Principle}}
\end{center}

Surely, it is very satisfying to have a rigorously relativistic energy functional because
this admits to predict the frequencies of spectral lines as precisely as the observational
data do require it in order to test the quality of the theory. However for the practical
applications it will mostly not be possible to solve exactly the relativistic eigenvalue
problem so that one is forced to resort to approximative solutions. But when the solutions
are not exact, it makes no sense to insert them into the exact (relativistic) energy
functional; but rather it is sufficient (and simpler) to use a similar approximation of
the energy functional, i.e.\ we have to look for its non-relativistic limit. In this sense
we must now deduce (from the relativistic predecessors) the non-relativistic
approximations of \emph{both} the corresponding RST field equations \emph{and} the energy
functional. Logically, both concepts must fit together in the sense that it does not
matter whether one first establishes the non-relativistic functional~($\tbbETT)$, say) and
then looks for the non-relativistic field equations as the extremal equations
of~$\tbbETT$; or whether one deduces the non-relativistic field equations directly from
their relativistic predecessors (\ref{eq:III.22a})-(\ref{eq:III.22c}), i.e.\ the following
approximation scheme must be commutative:
\label{commutative}

\bigskip
\begin{picture}(160,200)(0,0)
\setlength{\unitlength}{1mm}
\put(0,50){\framebox(50,18){$\displaystyle{\mbox{Relativistic} \atop \mbox{Field Equations}}$}}
\put(0.5,50.5){\framebox(49,17){}}
\put(85,50){\framebox(50,18){$\displaystyle{\mbox{Non-Relativistic} \atop \mbox{Field Equations}}$}}
\put(85.5,50.5){\framebox(49,17){}}
\put(0,0){\framebox(50,18){$\displaystyle{\mbox{Relativistic Principle} \atop \mbox{of Minimal Energy}}$}}
\put(0.5,0.5){\framebox(49,17){}}
\put(85,0){\framebox(50,18){$\displaystyle{\mbox{Non-Relativistic Principle} \atop \mbox{of Minimal Energy}}$}}
\put(85.5,0.5){\framebox(49,17){}}
\put(25,18){\vector(0,1){32}}
\put(110,18){\vector(0,1){32}}
\put(50,9){\vector(1,0){35}}
\put(50,59){\vector(1,0){35}}
\put(50,49.8){\makebox(35,18){${\mbox{\footnotesize non-relativistic} \atop \mbox{\footnotesize approximation}}$}}
\put(50,-0.2){\makebox(35,18){${\mbox{\footnotesize non-relativistic} \atop \mbox{\footnotesize approximation}}$}}
\put(16.2,13.7){\rotatebox{90}{\makebox(40,18){${\mbox{\footnotesize variational} \atop \mbox{\footnotesize procedure}}$}}}
\put(101.2,13.7){\rotatebox{90}{\makebox(40,18){${\mbox{\footnotesize variational} \atop \mbox{\footnotesize procedure}}$}}}
\end{picture}
\bigskip

The point here is that we have to establish the non-relativistic approximations in
such a way that the magnetic gauge covariance (or invariance, resp.) is properly
respected; this is not the case in the preceding paper~\cite{MaSo2} where the process of
linearization did spoil the strict magnetic gauge invariance/covariance.

\begin{center}
\emph{\large Non-Relativistic Eigenvalue Equations}
\end{center}

First let us deduce the non-relativistic energy eigenvalue equations directly from their
relativistic predecessors (\ref{eq:III.22a})-(\ref{eq:III.22c}). The method of
non-relativistic approximation consists here in resolving approximately the equations
for the ``positive'' Pauli spinors~$\app$ with respect to the ``negative'' Pauli
spinors~$\apm$ and substituting this back into the equation for the ``negative'' Pauli
spinors. Thus in the first step, the ``negative'' Pauli spinors do appear in the following
form~\cite{MaSo2}:
\begin{subequations}
  \begin{align}
    \label{eq:IV.36a}
    \ipm &\backsimeq \hR{I}\ipp\\*
    \label{eq:IV.36b}
    \iipm &\backsimeq \hR{II}\iipp + \hr{II}\iiipp\\*
    \label{eq:IV.36c}
    \iiipm &\backsimeq \hR{III}\iiipp + \hr{III}\iipp\ .
  \end{align}
\end{subequations}
This demonstrates again the exchange coupling of both identical particles~$(a=2,3)$ as
well as the corresponding decoupling of the first (non-identical) particle~$(a=1)$. The
resolvent operators, mediating the approximative transition to the non-relativistic regime,
are given by
\begin{subequations}
  \begin{align}
    \label{eq:IV.37a}
    \hR{I} &= \frac{i\hbar}{2\Mp c}\,\vec{\sigma}\sdot\left(\vec{\nabla}+i\vAI \right)\\*
    \label{eq:IV.37b}
    \hR{II} &= -\frac{i\hbar}{2\Me c}\,\vec{\sigma}\sdot\left(\vec{\nabla}+i\vAII\right)
    \quad , \quad\hr{II}=\frac{\hbar}{2\Me c}\,\vec{\sigma}\sdot\vec{B}\\*
    \label{eq:IV.37c}
    \hR{III} &= -\frac{i\hbar}{2\Me c}\,\vec{\sigma}\sdot\left(\vec{\nabla}+i\vAIII\right)
    \quad , \quad \hr{III}=\frac{\hbar}{2\Me c}\,\vec{\sigma}\sdot\vBstar\ .
  \end{align}
\end{subequations}

But now the crucial point is here that those non-relativistic approximations
(\ref{eq:IV.36a})-(\ref{eq:IV.36c}) must be covariant under the group of magnetic gauge
transformations! These transformations are defined for the magnetic
potentials~$\vAI,\vAII,\vAIII$ through the equations (\ref{eq:III.20a})-(\ref{eq:III.20c}),
furthermore for the exchange potential~$\{B_\mu\}=\{B_0,-\vec{B}\}$ through equation
(\ref{eq:III.9}), and finally for the Pauli spinors through
\begin{equation}
  \label{eq:IV.38}
  \sappm = e^{-i\alpha_a}\cdot\appm\ .
\end{equation}
Clearly, the latter transformation law is entailed on the two-component Pauli
spinors $\appm$ by the corresponding law (\ref{eq:III.6}) for the four-component Dirac
spinors~$\psi_a(\vec{r})$, namely just via the direct sum construction
(\ref{eq:III.21}). But the point with this group of magnetic gauge transformations is now
that those approximative relations (\ref{eq:IV.36a})-(\ref{eq:IV.36c}) must transform
covariantly under the magnetic gauge transformations. Indeed, it is easy to see that this
magnetic covariance requirement transcribes to the resolvent operators in the following
form:
\begin{subequations}
  \begin{align}
    \label{eq:IV.39a}
    e^{i\alpha_1}\cdot\shR{I}\cdot e^{-i\alpha_1} &= \hR{I}\\*
    \label{eq:IV.39b}
    e^{i\alpha_2}\cdot\shR{II}\cdot e^{-i\alpha_2} &= \hR{II}\quad,\quad
    e^{i\alpha_2}\cdot\shr{II}\cdot e^{-i\alpha_3}=\hr{II}\\* 
    \label{eq:IV.39c}
    e^{i\alpha_3}\cdot\shR{III}\cdot e^{-i\alpha_3} &= \hR{III}\quad,\quad
    e^{i\alpha_3}\cdot\shr{III}\cdot e^{-i\alpha_2}=
    \hr{III}\ .
  \end{align}
\end{subequations}
But this required transformation behavior of the resolvent operators can easily be
verified by simply evoking the corresponding transformation laws for the three-vector
potentials~$\vec{B},\vAI,\vAII,\vAIII$ as given by equations (\ref{eq:III.9}) and
(\ref{eq:III.20a})-(\ref{eq:III.20c}). Thus, the magnetic gauge covariance is
preserved in the first step.

The second step consists now in substituting those appropriate ``negative'' spinors $\apm$
(\ref{eq:IV.36a})-(\ref{eq:IV.36c}) back into the eigenvalue equations
(\ref{eq:III.22a})-(\ref{eq:III.22c}) for just the negative spinors.  This procedure then
leads us to the following Pauli-like equations for the ``positive'' spinors~$\app,\
a=1,2,3$
\begin{subequations}
  \begin{align}
    \label{eq:IV.40a}
    \hH{I}\ipp &= \EP{1}\cdot\ipp\\*
    \label{eq:IV.40b}
    \hH{II}\iipp + \hh{II}\iiipp &= \EP{2}\cdot\iipp\\*
    \label{eq:IV.40c}
    \hH{III}\iiipp + \hh{III}\iipp &= \EP{3}\cdot\iiipp\ .
  \end{align}
\end{subequations}
Here, the Pauli energy eigenvalues~$\EP{a}$ are non-relativistic approximations of the
former mass eigenvalues~$M_a$, i.e.
\begin{subequations}
  \begin{align}
    \label{eq:IV.41a}
    \EP{1} &\backsimeq -\left(\Mp+M_1\right)c^2\\*
    \label{eq:IV.41b}
    \EP{2} &\backsimeq \left(M_2-\Me\right)c^2\\*
    \label{eq:IV.41c}
    \EP{3} &\backsimeq \left(M_3-\Me\right)c^2\ ,
  \end{align}
\end{subequations}
furthermore the \emph{Pauli Hamiltonians}~$(\hat{\H})$ are given by
\begin{subequations}
  \begin{align}
    \label{eq:IV.42a}
    \hH{I} &=-\frac{\hbar^2}{2\Mp}\left[\left(\vec{\nabla}+i\vAI\right)^2 -
      \vec{\sigma}\sdot\vec{H}_{\mathrm{I}} \right] + \hbar c\; \IA_0\\*
    \label{eq:IV.42b}
    \hH{II} &=-\frac{\hbar^2}{2\Me}\left[\left(\vec{\nabla}+i\vAII\right)^2 -
      \vec{\sigma}\sdot\vec{H}_{\mathrm{II}} \right] - \hbar c\; \IIA_0\\*
    \label{eq:IV.42c}
    \hH{III} &=-\frac{\hbar^2}{2\Me}\left[\left(\vec{\nabla}+i\vAIII\right)^2 -
      \vec{\sigma}\sdot\vec{H}_{\mathrm{III}} \right] - \hbar c\; \IIIA_0\ ,
  \end{align}
\end{subequations}
and finally the \emph{exchange Hamiltonians}~$(\hat{\mathfrak{h}})$ are found to be of the
following form:
\begin{subequations}
  \begin{align}
    \label{eq:IV.43a}
    \hh{II} &= -\hbar c B_0 + \frac{\hbar^2}{2\Me}\left[\vec{\sigma}\sdot\vec{Y}
    -i\left(\vns\sdot\vec{B}\right)-2i\vec{B}\sdot\left(\vec{\nabla}+i\vAIII \right)\right]\\*
    \label{eq:IV.43b}
    \hh{III} &= -\hbar c \Bstar_0 + \frac{\hbar^2}{2\Me}\left[\vec{\sigma}\sdot\vYstar
    -i\left(\vns\sdot\vBstar\right)-2i\vBstar\sdot\left(\vec{\nabla}+i\vAII
    \right)\right]\ .
  \end{align}
\end{subequations}

Observe here that the present result (\ref{eq:IV.40a})-(\ref{eq:IV.40c}) for the
non-relativistic eigenvalue equations is the strictly gauge-covariant version of the
former results (III.74) and (III.76a)-(III.76b) of ref.~\cite{MaSo2}. Indeed, the present
Pauli Hamiltonians~$\hH{I,II,III}$ (\ref{eq:IV.42a})-(\ref{eq:IV.42c}) are exactly the
same as in the preceding paper~\cite{MaSo2}, see equations (III.75) and (III.77) of that
paper. The reason for this is that the former Pauli Hamiltonians of ref.~\cite{MaSo2} are
already invariant with respect to the magnetic gauge transformations:
\begin{subequations}
  \begin{align}
    \label{eq:IV.44a}
    e^{i\alpha_1}\,\shH{I} e^{-i\alpha_1} &= \hH{I}\\*
    \label{eq:IV.44b}
    e^{i\alpha_2}\,\shH{II} e^{-i\alpha_2} &= \hH{II}\\*
    \label{eq:IV.44c}
    e^{i\alpha_3}\,\shH{I} e^{-i\alpha_3} &= \hH{III}\ .
  \end{align}
\end{subequations}
Therefore the progress made here refers only to the exchange Hamiltonians~$\hh{II}$
and~$\hh{III}$ (\ref{eq:IV.43a})-(\ref{eq:IV.43b}). These objects appear now manifestly
gauge-invariant because they contain the gauge-covariant gradient operator~$\vns$
(\ref{eq:III.26}) rather than its counterpart~$\vec{\nabla}$ of ref.~\cite{MaSo2}. Thus
the magnetic covariance of the exchange Hamiltonians emerges in the following form:
\begin{subequations}
  \begin{align}
    \label{eq:IV.45a}
    e^{i\alpha_2}\,\shh{II} e^{-i\alpha_3} &= \hh{II}\\*
    \label{eq:IV.45b}
    e^{i\alpha_3}\,\shh{III} e^{-i\alpha_2} &= \hh{III}\ .    
  \end{align}
\end{subequations}

Summarizing, it has been demonstrated by explicit construction that it is possible to
deduce the non-relativistic eigenvalue equations from their relativistic originals under
the desired preservation of the magnetic gauge covariance. Now it should be a matter of
course that one wishes to see also the corresponding \emph{gauge-invariant} energy
functional~($\tbbETT$, say) whose extremal equations would just coincide with the present
non-relativistic eigenvalue equations (\ref{eq:IV.40a})-(\ref{eq:IV.40c}). Clearly, a
nearby guess suggests here to deduce this invariant functional~$\tbbETT$ from its
relativistic original~$\tETT$ (\ref{eq:IV.20}), again under preservation of the magnetic
gauge invariance. Moreover this non-relativistic functional~$\tbbETT$ is required to
produce also, through its extremalization with respect to the gauge fields, just the
non-relativistic version of the original gauge field equations
(\ref{eq:III.29a})-(\ref{eq:III.31b}).

\begin{center}
  \emph{\large Non-Relativistic Gauge Field Equations}
\end{center}

However, resorting to such a non-relativistic limit of the gauge field equations would
imply that one neglects (at least in lowest-order approximation) the fields of
\emph{magnetic} type altogether! The reason is that the magnetic fields couple to the
particle velocities which in the non-relativistic domain are negligibly small. On the
other hand, the present gauge invariance refers exclusively to the magnetic gauge
transformations; and therefore the neglection of all the magnetic-type fields would
eliminate also the magnetic gauge structure. Therefore in order to preserve this gauge
structure for the non-relativistic limit of the gauge field subsystem, one should neglect
only certain products of the magnetic-type fields \emph{provided this neglection does not
  spoil the magnetic gauge structure}. However such products of magnetic fields, being
well-suited for omission, do occur at almost all places of the gauge field subsystem, for
instance in connection with the equations (\ref{eq:III.23b})-(\ref{eq:III.23c}),
(\ref{eq:III.24b})-(\ref{eq:III.24c}) or (\ref{eq:III.31b}). Therefore, in order to not
pervert the essentially magnetic character of the gauge field configurations, we keep here
all the gauge field equations formally unchanged and do insert therein only the
non-relativistic limits for the densities~${}^{(a)}\!j_0,\vec{j}_a,h_0,\vec{h}$ (see
ref.~\cite{MaSo2}). Clearly, such a procedure implies that all the Poisson identities and
exchange identities do formally apply also for the present non-relativistic limit with
merely the constraints~$N,n$ being replaced by their non-relativistic versions~$\NN,\tilde{n}$.

Thus the last constraint which must be be considered here refers to the wave function
normalization (\ref{eq:III.49}). The non-relativistic version hereof consists simply in
using the non-relativistic approximation of the Dirac densities~$\akn$. The relativistic
density reads~\cite{MaSo2}
\begin{equation}
\label{eq:IV.46}
\akn = \apd\,\app + \amd\,\apm\ ,
\end{equation}
and since the non-relativistic limit consists in disregarding the ``negative'' Pauli
spinors $\apm$ one concludes for the non-relativistic approximation (~$\NND$, say)
\begin{equation}
  \label{eq:IV.47}
  \ND\Rightarrow\NND = \int d^3\vec{r}\;\;\apd\,\app -1\equiv 0\ .
\end{equation}
\smallskip
\begin{center}
  \emph{\large Non-Relativistic Energy Functional}
\end{center}

But now that the non-relativistic (however gauge-invariant) forms of all the constraints
have been revealed, one can easily write down the non-relativistic approximation
$\tbbETT$ of the original relativistic energy functional~$\tETT$ (\ref{eq:IV.20}):
\begin{equation}
  \begin{split}
  \label{eq:IV.48}
  \tbbETT &=
  \bbEIV+\sum\limits_{a=1}^3\lpa\,\cdot\,\NNDa + \lGe\,\cdot\,(\NNGe-\snGe) + \lGh\,\cdot\,(\NNGh-\snGh)\\*
  &+\lGg\,\cdot\,(\NNGg-\snGg)
  \equiv \bbEIV+\sum\limits_{a=1}^3\lpa\cdot\NNDa+\lG\cdot\left(\NNG-\snG\right)\ .
  \end{split}
\end{equation}
Here the three non-relativistic gauge field constraints with Lagrangean multipliers~$\lGe,\lGh,\lGg$ may be
collected again into the single constraint~$\NNG$ with Lagrangean multiplier~$\lG=-2$
\begin{subequations}
  \begin{align}
    \label{eq:IV.49a}
    \NNG\doteqdot\NNGe-\NNGh + \NNGg\\*
    \label{eq:IV.49b}
    \snG\doteqdot\snGe-\snGh+\snGg\ ,
  \end{align}
\end{subequations}
cf.\ the corresponding relativistic arrangement
(\ref{eq:IV.34a})-(\ref{eq:IV.34b}). Clearly, the principal structure of this
nonrelativistic functional (\ref{eq:IV.48}) is the same as for the relativistic
original~$\tETT$, namely the sum of a physical part (i.e.~$\bbEIV$) and of the
constraints. The physical part is of course the collection of all those physical
contributions which are already present also in the original relativistic version~$\Eiv$
(\ref{eq:IV.23}), however now in their non-relativistic approximation, i.e.
\begin{equation}
  \label{eq:IV.50}
  \bbEIV = \Ekin+\eER+\mER-\hEC+\gEC + \hbar c\int d^3\vec{r}\,\left(
   \vH{I}\sdot\vSm{I}+\vH{II}\sdot\vSm{II}+\vH{III}\sdot\vSm{III} \right)\ .
\end{equation}

It is true, this is the same form as in the preceding paper, see equation (\ref{eq:IV.21})
of ref.~\cite{MaSo2}; however, \emph{the invariance of any individual term with respect to the
  magnetic gauge transformations is now guaranteed!} Namely, the point here is that the
four gauge field contributions~$\eER,\mER,\hEC,\gEC$ are formally the same as their
relativistic originals (\ref{eq:IV.28})-(\ref{eq:IV.31}) if expressed in terms of the
corresponding field strengths~$\vec{E}_a,\vec{H}_a,\vec{X},\vec{Y}$; but observe here the
change of the magnetic contribution in the non-relativistic functional~$\tbbETT$
(\ref{eq:IV.48}) which no longer contains the magnetic constraint~$\NGm-\nGm$ of the
relativistic case (\ref{eq:IV.20}). Indeed this magnetic constraint is eliminated in
favour of the positive sign of~$\mER$ and of the emergence of the dipole interaction
energy~$(\sim \vec{H}\sdot\vec{S})$. Moreover, the gauge invariance of the magnetic dipole
energy is here manifest since the \emph{non-relativistic} Dirac dipole
densities~${}^{(\mathrm{m})}\vec{S}_a$ are defined by~\cite{MaSo2}
\begin{subequations}
  \begin{align}
    \label{eq:IV.51a}
    \vSm{1} &\doteqdot\frac{\hbar}{2\Mp c}\,\dvp{1}{+}\vec{\sigma}\,\vp{1}{+}\\*
    \label{eq:IV.51b}
    \vSm{2} &\doteqdot\frac{\hbar}{2\Me c}\,\dvp{2}{+}\vec{\sigma}\,\vp{2}{+}\\*
    \label{eq:IV.51c}
    \vSm{3} &\doteqdot\frac{\hbar}{2\Me c}\,\dvp{3}{+}\vec{\sigma}\,\vp{3}{+}\ .
  \end{align}
\end{subequations}
And finally, the total kinetic energy~$\Ekin$ in (\ref{eq:IV.50}) of course inherits  its
sum structure from its relativistic predecessor~$\Tkin$ (\ref{eq:IV.25}), i.e.
\begin{equation}
  \label{eq:IV.52}
  \Ekin = \sum\limits_{a=1}^3 \Ekin(a)\ ,
\end{equation}
with the manifestly gauge-invariant one-particle contributions
\begin{subequations}
  \begin{align}
    \label{eq:IV.53a}
    \Ekin(1) &= -\frac{\hbar^2}{2\Mp}\int d^3\vec{r}\;\;\dvp{1}{+}\left(\vec{\nabla}+i\vAI \right)^2\vp{1}{+}\\*
    \label{eq:IV.53b}
    \Ekin(2) &= -\frac{\hbar^2}{2\Me}\int d^3\vec{r}\;\;\dvp{2}{+}\left(\vec{\nabla}+i\vAII \right)^2\vp{2}{+}\\*
    \label{eq:IV.53c}
    \Ekin(3) &= -\frac{\hbar^2}{2\Me}\int d^3\vec{r}\;\;\dvp{3}{+}\left(\vec{\nabla}+i\vAIII\right)^2\vp{3}{+}\ .
  \end{align}
\end{subequations}

But now that the non-relativistic functional~$\tbbETT$ is precisely fixed in the form
(\ref{eq:IV.48}) it is a standard matter to deduce the non-relativistic RST field
equations through extremalization of that functional: The corresponding variational
procedure with respect to the Pauli spinors~$\appm$ (or~$\apmd$, resp.) yields the system
of gauge-covariant eigenvalue equations (\ref{eq:IV.40a})-(\ref{eq:IV.40c}) with the
Lagrangean multipliers
\begin{subequations}
  \begin{align}
    \label{eq:IV.54a}
    \lpa &= -\EP{a}\\*
    \label{eq:IV.54b}
    \lGe &= -\lGm = -\lGh = \lGg = -2\ ,
  \end{align}
\end{subequations}
see Appendix~B. And similarly, the extremalization of~$\tbbETT$ with respect to the gauge
fields~$\aAn,\vec{A}_a,B_0,\vec{B}$ (or their complex conjugates, resp.) lets emerge the
electromagnetic Poisson equations and their (gauge-covariant) exchange counterparts (see
Appendix~C). Thus the commutativity of the logical arrangement on p.~\pageref{commutative} is safely
validated which in turn supports the general logical consistency of RST. (The variational
deduction of the non-relativistic eigenvalue equations (\ref{eq:IV.40a})-(\ref{eq:IV.40c})
from the non-relativistic principle of minimal energy,~$\delta\tbbETT=0$, is a very instructive
exercise in dealing with the gauge-covariant structures, see Appendix~B. See also
Appendix~C for the variational deduction of the gauge field equations.)


\section{Ortho- and Para-Positronium}
\indent

As a practical demonstration for the usefulness of the constructed energy functional
$\tETT$, or its non-relativistic approximation $\tbbETT$, resp., one may consider now a
two-particle system ($N=2$) which is simple enough in order to avoid unnecessary
complications but, on the other hand, is sufficiently non-trivial in order to display the
physical correctness of the theoretical construction. Thus for the sake of demonstration,
we turn to positronium whose constituents (i.\,e. electron and positron) have identical
rest masses ($M_e=M_p\doteqdot M$) but different sign of charge so that both constituents
must count as non-identical particles. Consequently, these particles cannot feel the
exchange interactions, which entails the vanishing of all the exchange objects
($B_\mu\equiv0,\,G_{\mu\nu}\equiv0$, etc.). But this does not mean that the demonstration
becomes trivial because positronium is known to possess a sufficiently rich level
structure \cite{GR}. This is mainly due to the well-known ortho/para dichotomy which
occurs in a similar way also for muonium and hydrogen, albeit with different rest masses
\cite{CT}. Thus in the present context, positronium provides us with a test case for our
energy functional $\tETT$ (or $\tbbETT$, resp.) in a two-fold respect: {\bf (i)} one can
reveal the specific way in which the ortho/para dichotomy emerges in RST (in contrast to
the conventional theory), and {\bf (ii)} one can (at least qualitatively) discuss the
energy difference between certain ortho- and para-levels in order to compare the
corresponding RST predictions with those of the conventional theory and with the
observations.

Naturally, such a double program is to be carried through in two steps: In this section
(V.), we first study the emergence of the ortho/para dichotomy within the RST formalism
which is based upon the Whitney sum of one-particle bundles, in contrast to the
conventional formalism, which deals with tensor products of one-particle Hilbert
spaces. And in the next section (VI.) we apply the constructed RST energy functional
$\tETT$ (or $\tbbETT$, resp.) in order to discuss the origin of the energy difference
between the ortho- and para-levels. This energy difference appears to be caused not only
through different \emph{magnetic} interactions but also by the different anisotropy of the
\emph{electric} interaction potentials, which itself is due to the different pattern of
angular momentum inherent in the ortho- and para-configurations.

\vspace{4ex}
\begin{center}
  \emph{\textbf{1.\ Conventional Multiplet Structure}}
\end{center}

It is true, our model demonstration below (i.\,e. positronium) does {\em not}\/ consist of
identical fermions; but for the sake of comparison of the conventional and RST situations
it is very instructive to first consider a system of identical fermions (e.\,g.\ the two
helium electrons) from the {\em conventional}\/ viewpoint.

According to the conventional {\em spin-statistics theorem}\/, i.\,e. ``{\em Pauli
  principle}\/'' in popular terms \cite{CT}, any stationary bound system of identical
fermions (e.g.\ spin-$\frac{1}{2}$ particles) occupies a {\em skew-symmetric}\/ quantum
state ${}^{(-)}\Psi(\vec{r}_1,\vec{s}_1;\vec{r}_2,\vec{s}_2)$:
\begin{equation}
  \label{eq:V.1}
  {}^{(-)}\Psi(\vec{r}_1,\vec{s}_1;\vec{r}_2,\vec{s}_2)=-{}^{(-)}\Psi(\vec{r}_2,\vec{s}_2;\vec{r}_1,\vec{s}_1)\;.
\end{equation}
Analogously, the bosons (i.\,e. integer-spin particles) can occur only in {\em symmetric}\/ states
\begin{equation}
  \label{eq:V.2}
  {}^{(+)}\Psi(\vec{r}_1,\vec{s}_1;\vec{r}_2,\vec{s}_2)={}^{(+)}\Psi(\vec{r}_2,\vec{s}_2;\vec{r}_1,\vec{s}_1)\;.
\end{equation}
For a classification of such states one resorts to the assumption that their number is preserved when one thinks of the spin-spin interactions as being switched-off, with simultaneous neglection of the relativistic effects. Namely, the point with this hypothesis is that the wave functions \myrf{V.1} (or analogously also \myrf{V.2}) can then be factorized approximately into an orbital part $\psi(\vec{r}_1,\vec{r}_2)$ and a spin part $\chi(\vec{s}_1,\vec{s}_2)$:
\begin{equation}
  \label{eq:V.3}
  \Psi(\vec{r}_1,\vec{s}_1;\vec{r}_2,\vec{s}_2)=\psi(\vec{r}_1,\vec{r}_2)\otimes\chi(\vec{s}_1,\vec{s}_2)\;.
\end{equation}
But since the Pauli principle \myrf{V.1} demands that the total wave function must be antisymmetric under particle permutation, the factorization \myrf{V.3} admits exclusively the following combinations of symmetry properties for the orbital and spin parts
\begin{subequations}
  \begin{align}
  \label{eq:V.4a}
  \psi(\vec{r}_1,\vec{r}_2)=+\psi(\vec{r}_2,\vec{r}_1) &\Leftrightarrow\;\chi(\vec{s}_1,\vec{s}_2)=-\chi(\vec{s}_2,\vec{s}_1)\\
  \label{eq:V.4b}
  \psi(\vec{r}_1,\vec{r}_2)=-\psi(\vec{r}_2,\vec{r}_1) &\Leftrightarrow\;\chi(\vec{s}_1,\vec{s}_2)=+\chi(\vec{s}_2,\vec{s}_1)\;. 
  \end{align}
\end{subequations}

Here a closer inspection of the configuration space of a two-spin system shows that the antisymmetric spin configuration \myrf{V.4a} is actually a {\em singlet state}\/ $\chi_{[0,0]}$
\begin{equation}
  \label{eq:V.5}
  \chi_{[0,0]} =\frac{1}{\sqrt{2}}\left[\zeta_0^{\frac{1}{2},\frac{1}{2}}\otimes\zeta_0^{\frac{1}{2},-\frac{1}{2}}-\zeta_0^{\frac{1}{2},-\frac{1}{2}}\otimes\zeta_0^{\frac{1}{2},\frac{1}{2}}\right]\,.
\end{equation}
This antisymmetric decomposition of the spin part $\chi(\vec{s}_1,\vec{s}_2)$ refers to the standard spin basis $\big\{\zeta_0^{\frac{1}{2},\pm\frac{1}{2}}\big\}$ of the two-dimensional unitary space, i.\,e.
\begin{subequations}
  \begin{align}
  \label{eq:V.6a}
  \hat{\vec{s}}{\,}^2\zeta_0^{\frac{1}{2},\pm\frac{1}{2}} &=s(s+1)\left(\frac{\hbar}{2}\right)^2\zeta_0^{\frac{1}{2},\pm\frac{1}{2}}\\
  \label{eq:V.6b}
  \hat{s}_z\zeta_0^{\frac{1}{2},\pm\frac{1}{2}} &=\pm\frac{\hbar}{2}\,\zeta_0^{\frac{1}{2},\pm\frac{1}{2}}\\
  \label{eq:V.6c}
  \hat{\vec{l}}{\,}^2\zeta_0^{\frac{1}{2},\pm\frac{1}{2}} &=0\;,  
  \end{align}
\end{subequations}
where $\{\hat{\vec{s}}{\,}^2,\hat{s}_z,\hat{\vec{l}}{\,}^2\}$ are the corresponding one-particle operators for spin ($s=\frac{1}{2}$) and orbital angular momentum ($l=0,1,2,3,...)$, resp. Thus the antisymmetric singlet state $\chi_{[0,0]}$ has zero values for both the total spin $\hat{\vec{S}}$ ($=\hat{\vec{s}}_1\oplus\hat{\vec{s}}_2$) and its $z$-component $\hat{S}_z$ ($=\hat{s}_{z(1)}\oplus\hat{s}_{z(2)}$), i.\,e.
\begin{subequations}
  \begin{align}
  \label{eq:V.7a}
  \hat{\vec{S}}{}^2\chi_{[0,0]}&=\hbar^2\,S(S+1)\,\chi_{[0,0]}=0\\
  \label{eq:V.7b}
  \hat{S}_z\chi_{[0,0]}&=0\;.
  \end{align}
\end{subequations}

In contrast to this singlet structure, the symmetric spin parts \myrf{V.4b} do form a {\em triplet system}
\begin{subequations}
  \begin{align}
  \label{eq:V.8a}
  \chi_{[1,0]} &=\frac{1}{\sqrt{2}}\left[\zeta_0^{\frac{1}{2},\frac{1}{2}}\otimes\zeta_0^{\frac{1}{2},-\frac{1}{2}}+\zeta_0^{\frac{1}{2},-\frac{1}{2}}\otimes\zeta_0^{\frac{1}{2},\frac{1}{2}}\right]\\
  \label{eq:V.8b}
  \chi_{[1,1]} &=\zeta_0^{\frac{1}{2},\frac{1}{2}}\otimes\zeta_0^{\frac{1}{2},\frac{1}{2}}\\
  \label{eq:V.8c}
  \chi_{[1,-1]} &=\zeta_0^{\frac{1}{2},-\frac{1}{2}}\otimes\zeta_0^{\frac{1}{2},-\frac{1}{2}}
  \end{align}
\end{subequations}
such that
\begin{subequations}
  \begin{align}
  \label{eq:V.9a}
  \hat{\vec{S}}{}^2\chi_{[1,0]}&=\hbar^2\,S(S+1)\,\chi_{[1,0]}=2\hbar^2\,\chi_{[1,0]}\\
  \label{eq:V.9b}
  \hat{S}_z\chi_{[1,0]}&=0
  \end{align}
\end{subequations}
and
\begin{subequations}
  \begin{align}
  \label{eq:V.10a}
  \hat{\vec{S}}{}^2\chi_{[1,\pm1]}&=\hbar^2\,S(S+1)\,\chi_{[1,\pm1]}=2\hbar^2\,\chi_{[1,\pm1]}\\
  \label{eq:V.10b}
  \hat{S}_z\chi_{[1,\pm1]}&=\pm\hbar\,\chi_{[1,\pm1]}\;.
  \end{align}
\end{subequations}
Observe here that the singlet state $\chi_{[0,0]}$ \myrf{V.5} and the triplet state $\chi_{[1,0]}$ \myrf{V.8a} are degenerate with respect to the total spin component $\hat{S}_z$, cf. \myrf{V.7b} and \myrf{V.9b}, but this does not imply that the singlet state $\Psi_{\mathcal{P}}$ with symmetric orbital and antisymmetric spin part
\begin{equation}
  \begin{split}
  \label{eq:V.11}
  \Psi_{\mathcal{P}}\doteqdot\psi_+(\vec{r}_1,\vec{r}_2)\otimes\chi_{[0,0]}\\
  (\psi_+(\vec{r}_1,\vec{r}_2)=\psi_+(\vec{r}_2,\vec{r}_1))
  \end{split}
\end{equation}
has the same energy eigenvalue as the corresponding triplet state $\Psi_{\mathcal{O}}$. The reason is here that the orbital part $\psi_+(\vec{r}_1,\vec{r}_2)$ of the {\em para-state}\/ $\Psi_{\mathcal{P}}$ \myrf{V.11} is symmetric, whereas the orbital part $\psi_-(\vec{r}_1,\vec{r}_2)$ of $\Psi_{\mathcal{O}}$ as a member of the ortho-system is antisymmetric:
\begin{equation}
  \begin{split}
  \label{eq:V.12}
  \Psi_{\mathcal{O}}\doteqdot\psi_-(\vec{r}_1,\vec{r}_2)\otimes\chi_{[1,0]}\\
  (\psi_-(\vec{r}_1,\vec{r}_2)=-\psi_-(\vec{r}_2,\vec{r}_1))\;.
  \end{split}
\end{equation}
This symmetry arrangement implies namely that for the para-state $\Psi_{\mathcal{P}}$
\myrf{V.11} the two identical fermions tend to approach one another, whereas for the
ortho-state $\Psi_{\mathcal{O}}$ \myrf{V.12} they tend to retreat from one
another. Therefore one expects, since the electrostatic interaction of both identical
fermions is repulsive, that the para-state (i.\,e. the singlet $\Psi_{\mathcal{P}}$
\myrf{V.11}) has higher energy than the ortho-states (such as $\Psi_{\mathcal{O}}$
\myrf{V.12}) which themselves should constitute a {\em degenerate}\/ triplet. Possibly the
latter (electrostatic) degeneration will be eliminated as soon as one takes into account
also the magnetic interactions and the relativistic corrections. However, the typical
energy scale, being associated with these latter types of interactions ($\sim\as^2$), is
distinctly smaller than that of the typical electrostatic interactions ($\sim$ some few
electron volts \cite{CT}); and therefore the energy difference between triplet states
($\dSe$) and singlet states ($\eSo$) in an atom is mainly of {\em electrostatic}\/ origin!
For instance, the energy difference between the singlet state $\sf 1s\,2s\,\eSo$ and the
triplet state $\sf 1s\,2s\,\dSe$ in a helium atom amounts to (roughly) $0,8\,{\rm eV}$
\cite{CT}. Nevertheless, the helium groundstate $\sf 1s^2\,\eSo$ is of the para-type
$\Psi_{\mathcal{P}}$ \myrf{V.11} since it is only for this configuration that both
electrons can have the lowest possible principal quantum number ($n=1$).

However, this conventional picture of the multiplet structure and the associated energy eigenvalues does not apply to positronium, since the latter sytem does of course not consist of identical fermions. This implies that the Pauli principle is not valid here; and consequently that combination \myrf{V.4a}-\myrf{V.4b} of permutation symmetries, concerning the orbital and spin parts of the positronium wave function, does not apply here. Nevertheless, there also arises a conventional ortho/para-dichotomy for positronium, but with the associated energy difference between the singlet ($\eSo$) and triplet system ($\dSe$) being caused now by the {\em magnetic}\/ interactions! More concretely, the non-relativistic positronium groundstate will be of the form
\begin{equation}
  \label{eq:V.13}
  \Psi(\vec{r}_1,\vec{s}_1;\vec{r}_2,\vec{s}_2)=\tilde{\psi}(\vec{r}_1,\vec{r}_2)\otimes\left\{\begin{array}{l} {}^{\ds\chi_{[0,0]}}\\\hline\ds\chi_{[1,1]} \\ \ds\chi_{[1,0]} \\ \ds\chi_{[1,-1]}\end{array}\right.
\end{equation}
and thus is four-fold degenerate in this lowest order of approximation, since there is no
spin-induced symmetry restriction for the orbital factor
$\tilde{\psi}(\vec{r}_1,\vec{r}_2)$. But this four-fold degeneracy of the positronium
groundstate becomes eliminated in reality, namely mainly as a consequence of the spin-spin
interactions ($\sim\vec{s}_1\sdot\vec{s}_2$) of both non-identical fermions, see
ref. \cite{CT}. (For the classification of the positronium states see ref. \cite{GR}.)
However, the energy difference of the ortho-positronium state $\sf 1\,\dSe$ and the
para-positronium state $\sf 1\,\eSo$ amounts to only $8\cdot10^{-4}\,[{\rm eV}]$, see
ref. \cite{CT}; and this is thousand times smaller than the ortho/para splitting of the
electronic helium states quoted above.

Summarizing, the conventional theory seems able to predict the ortho/para splittings of
both the electric and magnetic type in excellent agreement with the observations; and thus
the question arises now: in which way can there emerge a comparable multiplet structure in
RST and what is the precise physical nature of the corresponding ortho/para energy
splitting?

\vspace{4ex}
\begin{center}
  \emph{\textbf{2.\ Ortho/Para Dichotomy in RST}}
\end{center}

Since the conventional theory and RST are based upon very different mathematical
structures (i.\,e. tensor products vs. Whitney sums) it should not come as a surprise that
the RST multiplets do arise in a totally other way. Recall here that in a conventional
entangled $N$-particle state $\Psi(\vec{r}_1,\vec{r}_2,...,\vec{r}_N)$ the individual
particles cannot be linked to single-particle states but rather do participate in that
common $N$-particle state $\Psi$ in a completely equivalent way. In contrast to this
democratic participation, any individual particle of an $N$-particle state $\Psi(\vec{r})$
does occupy in RST a well-defined one-particle state $\psi_a(\vec{r})$ ($a=1,...,N$) in
agreement with the Whitney sum construction
\begin{equation}
  \label{eq:V.14}
\Psi(\vec{r})=\psi_1(\vec{r})\oplus\psi_2(\vec{r})\oplus...\oplus\psi_N(\vec{r})\;.
\end{equation}

Anyone of these individual RST particles emits a unique electromagnetic potential ${A^a}_\mu$ ($a=1,...,N$) to be determined from the Maxwell equations, e.\,g. \myrf{II.12a}-\myrf{II.12c} for the case of a three-particle system. Additionally, any individual pair of {\em identical}\/ particles does cooperatively admit also an exchange potential ($B_\mu$) which is to be determined from the exchange part of the Maxwell equations, e.\,g. \myrf{II.12d} for the considered three-particle system. Subsequently we will inspect the positronium system ($N=2$) which is built up by two non-identical particles ($a=1,2$) so that the exchange potential is zero ($B_\mu\equiv0$) and the Maxwell equations \myrf{II.12a}-\myrf{II.12c} become simplified to the following form ($a=1,2$)
\begin{subequations}
  \begin{align}
    \label{eq:II.15a}
    {F^a}_{\mu\nu}\quad&=\nabla_\mu{A^a}_\nu-\nabla_\nu{A^a}_\mu \\*
    \label{eq:II.15b}
    \nabla^\mu {F^a}_{\mu\nu} &=4\pi\as\,{j^a}_\nu\;.
  \end{align}
\end{subequations}
For the stationary bound systems, the latter equation becomes simplified to the ordinary Poisson equation for the electric potentials $\aAr(\vec{r})$
\begin{equation}
  \label{eq:V.16}
  \Delta\aAr(\vec{r})=-4\pi\as\,\ajn\;,
\end{equation}
and analogously also for the magnetic three-vector potentials $\vec{A}_a(\vec{r})$
\begin{equation}
  \label{eq:V.17}
  \Delta\vec{A}_a(\vec{r})=-4\pi\as\,\vec{j}_a(\vec{r})\;.
\end{equation}
Clearly this is a special case of the general situation described by the equations \myrf{B.3}-\myrf{B.4b} and \myrf{B.7}-\myrf{B.8b} of the appendix.

Concerning now the corresponding matter equations, it seems reasonable to adopt the hypothesis that both particles do occupy physically equivalent one-particle quantum states since their rest masses are identical ($\leadsto$ putting $M_e=M_p\doteqdot M$) and their charges differ merely by sign. But obviously these equivalent states can occur in two forms; namely either the magnetic fields are parallel
\begin{subequations}
  \begin{align}
  \label{eq:V.18a}
  \mbox{\textbf{\emph{ortho-positronium:}}}\qquad\quad \vec{A}_1(\vec{r})&\equiv\vec{A}_2(\vec{r})\doteqdot\vec{A}_b(\vec{r})\\
  \label{eq:V.18b}
   \vec{H}_1(\vec{r})&\equiv\vec{H}_2(\vec{r})\doteqdot\vec{H}_b(\vec{r})\\
  \label{eq:V.18c}
  \vec{k}_1(\vec{r})&\equiv-\vec{k}_2(\vec{r})\doteqdot\vec{k}_b(\vec{r})\\
  \label{eq:V.18d}
  \vec{j}_1(\vec{r})&\equiv\vec{j}_2(\vec{r})\doteqdot\vec{j}_b(\vec{r})\equiv\vec{k}_b(\vec{r})\;;
  \end{align}
\end{subequations}
or they are anti-parallel
\begin{subequations}
  \begin{align}
  \label{eq:V.19a}
  \mbox{\textbf{\emph{para-positronium:}}}\qquad\quad \vec{A}_1(\vec{r})&\equiv-\vec{A}_2(\vec{r})\doteqdot\vec{A}_p(\vec{r})\\
  \label{eq:V.19b}
  \vec{H}_1(\vec{r})&\equiv-\vec{H}_2(\vec{r})\doteqdot\vec{H}_p(\vec{r})\\
  \label{eq:V.19c}
  \vec{k}_1(\vec{r})&\equiv\vec{k}_2(\vec{r})\doteqdot\vec{k}_p(\vec{r})\\
  \label{eq:V.19d}
  \vec{j}_1(\vec{r})&\equiv-\vec{j}_2(\vec{r})\doteqdot\vec{j}_p(\vec{r})\;.
  \end{align}
\end{subequations}
But in contrast to the magnetic objects, the corresponding electric fields must always differ in sign because both particles carry opposite charges:
\begin{subequations}
  \begin{align}
  \label{eq:V.20a}
  \iAn(\vec{r})&\equiv-\iiAn(\vec{r})\doteqdot\bpAo(\vec{r})\\
  \label{eq:V.20b}
  \vec{E}_1(\vec{r})&\equiv-\vec{E}_2(\vec{r})\doteqdot\vec{E}_{b/p}(\vec{r})=-\nabla\bpAo(\vec{r})\\
  \label{eq:V.20c}
  \ikn&\equiv\iikn\doteqdot\bpko(\vec{r})\\
  \label{eq:V.20d}
  \ijn(\vec{r})&\equiv-\iijn(\vec{r})\doteqdot\bpjo(\vec{r})\;.
  \end{align}
\end{subequations}
Therefore the electric Poisson equation \myrf{V.16} for the common electric potential $\bpA_0$ reads for both positronium configurations
\begin{equation}
  \label{eq:V.21}
  \Delta\bpA_0(\vec{r})=-4\pi\as\,\bpk_0(\vec{r})\;,
\end{equation}
and analogously the magnetic Poisson equation \myrf{V.17} becomes for the two different configurations
\begin{equation}
  \label{eq:V.22}
  \Delta\vec{A}_{b/p}=-4\pi\as\,\vec{k}_{b/p}(\vec{r})\;.
\end{equation}

But here the sources $\{\bpk_0(\vec{r});\vec{k}_{b/p}(\vec{r})\}$ of the electromagnetic
potentials \\ $\{\bpA_0(\vec{r}); \vec{A}_{b/p}(\vec{r})\}$ must now be built up by the matter
fields $\appm$ \myrf{III.21}. First, consider the Dirac density
$\akn=\bar{\psi}_a\gamma_0\psi_a$ \myrf{II.16}, which reads in terms of the Pauli spinors
$\appm$ as it is shown by equation \myrf{IV.46}. Consequently, the common form of the
charge density $\bpk_0(\vec{r})$ \myrf{V.20c} yields the following {\em common}\/
requirement for the Pauli spinors:
\begin{equation}
  \label{eq:V.23}
  \dvp{1}{+}(\vec{r})\vp{1}{+}(\vec{r})+\dvp{1}{-}(\vec{r})\vp{1}{-}(\vec{r}) \stackrel{!}{=}\dvp{2}{+}(\vec{r})\vp{2}{+}(\vec{r})+\dvp{2}{-}(\vec{r})\vp{2}{-}(\vec{r})\;.
\end{equation}
However, since the analogous requirements \myrf{V.18c} and \myrf{V.19c} for the Dirac {\em
  currents}\/ differ (or not) in sign, one is led to two different requirements for the
ortho- and para-case:
\begin{equation}
  \label{eq:V.24}
  \big\{\dvp{1}{+}\,\vec{\sigma}\,\vp{1}{-}+\dvp{1}{-}\,\vec{\sigma}\,\vp{1}{+}\big\} \stackrel{!}{=}\mp\big\{\dvp{2}{+}\,\vec{\sigma}\,\vp{2}{-}+\dvp{2}{-}\,\vec{\sigma}\,\vp{2}{+}\big\}
\end{equation}
where the upper/lower sign refers to the ortho/para-case, resp. The solutions for both requirements \myrf{V.23}-\myrf{V.24} are \cite{BeSo,BeSo2}\\
\begin{subequations}
  \begin{align}
  \label{eq:V.25a}
  \mbox{\bf ortho-positronium:\!}&\qquad\quad\begin{array}{r@{\;}l}\ds\ipp&\equiv\,-i\iipp\doteqdot\bpp\\
  \ipm&\equiv\,i\iipm\doteqdot\bpm\end{array}\\\nonumber\\
  \label{eq:V.25b}
  \mbox{\bf para-positronium:\;}&\qquad\quad\begin{array}{r@{\;}l}\ds\ipp&\equiv\, i\big(\hat{\vec{k}}\sdot\vec{\sigma}\big)\iipp\doteqdot\ppp\\
  \ipm&\equiv\,i\big(\hat{\vec{k}}\sdot\vec{\sigma}\big)\iipm\doteqdot\ppm\end{array}\\
  \nonumber
  &(\hat{\vec{k}}\doteqdot\frac{\vec{k}_p}{\|\vec{k}_p\|}\Rightarrow\hat{\vec{k}}\sdot\hat{\vec{k}}=1)\;.
  \end{align}
\end{subequations}

The most striking difference between the ortho- and para-states will surely refer to the energy $E_T$ being carried by those states. For a first tentative estimate of this energy difference of both types of states one may tend to the hypothesis that the magnetic forces are much smaller than their electric counterparts so that they can be treated as small perturbations. Thus one could perhaps first solve the RST eigenvalue problem under the exclusive action of the electric forces alone ({\em electrostatic approximation}\/) which would yield from equations \myrf{IV.20} and \myrf{IV.23} the corresponding total energy $\teETT$ of the positronium system as
\begin{equation}
  \label{eq:V.26}
  \tETT\Rightarrow\teETT=\eETiv=\big(\Z_{(1)}^2 + \Z_{(2)}^2\big)\cdot Mc^2 + 2\Tkin + \eER\;,
\end{equation}
provided all the constraints are validated. Observe here that (for different particles) all constraint terms referring to the exchange subsystem must vanish automatically because of $B_\mu\equiv0$:
\begin{equation}
  \label{eq:V.27}
  \NGh=\NGg=\nGe=\nGm=\nGh=\nGg\equiv0\;,
\end{equation}
while $\NGm$ \myrf{IV.5} is left unchanged and for $\NGe$ \myrf{IV.2} the last term is
zero. Furthermore, the rest masses $M_p,\,M_e$ of both particles are identical
($M_p=M_e\doteqdot M$) and are denoted simply by $M$.

Now it is of course very tempting to think that, when all the magnetic interactions are neglected, the ortho-states ($b$) and the para-states ($p$) will have the same energy $\eETiv$ \myrf{V.26}
\begin{equation}
  \label{eq:V.28}
  \eETiv\big|_b=\eETiv\big|_p\;,
\end{equation} 
and therefore it is only the magnetic energy term $\ERm$ \myrf{IV.29} which in lowest order of approximation is to be computed by means of the electrostatic solution and is then responsible for the energy difference between the ortho- and para-states
\begin{equation}
  \begin{split}
  \label{eq:V.29}
  \Delta\!^\textrm{(m)}E_\textrm{T}:=\Eiv\big|_b -\Eiv\big|_p &=\left(\eETiv\big|_b-\ERm\big|_b\right)-\left(\eETiv\big|_p-\ERm\big|_p\right)\\ &=-\left[\ERm\big|_b-\ERm\big|_p\right]\,.
  \end{split}
\end{equation} 
Observe here that this kind of reasoning is adopted also in the conventional approach \myrf{V.13} where the orbital factor $\tilde{\psi}(\vec{r}_1,\vec{r}_2)$ is determined by the electric forces alone and is thought to be the same for all four spin configurations (otherwise the groundstate degeneracy would not be four-fold). If those hypotheses like \myrf{V.28}-\myrf{V.29} were true also for the RST case, the ortho/para energy difference in lowest order would appear by means of $\ERm$ \myrf{IV.29} as
\begin{equation}
  \label{eq:V.30}
  \Delta\!^\textrm{(m)}E_\textrm{T} =-\frac{\hbar c}{4\pi\as}\int d^3\vec{r}\,\left\{\left[\vec{H}_1\sdot\vec{H}_2\right]_b-\left[\vec{H}_1\sdot\vec{H}_2\right]_p\right\}
\end{equation}
and thus by means of \myrf{V.18b} and \myrf{V.19b}
\begin{equation}
  \label{eq:V.31}
  \Delta\!^\textrm{(m)}E_\textrm{T} =-\frac{\hbar c}{4\pi\as}\int d^3\vec{r}\,\left[\vec{H}_b^2+\vec{H}_p^2\right]<0\;.
\end{equation}
This result says that the RST ortho-positronium (conventionally: $\eSo$) configuration has lower energy than the RST para-configuration (conventionally: $\dSe$) which is qualitatively in agreement with the observations (see the remarks below equation \myrf{V.13}).

Further information about the magnetic energy difference $\Delta\!^\textrm{(m)}E_\textrm{T}$ \myrf{V.31} is gained by observing that, in the absence of the exchange interactions, the magnetic fields $\vec{H}_{b/p}$ are simply the curls of the corresponding three-vector potentials $\vec{A}_{b/p}$, cf. \myrf{III.24a}-\myrf{III.24b}
\begin{equation}
  \label{eq:V.32}
  \vec{H}_{b/p}=\vec{\nabla}\times\vec{A}_{b/p}\;,
\end{equation}
and furthermore the curls of the magnetic field strengths are related to the Maxwellian three-currents $\vec{j}_{b/p}$ by
\begin{equation}
  \label{eq:V.33}
  \vec{\nabla}\times\vec{H}_{b/p}=-\Delta\vec{A}_{b/p}=4\pi\as\,\vec{j}_{b/p}\;.
\end{equation}
The combination of the latter two equations yields the standard solution of the vector potentials in terms of the Dirac currents $\vec{k}_{b/p}$ as
\begin{equation}
  \label{eq:V.34}
  \vec{A}_{b/p}=\as\int d^3\vec{r}\!\;'\,\frac{\vec{k}_{b/p}(\vec{r}\!\;')}{\|\vec{r}-\vec{r}\!\;'\|}\;.
\end{equation}
This result may now be used in combination with the magnetic Poisson identity \myrf{IV.4}-\myrf{IV.5} in order to recast the magnetic energy $\ERm$ into the following form:
\begin{equation}
  \label{eq:V.35}
  \ERm=\frac{\hbar c}{2}\int d^3\vec{r}\,\left[\vec{k}_1\sdot\vec{A}_2-\vec{k}_2\sdot\vec{A}_1\right]\,,
\end{equation}
i.\,e. concretely for both types of positronium configurations
\begin{equation}
  \label{eq:V.36}
  \ERm\big|_{b/p}=\pm\hbar c\int d^3\vec{r}\,\left(\vec{k}_{b/p}\sdot\vec{A}_{b/p}\right)\,.
\end{equation}
Thus it is possible to express the magnetic energy $\ERm$ completely in terms of the Dirac
currents as follows
\begin{equation}
  \label{eq:V.37}
  \ERm\big|_{b/p}=\pm\hbar c\as\iint d^3\vec{r}\,d^3\vec{r}\!\;'\,\frac{\vec{k}_{b/p}(\vec{r})\sdot\vec{k}_{b/p}(\vec{r}\!\;')}{\|\vec{r}-\vec{r}\!\;'\|}\;,
\end{equation}
and consequently the magnetic energy difference $\Delta\!^\textrm{(m)}E_\textrm{T}$ \myrf{V.29} reads in terms of the Dirac currents 
\begin{equation}
  \label{eq:V.38}
  \Delta\!^\textrm{(m)}E_\textrm{T}=-e^2\iint d^3\vec{r}\,d^3\vec{r}\!\;'\,\frac{\vec{k}_b(\vec{r})\sdot\vec{k}_b(\vec{r}\!\;')+\vec{k}_p(\vec{r})\sdot\vec{k}_p(\vec{r}\!\;')}{\|\vec{r}-\vec{r}\!\;'\|}\;.
\end{equation}

This result presents now a very critical point for the intended estimate of the ortho/para energy difference. Namely, at first glance it seems very tempting to think that, because of the smallness of the magnetic interactions as compared to their electric counterparts, the neglection of the magnetic forces will induce only a negligibly small change of the Dirac currents $\vec{k}_a(\vec{r})$ ($a=1,2$) as they are obtained by the electrostatic approximation, so that one should be allowed to equate approximately the ortho-and para-currents, i.\,e.
\begin{equation}
  \label{eq:V.39}
  \vec{k}_b(\vec{r})\simeq\vec{k}_p(\vec{r})\doteqdot\vec{k}_*(\vec{r})\;.
\end{equation}
Furthermore, this common current $\vec{k}_*(\vec{r})$ could be eventually taken from the
solution of the {\em electrostatic}\/ (and therefore truncated) RST eigenvalue
problem. Indeed, such a philosophy was put forward in two preceding papers
\cite{MaSo3,BeSo2}; and by such presumptions the ortho/para energy difference
$\Delta\!^\textrm{(m)}E_\textrm{T}$ \myrf{V.38} would be finally found in the following
form
\begin{equation}
  \label{eq:V.40}
  \Delta\!^\textrm{(m)}E_\textrm{T}\Rightarrow-2e^2\iint d^3\vec{r}\,d^3\vec{r}\!\;'\,\frac{\vec{k}_*(\vec{r})\sdot\vec{k}_*(\vec{r}\!\;')}{\|\vec{r}-\vec{r}\!\;'\|}\;,
\end{equation}
see equation (5.113) of ref. \cite{BeSo2}. (Unfortunately, in ref. \cite{BeSo2} the notations for the Dirac currents $\vec{k}_b$ and $\vec{k}_p$ are interchanged relative to the present definitions \myrf{V.18c} and \myrf{V.19c}).

However, serious objections against such kind of reasoning seem adequate because it is highly questionable whether the Dirac currents $\vec{k}_a$ ($a=1,2$) are really (or almost) identical for the ortho- and para-system, even in the electrostatic approximation! The crucial point in this context is namely that two strongly interacting fermions do form a bosonic system when being considered as a whole, and in RST this bosonic character of the compound system becomes transferred to the individual particles which as a consequence do then occupy {\em bosonic one-particle states}\/. In general, such exotic RST states carry then an {\em integer}\/ eigenvalue of total angular momentum $\hat{J}_z=\hat{J}_z^{(+)}\oplus\hat{J}_z^{(-)}$ (the $z$-axis is adopted to be the symmetry axis of the two-particle system):
\begin{equation}
  \begin{split}
  \label{eq:V.41}
  \hat{\mathcal{J}}_z\,\psi_a(\vec{r}) &=\big(\hat{J}_z^{(+)\,}\app\big)\oplus\big(\hat{J}_z^{(-)\,}\apm\big)=j_z\,\psi_a(\vec{r})\\
  &{}\qquad\qquad j_z=0,\pm1,\pm2,...
  \end{split}
\end{equation}

But since the one-particle wave functions $\psi_a(\vec{r})$ are Dirac four-spinors, this
integrity condition \myrf{V.41} can hold only at the price of non-uniqueness of the
spinors $\psi_a(\vec{r})$, see the discussion of this effect in ref.s \cite{MaSo2,BeSo2}.
Thus for each one of the two particles of the positronium groundstate one expects $j_z=0$
($\leadsto$ singlet) for the para-state and $j_z=0,\pm1$ ($\leadsto$ triplet) for the
ortho-system. Now it is well-known in (conventional) quantum mechanics that wave
functions~$\psi_a$ with different eigenvalues of angular momentum can display a rather
different spatial pattern and this difference will then be transferred also to the
currents $\vec{k}_a$ which are (more or less uniquely) determined by the wave functions
$\psi_a$, see the Dirac currents \myrf{II.16}. Therefore one concludes that, in contrast
to the equality of the densities $\akn$ \myrf{V.20c}, the presumed equality \myrf{V.39} of
ortho- and para-currents $\vec{k}_b$ and $\vec{k}_p$ can not be correct in general, not
even in the electrostatic approximation, cf.\ fig.1 vs.\ fig.~6; and as a consequence the
result (\ref{eq:V.38}) is to be prefered in place of (\ref{eq:V.46}).
Clearly, this conclusion forces us now to inspect more rigorously the spatial pattern of
the ortho/para wave functions $\psi_a(\vec{r})$ and of the associated curents
$\vec{k}_a(\vec{r})$.

\vspace{4ex}
\begin{center}
  \emph{\textbf{3.\ Positronium Eigenvalue Problem}}
\end{center}

The postulated positronium constraints \myrf{V.18a}-\myrf{V.20d} for both gauge field
modes ($a=1,2$) imply the corresponding constraints \myrf{V.23}-\myrf{V.24} for the matter
modes $\appm$ ($a=1,2$) so that both kinds of constraints are consistent with each other
and thus admit to deal with only one single matter field $\varphi_\pm$ for the considered
two-particle system: $\bppm$ \myrf{V.25a} for the ortho-configuration and $\pppm$
\myrf{V.25b} for the para-configuration. But with this arrangement there arises now a
serious problem; namely the question whether {\em each}\/ of these two links
\myrf{V.25a}-\myrf{V.25b} is additionally consistent also with the two-particle eigenvalue
equations \myrf{III.22a}-\myrf{III.22b}? The latter consistency requirement says that any
of the two equations \myrf{III.22a} and \myrf{III.22b} must collapse to the same equation
for the Pauli spinors $\appm$, namely either for $\bppm$ in the ortho-case or for $\pppm$
in the para-case. Clearly, the situation with ortho-positronium \myrf{V.25a} is simpler
and an explicit calculation yields for the one residual Pauli pair of equations \cite{BeSo}:
\begin{equation}
  \begin{split}
  \label{eq:V.42}
  i\,\vec{\sigma}\sdot\big(\vec{\nabla}+i\vec{A}_b\big)\bppm-\bAn\cdot\bpmp&=\frac{M_*\pm M}{\hbar}\,c\cdot\bpmp\\
  &\hspace{-9em}\mbox{\textbf{(\emph{ortho-positronium}\/)}}\;.
  \end{split}
\end{equation}
Here both mass eigenvalues $M_a$ are assumed to differ merely in sign and the rest masses of electron ($M_e$) and positron ($M_p$) are equal
\begin{subequations}
  \begin{align}
  \label{eq:V.43a}
  M_1&=-M_2\doteqdot-M_*\\
  \label{eq:V.43b}
  M_e&=M_p\doteqdot M\;.
  \end{align}
\end{subequations}
However, the situation with para-positronium \myrf{V.25b} is somewhat more intricate but
nevertheless admits also both Pauli pairs of equations \myrf{III.22a}-\myrf{III.22b} to
collapse onto one single pair:
\begin{equation}
  \begin{split}
  \label{eq:V.44}
  i\,\vec{\sigma}\sdot\big(\vec{\nabla}-i\vec{A}_p\big)\pppm-\pAo\cdot\ppmp&=\frac{M_*\pm M}{\hbar}\,c\cdot\ppmp\\
  &\hspace{-8.8em}\mbox{\textbf{(\emph{para-positronium}\/)}}\;.
  \end{split}
\end{equation}
The deduction of this para-equation is not only more difficult than its ortho-counterpart \myrf{V.42} but it yields also an additional constraint upon the para-spinors $\pppm$ \myrf{V.25b}, i.\,e.
\begin{equation}
  \label{eq:V.45}
  \hat{J}_z^{(\pm)}\,\pppm=0\;,
\end{equation}
which itself is a special case of equation \myrf{V.41}
\begin{equation}
  \label{eq:V.46}
  \hat{\mathcal{J}}_z\,\psi_a(\vec{r})=0\;.
  \end{equation}
For both cases \myrf{V.42} and \myrf{V.44} it was assumed here that both the three-vector potentials $\vec{A}_a$ and the Dirac currents $\vec{k}_a$ are of azimuthal character, i.\,e.
\begin{subequations}
  \begin{align}
  \label{eq:V.47a}
  \vec{A}_a&=\aA_\phi\cdot\vec{\rm e}_\phi\\
  \label{eq:V.47b}
  \vec{k}_a&=\ak_\phi\cdot\vec{\rm e}_\phi\;.
  \end{align}
\end{subequations}

But observe that the ortho-case \myrf{V.42} must not be associated with such a restriction
as \myrf{V.46}! Moreover, it may seem that the definitions \myrf{V.18a}-\myrf{V.19d} for
the ortho- and para-configurations are rather arbitrary; but actually we do assume here
that such configurations are the only localized solutions of the two-particle system
\myrf{III.22a}-\myrf{III.22b} {\em if both rest masses are identical}\/
(i.\,e. $M_e=M_p\doteqdot M$). Clearly, the proof of this hypothesis needs an extra
discussion. If the hypothesis turned out to be wrong, the RST predictions would come in
conflict with the observational data which yield a {\em one-particle}\/ spectrum for
positronium \cite{GR}. Observe also that, when the magnetic interactions are neglected
($\vec{A}_b=\vec{A}_p\equiv0$), the ortho-equation \myrf{V.42} becomes identical to the
para-equation \myrf{V.44}; but since the angular-momentum restriction \myrf{V.45} does
hold only for the para-case (also in the electrostatic approximation) one expects that
ortho-positronium has a richer spectrum (see below).

The vanishing eigenvalue of total angular momentum $\hat{J}_z$ \myrf{V.46} is a very
unorthodox feature of a Dirac particle's quantum state because one usually expects
half-integer eigenvalues. However, as mentioned above, the bosonic character of a
two-fermion system is transferred in RST to the individual constituents so that they can
{\em individually}\/ acquire bosonic properties. This amazing effect may be elaborated
in RST by an adequate choice of a spinor basis relative to which all the Pauli spinors
$\appm$ are then to be decomposed. Our choice of such a basis does not directly refer to
the spinors $\zeta^{j,m}_l$ \myrf{V.6a}-\myrf{V.6c} but rather to a certain modification
of them \cite{BeSo2}
\begin{subequations}
  \begin{align}
  \label{eq:V.48a}
  \wop&={\rm e}^{-i\bbar\phi}\cdot\zeta^{\frac{1}{2},\frac{1}{2}}_0\\
  \label{eq:V.48b}
  \wom&={\rm e}^{i\bbar\phi}\cdot\zeta^{\frac{1}{2},-\frac{1}{2}}_0\\
  \label{eq:V.48c}
  \wep&={\rm e}^{-i\bbar\phi}\cdot\zeta^{\frac{1}{2},\frac{1}{2}}_1\\
  \label{eq:V.48d}
  \wem&={\rm e}^{i\bbar\phi}\cdot\zeta^{\frac{1}{2},-\frac{1}{2}}_1\;,
  \end{align}
\end{subequations}
where the {\em boson number}\/ $\bbar$ is a real number with range $-\frac{1}{2}\leq\bbar\leq\frac{1}{2}$. This number measures the bosonic character of a Dirac particle which becomes more obvious from the action of $\hat{J}^{(\pm)}_z$ upon those basis spinors
\begin{subequations}
  \begin{align}
  \label{eq:V.49a}
  \hat{J}_z^{(+)}\,\wop&=-\big(\bbar-\frac{1}{2}\big)\hbar\cdot\wop\\
  \label{eq:V.49b}
  \hat{J}_z^{(+)}\,\wom&=\big(\bbar-\frac{1}{2}\big)\hbar\cdot\wom\\
  \label{eq:V.49c}
  \hat{J}_z^{(-)}\,\wep&=-\big(\bbar-\frac{1}{2}\big)\hbar\cdot\wep\\
  \label{eq:V.49d}
  \hat{J}_z^{(-)}\,\wem&=\big(\bbar-\frac{1}{2}\big)\hbar\cdot\wem\;.
  \end{align}
\end{subequations}
Whenever the boson number $\bbar$ is zero, the new basis $\{\woepm\}$ \myrf{V.48a}-\myrf{V.48d} coincides with the original basis $\{\zeta^{\frac{1}{2},\pm\frac{1}{2}}_{0,1}\}$ \myrf{V.6a}-\myrf{V.6c} and one deals then with a purely fermionic basis, since the eigenvalues $j_z$ of $\hat{J}^{(\pm)}_z$ are half-integer. However, for $\bbar=\pm\frac{1}{2}$ the eigenvalues of $\hat{J}_z$ are integer and one has a purely bosonic basis (the case $-\frac{1}{2}<\bbar<\frac{1}{2}$ with $\bbar\neq0$ is something between a fermionic and bosonic basis and needs an extra discussion).

But once a spinor basis has been selected in such a way, one can decompose the Pauli spinors $\bpppm$ for the ortho- and para-system in the usual way
\begin{subequations}
  \begin{align}
  \label{eq:V.50a}
  \bpphip&=\bpMRp\cdot\wop+\bpMSp\cdot\wom\\
  \label{eq:V.50b}
  \bpphim&=-i\left\{\bpMRm\cdot\wep+\bpMSm\cdot\wem\right\}\,.
  \end{align}
\end{subequations}
Strictly speaking, such a spinor coordinatization should first refer to the original single-particle states $\appm$ ($a=1,2$); i.\,e. one puts
\begin{subequations}
  \begin{align}
  \label{eq:V.51a}
  \app&=\aMRp\cdot\wop+\aMSp\cdot\wom\\
  \label{eq:V.51b}
  \apm&=-i\left\{\aMRm\cdot\wep+\aMSm\cdot\wem\right\}\,,
  \end{align}
\end{subequations}
but on account of the ortho- and para-identifications \myrf{V.25a}-\myrf{V.25b} the wave amplitudes $\aMRpm,\,\aMSpm$ are also to be identified in a certain way, namely for \emph{\textbf{ortho-positronium}}~\myrf{V.25a}
\begin{subequations}
  \begin{align}
  \label{eq:V.52a}
  \eMRp&\equiv-i\zMRp\doteqdot\bMRp\\
  \label{eq:V.52b}
  \eMSp&\equiv-i\zMSp\doteqdot\bMSp\\
  \label{eq:V.52c}
  \eMRm&\equiv i\zMRm\doteqdot\bMRm\\
  \label{eq:V.52d}
  \eMSm&\equiv i\zMSm\doteqdot\bMSm
  \end{align}
\end{subequations}
and similarly in the case of \emph{\textbf{para-positronium}}\/ \myrf{V.25b} for boson number $\bbar=\frac{1}{2}$:
\begin{subequations}
  \begin{align}
  \label{eq:V.53a}
  \eMRp&\equiv\zMSp\doteqdot\pMRp\\
  \label{eq:V.53b}
  \eMSp&\equiv-\zMRp\doteqdot\pMSp\\
  \label{eq:V.53c}
  \eMRm&\equiv-\zMSm\doteqdot\pMRm\\
  \label{eq:V.53d}
  \eMSm&\equiv\zMRm\doteqdot\pMSm\;.
  \end{align}
\end{subequations}
Observe here for the latter case of para-positronium that the requirement of vanishing
angular momentum \myrf{V.46} can be attained in the simplest way by {\bf (i)} choosing for
the boson number $\bbar$ the special value $\bbar=\frac{1}{2}$, which puts the eigenvalues
$j_z$ of $\hat{J}_z$ in \myrf{V.41} to zero ($j_z\Rightarrow0$), and {\bf (ii)} adopting
the wave amplitudes $\pMRpm,\,\pMSpm$ as being independent of the azimuthal angle $\phi$,
i.\,e.
\begin{subequations}
  \begin{align}
  \label{eq:V.54a}
  \pMRpm(r,\vartheta,\phi)\Rightarrow\pRpm(r,\vartheta)\\
  \label{eq:V.54b}
  \pMSpm(r,\vartheta,\phi)\Rightarrow\pSpm(r,\vartheta)\;.
  \end{align}
\end{subequations}

The parametrizations \myrf{V.51a}-\myrf{V.51b} of the Pauli spinors $\bpppm$ in terms of
the (complex-valued) wave amplitudes $\bpMRpm,\,\bpMSpm$ can now be inserted in the
two-particle Pauli eigenvalue equations for ortho-positronium \myrf{V.42} and
para-positronium \myrf{V.44} in order to yield the corresponding eigenvalue equations for
these wave amplitudes; i.\,e. for \emph{\textbf{ortho-positronium}}
\begin{subequations}
  \begin{align}
  \label{eq:V.55a}
  \frac{\partial\,\bMRp}{\partial r} +\frac{i}{r}\cdot\frac{\partial\,\bMRp}{\partial\phi}+\frac{\bbar}{r}\cdot\bMRp-\bAo\cdot\bMRm -\sin\vartheta\,\bAphi\cdot\bMRp\hspace{-9em}&\hspace{9em}\nonumber\\
  {}+{\rm e}^{2i(\bbar-\frac{1}{2})\phi}\cdot\Bigg\{\frac{1}{r}\cdot\frac{\partial\,\bMSp}{\partial\vartheta} +\frac{\cot\vartheta}{r}\,\big[\bbar\cdot\bMSp-i\,\frac{\partial\,\bMSp}{\partial\phi}\big]\hspace{-1em}&\nonumber\\
  +\cos\vartheta\,\bAphi\cdot\bMSp\Bigg\}&=\frac{M+M_*}{\hbar}\,c\cdot\bMRm
  \\\nonumber\\
  \label{eq:V.55b}
  \frac{\partial\,\bMSp}{\partial r} -\frac{i}{r}\cdot\frac{\partial\,\bMSp}{\partial\phi}+\frac{\bbar}{r}\cdot\bMSp-\bAo\cdot\bMSm +\sin\vartheta\,\bAphi\cdot\bMSp\hspace{-7em}&\hspace{7em}\nonumber\\
  {}-{\rm e}^{-2i(\bbar-\frac{1}{2})\phi}\cdot\Bigg\{\frac{1}{r}\cdot\frac{\partial\,\bMRp}{\partial\vartheta} +\frac{\cot\vartheta}{r}\,\big[\bbar\cdot\bMRp+i\,\frac{\partial\,\bMRp}{\partial\phi}\big]\hspace{-2.8em}&\nonumber\\
  -\cos\vartheta\,\bAphi\cdot\bMRp\Bigg\}&=\frac{M+M_*}{\hbar}\,c\cdot\bMSm
  \\\nonumber\\
  \label{eq:V.55c}
  \frac{\partial\,\bMRm}{\partial r} -\frac{i}{r}\cdot\frac{\partial\,\bMRm}{\partial\phi}+\frac{2-\bbar}{r}\cdot\bMRm+\bAo\cdot\bMRp +\sin\vartheta\,\bAphi\cdot\bMRm\hspace{-11em}&\hspace{11em}\nonumber\\
  {}-{\rm e}^{2i(\bbar-\frac{1}{2})\phi}\cdot\Bigg\{\frac{1}{r}\cdot\frac{\partial\,\bMSm}{\partial\vartheta} +\frac{\cot\vartheta}{r}\,\big[\bbar\cdot\bMSm-i\,\frac{\partial\,\bMSm}{\partial\phi}\big]\hspace{-1em}&\nonumber\\
  +\cos\vartheta\,\bAphi\cdot\bMSm\Bigg\}&=\frac{M-M_*}{\hbar}\,c\cdot\bMRp
  \\\nonumber\\
  \label{eq:V.55d}
  \frac{\partial\,\bMSm}{\partial r} +\frac{i}{r}\cdot\frac{\partial\,\bMSm}{\partial\phi}+\frac{2-\bbar}{r}\cdot\bMSm+\bAo\cdot\bMSp -\sin\vartheta\,\bAphi\cdot\bMSm\hspace{-9.5em}&\hspace{9.5em}\nonumber\\
  {}+{\rm e}^{-2i(\bbar-\frac{1}{2})\phi}\cdot\Bigg\{\frac{1}{r}\cdot\frac{\partial\,\bMRm}{\partial\vartheta} +\frac{\cot\vartheta}{r}\,\big[\bbar\cdot\bMRm+i\,\frac{\partial\,\bMRm}{\partial\phi}\big]\hspace{-2.5em}&\nonumber\\ -\cos\vartheta\,\bAphi\cdot\bMRm\Bigg\}&=\frac{M-M_*}{\hbar}\,c\cdot\bMSp\;.
  \end{align}
\end{subequations}
Here, the value of the boson number $\bbar$ has been left unspecified for the sake of generality; but it is clear from equations \myrf{V.49a}-\myrf{V.49d} that the right value for {\em ortho-positronium}\/ will be $\bbar=-\frac{1}{2}$.

Quite analogously, for {\em para-positronium} one will adopt the value
$\bbar=\frac{1}{2}$, which puts the eigenvalues of $\hat{J}_z$ to zero, cf. \myrf{V.45};
and additionally one adopts the wave amplitudes $\aMRpm,\,\aMSpm$
\myrf{V.54a}-\myrf{V.54b} to be real-valued functions ($\aRpm,\,\aSpm$) over
three-space. For this arrangement, the two eigenvalue equations
 (\ref{eq:V.44}) read then in terms of those (real-valued) wave amplitudes
$\pRpm,\,\pSpm$ for \emph{\textbf{para-positronium}}
\begin{subequations}
  \begin{align}
  \frac{\partial\pRp}{\partial r}+\frac{1}{2r}\cdot\pRp-\pAo\cdot\pRm +\pAphi\left[\sin\vartheta\cdot\pRp-\cos\vartheta\cdot\pSp\right]+\frac{1}{r}\frac{\partial\pSp}{\partial\vartheta}&\nonumber\\\nonumber\\ 
  \label{eq:V.56a}
+\frac{\cot\vartheta}{2r}\cdot\pSp=\frac{M+M_*}{\hbar}\,c&\cdot\pRm\\\nonumber\\
  \frac{\partial\pSp}{\partial r}+\frac{1}{2r}\cdot\pSp-\pAo\cdot\pSm -\pAphi\left[\cos\vartheta\cdot\pRp+\sin\vartheta\cdot\pSp\right]-\frac{1}{r}\frac{\partial\pRp}{\partial\vartheta}&\nonumber\\\nonumber\\
  \label{eq:V.56b}
-\frac{\cot\vartheta}{2r}\cdot\pRp=\frac{M+M_*}{\hbar}\,c&\cdot\pSm\\\nonumber\\
  \frac{\partial\pRm}{\partial r}+\frac{3}{2r}\cdot\pRm+\pAo\cdot\pRp -\pAphi\left[\sin\vartheta\cdot\pRm-\cos\vartheta\cdot\pSm\right]-\frac{1}{r}\frac{\partial\pSm}{\partial\vartheta}&\nonumber\\\nonumber\\
  \label{eq:V.56c}
-\frac{\cot\vartheta}{2r}\cdot\pSm=\frac{M-M_*}{\hbar}\,c&\cdot\pRp\\\nonumber\\
  \frac{\partial\pSm}{\partial r}+\frac{3}{2r}\cdot\pSm+\pAo\cdot\pSp +\pAphi\left[\cos\vartheta\cdot\pRm+\sin\vartheta\cdot\pSm\right]+\frac{1}{r}\frac{\partial\pRm}{\partial\vartheta}&\nonumber\\\nonumber\\
  \label{eq:V.56d}
+\frac{\cot\vartheta}{2r}\cdot\pRm=\frac{M-M_*}{\hbar}\,c&\cdot\pSp\;.
  \end{align}
\end{subequations}

Both eigenvalue systems, i.\,e. the ortho-system \myrf{V.55a}-\myrf{V.55d} and the para-system \myrf{V.56a}-\myrf{V.56d} provide now the basis for a thorough inspection of the RST multiplet and degeneracy structure. Observe here that the main feature of this structure is the ortho/para dichotomy for which RST supplies a very natural basis, namely the \mbox{(anti-)}parallelity of the magnetic fields. In contrast to this rather dynamical criterion, the conventional theory puts this ortho/para dichotomy on the basis of the composition rule for the angular momenta of both particles which is rather a kinematical criterion than a dynamical one! In this context it is worthwile to reconsider the kinematical consequences of the dynamical requirement concerning the \mbox{(anti-)}parallelity of the magnetic fields. Clearly, the parallelity or antiparallelity of the magnetic field strengths $\vec{H}_a(\vec{r})$ ($a=1,2$) entails the antiparallelity or parallelity of the corresponding Dirac currents $\vec{k}_a(\vec{r})$ (or Maxwell currents $\vec{j}_a$, resp.) because the latter are the curls of the magnetic fields, cf. \myrf{V.33}. But on the other hand, these currents are themselves generated by the wave functions $\psi_a(\vec{r})$, cf. \myrf{II.17a}-\myrf{II.17e}; and therefore the currents can transcribe those \mbox{(anti-)}parallelity requirements for the field strengths $\vec{H}_a$ to the wave functions, i.\,e. ultimately to the wave amplitudes $\aMRpm,\,\aMSpm$. More concretely, when the Dirac currents $\vec{k}_a$ are decomposed with respect to the basis triad $\{\vec{\rm e}_r,\vec{\rm e}_\vartheta,\vec{\rm e}_\phi\}$ of the spherical polar coordinates $\{r,\,\vartheta,\,\phi\}$
\begin{equation}
  \label{eq:V.56'}
  \vec{k}_a=\ak_r\cdot\vec{\rm e}_r+\ak_\vartheta\cdot\vec{\rm e}_\vartheta+\ak_\phi\cdot\vec{\rm e}_\phi\;,
\end{equation}
the current components $\ak_r,\,\ak_\vartheta,\,\ak_\phi$ are linked to the wave amplitudes through
\begin{subequations}
  \begin{align}
  \label{eq:V.57a}
  \ak_r&=\frac{i}{4\pi}\left\{\aMRpS\cdot\aMRm+\aMSpS\cdot\aMSm-\aMRmS\cdot\aMRp-\aMSmS\cdot\aMSp\right\}\\
  \label{eq:V.57b}
  \ak_\vartheta&=-\frac{i}{4\pi}\left\{{\rm e}^{2i(\bbar-\frac{1}{2})\phi}\cdot\MCa-{\rm e}^{-2i(\bbar-\frac{1}{2})\phi}\cdot\MCaS\right\}\\
  &{}\hspace{8em}(\MCa\doteqdot\aMRpS\cdot\aMSm+\aMRmS\cdot\aMSp)\nonumber\\
  \label{eq:V.57c}
  \ak_\phi&= \frac{\sin\vartheta}{4\pi}\left\{\aMRpS\cdot\aMRm+\aMRmS\cdot\aMRp-\aMSpS\cdot\aMSm-\aMSmS\cdot\aMSp\right\}\nonumber\\
  &\hspace{1.5em}-\frac{\cos\vartheta}{4\pi}\left\{{\rm e}^{2i(\bbar-\frac{1}{2})\phi}\cdot\MCa+{\rm e}^{-2i(\bbar-\frac{1}{2})\phi}\cdot\MCaS\right\}\;.
  \end{align}
\end{subequations}

Now one can test here whether the ortho-antiparallelity \myrf{V.18c} and the para-parallelity \myrf{V.19c} of the Dirac currents $\vec{k}_a(\vec{r})$ are really validated by both the ortho-identifications \myrf{V.52a}-\myrf{V.52d} and the para-identifications \myrf{V.53a}-\myrf{V.53d}. For the latter para-case one finds
\begin{subequations}
  \begin{align}
  \label{eq:V.58a}
  \ek_r&=-\zk_r\\
  \label{eq:V.58b}
  \ek_\vartheta&=-\zk_\vartheta\\
  \label{eq:V.58c}
  \ek_\phi&=+\zk_\phi\;.
  \end{align}
\end{subequations}
But obviously it is only the azimuthal component \myrf{V.58c} which is consistent with the para-identification \myrf{V.19c}, but not the radial and longitudinal components \myrf{V.58a}-\myrf{V.58b}. If one therefore insists on the {\em three-vector}\/ identification \myrf{V.19c}, one is forced to let vanish the radial and longitudinal components of the Dirac para-current $\vec{k}_p$:
\begin{subequations}
  \begin{align}
  \label{eq:V.59a}
  \ek_r&\equiv\zk_r\doteqdot\pk_r\stackrel{!}{\equiv}0\\
  \label{eq:V.59b}
  \ek_\vartheta&\equiv\zk_\vartheta\doteqdot\pk_\vartheta\stackrel{!}{\equiv}0\;.
  \end{align}
\end{subequations}
As a result, the Dirac para-current $\vec{k}_p$ must encircle the $z$-axis in the planes $z=\mbox{const}$
\begin{equation}
  \label{eq:V.60}
  \vec{k}_p=\pk_\phi\cdot\vec{\rm e}_\phi\;,
\end{equation}
cf. the anticipation \myrf{V.47b}. Moreover it is easy to see that both requirements of vanishing current components \myrf{V.59a}-\myrf{V.59b} can be satisfied for $\bbar=\frac{1}{2}$ by simply taking all the wave amplitudes $\aMRpm,\,\aMSpm$ to be real-valued ($\aMRpm\Rightarrow\aRpm;\,\aMSpm\Rightarrow\aSpm$), as it was already anticipated for the para-system \myrf{V.56a}-\myrf{V.56d}.

However, concerning the ortho-system, one finds that the current components \myrf{V.56'} for the ortho-value of $\bbar$ (i.\,e. $\bbar=-\frac{1}{2}$), together with the ortho-identifications \myrf{V.52a}-\myrf{V.52d}, appear in the following form
\begin{subequations}
  \begin{align}
  \label{eq:V.61a}
  \ek_r&=-\zk_r\\
  \label{eq:V.61b}
  \ek_\vartheta&=-\zk_\vartheta\\
  \label{eq:V.61c}
  \ek_\phi&=-\zk_\phi\;.
  \end{align}
\end{subequations}
This means that the ortho-antiparallelity of the Dirac currents \myrf{V.18c} can always be
attained by the present ortho-parametrization of the wave functions $\psi_a(\vec{r})$,
with no additional restriction as in the preceding para-situation. Fortunately, this
admits to evoke here the toroidal configuration
($k_\phi\equiv0,\,k_\vartheta\neq0,\,k_r\neq0$) of ref. \cite{GBMS} which may be adopted
here in order to provide a representation for that ortho-state with vanishing $j_z$
($=0$). However we will subsequently restict ourselves to that rotational symmetry around
the $z$-axis as given by equations \myrf{V.47a}-\myrf{V.47b}, for the discussion of both
the para- and the ortho-system. In this sense, we specialize down the general Poisson
equation \myrf{V.22} for the vector potentials $\vec{A}_{b/p}$ to their $SO(2)$ symmetric
form
\begin{equation}
  \label{eq:V.62}
  \Delta\bpA_\phi-\frac{\bpA_\phi}{r^2\sin^2\vartheta}=-4\pi\as\cdot\bpk_\phi
\end{equation}
which together with its electric counterpart \myrf{V.21} completes the present eigenvalue equations of the ortho- and para-type to a closed system.

\section{Energy Difference of Ortho- and Para-States}
\indent

Once the emergence of the ortho/para dichotomy in RST has been clarified, there naturally
arises the question of degeneracy of the corresponding ortho- and para-states. But
here it should be clear that, corresponding to the present state of the art, this question
can be discussed only from a rather qualitative point of view because exact (analytic or
numerical) solutions of the RST two-particle eigenvalue problem are presently not
known. Nevertheless, the knowledge of the exact form of the RST energy functional~$\tETT$
(\ref{eq:IV.20}) admits to draw same general conclusions concerning this question of
degeneracy.

First, it will be necessary to define some order of approximation which corresponds to the
conventional situation (\ref{eq:V.13}), where the singlet and triplet states carry the
same energy. As a matter of course, this first approximation step will be essentially the
same in RST as in the conventional theory, namely the neglection of the magnetic
interactions ($\leadsto$ \emph{``electrostatic approximation''}). This neglection of the
magnetic forces, however, does not yet entail the mathematical equivalence of the
eigenvalue equations for the ortho- and para- configurations; indeed, there persists a
crucial kinematical difference: whereas the current of the para-states ($\leadsto$
\emph{``para-current''}~$\vec{k}_p$) must always encircle the axis of rotational symmetry
(i.e.\ the z-axis) and thus realizes a cylindrical structure, see equation \myrf{V.60}, the
\emph{``ortho-current''}~$\vec{k}_b$ can obey this cylindrical geometry only for
non-vanishing angular momentum~$(j_z=\pm 1)$, but for vanishing angular momentum~$(j_z=0)$ it
is required to undergo a \emph{toroidal} geometry (previously discussed in
ref.~\cite{GBMS}). In the present paper, we restrict ourselves to the cylindrical structure
(i.e.~$j_z=0$ for the para-states and~$j_z=\pm1$ for the ortho-states); but even within
this subclass of the cylindrical configurations the link between the charge
densities~$\bpko$ and the wave amplitudes~$\bpMRpm,\bpMSpm$ looks rather different: for the
para-configurations, the essential angular variability of~$\pko(r)$ goes
like~$(\sin\vartheta)^{-1}$, cf.\ (\ref{eq:VI.17}) below; but for the ortho-densities~$\bkn$
there occurs a certain subdichotomy, see figure~5, namely in such a way that both forms of
ortho-matter with density~$\bkn$ are pressed off the rotational axis upon which they
become zero~$({}^{(b)}k_0 \big|_{\vartheta=0,\pi}=0)$.

Clearly, this specific distribution of ortho- and para-matter is responsible for the
difference in angular momentum (i.e.~$j_z=0$ for the para-distribution being concentrated
close to the rotational axis~$(\sim(\sin\vartheta)^{-1})$; and~$j_z=\pm1$ for the
ortho-distribution concentrated off the z-axis~$(\sim \sin\vartheta)$). However such a
redistribution of the charge densities does imply also a corresponding alteration of the
electrostatic interaction energy~$\eER$ so that the ortho- and para-configurations would
always carry different energy~$\ET$ and the ortho/para degeneracy would be broken. Thus
one is forced to conclude that, in order to encounter the degeneracy phenomenon in RST, it
is not sufficient to neglect the magnetic interactions as is the case with the
conventional theory, see the discussion of equation \myrf{V.13}. Additionally, one has to
resort to a further approximation, i.\,e. the {\em spherically symmetric approximation}\/,
which consists in assuming the electrostatic interaction potential $\bpAo(\vec{r})$ (not
the wave amplitudes!) to be spherically symmetric:
$\bpAo(\vec{r})\Rightarrow\bpAo(r)\;(r=|\vec{r}|)$. And indeed, it is just this assumption
which yields the complete mathematical equivalence of the eigenvalue problems for the
ortho- and para-configurations, cf. the para-eigenvalue equations
\myrf{VI.23a}-\myrf{VI.23b} to their ortho-counterparts \myrf{VI.115a}-\myrf{VI.115b}. The
(approximately) common energy spectrum is sketched in fig.~2. The RST energy levels of
this common spectrum do agree with their conventional counterparts (\ref{eq:VI.88}) with
deviations of magnitude of some 10\% (see the table on p.~\pageref{tabE}). This is a
remarkable result because it is based upon a very simple trial function~$\tilde{\Phi}(r)$
(\ref{eq:VI.55}) with only two variational parameters!

\vspace{4ex}
\begin{center}
  \emph{\textbf{1.\ Para-Positronium}}
\end{center}

As a matter of course, we will first turn to the simplest field configuration which is
surely the para-system with its singlet character ($\bbar=\frac{1}{2}$). The corresponding
energy eigenvalue problem is defined by the mass eigenvalue equations
\myrf{V.56a}-\myrf{V.56d} together with the Poisson equations \myrf{V.21} and \myrf{V.62}
for the electric potential $\pAo(r,\vartheta)$ and the azimuthal component
$\pAphi(r,\vartheta)$ of the three-vector potential $\vec{A}_p(\vec{r})$, cf. equation
\myrf{V.22}. Here it should be self-evident that it is very hard (if not impossible) to
find exact solutions for the para-system so that one is forced to resort to an
approximative treatment. Especially for the para-groundstate one expects the highest
possible symmetry of the solutions; but this is surely not the $SO(3)$ symmetry with
respect to the origin ($r=0$) because this would imply hedgehog-like three-vector fields
$\vec{E}_p(\vec{r})$ and $\vec{H}_p(\vec{r}),\,\vec{k}_p(\vec{r})$. This could be accepted
for the electric field strength $\vec{E}_p(\vec{r})$ ($\simeq\vec{r}$), but not for its
magnetic counterpart $\vec{H}_p(\vec{r})$ because the latter would imply the existence of
magnetic monopoles (to be excluded here).

\vspace{3ex}
\begin{center}
\emph{\large Electrostatic Approximation}
\end{center}

Hence for a first rough estimate of the para-levels one may resort to the {\em
  electrostatic approximation}\/ where all objects of the magnetic type are to be
neglected. Here the hypothesis is that the \emph{magnetic} forces are much weaker (than
their \emph{electric} counterparts) and therefore they can hardly influence the charge
densities $\ajn$ whose spatial distribution will be dominated by the \emph{electric}
forces. Since the corresponding electrostatic interaction energy ($\ERe$) is itself also
much greater than its magnetic counterpart ($\ERm$), the influence of the magnetic
interactions on the total energy $\ET$ appears therefore to be negligibly small. Clearly,
such a picture of the magnitude of the intra-atomic gauge forces seems to provide the
legitimation for neglecting the magnetic interactions altogether. However, there may exist
certain regions in three-space (here the $z$-axis) where the magnetic forces become
(almost) infinite and thus enforce a distinct redistribution of the electric charge
densities $\ajn$ so that the electrostatic energy $\ERe$ may become modified
considerably. Evidently, for such a situation the electrostatic approximation is not
adequate; but we will assume here that this magnetically induced shift of the electric
energy $\ERe$ will not spoil too much the results. However, it may well be possible that
through the omission of such an energy shift the lower bound of the original energy
functional $\tET$ becomes underestimated by the electrostatic approximation (see the
discussion of the positronium groundstate in ref. \cite{MaSo2}).
 
In any case, by this neglection of magnetism the original para-system \myrf{V.56a}-\myrf{V.56d} reduces to
\begin{subequations}
  \begin{align}
  \frac{\partial\ptRp}{\partial r}+\frac{1}{r}\frac{\partial\ptSp}{\partial\vartheta}-\pAo\cdot\ptRm   \label{eq:VI.1a}
=\frac{M+M_*}{\hbar}\,c&\cdot\ptRm\\\nonumber\\
  \frac{\partial\ptSp}{\partial r}-\frac{1}{r}\frac{\partial\ptRp}{\partial\vartheta}-\pAo\cdot\ptSm   \label{eq:VI.1b}
=\frac{M+M_*}{\hbar}\,c&\cdot\ptSm\\\nonumber\\
  \frac{\partial\ptRm}{\partial r}+\frac{1}{r}\cdot\ptRm-\frac{1}{r}\frac{\partial\ptSm}{\partial\vartheta}+\pAo\cdot\ptRp   \label{eq:VI.1c}
=\frac{M-M_*}{\hbar}\,c&\cdot\ptRp\\\nonumber\\
  \frac{\partial\ptSm}{\partial r}+\frac{1}{r}\cdot\ptSm+\frac{1}{r}\frac{\partial\ptRm}{\partial\vartheta}+\pAo\cdot\ptSp   \label{eq:VI.1d}
=\frac{M-M_*}{\hbar}\,c&\cdot\ptSp\;.
  \end{align}
\end{subequations}
Here the modified wave amplitudes $\ptRpm,\,\ptSpm$ are defined in terms of the original
real-valued amplitudes $\pMRpm$ ($\Rightarrow\pRpm$) and $\pMSpm$ ($\Rightarrow\pSpm$)
\myrf{V.54a}-\myrf{V.54b} through
\begin{subequations}
  \begin{align}
  \label{eq:VI.2a}
  \ptRpm&\doteqdot\sqrt{r\sin\vartheta}\cdot\pRpm\\
  \label{eq:VI.2b}
  \ptSpm&\doteqdot\sqrt{r\sin\vartheta}\cdot\pSpm\;.
  \end{align}
\end{subequations}

Being satisfied for the moment with that electrostatic approximation, the relativistic energy functional $\tETT$ \myrf{IV.20} becomes simplified for the two {\em non-identical}\/ particles which, however, must be assumed to occupy two {\em physically equivalent quantum states}\/ \myrf{V.25b}. This implies that both mass renormalization factors $\Z_{(a)}$ become identical
\begin{equation}
  \label{eq:VI.3}
  \Z_{(1)}^2=\Z_{(2)}^2\doteqdot\tZp^2
\end{equation}
with
\begin{equation}
  \begin{split}
  \label{eq:VI.4}
  \tZp^2=\frac{1}{2}\int d^2\vec{r}&\, \left\{\big(\ptRp\big)^2+\big(\ptSp\big)^2-\big(\ptRm\big)^2-\big(\ptSm\big)^2\right\}\\
  &(d^2\vec{r}\doteqdot r\,dr\,d\vartheta)\;.
  \end{split}
\end{equation}
Furthermore, the adopted physical equivalence of both particle states implies also that
the kinetic energies $\Tkina$ \myrf{IV.26}-\myrf{IV.27} must turn out to be identical
\begin{equation}
  \label{eq:VI.5}
  \Tkine=\Tkinz\doteqdot {}^{(p)}\tTkin\;.
\end{equation}
Here the common one-particle kinetic energy $\tTkin$ splits up into a radial part ($\tilde{T}_r$) and a longitudinal part ($\tilde{T}_\vartheta$)
\begin{equation}
  \label{eq:VI.6}
  {}^{(p)\!}\tTkin={}^{(p)\!}\tilde{T}_{(r)}+{}^{(p)\!}\tilde{T}_{(\vartheta)}
\end{equation}
with the radial part being given by
\begin{equation}
  \begin{split}
  \label{eq:VI.7}
  {}^{(p)\!}\tilde{T}_{(r)}=\frac{\hbar c}{4}\int d^2\vec{r}\,
  \left\{\ptRm\cdot\frac{\partial\ptRp}{\partial r}-\ptRp\cdot\frac{\partial\ptRm}{\partial r} -\frac{\ptRp\cdot\ptRm}{r}+\ptSm\cdot\frac{\partial\ptSp}{\partial r}\right.\\
  \left.-\ptSp\cdot\frac{\partial\ptSm}{\partial r}-\frac{\ptSp\cdot\ptSm}{r}\right\}
  \end{split}
\end{equation}
and the longitudinal part by
\begin{equation}
  \label{eq:VI.8}
  {}^{(p)\!}\tilde{T}_{(\vartheta)}=\frac{\hbar c}{4}\int\frac{d^2\vec{r}}{r}\,
  \left\{\ptRm\cdot\frac{\partial\ptSp}{\partial\vartheta}-\ptSp\cdot\frac{\partial\ptRm}{\partial\vartheta} +\ptRp\cdot\frac{\partial\ptSm}{\partial\vartheta}
  -\ptSm\cdot\frac{\partial\ptRp}{\partial\vartheta}\right\}\,.
\end{equation}
And finally, the electric constraint $\NGe$ \myrf{IV.1}-\myrf{IV.2} is simplified to $\tNGe$
\begin{equation}
  \label{eq:VI.9}
  \tNGe=\ERe-\tMe c^2\equiv0
\end{equation}
with the electrostatic interaction energy $\ERe$ being given by
\begin{equation}
  \label{eq:VI.10}
  \ERe=-\frac{\hbar c}{4\pi\as}\int d^3\vec{r}\;\|\vec{\nabla}\pAo\|^2
\end{equation}
and its mass equivalent by
\begin{equation}
  \label{eq:VI.11}
  \tMe c^2=-\frac{\hbar c}{2}\int d^2\vec{r}\;\pAo\left\{\big(\ptRp\big)^2+\big(\ptSp\big)^2+\big(\ptRm\big)^2+\big(\ptSm\big)^2\right\}\,.
\end{equation}

Thus, putting now together all energy contributions yields for the relativistic energy functional $\tETT$ \myrf{IV.20} with \myrf{IV.23} in the {\em electrostatic approximation}
\begin{equation}
  \label{eq:VI.12}
  \tETT\Rightarrow\tilde{E}_{\rm\{T\}}=2Mc^2\cdot\tZp^2+4{}^{(p)\!}\tTkin+\ERe+2\lD\cdot\tND+\lGe\cdot\tNGe\;.
\end{equation}
Here the Lagrangean multiplier $\lD$ is given by the sum of the individual multipliers
\begin{equation}
  \label{eq:VI.13}
  \lD\doteqdot\lDe+\lDz=(M_1-M_2)c^2=-2M_*c^2\;,
\end{equation}
cf. \myrf{IV.22a}; and both normalization constraints $\NDa$ \myrf{III.49} collapse to a single one just on account of the physical equivalence of both quantum states
\begin{equation}
  \label{eq:VI.14}
  \NDe=\NDz\doteqdot\tND
\end{equation}
with
\begin{equation}
  \label{eq:VI.15}
  \tND\doteqdot\frac{1}{2}\int d^2\vec{r}\,\left\{\big(\ptRp\big)^2+\big(\ptSp\big)^2+\big(\ptRm\big)^2+\big(\ptSm\big)^2\right\}-1\equiv0\;.
\end{equation}

Indeed, after all it is now a very satisfying consistency check to deduce the relativistic para-system \myrf{VI.1a}-\myrf{VI.1d} from the {\em principle of minimal energy}\/ \myrf{IV.21} with the functional $\tETT$ being given by its present electrostatic approximation $\tilde{E}_{\rm\{T\}}$ \myrf{VI.12}. But for the subsequent construction of (approximate) solutions one should keep in mind those boundary conditions which are needed for the variational deduction:
\begin{subequations}
  \begin{align}
  \label{eq:VI.16a}
  \lim_{r\rightarrow0}\left(r\cdot\ptRp\cdot\ptRm\right)=\lim_{r\rightarrow0}\left(r\cdot\ptSp\cdot\ptSm\right)&=0\\
  \label{eq:VI.16b}
  \left[\ptRp\cdot\ptSm\right]_{\vartheta=\pi}-\left[\ptRp\cdot\ptSm\right]_{\vartheta=0}&=0\\
  \label{eq:VI.16c}
  \left[\ptRm\cdot\ptSp\right]_{\vartheta=\pi}-\left[\ptRm\cdot\ptSp\right]_{\vartheta=0}&=0\;.
  \end{align}
\end{subequations}
Finally, in order to close the truncated eigenvalue system \myrf{VI.1a}-\myrf{VI.1d}, one has to add only the electric Poisson equation \myrf{V.21} for the potential $\pAo$
\begin{equation}
  \label{eq:VI.17}
  \Delta\pAo=-4\pi\as\,\pko=-\as\,\frac{\big(\ptRp\big)^2+\big(\ptSp\big)^2+\big(\ptRm\big)^2+\big(\ptSm\big)^2}{r\sin\vartheta}
\end{equation}
since the azimuthal magnetic potential $\pAphi$ is to be neglected in the electrostatic approximation. Of course in the true spirit of the {\em principle of minimal energy}\/, this Poisson equation \myrf{VI.17} must be deducible also from the approximative energy functional $\tilde{E}_{\rm\{T\}}$ \myrf{VI.12}, namely through its extremalization with respect to the electric potential $\pAo$!
\clearpage
\vspace{3ex}
\begin{center}
\emph{\large Spherically Symmetric Approximation}
\end{center}

Despite the simplifying effect of the electrostatic approximation, the resulting
eigenvalue system \myrf{VI.1a}-\myrf{VI.1d} plus the Poisson equation \myrf{VI.17} is
still a too complicated system of coupled equations in order to find the exact solutions
thereof. Consequently, it becomes necessary to think of a further approximation step which
should refer to the gauge potential $\pAo(r,\vartheta)$ as a solution of the electrostatic
Poisson equation \myrf{VI.17}. Indeed, what obviously does complicate things with that
equation is the fact that its solutions $\pAo$ will depend in general on both variables
$r,\,\vartheta$ even if all the wave amplitudes $\ptRpm,\,\ptSpm$ would be assumed to be
spherically symmetric (i.\,e. not being dependent on the spherical polar angle
$\vartheta$). Therefore it suggests itself to adopt also the electric potential
$\pAo(r,\vartheta)$ to depend exclusively on the radial variable $r$ but not on the angle
$\vartheta$, i.\,e. the approximation is the following (``\/{\em spherically symmetric
  approximation}\/'')
\begin{equation}
  \label{eq:VI.18}
  \pAo(r,\vartheta)\Rightarrow\ppAo(r)\;.
\end{equation}

But observe here that our present assumption of spherical symmetry \myrf{VI.18} does
concern only the gauge potential $\pAo$, not the wave amplitudes $\ptRpm,\,\ptSpm$
themselves! Indeed, we admit these amplitudes to depend on the angle $\vartheta$ in the
following way:
\begin{subequations}
  \begin{align}
  \label{eq:VI.19a}
  \ptRpm(r,\vartheta)&=\sRpm(r)\cdot f_R(\vartheta)\\
  \label{eq:VI.19b}
  \ptSpm(r,\vartheta)&=\sSpm(r)\cdot f_S(\vartheta)\;.
  \end{align}
\end{subequations}
The angular functions $f_R(\vartheta)$ and $f_S(\vartheta)$ of this product ansatz must now be determined in such a way that the four partial differential equations \myrf{VI.1a}-\myrf{VI.1d} become reduced to ordinary differential equations with respect to the radial variable $r$. It is easy to see that this goal can be attained by demanding that both angular functions $f_R(\vartheta)$ and $f_S(\vartheta)$ obey the following coupled system
\begin{subequations}
  \begin{align}
  \label{eq:VI.20a}
  \frac{df_R(\vartheta)}{d\vartheta}&=\lP\cdot f_S(\vartheta)\\
  \label{eq:VI.20b}
  \frac{df_S(\vartheta)}{d\vartheta}&=-\lP\cdot f_R(\vartheta)\;,
  \end{align}
\end{subequations}
where $\lP$ is some real-valued constant.

Actually, by this demand the original system \myrf{VI.1a}-\myrf{VI.1d} is simplified to the following coupled set of ordinary differential equations:
\begin{subequations}
  \begin{align}
  \label{eq:VI.21a}
  \frac{d\sRp(r)}{dr}-\frac{\lP}{r}\cdot\sSp(r)-\ppAo(r)\cdot\sRm(r)&=\frac{M+M_*}{\hbar}\,c\cdot\sRm(r)\\
  \label{eq:VI.21b}
  \frac{d\sSp(r)}{dr}-\frac{\lP}{r}\cdot\sRp(r)-\ppAo(r)\cdot\sSm(r)&=\frac{M+M_*}{\hbar}\,c\cdot\sSm(r)\\
  \label{eq:VI.21c}
  \frac{d\sRm(r)}{dr}+\frac{1}{r}\cdot\sRm(r)+\frac{\lP}{r}\cdot\sSm(r)+\ppAo(r)\cdot\sRp(r) &=\frac{M-M_*}{\hbar}\,c\cdot\sRp(r)\\
  \label{eq:VI.21d}
  \frac{d\sSm(r)}{dr}+\frac{1}{r}\cdot\sSm(r)+\frac{\lP}{r}\cdot\sRm(r)+\ppAo(r)\cdot\sSp(r) &=\frac{M-M_*}{\hbar}\,c\cdot\sSp(r)\;.
  \end{align}
\end{subequations}
Clearly, such a system of ordinary differential equations needs further
discussion~\cite{MaSo2} since the {\em same}\/ eigenvalue $M_*$ emerges in each equation!
In view of this fact one will suppose that the four solutions $\sRpm(r),\,\sSpm(r)$ can
not be too different and thus one tries the identifications
\begin{equation}
  \label{eq:VI.22}
  \sRpm(r)\equiv\sSpm(r)\doteqdot\tPhipm(r)\;.
\end{equation}
And indeed, by this assumption the four eigenvalue equations \myrf{VI.21a}-\myrf{VI.21d} become contracted without any contradiction to only two equations
\begin{subequations}
  \begin{align}
  \label{eq:VI.23a}
  \frac{d\tPhip(r)}{dr}-\frac{\lP}{r}\cdot\tPhip(r)-\ppAo(r)\cdot\tPhim(r)&=\frac{M+M_*}{\hbar}\,c\cdot\tPhim(r)\\
  \label{eq:VI.23b}
  \frac{d\tPhim(r)}{dr}+\frac{\lP+1}{r}\cdot\tPhim(r)+\ppAo(r)\cdot\tPhip(r) &=\frac{M-M_*}{\hbar}\,c\cdot\tPhip(r)\;.
  \end{align}
\end{subequations}
These two equations, together with the spherically symmetric approximation of the electric Poisson equation \myrf{VI.17}, constitutes now the eigenvalue problem of para-positronium {\em in the spherically symmetric and electrostatic approximation}\/.

The spherically symmetric form of the Poisson equation will readily be discussed, but first one should think about the possible values of the constant $\lP$ appearing in the angular system \myrf{VI.20a}-\myrf{VI.20b}. For this purpose, one converts this first-order system \myrf{VI.20a}-\myrf{VI.20b} into a second-order system by differentiating once more
\begin{subequations}
  \begin{align}
  \label{eq:VI.24a}
  \frac{d^2f_R(\vartheta)}{d\vartheta^2}&=-\lP^2\cdot f_R(\vartheta)\\
  \label{eq:VI.24b}
  \frac{d^2f_S(\vartheta)}{d\vartheta^2}&=-\lP^2\cdot f_S(\vartheta)\;.
  \end{align}
\end{subequations}
The solutions of this standard problem are well known and are written here as
\begin{subequations}
  \begin{align}
  \label{eq:VI.25a}
  f_R(\vartheta)&=\cos(\lP\cdot\vartheta)\\
  \label{eq:VI.25b}
  f_S(\vartheta)&=-\sin(\lP\cdot\vartheta)\;,
  \end{align}
\end{subequations}
or alternatively
\begin{subequations}
  \begin{align}
  \label{eq:VI.26a}
  f_R(\vartheta)&=\sin(\lP\cdot\vartheta)\\
  \label{eq:VI.26b}
  f_S(\vartheta)&=\cos(\lP\cdot\vartheta)\;.
  \end{align}
\end{subequations}

But with the general shape of the wave functions being known, i.\,e.
\begin{subequations}
  \begin{align}
  \label{eq:VI.27a}
  \ptRpm(r,\vartheta)&=\tPhipm(r)\cdot f_R(\vartheta)\\
  \label{eq:VI.27b}
  \ptSpm(r,\vartheta)&=\tPhipm(r)\cdot f_S(\vartheta)\;,
  \end{align}
\end{subequations}
one can reconsider the former boundary conditions \myrf{VI.16b}-\myrf{VI.16c} which yields the requirement
\begin{equation}
  \label{eq:VI.28}
  \sin(2\pi\lP)\stackrel{!}{=}0\;,
\end{equation}
and this can be obeyed by fixing the value of $\lP$ by the \mbox{(half-)}integers
\begin{equation}
  \label{eq:VI.29}
  \lP=0,\left(\frac{1}{2}\right),1,\left(\frac{3}{2}\right),2,\left(\frac{5}{2}\right),3\ldots
\end{equation}

In this context it is also interesting to remark that, through this disposal \myrf{VI.29}
of $\lP$, the para-current $\vec{k}_p$ \myrf{V.60} remains finite on the $z$-axis
(i.\,e. $\vartheta=0$ and $\vartheta=\pi$). Indeed, substituting the present product form
\myrf{VI.27a}-\myrf{VI.27b} of the wave amplitudes back into the general form \myrf{V.57c}
of the Dirac current, one finds
\begin{equation}
  \label{eq:VI.30}
  \pk_\phi(r,\vartheta)=\pm\frac{\tPhip(r)\cdot\tPhim(r)}{2\pi r}\cdot\frac{\sin\big[(2\lP+1)\cdot\vartheta\big]}{\sin\vartheta}\;.
\end{equation}
(The positive/negative sign refers here to the cases (VI.25/26)). This result says that the para-current $\vec{k}_p$ remains finite but nevertheless is singular on the $z$-axis; however, as we will readily see, such kind of singularity does not spoil the energy functional, even if the (presently neglected) magnetic interactions are included. The alternating flow directions of the para-current $\vec{k}_p$ around the $z$-axis are sketched in fig.~1.

The energy functional $\tilde{E}_{\{T\}}$ \myrf{VI.12} is needed now in the context of the spherically symmetric approximation because it is not obvious how the corresponding approximate form of the electric Poisson equation \myrf{VI.17} should look like. In order to deduce this from the {\em principle of minimal energy}\/, one substitutes those spherically symmetric ans\"atze for the potential $\pAo$ \myrf{VI.18} and for the wave functions \myrf{VI.19a}-\myrf{VI.19b} into the electrostatic approximation $\tEsT$ \myrf{VI.12} of the original energy functional $\tETT$ and thereby finds the following results for its constituents:
\begin{enumerate}[\bf (i)]
\item the {\em mass renormalization factor}\/ $\tilde{\Z}_\mathcal{P}^2$ \myrf{VI.4} emerges as
\begin{equation}
  \label{eq:VI.31}
  \tilde{\Z}_\mathcal{P}^2\Rightarrow\tilde{\Z}_\Phi^2=\frac{\pi}{2}\int_0^\infty dr\,r\left\{\big(\tPhip(r)\big)^2-\big(\tPhim(r)\big)^2\right\}
\end{equation}
\item next, the radial and longitudinal kinetic energies ${}^{(p)}\tilde{T}_{(r)}$  \myrf{VI.7} and ${}^{(p)}\tilde{T}_{(\vartheta)}$ \myrf{VI.8} become simplified to
\begin{subequations}
  \begin{align}
  \label{eq:VI.32a}
  {}^{(p)}\tilde{T}_{(r)}&\Rightarrow{}^{[p]}\tilde{T}_{(r)}=\frac{\pi}{4}\,\hbar c\int_0^\infty dr\,r\,\left\{\tPhim(r)\cdot\frac{d\tPhip(r)}{dr}-\tPhip(r)\cdot\frac{d\tPhim(r)}{dr}-\frac{\tPhip(r)\cdot\tPhim(r)}{r}\right\}\\
  \label{eq:VI.32b}
  {}^{(p)}\tilde{T}_{(\vartheta)}&\Rightarrow{}^{[p]}\tilde{T}_{(\vartheta)}=-\frac{\pi}{2}\,\hbar c\lP\int_0^\infty dr\;\tPhip(r)\cdot\tPhim(r)
  \end{align}
\end{subequations}
\item the electric Poisson constraint $\tNGe$ \myrf{VI.9} is left formally unchanged, i.\,e.
\begin{equation}
  \label{eq:VI.33}
  \tNGe\Rightarrow\tNGee=\ERee-\tMee c^2\equiv0,
\end{equation}
with the electrostatic interaction energy $\ERe$ \myrf{VI.10} appearing now as
\begin{equation}
  \label{eq:VI.34}
  \ERe\Rightarrow\ERee=-\frac{\hbar c}{\as}\int_0^\infty dr\,\left(r\cdot\frac{d\ppAn}{dr}\right)^2
\end{equation}
and analogously its mass equivalent \myrf{VI.11}
\begin{equation}
  \label{eq:VI.35}
  \tMe c^2\Rightarrow\tMee c^2=-\frac{\pi}{2}\,\hbar c\int_0^\infty dr\,r\,\ppAn(r)\cdot\left\{\left(\tPhip(r)\right)^2+\left(\tPhim(r)\right)^2\right\}
\end{equation}
\item and finally the normalization constraint \myrf{VI.14} reads now
\begin{equation}
 \label{eq:VI.36}
 \tND\Rightarrow\tNPhi\doteqdot\frac{\pi}{2}\int_0^\infty dr\,r\,\left\{\left(\tPhip(r)\right)^2+\left(\tPhim(r)\right)^2\right\}-1\equiv0\;.
\end{equation}
\end{enumerate}

Putting now together all four contributions {\bf (i)}-{\bf (iv)} yields the desired spherically symmetric approximation ($\tilde{E}_{[\Phi]}$, say) of the electrostatic approximation $\tEsT$ \myrf{VI.12} as
\begin{equation}
  \label{eq:VI.37}
  \tilde{E}_{[\Phi]}= 2\,Mc^2\cdot\tilde{\Z}_\Phi^2+4{}^{[p]}\tilde{T}_\textrm{kin}+\ERee+2\lD\cdot\tNPhi+\lGe\cdot\tNGee\;.
\end{equation}
Here it is again a nice exercise to convince oneself of the fact that the spherically symmetric eigenvalue equations \myrf{VI.23a}-\myrf{VI.23b} actually do arise as the extremal equations due to the present functional $\tilde{E}_{[\Phi]}$ \myrf{VI.37}! But furthermore, the extremalization of $\tilde{E}_{[\Phi]}$ with respect to the electric potential $\ppAn$ yields the wanted spherically symmetric form of the electric Poisson equation \myrf{VI.17}:
\begin{equation}
  \label{eq:VI.38}
  \frac{d^2\ppAn}{dr^2}+\frac{2}{r}\frac{d\ppAn}{dr}= -\frac{\pi}{2}\,\as\,\frac{\left(\tPhip(r)\right)^2+\left(\tPhim(r)\right)^2}{r}\;,
\end{equation}
with the standard solution
\begin{equation}
  \label{eq:VI.39}
  \ppAn(r)=\frac{\as}{8}\int\frac{d^3\vec{r}\!\;'\,}{r'}\, \frac{\left(\tPhip(r')\right)^2+\left(\tPhim(r')\right)^2}{\|\vec{r}-\vec{r}\!\;'\,\|}\;.
\end{equation}
As a consistency check, one easily deduces hereof the asymptotic Coulomb form \myrf{III.60} of the electric potential $\ppAn(r)$
\begin{equation}
  \label{eq:VI.40}
  \lim_{r\rightarrow\infty}\ppAn(r)=\frac{\as}{r}
\end{equation}
by simply regarding the normalization condition in its spherically symmetric form \myrf{VI.36}.

Summarizing, the combination of the electrostatic and spherically symmetric approximation
consists of the mass eigenvalue equations \myrf{VI.23a}-\myrf{VI.23b} plus the electric
Poisson equation \myrf{VI.38}. If the (exact or approximate) solutions of this eigenvalue
problem could be attained, one would substitute these solutions into the associated energy
functional $\tEPhi$ \myrf{VI.37} and would thus obtain the corresponding energy spectrum
(i.\,e. the singlet spectrum) of para-positronium. But although the corresponding energy
levels would be still of relativistic nature, the corresponding relativistic effects would
not be of much help! The reason is that the {\em neglected}\/ magnetic and anisotropic
effects are in general of the same order of magnitude as the relativistic effects, so that
the numerical usefulness would be spoiled in any case. Therefore it is more meaningful to
finally do the last step of the approximations, namely by turning to the non-relativistic
domain.

\vspace{3ex}
\begin{center}
\emph{\large Non-Relativistic Approximation}
\end{center}

Though the two preceding approximations (i.\,e. the electrostatic and the spherically
symmetric one) did considerably simplify the positronium eigenvalue problem, a further
rather technical simplification (i.\,e. the transition to the non-relativistic domain) is
highly welcome when one is satisfied with a first rough estimate of the positronium
spectrum. The point of departure for such a purpose is the eigenvalue system
\myrf{VI.23a}-\myrf{VI.23b} in combination with the Poisson equation \myrf{VI.38} for the
electric interaction potential $\ppAn$ which has already been assumed to be spherically
symmetric. Now in order to deduce the non-relativistic approximation of those eigenvalue
equations \myrf{VI.23a}-\myrf{VI.23b} one resolves (approximately) the first equation
\myrf{VI.23a} with respect to the ``negative'' wave amplitude $\tPhim(r)$ in terms of the
``positive'' amplitude $\tPhip(r)$, i.\,e. one puts
\begin{equation}
  \label{eq:VI.41}
  \tPhim(r)\simeq\frac{\hbar}{2Mc}\left\{\frac{d\tPhip(r)}{dr}-\frac{\lP}{r}\cdot\tPhip(r)\right\}\,,
\end{equation}
and then one substitutes this into the second equation \myrf{VI.23b}. This then yields the following Schr\"odinger-like equation for the non-relativistic approximation ($\tPhi(r)$, say) of the ``positive'' amplitude $\tPhip(r)$:
\begin{equation}
  \label{eq:VI.42}
  -\frac{\hbar^2}{2M}\left\{\frac{d^2\tPhi(r)}{dr^2}+\frac{1}{r}\frac{d\tPhi(r)}{dr}\right\} +\frac{\hbar^2}{2M}\frac{\lP^2}{r^2}\,\tPhi(r)-\hbar c\ppAn(r)\cdot\tPhi(r)=E_*\cdot\tPhi(r)
\end{equation}
where the Schr\"odinger eigenvalue $E_*$ emerges as the difference of the mass eigenvalue~$M_*$ and the rest mass $M$
\begin{equation}
  \label{eq:VI.43}
  E_*\doteqdot(M_*-M)c^2\;.
\end{equation}

Finally, in order to close this eigenvalue equation \myrf{VI.42}, one has to add the non-relativistic approximation of the electric Poisson equation \myrf{VI.38}. This, however, is a simple thing because the general form of the Poisson equation survives the transition to the non-relativistic domain so that one merely has to resort to the non-relativistic approximation for the charge density $\pko(r)$. But since the latter approximation merely consists in neglecting the ``negative'' wave amplitude $\tPhim(r)$, the non-relativistic approximation of the electric Poisson equation \myrf{VI.38} is immediately written down as
\begin{equation}
  \label{eq:VI.44}
  \frac{d^2\ppAn}{dr^2}+\frac{2}{r}\frac{d\ppAn}{dr}= -\frac{\pi}{2}\,\as\,\frac{\left(\tPhi(r)\right)^2}{r}\;.
\end{equation}
Thus the non-relativistic eigenvalue problem for para-positronium ultimately consists of both equations \myrf{VI.42} and \myrf{VI.44}.

However, notice here that the ``Schr\"odinger eigenvalue'' $E_*$ \myrf{VI.42}-\myrf{VI.43} is a one-particle quantity and therefore cannot be identified with the (non-relativistic) RST energy ($\EEPhi$, say) of our two-particle problem. Rather, in order to get this two-particle energy $\EEPhi$ due to the solutions of the present one-particle problem \myrf{VI.42}-\myrf{VI.44}, one has to substitute these solutions into the non-relativistic descendant ($\tEEPhi$, say) of its relativistic predecessor $\tEPhi$ \myrf{VI.37}. Thus it becomes now necessary to deduce that non-relativistic functional $\tEEPhi$ which then simultaneously provides us also with the possibility to check whether the non-relativistic system of eigenvalue and Poisson equations \myrf{VI.42}-\myrf{VI.44} does really represent the extremal equations of the wanted non-relativistic functional $\tEEPhi$.

But of course, the desired non-relativistic version $\tEEPhi$ of the relativistic $\tEPhi$ \myrf{VI.37} will simply consist of the non-relativistic constituents of $\tEPhi$ as they are collected by equations \myrf{VI.31}-\myrf{VI.36}. Therefore it is sufficient to simply list up those non-relativistic versions which come about by eliminating the ``negative'' wave amplitude $\tPhim(r)$ by means of the equation \myrf{VI.41}:
\begin{enumerate}[\bf (i)]
\item {\em mass renormalization \myrf{VI.31}:}
\begin{equation}
  \label{eq:VI.45}
  2\,Mc^2\tilde{\Z}_\Phi^2\Rightarrow 2\,Mc^2-2\big({}^{(r)}E_\textrm{kin}+{}^{[\vartheta]}E_\textrm{kin}\big)+2\pi\,\frac{\hbar^2}{2M}\,\lP\int_0^\infty dr\;\tPhi(r)\cdot\frac{d\tPhi(r)}{dr}
\end{equation}
\item {\em radial and longitudinal kinetic energy \myrf{VI.32a}-\myrf{VI.32b}:}
\begin{subequations}
  \begin{align}
  \label{eq:VI.46a}
  {}^{[p]}\tilde{T}_{(r)}&\Rightarrow\frac{\pi}{2}\frac{\hbar^2}{2M}\int_0^\infty dr\,r\,\left(\frac{d\tPhi(r)}{dr}\right)^2\doteqdot{}^{(r)}E_\textrm{kin}\\
  \label{eq:VI.46b}
  {}^{[p]}\tilde{T}_{(\vartheta)}&\Rightarrow {}^{[\vartheta]}E_\textrm{kin}-\frac{\pi}{2}\frac{\hbar^2}{2M}\,\lP\int_0^\infty dr\;\tPhi(r)\cdot\frac{d\tPhi(r)}{dr}\\
  \label{eq:VI.46c}
  &\big(\,{}^{[\vartheta]}E_\textrm{kin}\doteqdot\frac{\pi}{2}\frac{\hbar^2}{2M}\,\lP^2\int_0^\infty dr\,\frac{\left(\tPhi(r)\right)^2}{r}\,\big)
  \end{align}
\end{subequations}
\item {\em electric Poisson identity \myrf{VI.33}-\myrf{VI.35}:}
\begin{equation}
  \label{eq:VI.47}
  \tNGee\Rightarrow\tNNGee\doteqdot\ERee-\tMMee c^2=0
\end{equation}
\begin{equation}
  \label{eq:VI.48}
  \tMee c^2\Rightarrow\tMMee c^2\doteqdot-\frac{\pi}{2}\,\hbar c\int_0^\infty dr\,r\;\ppAn\cdot\left(\tPhi(r)\right)^2
\end{equation}
\item {\em normalization condition \myrf{VI.36}:}
\begin{equation}
  \label{eq:VI.49}
  \tilde{N}_\Phi\Rightarrow\tNNPhi\doteqdot\frac{\pi}{2}\int_0^\infty dr\,r\,\left(\tPhi(r)\right)^2-1=0\;.
\end{equation}
\end{enumerate}
Collecting now all these partial energies (with omission of the rest mass $2\,Mc^2$) into the desired functional $\tEEPhi$ as the non-relativistic limit of $\tilde{E}_{[\Phi]}$ one arrives at the following result
\begin{equation}
  \label{eq:VI.50}
  \tEPhi\Rightarrow\tEEPhi= 2\big({}^{(r)}E_\textrm{kin}+{}^{[\vartheta]}E_\textrm{kin}\big)+\ERee+2\lS\cdot\tNNPhi+\lGe\cdot\tNNGee\;,
\end{equation}
where the Lagrangean multiplier $\lS$ is related to the one-particle Schr\"odinger eigenvalue $E_*$ \myrf{VI.43} through
\begin{equation}
  \label{eq:VI.51}
  \lS=-E_*\;.
\end{equation}
Indeed this is a very satisfying result because it provides an instructive exercise to
verify that the non-relativistic eigenvalue system \myrf{VI.42}-\myrf{VI.44} actually
consists of just the extremal equations due to the functional $\tEEPhi$ \myrf{VI.50}!
Observe also that for $\lP=1$ the present functional $\tEEPhi$ \myrf{VI.50} becomes
identical to the functional $\tilde{\mathbb{E}}_{[T]}^{(0)}$ of ref. \cite{MaSo2} (see
there the equations (V.109)-(V.113), which served to treat the first excited state ${\sf
  2\,^1P_1}$ of positronium).

This non-relativistic functional $\tEEPhi$ owns now some interesting properties which are worth while to be considered in some detail. First, notice that both the mass renormalization factor $\tilde{\Z}_\Phi^2$ \myrf{VI.45} and the longitudinal kinetic energy ${}^{[p]}\tilde{T}_{(\vartheta)}$ \myrf{VI.46b} contain the same boundary term, albeit only up to sign
\begin{equation}
  \label{eq:VI.52}
  2\pi\frac{\hbar^2}{2M}\,\lP\int_0^\infty dr\;\tPhi(r)\cdot\frac{d\tPhi(r)}{dr}=-\pi\,\frac{\hbar^2}{2M}\,\lP\cdot\big(\tPhi(0)\big)^2\;.
\end{equation}
It is true, this term cancels for the construction of the non-relatvistic functional $\tEEPhi$ \myrf{VI.50}; but if one wishes to insist on both non-relativistic contributions \myrf{VI.45} and \myrf{VI.46b} being individually independent of boundary terms, one has to demand that the wave amplitude $\tPhi(r)$ must vanish at the origin (i.\,e. $\tPhi(0)=0$) whenever the quantum number of angular momentum $\lP$ is non-zero, cf. \myrf{VI.29}. Or the other way round: it is exclusively for $\lP=0$ that one can admit $\tPhi(0)\neq0$; this circumstance must be regarded also for the construction of variational solutions (see below).

Next, observe that the mass renormalization term \myrf{VI.45} contains minus twice the
one-particle kinetic energy $E_\textrm{kin}$
($={}^{(r)}E_\textrm{kin}+{}^{[\vartheta]}E_\textrm{kin}$), but since the one-particle
kinetic energy ${}^{[p]}\tilde{T}_\textrm{kin}$ enters the relativistic functional
$\tEPhi$ \myrf{VI.37} with a pre-factor of four, the combination of both effects yields
the double one-particle energy $E_\textrm{kin}$
($={}^{(r)}E_\textrm{kin}+{}^{[\vartheta]}E_\textrm{kin}$) for the non-relativistic
functional $\tEEPhi$ \myrf{VI.50}, as it must naturally be expected for a two-particle
system. This effect has been observed already in some of the preceding treatises
\cite{MaSo3}. Summarizing, the non-relativistic functional $\tEEPhi$ \myrf{VI.50} appears
physically plausible and logically consistent so that its practical usefulness may now be
tested in a concrete example.

\vspace{3ex}
\begin{center}
\emph{\large Non-Relativistic Para-Spectrum}
\end{center}

The non-relativistic functional $\tEEPhi$ \myrf{VI.50} may now be taken as the point of departure for building up an approximation formalism for the non-relativistic energy level system of para-positronium. For this purpose, one will select some trial function $\tPhi(r)$, obeying the normalization condition \myrf{VI.49}; and furthermore one will then solve the electric Poisson equation \myrf{VI.44} for the potential $\ppAn(r)$ which is due to the selected trial function $\tPhi(r)$. As a consequence, the electric Poisson identity \myrf{VI.47} will thus be obeyed automatically and therefore the functional $\tEEPhi$ \myrf{VI.50} becomes reduced to its physical part $\EEivPhi$ \myrf{IV.23} which reads for the present situation
\begin{equation}
  \label{eq:VI.53}
  \EEivPhi=2\big({}^{(r)}E_\textrm{kin}+{}^{[\vartheta]}E_\textrm{kin}\big)+\ERee\;.
\end{equation}
This yields essentially an ordinary function $\EEivbnk$ of the variational parameters $\{\beta,\nu_k\}$ being contained in the selected trial function $\tPhi(r)$; and thus the extremal values of this energy function $\EEivbnk$, which are to be determined by
\begin{subequations}
  \begin{align}
  \label{eq:VI.54a}
  \frac{\partial\EEivbnk}{\partial\beta}=0\\
  \label{eq:VI.54b}
  \frac{\partial\EEivbnk}{\partial\nu_k}=0\;,
  \end{align}
\end{subequations}
will then represent approximately the desired energy levels of para-positronium.

The general structure of this level system is then obviously the following: since the
energy function $\EEivbnk$ depends also on the quantum number $\lP$ \myrf{VI.46c} of
angular momentum, one expects for any fixed value of $\lP=0,1,2,3,\ldots$ a finite number
of stationary points \myrf{VI.54a}-\myrf{VI.54b} of the function $\EEivbnk$. This latter
set of excited states for any fixed $\lP$ may be enumerated by the integers $\nP$ with
$\nP=\lP+1,\lP+2,\lP+3,...$ so that any para-level is specified by the two quantum numbers
$\{\nP,\lP\}$, see fig.~2. Naturally, for such a kind of level system there must arise the
question of {\em degeneracy}\/, namely whether or not the energy of the levels due to the
same $\nP$ but different $\lP$ turns out to be the same? Clearly, this question could be
answered with sureness only if the exact solutions of the eigenvalue problem
\myrf{VI.42}-\myrf{VI.44} could be found which of course is not possible; the subsequent
approximative treatment of the para-level system with only two variational parameters can
yield only the ``groundstates''~$(\nP=\lP+1)$ for any~$\lP$.

As a simple example of the proposed approximation scheme we select a trial function
$\tPhi(r)$ which contains only two (real-valued) variational parameters $\beta$ and $\nu$,
i.\,e. we try
\begin{equation}
  \label{eq:VI.55}
  \tPhi(r)=\Phi_*r^\nu{\rm e}^{-\beta r}\;.
\end{equation}
Here, the normalization constant $\Phi_*$ must be fixed by the constraint of wave function normalization \myrf{VI.49} which yields
\begin{equation}
  \label{eq:VI.56}
  \Phi_*^2=\frac{2}{\pi}\cdot\frac{(2\beta)^{2\nu+2}}{\Gamma(2\nu+2)}\;,
\end{equation}
so that our trial function $\tPhi(r)$ \myrf{VI.55} may be recast to a dimensionless form $\tPhi(y)$
\begin{equation}
  \label{eq:VI.57}
  \tPhi(r)=\sqrt{\frac{2}{\pi}}\cdot\frac{2\beta}{\sqrt{\Gamma(2\nu+2)}}\,\tPhi(y)
\end{equation}
with
\begin{equation}
  \label{eq:VI.58}
  \tPhi(y)=y^\nu{\rm e}^{-\frac{y}{2}}\;.
\end{equation}
Here the dimensionless variable $y$ is defined through
\begin{equation}
  \label{eq:VI.59}
  y=2\beta r\;,
\end{equation}
and the dimensionless form of the normalization condition \myrf{VI.49} appears now as
\begin{equation}
  \label{eq:VI.60}
  \frac{1}{\Gamma(2\nu+2)}\int_0^\infty dy\;y^{2\nu+1}{\rm e}^{-y}=1\;.
\end{equation}

The next step must now refer to the calculation of the first term of the functional $\EEivPhi$ \myrf{VI.53}, i.\,e. the kinetic energy $E_\textrm{kin}$ ($={}^{(r)}E_\textrm{kin}+{}^{[\vartheta]}E_\textrm{kin}$). The radial part hereof is defined quite generally by equation \myrf{VI.46a}, which adopts here for the chosen trial function $\tPhi(r)$ \myrf{VI.57}-\myrf{VI.60} the following form
\begin{equation}
  \label{eq:VI.61}
  {}^{(r)}E_\textrm{kin}=\frac{\hbar^2}{2M}\,(2\beta)^2\cdot{}^{(r)}\varepsilon_\textrm{kin}(\nu)\;,
\end{equation}
with the (dimensionless) kinetic function ${}^{(r)}\varepsilon_\textrm{kin}(\nu)$ for the radial motion being given by
\begin{equation}
  \label{eq:VI.62}
  {}^{(r)}\varepsilon_\textrm{kin}(\nu)\doteqdot\frac{1}{\Gamma(2\nu+2)}\int_0^\infty dy\,\left(\frac{d\tPhi(y)}{dy}\right)^2=\frac{1}{4(2\nu+1)}\;.
\end{equation}
Analogously, the longitudinal part ${}^{[\vartheta]}E_\textrm{kin}$ \myrf{VI.46c} of the kinetic energy is found to look like
\begin{equation}
  \label{eq:VI.63}
  {}^{[\vartheta]}E_\textrm{kin}=\frac{\hbar^2}{2M}\,(2\beta)^2\lP^2\cdot{}^{[\vartheta]}\varepsilon_\textrm{kin}(\nu)
\end{equation}
where the kinetic function ${}^{[\vartheta]}\varepsilon_\textrm{kin}(\nu)$ for the longitudinal motion turns out as
\begin{equation}
  \label{eq:VI.64}
  {}^{[\vartheta]}\varepsilon_\textrm{kin}(\nu)\doteqdot\frac{1}{\Gamma(2\nu+2)}\int_0^\infty\frac{dy}{y}\; \tPhi(y)^2=\frac{1}{2\nu(2\nu+1)}\;.
\end{equation}
Thus the total kinetic energy $E_\textrm{kin}$ is found as
\begin{equation}
  \begin{split}
  \label{eq:VI.65}
  E_\textrm{kin}={}^{(r)}E_\textrm{kin}+{}^{[\vartheta]}E_\textrm{kin}
  &=\frac{\hbar^2}{2M}\,(2\beta)^2\big\{{}^{(r)}\varepsilon_\textrm{kin}(\nu) +\lP^2\cdot{}^{[\vartheta]}\varepsilon_\textrm{kin}(\nu)\big\}\\
  &\doteqdot\frac{e^2}{2\aB}\cdot(2\aB\beta)^2\cdot\varepsilon_\textrm{kin}(\nu)\\
  (\aB=\frac{\hbar^2}{Me^2}&\;...\;\mbox{Bohr radius})
  \end{split}
\end{equation}
with the {\em kinetic function}\/ $\varepsilon_\textrm{kin}(\nu)$ for the total motion being given simply by the sum of its radial and longitudinal parts, i.\,e.
\begin{equation}
  \label{eq:VI.66}
  \varepsilon_\textrm{kin}(\nu)= {}^{(r)}\varepsilon_\textrm{kin}(\nu)+\lP^2\cdot{}^{[\vartheta]}\varepsilon_\textrm{kin}(\nu)= \frac{1}{2\nu+1}\left(\frac{1}{4}+\frac{\lP^2}{2\nu}\right)\,.
\end{equation}

Let us remark here that this result \myrf{VI.65}-\myrf{VI.66} for the kinetic energy $E_\textrm{kin}$ has been discussed extensively for the special cases $\lP=0,1$ in two preceding papers: putting first $\lP=0$ and then $\nu=0$, one has from \myrf{VI.66} $\varepsilon_\textrm{kin}\Rightarrow\frac{1}{4}$ and thus from \myrf{VI.65}
\begin{equation}
  \label{eq.VI.67}
  E_\textrm{kin}\Rightarrow\frac{e^2}{2\aB}\,(\aB\beta)^2
\end{equation}
which is just the equation (IV.22) of ref. \cite{MaSo} (hint: rescale here
$\aB\beta\Rightarrow\beta$ and put there $T_\mathbb{N}\Rightarrow0$ for
$\mathbb{N}=0$). Furthermore, putting here $\lP=1$ lets the present equations
\myrf{VI.65}-\myrf{VI.66} collapse to the equations (V.127)-(V.128) of
ref. \cite{MaSo2}. This provides us with the possibility to compare our subsequent results to
those of the two precedent papers.

The present calculations of the value of the kinetic energy functional
$E_\textrm{kin}={}^{(r)}E_\textrm{kin}+{}^{[\vartheta]}E_\textrm{kin}$ upon the selected
class of two-parameter trial functions $\tPhi(r)$ \myrf{VI.55} represents a somewhat
fortunate situation because the dependence of the kinetic energy on the two variational
parameters $\beta$ and $\nu$ could be expressed by very simple elementary functions
(e.\,g. $E_\textrm{kin}\sim\beta^2$, see equation \myrf{VI.65}). Such a happy chance
does not apply to the electrostatic interaction energy $\ERee$ \myrf{VI.34} which is the
second constituent of the energy functional $\EEivPhi$ \myrf{VI.53} under
consideration. Indeed, a comparable simplification occurs here only for (half-)integer
values of the variational parameter $\nu$ (i.\,e. $2\nu=0,1,2,3,...$);but for the
extremalization one wishes to deal with a continuous variable $\nu$, and this suggests to
inspect the electric Poisson identity somewhat more thoroughly, see appendix D.

Recall here that, just on account of this identity, the interaction energy $\ERee$ \myrf{VI.34}
\begin{eqnarray}
  \label{eq:VI.68}
  \ERee&=&-\frac{\hbar c}{\as}\int_0^\infty dr\,\left(r\cdot\frac{d\ppAo(r)}{dr}\right)^2 \equiv-e^2\,(2\beta)\int_0^\infty dy\;y^2\cdot\left(\frac{d\tilde{a}_\nu(y)}{dy}\right)^2\\
  \nonumber &&\hspace{5em}\big(\mbox{putting}\;\;\ppAo(r)\doteqdot2\beta\as\cdot\tilde{a}_\nu(y)\big)
\end{eqnarray}
must equal its mass equivalent $\tMMee c^2$ \myrf{VI.48}, i.\,e.
\begin{equation}
  \label{eq:VI.69}
  \tMMee c^2=-\frac{\pi}{2}\,\hbar c\int_0^\infty dr\;r\,\ppAo(r)\cdot\tilde{\Phi}(r)^2\equiv-e^2\,(2\beta)\,\frac{1}{\Gamma(2\nu+2)}\int_0^\infty dy\;y\,\tilde{a}_\nu(y)\cdot\tilde{\Phi}(y)^2\;.
\end{equation}
Consequently, the electric Poisson identity reads in dimensionless form
\begin{equation}
  \label{eq:VI.70}
  \int_0^\infty dy\;y^2\cdot\left(\frac{d\tilde{a}_\nu(y)}{dy}\right)^2=\frac{1}{\Gamma(2\nu+2)}\int_0^\infty dy\;y\,\tilde{a}_\nu(y)\cdot\tilde{\Phi}(y)^2\doteqdot\varepsilon_\textrm{pot}(\nu)
\end{equation}
which may be verified also directly through integration by parts with use of the dimensionless form of the Poisson equation \myrf{VI.44}
\begin{equation}
  \label{eq:VI.71}
  \frac{d^2\tilde{a}_\nu(y)}{dy^2}+\frac{2}{y}\frac{d\tilde{a}_\nu(y)}{dy}=-{\rm e}^{-y}\cdot\frac{y^{2\nu-1}}{\Gamma(2\nu+2)}\;.
\end{equation}

The problem is now to determine the {\em potential function}\/
$\varepsilon_\textrm{pot}(\nu)$ by one or the other of the two possibilities \myrf{VI.70},
not only for (half-)integer values of the variational parameter $\nu$ (as it has been done
in the preceding paper \cite{MaSo2}) but by treating $\nu$ as a truly continuous
parameter. Clearly it is only for this case that the extremalizing condition \myrf{VI.54b}
can be meaningful, with the energy function $\EEivbn$ being given by
\begin{equation}
  \label{eq:VI.72}
  \EEivbn=2E_\textrm{kin}+\ERee= \frac{e^2}{\aB}\left\{(2\aB\beta)^2\cdot\varepsilon_\textrm{kin}(\nu)-(2\aB\beta)\cdot\varepsilon_\textrm{pot}(\nu)\right\}\,
\end{equation}
where the energy scale is determined by the well-known {\em atomic unit}\/ (a.u.)
\begin{equation}
  \label{eq:VI.73}
  1\,{\rm a.u.}\doteqdot\frac{e^2}{2\aB}=13.605...\,[{\rm eV}]\;.
\end{equation}
The specific form of the present energy function $\EEivbn$ \myrf{VI.72} has been deduced
for the first time in the preceding papers (see equation (V.138) of ref. \cite{MaSo2}); and
it has already been pointed out that the well-known {\em virial theorem}\/ \cite{CT} is
implied by that type of energy function \myrf{VI.72}. Indeed, this can easily be verified
by simply evoking the first equilibrium condition \myrf{VI.54a} which yields for the
present situation \myrf{VI.72}
\begin{equation}
  \label{eq:VI.74}
  \frac{\partial\EEivbn}{\partial\beta}= 4e^2\left\{(2\aB\beta)\cdot\varepsilon_\textrm{kin}(\nu)-\frac{1}{2}\varepsilon_\textrm{pot}(\nu)\right\}\stackrel{!}{=}0\;.
\end{equation}
This says that the energy-minimizing value ($\beta_*$) of $\beta$ is given (for arbitrary $\nu$) by
\begin{equation}
  \label{eq:VI.75}
  2\aB\beta_*=\frac{\varepsilon_\textrm{pot}(\nu)}{2\varepsilon_\textrm{kin}(\nu)}
\end{equation}
which then yields for the kinetic energy \myrf{VI.65}
\begin{equation}
  \label{eq:VI.76}
  E_\textrm{kin}(\beta_*,\nu)\Rightarrow E_\textrm{kin}(\nu)=\frac{e^2}{2\aB}\cdot\frac{\varepsilon_\textrm{pot}^2(\nu)}{4\varepsilon_\textrm{kin}(\nu)}
\end{equation}
and analogously for the interaction energy \myrf{VI.68} or \myrf{VI.69}, resp.,
\begin{equation}
  \label{eq:VI.77}
  \ERee=\tMMee c^2\Rightarrow\EE_\textrm{pot}(\nu)=-\frac{e^2}{2\aB}\cdot\frac{\varepsilon_\textrm{pot}^2(\nu)}{\varepsilon_\textrm{kin}(\nu)}\;.
\end{equation}

Consequently, the kinetic energy $2E_\textrm{kin}(\nu)$ \myrf{VI.76} at the equilibrium value of $\beta$ equals always (minus) half of the interaction energy $\ERee$, i.\,e.
\begin{equation}
  \label{eq:VI.78}
  2\cdot E_\textrm{kin}(\nu)=-\frac{1}{2}\,\ERee=-\frac{1}{2}\,\tMMee c^2\;,
\end{equation}
which in the conventional theory holds for the Coulomb-like interactions and is generally
known as the {\em ``virial theorem''}\/ \cite{CT,GBMS}. But the amazing point is here that
this virial theorem \myrf{VI.78} does hold also in RST quite generally for the
electrostatic approximation where the interaction potential $\ppAo(r)$
\begin{equation}
  \label{eq:VI.79}
  \ppAo(r)=\frac{\as}{8}\int\frac{d^3\vec{r}\,'}{r'}\,\frac{\tilde{\Phi}(r')}{\|\vec{r}-\vec{r}\,'\|}\; \xrightarrow{(r\rightarrow\infty)}\;\frac{\as}{r}\;,
\end{equation}
as the solution of the Poisson equation \myrf{VI.44}, adopts the Coulomb form only in the asymptotic region ($r\rightarrow\infty$), {\em but not for finite distance $r$}\/! Notice also that the RST virial theorem holds for arbitrary values of the variational parameter $\nu$, not only for that value ($\nu_*$, say) which further minimalizes the energy function $\EEivbn$ \myrf{VI.72} beyond the minimalization through the first variable $\beta$.

This final step of energy minimalization is described now by the second equilibrium condition \myrf{VI.54b} which then reads for our present energy function $\EEivbn$ \myrf{VI.72}
\begin{equation}
  \label{eq:VI.80}
  \frac{\partial  {\mathbb{E}^\textrm{(IV)}(\beta_*,\nu)}}{\partial\nu}= \frac{e^2}{\aB}\left\{(2\aB\beta_*)^2\cdot\frac{\partial\varepsilon_\textrm{kin}(\nu)}{\partial\nu}-(2\aB\beta_*)\cdot\frac{\partial\varepsilon_\textrm{pot}(\nu)}{\partial\nu}\right\}\stackrel{!}{=}0
\end{equation}
and thus would evidently require also the {\em explicit}\/ knowledge of the potential function $\varepsilon_\textrm{pot}(\nu)$ \myrf{VI.70}, analogously to the kinetic function $\varepsilon_\textrm{kin}(\nu)$ \myrf{VI.66}. However, the determination of the analytic form of the potential function $\varepsilon_\textrm{pot}(\nu)$ is a difficult problem so that we have to be satisfied here with a power series representation (see appendix D).

If the interaction energy of both positronium constituents is interpreted as the energy $\ERee$ \myrf{VI.68} being concentrated in the gauge field $\ppAo(r)$, then the potential function $\varepsilon_\textrm{pot}(\nu)$ is found as
\begin{equation}
  \label{eq:VI.81}
  \varepsilon_\textrm{pot}(\nu)=\int_0^\infty dy\;y^2\cdot\left(\frac{d\tilde{a}_\nu(y)}{dy}\right)^2= \frac{1}{2^{4\nu+3}}\sum_{m,n=0}^\infty\frac{1}{2^{m+n}}\cdot\frac{\Gamma(4\nu+3+m+n)}{\Gamma(2\nu+3+m)\cdot\Gamma(2\nu+3+n)}\;,
\end{equation}
cf. equation \myrf{D8}.
On the other hand, if one prefers to interprete the physical interaction energy as the mass equivalent $\tMMee c^2$ \myrf{VI.48} then the potential function $\varepsilon_\textrm{pot}(\nu)$ is found to be of the form
\begin{eqnarray}
  \varepsilon_\textrm{pot}(\nu)&=&\frac{1}{\Gamma(2\nu+2)}\int_0^\infty dy\;y\,\tilde{a}_\nu(y)\cdot\tilde{\Phi}(y)^2\nonumber\\
  \label{eq:VI.82}
  &=& \frac{1}{2\nu+1}\left\{1-\frac{1}{2^{4\nu+2}}\sum_{n=0}^\infty\frac{n}{2^n}\cdot\frac{\Gamma(4\nu+2+n)}{\Gamma(2\nu+2)\cdot\Gamma(2\nu+2+n)}\right\}\,,
\end{eqnarray}
cf. equation \myrf{D18}. Consequently, the electric Poisson identity implies the numerical
equality of both potential functions $\varepsilon_\textrm{pot}(\nu)$ \myrf{VI.81} and
\myrf{VI.82} which may be checked explicitly by selecting some integers for the
variational parameter $2\nu$, see the table in appendix D and fig.~3.

The desired minimal value ($\EE^\textrm{(IV)}_*$, say) of the energy function $\EEivbn$
\myrf{VI.72} is found by resolving both equilibrium conditions \myrf{VI.74} and
\myrf{VI.80} for the equilibrium values $\beta_*$ and $\nu_*$ and substituting this into
the energy function such that
\begin{equation}
  \label{eq:VI.83}
  \EE^\textrm{(IV)}_*=\EEivbn\Big|{}_{{}_{\beta=\beta_*\atop\nu=\nu_*}}= \frac{e^2}{\aB}\left\{(2\aB\beta_*)^2\cdot\varepsilon_\textrm{kin}(\nu_*)-(2\aB\beta_*)\cdot\varepsilon_\textrm{pot}(\nu_*)\right\}\,.
\end{equation}
However, in contrast to the first equilibrium condition \myrf{VI.74}, the second equilibrium condition \myrf{VI.80} is hard (if not impossible) to solve exactly for the equilibrium values $\beta_*,\,\nu_*$; and therefore we have to resort to an approximative solution of the extremalization problem defined by those equations \myrf{VI.74} and \myrf{VI.80}. To this end, one substitutes both the kinetic energy $E_\textrm{kin}(\beta_*,\nu)$ \myrf{VI.76} and the potential energy $\ERee\doteqdot\EE_\textrm{pot}(\nu)$ \myrf{VI.77} into the energy function $\EEivbn$ \myrf{VI.72} which yields a resulting function $\EET(\nu)$ ($\doteqdot\EE^\textrm{(IV)}(\beta_*,\nu)$)
\begin{eqnarray}
  \nonumber\EET(\nu)&=&2E_\textrm{kin}(\nu)+\EE_\textrm{pot}(\nu)\\
  \label{eq:VI.84}
  &=&\frac{e^2}{2\aB}\left\{\frac{\varepsilon_\textrm{pot}^2(\nu)}{2\varepsilon_\textrm{kin}(\nu)} -\frac{\varepsilon_\textrm{pot}^2(\nu)}{\varepsilon_\textrm{kin}(\nu)}\right\}\;=\;-\frac{e^2}{4\aB}\,\frac{\varepsilon_\textrm{pot}^2(\nu)}{\varepsilon_\textrm{kin}(\nu)}\;.
\end{eqnarray}
Recalling here the fact that the conventional non-relativistic groundstate energy ($E_\textrm{conv}$) of positronium is given by
\begin{equation}
  \label{eq:VI.85}
  E_\textrm{conv}=-\frac{e^2}{4\aB}=-6.802\ldots\,[{\rm eV}]
\end{equation}
the present RST energy function $\EET(\nu)$ \myrf{VI.84} does appear as
\begin{equation}
  \label{eq:VI.86}
  \EET(\nu)=E_\textrm{conv}\cdot S_\mathcal{P}(\nu)
\end{equation}
with the {\em spectral function}\/ $S_\mathcal{P}$ being defined by
\begin{equation}
  \label{eq:VI.87}
  S_\mathcal{P}(\nu)=\frac{\varepsilon_\textrm{pot}^2(\nu)}{\varepsilon_\textrm{kin}(\nu)}\;.
\end{equation}

Now according to the {\em principle of minimal energy}\/, the extremal points of the
spectral function $S_\mathcal{P}(\nu)$ determine an approximate solution of the original
eigenvalue problem defined by the former equations \myrf{VI.42}-\myrf{VI.44} where the
extremal values $\beta_*$ and $\nu_*$ do then fix the corresponding wave function
$\tilde{\Phi}(r)$ \myrf{VI.55}, potential $\ppAo(r)$ \myrf{VI.79} and associated energc
$\EE^\textrm{(IV)}_*$ \myrf{VI.83}. But on behalf of the simplicity of the chosen trial
function $\tilde{\Phi}(r)$ \myrf{VI.55} one will not expect to obtain by this procedure
the totality of para-states; but rather one will find the state of lowest energy for any
admitted value of the quantum number $\lP$ of angular momentum, see fig.~2. By the former
convention, this state has principal quantum number $\nP=\lP+1$. By comparison with the
conventional theory (see ref. \cite{MaSo}), the conventional spectrum
\begin{eqnarray}
  \label{eq:VI.88}
  \EE^{(n)}_\textrm{conv}&=&-\frac{e^2}{4\aB}\cdot\frac{1}{n^2}\\
  \nonumber
  (n&=&1,2,3,4,\,...)
\end{eqnarray}
would obviously be reproduced by the present RST approximation method if the minimal values of the spectral function $S_\mathcal{P}$ \myrf{VI.87} turned out for any prescribed $\lP$ as
\begin{equation}
  \label{eq:VI.89}
  \S_\mathcal{P}\Big|_\textrm{min}=\frac{1}{\nP^2}=\frac{1}{(\lP+1)^2}\;.
\end{equation}

In a preceding paper \cite{MaSo3}, the conventional groundstate energy
$\EE^{(n=1)}_\textrm{conv}$ \myrf{VI.88} has {\em exactly}\/ been reproduced for $\lP=0$
by choosing $\nu=0$ and extremalizing the energy function merely with respect to the
remaining variational parameter $\beta$. But clearly, since we are dealing here with an
approximation method, the true RST groundstate prediction (in the electrostatic and
spherically symmetric approximation) must be somewhat lower than the conventional value
\myrf{VI.85}, see the discussion of this in ref. \cite{MaSo}. In a subsequent paper
\cite{MaSo2}, the first excited positronium state ($\lP=1$) has been treated by means of
the present approximation method with two variational parameters $\beta$ and $\nu$; and
the corresponding RST prediction did approach the conventional value
$\EE^{(n=2)}_\textrm{conv}$ \myrf{VI.88} up to some few percents. However, this result
came about by approximating the spectral function $S_\mathcal{P}(\nu)$ \myrf{VI.87} by its
values upon the (half-)integers $\nu=0,\frac{1}{2},1,\frac{3}{2},\,...$ together with the
help of an interpolating polynomial. But clearly, the reliability of the RST predictions
can be improved now by numerically calculating the spectral function $S_\mathcal{P}(\nu)$
\myrf{VI.87} for \emph{continuous} parameter $\nu$ and thus looking for its minimal values for
any $\lP\;(=0,1,2,3,\ldots)$, see fig.~4.

Perhaps the most remarkable result of the present calculations refers to the positronium
groundstate (i.e.~$\nP=0 \Leftrightarrow \lP=1$). Here the conventional value
(\ref{eq:VI.85}) is exactly reproduced also with the present two-parameter trial
function~$\tPhi(r)$ (\ref{eq:VI.55}). This is a somewhat amazing result because the same
prediction could be obtained in a preceding paper~\cite{MaSo3} with only a
\emph{one-parameter} trial function which emerges from the present two-parameter trial
function (\ref{eq:VI.55}) by putting one of the two variational parameters to zero
(i.e.~$\nu\rightarrow 0$). Indeed, both coinciding predictions are consistent because, in
the present two-parameter approach, the groundstate energy emerges as a \emph{boundary
  minimum}, namely just for~$\nu=0$ (see fig.~4).

Concerning the energy of the excited states~$(\nP=2,3,4,5,6)$ the RST predictions are
found to differ from the conventional values~$\Ec^{(n)}$ (\ref{eq:VI.88}) by roughly 10\%
(see the table below). This difference may be attributed partly to the use of the
spherically symmetric approximation and partly to the use of an imperfect trial
function~$\tPhi(r)$ (\ref{eq:VI.55}) with only two variational parameters~$\beta$
and~$\nu$. A better trial function~$\tPhi(r)$ would surely shift the RST predictions
towards its conventional counterparts~$\Ec^{(n)}$ since the RST energy
functional~$\tEEPhi$ (\ref{eq:VI.50}) is bounded from below for any chosen value~$\lP$ of
orbital angular momentum (strictly speaking, the quantum number~$\lP$ emerges only in the
context of the product ansatz (\ref{eq:VI.19a})-(\ref{eq:VI.20b}); but it is assumed here
that the hypothetical exact solutions of the RST eigenvalue problem can also be enumerated
by some quantum number of orbital  angular momentum which then is expected to have the
same classifying effect upon the set of solutions as is the case with the
present~$\lP$). On the other hand, for the purpose of obtaining more exact solutions to
the RST eigenvalue problem one should go beyond the spherically symmetric approximation of
the interaction potential~$\pAo(r,\vartheta)$, cf.\ (\ref{eq:VI.18}); and whether or not
this also shifts the RST predictions towards their conventional counterparts appears to be
unclear at the moment.

\begin{flushleft}
  \begin{tabular}{|c||c|c|c|c|}
  \hline
  $\nP\ (=\lP+1)$ & $\Ec^{(n)}$\ [eV], (\ref{eq:VI.88}) & $\EET(\nu_*)$\ [eV],  (\ref{eq:VI.84}) &
  $\nu_*$ & $\ds\frac{\EET(\nu_*)-\Ec^{(n)}}{|\Ec^{(n)}|}\ [\%]$ \\
  \hline\hline
  1 & $-6,802\ldots$ & $-6,802\ldots$ & 0 & 0\\ \hline
  2 & $-1,700\ldots$ & $-1,550\ldots$ & $1,794\ldots$ &   8,8\\  \hline
  3 & $-0,755\ldots$ & $-0,669\ldots$ & $3,752\ldots$ &  11,4\\ \hline
  4 & $-0,425\ldots$ & $-0,373\ldots$ & $5,874\ldots$ &  12,2\\ \hline
  5 & $-0,272\ldots$ & $-0,239\ldots$ & $8,130\ldots$ &  12,1\\ \hline
  6 & $-0,188\ldots$ & $-0,166\ldots$ & $10,504\ldots$ & 11,7\\ \hline
 10 & $-0,06802\ldots$ & $-0,06068\ldots$ & $20,953\ldots$ & 10,7\\ \hline
 20 & $-0,01700\ldots$ & $-0,01554\ldots$ & $51,919\ldots$ & 8,5\\ \hline
 50 & $-0,00272\ldots$ & $-0,00256\ldots$ & $172,161\ldots$ & 5,7\\ \hline
  \hline
  \end{tabular}
  \label{tabE}
\end{flushleft}
Observe here that the relative deviation of the conventional and RST predictions (last
column) decreases for increasing principal quantum number~$\nP$ (e.g.\ 3,7\% for~$\nP=100$)!
\vspace{4ex}
\begin{center}
  \emph{\textbf{2.\ Ortho-Positronium}}
\end{center}

The discussion of ortho-positronium will be carried through along a certain path of
arguments very similar to those which were presented for para-positronium in the preceding
subsection. Especially, we will be satisfied again with the electrostatic, spherically
symmetric, and non-relativistic approximation. The consideration of the magnetic
interactions will be deferred to the next subsection when both the para- and ortho-level
system will be sufficiently worked out in the above mentioned approximation which then
will serve as a first rough estimate of the energy difference of the ortho- and
para-versions of the positronium states.
\pagebreak
\vspace{3ex}
\begin{center}
\emph{\large Eigenvalue Problem for the Ortho-States}
\end{center}

The point of departure is of course the ortho-system \myrf{V.55a}-\myrf{V.55d} for which
the boson number $\bbar$ must be put to minus the value of the para-configuration,
i.\,e. $\bbar\Rightarrow-\frac{1}{2}$. Furthermore, the electrostatic approximation
requires to neglect the magnetic potential $\bAphi$. With these two specializing arguments
the ortho-system \myrf{V.55a}-\myrf{V.55d} becomes simplified to the following system:
\begin{subequations}
  \begin{align}
  \label{eq:VI.90a}
  \frac{\partial\,\bMRp}{\partial r} +\frac{i}{r}\cdot\frac{\partial\,\bMRp}{\partial\phi}-\frac{1}{2r}\cdot\bMRp-\bAo\cdot\bMRm\hspace{4em}&\nonumber\\
  {}+{\rm e}^{-2i\phi}\cdot\Bigg\{\frac{1}{r}\cdot\frac{\partial\,\bMSp}{\partial\vartheta} +\frac{\cot\vartheta}{r}\,\big[-\frac{1}{2}\,\bMSp-i\,\frac{\partial\,\bMSp}{\partial\phi}\big]\Bigg\}&=\frac{M+M_*}{\hbar}\,c\cdot\bMRm
  \\\nonumber\\
  \label{eq:VI.90b}
  \frac{\partial\,\bMSp}{\partial r} -\frac{i}{r}\cdot\frac{\partial\,\bMSp}{\partial\phi}-\frac{1}{2r}\cdot\bMSp-\bAo\cdot\bMSm \hspace{5em}&\nonumber\\
  {}-{\rm e}^{2i\phi}\cdot\Bigg\{\frac{1}{r}\cdot\frac{\partial\,\bMRp}{\partial\vartheta} +\frac{\cot\vartheta}{r}\,\big[-\frac{1}{2}\,\bMRp+i\,\frac{\partial\,\bMRp}{\partial\phi}\big]\Bigg\}&=\frac{M+M_*}{\hbar}\,c\cdot\bMSm
  \\\nonumber\\
  \label{eq:VI.90c}
  \frac{\partial\,\bMRm}{\partial r} -\frac{i}{r}\cdot\frac{\partial\,\bMRm}{\partial\phi}+\frac{5}{2r}\cdot\bMRm+\bAo\cdot\bMRp \hspace{4em}&\nonumber\\
  {}-{\rm e}^{-2i\phi}\cdot\Bigg\{\frac{1}{r}\cdot\frac{\partial\,\bMSm}{\partial\vartheta} +\frac{\cot\vartheta}{r}\,\big[-\frac{1}{2}\,\bMSm-i\,\frac{\partial\,\bMSm}{\partial\phi}\big]\Bigg\}&=\frac{M-M_*}{\hbar}\,c\cdot\bMRp
  \\\nonumber\\
  \label{eq:VI.90d}
  \frac{\partial\,\bMSm}{\partial r} +\frac{i}{r}\cdot\frac{\partial\,\bMSm}{\partial\phi}+\frac{5}{2r}\cdot\bMSm+\bAo\cdot\bMSp \hspace{5em}&\nonumber\\
  {}+{\rm e}^{2i\phi}\cdot\Bigg\{\frac{1}{r}\cdot\frac{\partial\,\bMRm}{\partial\vartheta} +\frac{\cot\vartheta}{r}\,\big[-\frac{1}{2}\,\bMRm+i\,\frac{\partial\,\bMRm}{\partial\phi}\big]\Bigg\}&=\frac{M-M_*}{\hbar}\,c\cdot\bMSp\;.
  \end{align}
\end{subequations}
This eigenvalue system is to be complemented again by the electric Poisson equation \myrf{V.21}
\begin{equation}
  \label{eq:VI.91}
  \Delta\bAo=-\as\cdot\left\{\bMRpS\cdot\bMRp+\bMSpS\cdot\bMSp+\bMRmS\cdot\bMRm+\bMSmS\cdot\bMSm\right\}\,.
\end{equation}

The crucial point with the present ortho-system \myrf{VI.90a}-\myrf{VI.90d} refers of
course to the emergence of its triplet character, i.\,e. one wishes to see in what way
there do arise from the present ortho-system three different types of solutions (with
$j_z=1,0,-1$) which have in common the same energy albeit only for the special case of
neglection of the magnetic interactions. Furthermore with the latter presumptions, the
common energy of these ortho-states should be the same as the energy due to the
corresponding para-states, see the discussion of the analogous situation in the
conventional theory (below equation \myrf{V.13}).

\vspace{3ex}
\begin{center}
\emph{\large Vanishing Angular Momentum $j_z=0$}
\end{center}

Naturally, it is very tempting (but not correct) to think that the RST counterparts of the conventional triplet states with $S=1$, $S_z=0$ are included in the set of solutions of the ortho-eigenvalue system \myrf{VI.90a}-\myrf{VI.90d}, namely in such a way that both kinds of wave amplitudes $\bMRpm$ and $\bMSpm$ do occur in a symmetric way, just in order to have a vanishing $z$-component ($j_z=0$) of total angular momentum $\hat{J}_z$. In this sense, one tends to try the following ansatz for the complex-valued amplitudes $\bMRpm(\vec{r})$ and $\bMSpm(\vec{r})$:
\begin{subequations}
  \begin{align}
  \label{eq:VI.92a}
  \bMRpm(r,\vartheta,\phi)&={\rm e}^{-i\phi}\cdot\bRpm(r,\vartheta)\\
  \label{eq:VI.92b}
  \bMSpm(r,\vartheta,\phi)&={\rm e}^{i\phi}\cdot\bSpm(r,\vartheta)\;,
  \end{align}
\end{subequations}
where the reduced wave amplitudes $\bRpm(r,\vartheta)$ and $\bSpm(r,\vartheta)$ are adopted to be real-valued and independent of the azimuthal angle $\phi$. Inserting this ansatz into the present ortho-system \myrf{VI.90a}-\myrf{VI.90d} yields the following system for the real-valued amplitudes $\bRpm,\,\bSpm$, which for the sake of convenience may be transformed to $\btRpm,\,\btSpm$ according to
\begin{subequations}
  \begin{align}
  \label{eq:VI.93a}
  \btRpm&=\sqrt{r\sin\vartheta}\cdot\bRpm\\
  \label{eq:VI.93b}
  \btSpm&=\sqrt{r\sin\vartheta}\cdot\bSpm\;.
  \end{align}
\end{subequations}
This lets reappear the eigenvalue system \myrf{VI.90a}-\myrf{VI.90d} in the following form:
\begin{subequations}
  \begin{align}
  \label{eq:VI.94a}
  \frac{\partial\,\btRp}{\partial r}+\frac{1}{r}\cdot\frac{\partial\,\btSp}{\partial\vartheta}-\bAo\cdot\btRm&= \frac{M+M_*}{\hbar}\,c\cdot\btRm\\\nonumber\\
  \label{eq:VI.94b}
  \frac{\partial\,\btSp}{\partial r}-\frac{1}{r}\cdot\frac{\partial\,\btRp}{\partial\vartheta}-\bAo\cdot\btSm&= \frac{M+M_*}{\hbar}\,c\cdot\btSm\\\nonumber\\
  \label{eq:VI.94c}
  \frac{\partial\,\btRm}{\partial r}+\frac{1}{r}\cdot\btRm-\frac{1}{r}\cdot\frac{\partial\,\btSm}{\partial\vartheta}+\bAo\cdot\btRp&= \frac{M-M_*}{\hbar}\,c\cdot\btRp\\\nonumber\\
  \label{eq:VI.94d}
  \frac{\partial\,\btSm}{\partial r}+\frac{1}{r}\cdot\btSm+\frac{1}{r}\cdot\frac{\partial\,\btRm}{\partial\vartheta}+\bAo\cdot\btSp&= \frac{M-M_*}{\hbar}\,c\cdot\btSp\;.
  \end{align}
\end{subequations}
However, this special subcase of the ortho-system \myrf{VI.90a}-\myrf{VI.90d} is exactly identical to the para-system \myrf{VI.1a}-\myrf{VI.1d}; and therefore the present ortho-system \myrf{VI.94a}-\myrf{VI.94d} does actually not describe a new quantum state of the ortho-type!

The reason for this ``failure'' of the ortho-ansatz \myrf{VI.92a}-\myrf{VI.92b} is the following: If that ansatz is resubstituted into the Pauli spinors $\bppm$ \myrf{V.50a}-\myrf{V.50b} with the basis spinors $\woepm$ being given by \myrf{V.48a}-\myrf{V.48d} for boson number $\bbar=-\frac{1}{2}$, then one will ultimately end up with the following decomposition of the proposed Pauli spinors for the ortho-system
\begin{subequations}
  \begin{align}
  \label{eq:VI.95a}
  {}^{(b)}\varphi_{+}&=\bRp\cdot\wop+\bSp\cdot\wom\\
  \label{eq:VI.95b}
  {}^{(b)}\varphi_{-}&=-i\left\{\bRm\cdot\wep+\bSm\cdot\wem\right\}\,
  \end{align}
\end{subequations}
where now the basis spinors $\woepm$ are those for $\bbar=+\frac{1}{2}$, i.\,e. one recovers merely the para-case being described above the para-system \myrf{V.56a}-\myrf{V.56d}. In other words, the ortho-ansatz \myrf{VI.92a}-\myrf{VI.92b} for the wave amplitudes rescinds the transition from the para-basis ($\bbar=\frac{1}{2}$) to the ortho-basis ($\bbar=-\frac{1}{2}$). It is very instructive to reformulate this effect also in terms of angular momentum: The action of $\hat{J}_z$ on the ortho-spinors \myrf{V.50a}-\myrf{V.50b} reads explicitly
\begin{subequations}
  \begin{align}
  \label{eq:VI.96a}
  \hat{J}^{(+)}_z\,^{(b)}\varphi_{+}&= \big(\hat{L}_z\bMRp\big)\cdot\wop+\big(\hat{L}_z\bMSp\big)\cdot\wom+\bMRp\cdot\hat{J}^{(+)}_z\wop+\bMSp\cdot\hat{J}^{(+)}_z\wom\\
  \label{eq:VI.96b}
  \hat{J}^{(-)}_z\,^{(b)}\varphi_{-}&= -i\left\{\big(\hat{L}_z\bMRm\big)\cdot\wep+\big(\hat{L}_z\bMSm\big)\cdot\wem+\bMRm\cdot\hat{J}^{(-)}_z\wep+\bMSm\cdot\hat{J}^{(-)}_z\wem\right\}\,,
  \end{align}
\end{subequations}
where the ortho-form ($\bbar=-\frac{1}{2}$) of the basis spinors $\woepm$ \myrf{V.48a}-\myrf{V.48d} is to be used.
But here one concludes for the ortho-value (i.\,e. $\bbar=-\frac{1}{2}$) of the boson number $\bbar$ from the eigenvalue equations of angular momentum \myrf{V.49a}-\myrf{V.49b}
\begin{subequations}
  \begin{align}
  \label{eq:V.97a}
  \hat{J}_z^{(+)}\,\wop&=\hbar\cdot\wop\\
  \label{eq:V.97b}
  \hat{J}_z^{(+)}\,\wom&=-\hbar\cdot\wom\\
  \label{eq:V.97c}
  \hat{J}_z^{(-)}\,\wep&=\hbar\cdot\wep\\
  \label{eq:V.97d}
  \hat{J}_z^{(-)}\,\wem&=-\hbar\cdot\wem\;,
  \end{align}
\end{subequations}
and similarly for the wave amplitudes $\bMRpm,\,\bMSpm$ \myrf{VI.92a}-\myrf{VI.92b}
\begin{subequations}
  \begin{align}
  \label{eq:VI.98a}
  \hat{L}_z\bMRpm&=-\hbar\cdot\bMRpm\\
  \label{eq:VI.98b}
  \hat{L}_z\bMSpm&=\hbar\cdot\bMSpm\;.
  \end{align}
\end{subequations}
Obviously, both the spinor basis $\woepm$ and the wave amplitudes $\bMRpm,\,\bMSpm$ carry the expexted quantum of angular momentum, but if this result is inserted into the relations for the Pauli spinors \myrf{VI.96a}-\myrf{VI.96b} one recovers the {\em para-situation}\/
\begin{subequations}
  \begin{align}
  \label{eq:VI.99a}
  \hat{J}_z^{(+)}\,{}^{(b)}\varphi_{+}&=0\\
  \label{eq:VI.99b}
  \hat{J}_z^{(-)}\,{}^{(b)}\varphi_{-}&=0\;.
  \end{align}
\end{subequations}
Thus the result is that the orbital angular momentum of the wave amplitudes compensates the (total) angular momentum of the spinor basis so that the Dirac spinor field $\psi_\mathcal{O}$
\begin{equation}
  \label{eq:VI.100}
  \psi_\mathcal{O}(\vec{r})\doteqdot{}^{(b)}\varphi_{+}(\vec{r})\oplus{}^{(b)}\varphi_{-}(\vec{r})
\end{equation}
would carry here vanishing angular momentum:
\begin{equation}
  \label{eq:VI.101}
  \hat{\mathcal{J}}\!_z\,\psi_\mathcal{O}= \hat{J}_z^{(+)}\,{}^{(b)}\varphi_{+}\oplus\hat{J}_z^{(-)}\,{}^{(b)}\varphi_{-}\equiv0\;.
\end{equation}
In contrast to this, for the para-state
$\psi_\mathcal{P}={}^{(p)}\varphi_{+}\oplus{}^{(p)}\varphi_{-}$ neither the spinor basis
nor the wave amplitudes do carry any angular momentum; but the net effect~$(j_z=0)$ is the same.

Summarizing, the triplet states with $j_z=0$ cannot be described by those field equations
where the associated current $\vec{j}(\vec{r})$ (or $\vec{k}(\vec{r})$, resp.) encircles
the $z$-axis, as shown by equation \myrf{V.47b}. Rather, one must evoke here the toroidal
geometry which has already been studied in a previous paper \cite{GBMS}. This, however,
requires an extra discussion not to be presented in this paper. For the subsequent
discussion of the ortho-states we will restrict ourselves to the RST counterparts of
those conventional states with $j_z=\pm1$, cf. \myrf{V.10a}-\myrf{V.10b}.

\vspace{3ex}
\begin{center}
\emph{\large Ortho-States with $j_z=-1$}
\end{center}

The states with definite orientation of their total angular momentum $J_z$ along the (positive or negative) $z$-axis can intuitively be associated with an ortho-current $\vec{k}_b$ which encircles the $z$-axis in the usual way
\begin{equation}
  \label{eq:VI.102}
  \vec{k}_b={}^{(b)\!}k_\phi\,\vec{e}_\phi\;,
\end{equation}
cf. the analogous para-case \myrf{V.60}. For analyzing this ortho-situation, one has to
restart from the original eigenvalue system \myrf{VI.90a}-\myrf{VI.90d} for the
ortho-states, but now avoiding such an ansatz for the wave amplitudes like
\myrf{VI.92a}-\myrf{VI.92b} which turned out to be adequate for the para-states. Clearly,
the wanted ansatz for those ortho-states with definite orientation along the $z$-axis
must respect the fact that either the positive ($z>0$) or the negative ($z<0$) direction
is singled out; and therefore both wave amplitudes $\bMRpm$ and $\bMSpm$ cannot enter the
description of a triplet state in such a symmetric way as it was the case for the singlet
states, cf. \myrf{V.54a}-\myrf{V.54b}. Thus we try now for the treatment of
ortho-positronium the following ansatz for the wave amplitudes $\bMRpm,\,\bMSpm$
\myrf{V.50a}-\myrf{V.50b}:
\begin{subequations}
  \begin{align}
  \label{eq:VI.103a}
  \bMRpm(r,\vartheta,\phi)&=\frac{{\rm e}^{-2i\phi}}{\sin\vartheta\sqrt{r\sin\vartheta}}\cdot\btRpm(r,\vartheta)\\
  \label{eq:VI.103b}
  \bMSpm(r,\vartheta,\phi)&=\sqrt{\frac{\sin\vartheta}{r}}\cdot\btSpm(r,\vartheta)\;,
  \end{align}
\end{subequations}
where the new amplitudes $\btRpm$ and $\btSpm$ do not depend upon the azimuthal angle $\phi$. Obviously, the present ortho-ansatz \myrf{VI.103a}-\myrf{VI.103b} is the analogue of the para-case \myrf{VI.2a}-\myrf{VI.2b}.

Following now further the line of those arguments which were put forward for the para-configurations in order to reformulate the eigenvalue equations in terms of the new amplitudes $\ptRpm,\,\ptSpm$ (cf. \myrf{VI.1a}-\myrf{VI.1d}), one substitutes the present ansatz \myrf{VI.103a}-\myrf{VI.103b} into the original ortho-system of eigenvalue equations \myrf{VI.90a}-\myrf{VI.90d} which then reappears in the following form: 
\begin{subequations}
  \begin{align}
  \label{eq:VI.104a}
  \frac{\partial\btRp}{\partial r}+\frac{1}{r}\cdot\btRp+\frac{\sin^2\vartheta}{r}\cdot\frac{\partial\btSp}{\partial\vartheta}-\bAo\cdot\btRm&=\frac{M+M_*}{\hbar}\,c\cdot\btRm\\\nonumber\\
  \label{eq:VI.104b}
  \frac{\partial\btSp}{\partial r}-\frac{1}{r}\cdot\btSp-\frac{1}{r\sin^2\vartheta}\cdot\frac{\partial\btRp}{\partial\vartheta}-\bAo\cdot\btSm&=\frac{M+M_*}{\hbar}\,c\cdot\btSm\\\nonumber\\
  \label{eq:VI.104c}
  \frac{\partial\btRm}{\partial r}-\frac{\sin^2\vartheta}{r}\cdot\frac{\partial\btSm}{\partial\vartheta}+\bAo\cdot\btRp&=\frac{M-M_*}{\hbar}\,c\cdot\btRp\\\nonumber\\
  \label{eq:VI.104d}
  \frac{\partial\btSm}{\partial r}+\frac{2}{r}\cdot\btSm+\frac{1}{r\sin^2\vartheta}\cdot\frac{\partial\btRm}{\partial\vartheta}+\bAo\cdot\btSp&=\frac{M-M_*}{\hbar}\,c\cdot\btSp\;.
  \end{align}
\end{subequations}
Indeed, this is just the ortho-counterpart of the former para-system \myrf{VI.1a}-\myrf{VI.1d}; and therefore it suggests itself to look here also for some product ansatz similar to the para-case \myrf{VI.19a}-\myrf{VI.19b}; i.\,e. we try here again the {\em spherically symmetric approximation}\/ (in its ortho-form)
\begin{subequations}
  \begin{align}
  \label{eq:VI.105a}
  \btRpm(r,\vartheta)&=\ssRpm(r)\cdot g_R(\vartheta)\\
  \label{eq:VI.105b}
  \btSpm(r,\vartheta)&=\ssSpm(r)\cdot g_S(\vartheta)\;.
  \end{align}
\end{subequations}
For the angular functions $g_R(\vartheta)$ and $g_S(\vartheta)$ we postulate the coupled first-order equations
\begin{subequations}
  \begin{align}
  \label{eq:VI.106a}
  \sin^2\vartheta\cdot\frac{dg_S(\vartheta)}{d\vartheta}&=\dlO\cdot g_R(\vartheta)\\
  \label{eq:VI.106b}
  \frac{1}{\sin^2\vartheta}\cdot\frac{dg_R(\vartheta)}{d\vartheta}&=\ddlO\cdot g_S(\vartheta)\\
  \nonumber
  (\dlO,\ddlO&={\rm const.})\ ,
  \end{align}
\end{subequations}
since by this choice the original ortho-system \myrf{VI.104a}-\myrf{VI.104d} becomes neatly reduced to a system of {\em ordinary}\/ differential equations
\begin{subequations}
  \begin{align}
  \label{eq:VI.107a}
  \frac{d\ssRp(r)}{dr}+\frac{1}{r}\cdot\ssRp(r)+\frac{\dlO}{r}\cdot\ssSp(r)-\bbAo(r)\cdot\ssRm(r)&= \frac{M+M_*}{\hbar}\,c\cdot\ssRm(r)\\
  \label{eq:VI.107b}
  \frac{d\ssSp(r)}{dr}-\frac{1}{r}\cdot\ssSp(r)-\frac{\ddlO}{r}\cdot\ssRp(r)-\bbAo(r)\cdot\ssSm(r)&= \frac{M+M_*}{\hbar}\,c\cdot\ssSm(r)\\
  \label{eq:VI.107c}
  \frac{d\ssRm(r)}{dr}-\frac{\dlO}{r}\cdot\ssSm(r)+\bbAo(r)\cdot\ssRp(r)&=\frac{M-M_*}{\hbar}\,c\cdot\ssRp(r)\\
  \label{eq:VI.107d}
  \frac{d\ssSm(r)}{dr}+\frac{2}{r}\cdot\ssSm(r)+\frac{\ddlO}{r}\cdot\ssRm(r)+\bbAo(r)\cdot\ssSp(r)&= \frac{M-M_*}{\hbar}\,c\cdot\ssSp(r)\;.
  \end{align}
\end{subequations}
Evidently, this system is again the ortho-counterpart of the para-case \myrf{VI.21a}-\myrf{VI.21d} where the role of the para-number $\lP$ is now played by the two ortho-numbers $\dlO$ and $\ddlO$. In order to clarify their meaning, we defer for a moment the discussion of the present ortho-system \myrf{VI.107a}-\myrf{VI.107d} and first analyze the angular system \myrf{VI.106a}-\myrf{VI.106b} which, of course, is the ortho-counterpart of the former para-system \myrf{VI.20a}-\myrf{VI.20b} but is considerably more complicated.

In order to see more clearly the type of solutions admitted by that angular system \myrf{VI.106a}-\myrf{VI.106b} one differentiates once more and thus deduces a {\em decoupled}\/ second-order system (as the ortho-analogue of the para-case \myrf{VI.24a}-\myrf{VI.24b}):
\begin{subequations}
  \begin{align}
  \label{eq:VI.108a}
  \frac{d^2g_S(\vartheta)}{d\vartheta^2}+2\cot\vartheta\cdot\frac{dg_S(\vartheta)}{d\vartheta}&= (\dlO\ddlO)\cdot g_S(\vartheta)\\
  \label{eq:VI.108b}
  \frac{d^2g_R(\vartheta)}{d\vartheta^2}-2\cot\vartheta\cdot\frac{dg_R(\vartheta)}{d\vartheta}&= (\dlO\ddlO)\cdot g_R(\vartheta)\;.
  \end{align}
\end{subequations}
Sometimes it may be rather instructive to first consider an unphysical solution of the
problem since this lets one truly appreciate the elaboration of the physically correct
case. Such an unphysical solution can easily be guessed and is given by
\begin{subequations}
  \begin{align}
  \label{eq:VI.109a}
  g_S(\vartheta)&\Rightarrow\frac{1}{\sin\vartheta}\\
  \label{eq:VI.109b}
  g_R(\vartheta)&\Rightarrow\cos\vartheta
  \end{align}
\end{subequations}
where the constants $\dlO,\,\ddlO$ must both be fixed by
\begin{subequations}
  \begin{align}
  \label{eq:VI.110a}
  \dlO&\Rightarrow-1\\
  \label{eq:VI.110b}
  \ddlO&\Rightarrow-1\;.
  \end{align}
\end{subequations}
However, this solution must be rejected because the wave amplitudes \myrf{VI.103a}-\myrf{VI.103b} would adopt the following form
\begin{subequations}
  \begin{align}
  \label{eq:VI.111a}
  \bMRpm(r,\vartheta,\phi)&=\frac{{\rm e}^{-2i\phi}}{\sin\vartheta\sqrt{r\sin\vartheta}}\cdot g_R(\vartheta)\cdot\ssRpm(r)\Rightarrow\frac{\cos\vartheta\cdot{\rm e}^{-2i\phi}}{\sin\vartheta\sqrt{r\sin\vartheta}}\cdot\ssRpm(r)\\
  \label{eq:VI.111b}
  \bMSpm(r,\vartheta,\phi)&=\sqrt{\frac{\sin\vartheta}{r}}\cdot g_S(\vartheta)\cdot\ssSpm(r)\Rightarrow\frac{1}{\sqrt{r\sin\vartheta}}\cdot\ssSpm(r)\;.
  \end{align}
\end{subequations}
Indeed, since the wave amplitude $\bMRpm$ \myrf{VI.111a} diverges like
$\sim(\sin\vartheta)^{-\frac{3}{2}}$ on the $z$-axis, the normalization condition
\myrf{III.49} cannot be satisfied! And clearly it makes no sense to deal with
non-normalizable wave functions. The normalizable solutions of the angular system
\myrf{VI.106a}-\myrf{VI.106b}, or \myrf{VI.108a}-\myrf{VI.108b}, resp., are worked out in
detail in appendix E.

For a better understanding of the ortho-system it is very instructive to elucidate the meaning of those constants $\dlO,\,\ddlO$ emerging in the angular systems \myrf{VI.106a}-\myrf{VI.106b} or \myrf{VI.108a}-\myrf{VI.108b}, resp. To this end, one passes over to the non-relativistic limit of the one-dimensional eigenvalue system \myrf{VI.107a}-\myrf{VI.107d}. Here the usual procedure consists in first identifying both ``positive'' and both ``negative'' amplitudes with each other, i.\,e. one puts
\begin{equation}
  \label{eq:VI.112}
  \ssRpm(r)\equiv\ssSpm(r)\doteqdot\tOpm(r)\;,
\end{equation}
so that the radial eigenvalue system becomes reduced to only two equations for $\tOpm(r)$. However, this reduction can be performed in {\em two}\/ ways. Here the first version is the following:
\begin{subequations}
  \begin{align}
  \label{eq:VI.113a}
  \frac{d\tOp(r)}{dr}+\frac{\lO}{r}\cdot\tOp(r)-\bbAo(r)\cdot\tOm(r)=\frac{M+M_*}{\hbar}\,c\cdot\tOm(r)\\
  \label{eq:VI.113b}
  \frac{d\tOm(r)}{dr}-\frac{\lO-1}{r}\cdot\tOm(r)+\bbAo(r)\cdot\tOp(r)=\frac{M-M_*}{\hbar}\,c\cdot\tOp(r)\;,
  \end{align}
\end{subequations}
which is valid when the introduced {\em ortho-number}\/ $\lO$ of orbital angular momentum is linked to the angular constants $\dlO$ and $\ddlO$ by
\begin{subequations}
  \begin{align}
  \label{eq:VI.114a}
  \dlO&=\lO-1\\
  \label{eq:VI.114b}
  \ddlO&=-(\lO+1)\;.
  \end{align}
\end{subequations}
The second version looks slightly different
\begin{subequations}
  \begin{align}
  \label{eq:VI.115a}
  \frac{d\tOp(r)}{dr}-\frac{\lO}{r}\cdot\tOp(r)-\bbAo(r)\cdot\tOm(r)=\frac{M+M_*}{\hbar}\,c\cdot\tOm(r)\\
  \label{eq:VI.115b}
  \frac{d\tOm(r)}{dr}+\frac{\lO+1}{r}\cdot\tOm(r)+\bbAo(r)\cdot\tOp(r)=\frac{M-M_*}{\hbar}\,c\cdot\tOp(r)
  \end{align}
\end{subequations}
where the link of the ortho-number $\lO$ and the angular constants $\dlO,\,\ddlO$ reads now
\begin{subequations}
  \begin{align}
  \label{eq:VI.116a}
  \dlO&=-(\lO+1)\\
  \label{eq:VI.116b}
  \ddlO&=\lO-1\;.
  \end{align}
\end{subequations}
Observe here that {\em both}\/ links \myrf{VI.114a}-\myrf{VI.114b} \emph{and}
\myrf{VI.116a}-\myrf{VI.116b} entail a common form of link, namely
\begin{subequations}
  \begin{align}
  \label{eq:VI.117a}
  \dlO\cdot\ddlO&=1-\lO^2\\
  \label{eq:VI.117b}
  \dlO+\ddlO&=-2\;,
  \end{align}
\end{subequations}
and this is closely related to the posible values of $\lO$ ($\Rightarrow2,4,6,8,\,...$), see appendix E.

Clearly, the present ortho-number $\lO$ is the ortho-counterpart of the para-number $\lP$ defined through \myrf{VI.20a}-\myrf{VI.20b}; and this ortho/para analogy refers also to the pair of radial functions $\tilde{\Phi}_\pm(r)$ \myrf{VI.22} and $\tilde{\Omega}_\pm(r)$ \myrf{VI.112}. However, both corresponding eigenvalue systems, i.\,e. the former para-case \myrf{VI.23a}-\myrf{VI.23b} and the present ortho-cases \myrf{VI.113a}-\myrf{VI.113b} plus \myrf{VI.115a}-\myrf{VI.115b}, own a rather different degree of complexity so that the treatment of the ortho-case requires an extra discussion (see appendix E). On the other hand, the relative intricacy of the ortho-case disappears with respect to the eigenvalue equations if one writes down their non-relativistic approximation which is {\em shared by both}\/ versions \myrf{VI.113a}-\myrf{VI.113b} and \myrf{VI.115a}-\myrf{VI.115b}:
\begin{equation}
  \label{eq:VI.118}
  -\frac{\hbar^2}{2M}\left(\frac{d^2\tO(r)}{dr^2}+\frac{1}{r}\frac{d\tO(r)}{dr}\right) +\frac{\hbar^2}{2M}\frac{\lO^2}{r^2}\cdot\tO(r)-\hbar c\,\bbAo(r)\cdot\tO(r)=E_*\cdot\tO(r).
\end{equation}
Clearly, since the ortho-number $\lO$ appears here in quadratic form ($\sim\lO^2$) both
relativistic versions of the eigenvalue equations must have the same non-relativistic
limit \myrf{VI.118}; namely observe that the change of sign $\lO\rightarrow-\lO$ leaves
invariant the non-relativistic equation \myrf{VI.118} but interchanges both relativistic
versions
\myrf{VI.113a}/\myrf{VI.113b}$\leftrightarrow$\linebreak\myrf{VI.115a}/\myrf{VI.115b}. Furthermore,
the interpretation of the ortho-number $\lO$ as a quantum number of orbital angular
momentum seems legitimated by its appearance in connection with the centrifugal term
($\sim\lO^2$) in the Schr\"odinger-like equation \myrf{VI.118}, cf. the para-counterpart
\myrf{VI.42}.

Finally, it is also instructive to convince oneself of the fact that our improved ansatz
\myrf{VI.111a}-\myrf{VI.111b} (with general $g_R(\vartheta)$ and $g_S(\vartheta)$) for the
wave amplitudes $\bMRpm,\,\bMSpm$ actually equips now the Dirac spinor $\psi_\mathcal{O}$
\myrf{VI.100} with the desired \emph{one} quantum of angular momentum, i.\,e.
\begin{equation}
  \label{eq:VI.119}
  \hat{\mathcal{J}}\!_z\,\psi_\mathcal{O}=-\hbar\,\psi_\mathcal{O}\;,
\end{equation}
in contrast to the spoiled result \myrf{VI.101}! The present nice result for the ortho-wave function $\psi_\mathcal{O}$ comes about by observing its Whitney sum structure (cf. \myrf{VI.100}) with
\begin{eqnarray}
  \nonumber
  \hat{J}^{(+)}_z\,\bpp&=& (\hat{L}_z\bMRp)\cdot\wop+\bMRp\cdot(\hat{J}^{(+)}_z\,\wop)+(\hat{L}_z\bMSp)\cdot\wom+\bMSp\cdot(\hat{J}^{(+)}_z\,\wom)\\
  \label{eq:VI.120}
  &=&-\hbar\cdot\bpp\;.
\end{eqnarray}
Here it is especially interesting to observe that the general wave amplitudes $\bMRpm$ \myrf{VI.111a} carry two quanta of orbital angular momentum, i.\,e.
\begin{equation}
  \label{eq:VI.121}
  \hat{L}_z\,\bMRpm=-2\hbar\,\bMRpm\;,
\end{equation}
which however is partially compensated by the one quantum being carried by the spinor basis $\wopm$
\begin{equation}
  \label{eq:VI.122}
  \hat{J}_z^{(+)}\,\wopm=\pm\hbar\cdot\wopm\;.
\end{equation}
On the other hand, the amplitude field $\bMSpm$ \myrf{VI.111b} carries no angular momentum at all
\begin{equation}
  \label{eq:VI.123}
  \hat{L}_z\,\bMSpm=0\;,
\end{equation}
whereas the spinor basis $\wepm$ contributes one quantum
\begin{equation}
  \label{eq:VI.124}
  \hat{J}_z^{(-)}\,\wepm=\pm\hbar\cdot\wepm\;,
\end{equation}
cf. \myrf{V.49a}-\myrf{V.49d}. Thus the negative Pauli spinor carries also one quantum of total angular momentum
\begin{equation}
  \label{eq:VI.125}
  \hat{J}^{(-)}_z\,\bpm= -\hbar\,\bpm\;,
\end{equation}
so that both results \myrf{VI.120} and \myrf{VI.125} actually add up to the claimed result
\myrf{VI.119} for the Whitney sum $\psi_\mathcal{O}$ \myrf{VI.119}.

\vspace{3ex}
\begin{center}
\emph{\large Energy Functional for the Ortho-States}
\end{center}

The ortho-eigenvalue system (either in its relativistic form \myrf{VI.115a}-\myrf{VI.115b}
or in its non-relativistic approximation \myrf{VI.118}) appears to be very similar to its
para-counterpart \myrf{VI.23a}-\myrf{VI.23b}, or \myrf{VI.42}, resp.; but this should not
lead one astray to think that such a formal similarity is sufficient in order to ensure
the numerical equality of the corresponding energy spectra (due to the ortho- and
para-types). In fact, for such a numerical equality (i.e.\ the \emph{ortho/para
  degeneracy}) it is necessary that two additional conditions must be satisfied; namely
the numerical equality must also refer to {\bf (i)} the interaction potentials
(i.\,e. $\bbAo\Leftrightarrow\ppAo$) and {\bf (ii)} to the energy functionals
($\tEEPhi\Leftrightarrow\tEEO$). This means that, for the sake of comparison, we have to
set up now the ortho-functionals $\tEO,\,\tEEO$ as the ortho-counterparts of the
para-functionals $\tEPhi$ \myrf{VI.37} and $\tEEPhi$ \myrf{VI.50}, resp.

Turning here first to the relativistic form ($\tEO$), one restarts with the proposed
ortho-ansatz \myrf{VI.111a}-\myrf{VI.111b}, with general angular functions
$g_R(\vartheta),\,g_S(\vartheta)$ and under observation of the ortho-identification
\myrf{VI.112}; and one substitutes this ansatz now into the original energy functional
$\tETT$ \myrf{IV.20} with vanishing magnetic and exchange objects. Thus one obtains the
desired ortho-counterpart $\tEO$ of the para-functional $\tEPhi$ \myrf{VI.37} in the
following form
\begin{equation}
  \label{eq:VI.126}
  \tEO=2Mc^2\cdot\tZO^2+4\btTkin+\ERee+2\lD\cdot\tNO+\lGe\cdot\tNGee\;.
\end{equation}
The problem is now to adapt all the individual contributions to the peculiarities of the ortho-configurations. First, consider the mass renormalization factor $\Z_\mathcal{O}^2$
\begin{equation}
  \label{eq:VI.127}
  \Z_\mathcal{O}^2\doteqdot\int d^3\vec{r}\;\bar{\psi}_\mathcal{O}(\vec{r})\,\psi_\mathcal{O}(\vec{r})= \int d^3\vec{r}\,\left\{{}^{(b)}\varphi_{(+)}^\dagger(\vec{r})\,{}^{(b)}\varphi_{(+)}(\vec{r})-{}^{(b)}\varphi_{(-)}^\dagger(\vec{r})\,{}^{(b)}\varphi_{(-)}(\vec{r})\right\}
\end{equation}
which reads in terms of the wave amplitudes $\bMRpm,\,\bMSpm$ \myrf{V.50a}-\myrf{V.50b}
\begin{equation}
  \label{eq:VI.128}
  \Z_\mathcal{O}^2= \int\frac{d^3\vec{r}}{4\pi}\,\left\{\bMRpS\cdot\bMRp-\bMRmS\cdot\bMRm+\bMSpS\cdot\bMSp-\bMSmS\cdot\bMSm\right\}\,.
\end{equation}
But substituting here our general ortho-ansatz \myrf{VI.111a}-\myrf{VI.111b} lets reappear the mass renormalization factor $\Z_\mathcal{O}^2$ in the following factorized form
\begin{eqnarray}
  \nonumber
  \Z_\mathcal{O}^2\Rightarrow\tZO^2&=& \int\frac{d\Omega}{4\pi}\,\left\{\frac{g_R^2(\vartheta)}{\sin^3\vartheta}+\sin\vartheta\cdot g_S^2(\vartheta)\right\}\cdot\int dr\,r^2\,\frac{\tOp^2(r)-\tOm^2(r)}{r}\\
  \label{eq:VI.129}
  &&\qquad{{}\atop\displaystyle(d\Omega=\sin\vartheta\,d\vartheta\,d\phi)\;.}
\end{eqnarray}
This should be compared to the wave function normalization and thus suggests to establish separate normalization conditions for the angular and for the radial parts of the wave amplitude, cf. \myrf{III.49} and \myrf{IV.46}
\begin{eqnarray}
  \nonumber
  1&\stackrel{!}{=}&\int d^3\vec{r}\;{}^{(b)}k_0(\vec{r})=\int d^3\vec{r}\,\left\{{}^{(b)}\varphi_{+}^\dagger(\vec{r})\,{}^{(b)}\varphi_{+}(\vec{r})+{}^{(b)}\varphi_{-}^\dagger(\vec{r})\,{}^{(b)}\varphi_{-}(\vec{r})\right\}\\
  \label{eq:VI.130}
  &=& \int\frac{d^3\vec{r}}{4\pi}\,\left\{\frac{g_R^2(\vartheta)}{\sin^3\vartheta}+\sin\vartheta\cdot g_S^2(\vartheta)\right\}\cdot\frac{\tOp^2(r)+\tOm^2(r)}{r}\;,
\end{eqnarray}
i.\,e. we satisfy this normalization condition by putting separately
\begin{subequations}
  \begin{align}
  \label{eq:VI.131a}
  \int\frac{d\Omega}{4\pi}\,\left\{\frac{g_R^2(\vartheta)}{\sin^3\vartheta}+\sin\vartheta\cdot g_S^2(\vartheta)\right\}&\stackrel{!}{=}1\\
  \label{eq:VI.131b}
  \tNO\doteqdot\int dr\,r^2\,\frac{\tOp^2(r)+\tOm^2(r)}{r}-1&\equiv0\;.
  \end{align}
\end{subequations}
With this convention, the mass renormalization factor $\tZ_\Omega^2$ \myrf{VI.129} becomes reduced to
\begin{equation}
  \label{eq:VI.132}
  \tZO^2=\int dr\,r^2\,\frac{\tOp^2(r)-\tOm^2(r)}{r}\;.
\end{equation}
Clearly, this is again the ortho-counterpart of the para-case \myrf{VI.31}.

Concerning now the kinetic energy $\btTkin$, i.\,e. the second contribution to the energy functional $\tEO$ \myrf{VI.126}, one is forced to plunge into some technical details which are presented in appendix F. The result is, cf. equation \myrf{F14}
\begin{equation}
  \label{eq:VI.133}
  \bbtTkin=-\frac{\hbar c}{2}\int dr\,r^2\,\left\{\frac{\tOp(r)}{r}\frac{d\tOm(r)}{dr}-\frac{\tOm(r)}{r}\frac{d\tOp(r)}{dr}+(1\mp2\lO)\cdot\frac{\tOp(r)\cdot\tOm(r)}{r^2}\right\}\,.
\end{equation}
Furthermore, the electrostatic interaction energy $\ERee$ in the spherically symmetric approximation is the same as for the para-configurations, cf. \myrf{VI.34}
\begin{equation}
  \label{eq:VI.134}
  \ERee=-\frac{\hbar c}{\as}\int_0^\infty dr\,\left(r\cdot\frac{d\bbAo(r)}{dr}\right)^2\,, 
\end{equation}
whereas the ortho-counterpart of the electrostatic Poisson identity \myrf{VI.35} requires to restart from the general form of the mass equivalent $\Mee c^2$
\begin{equation}
  \label{eq:VI.135}
  \Mee c^2\doteqdot-\hbar c\int d^3\vec{r}\;{}^{(b)}k_0(\vec{r})\cdot\bAo(\vec{r})\;.
\end{equation}
But by observation of the ortho-density ${}^{(b)}k_0(\vec{r})$ \myrf{VI.130} this object factorizes again to
\begin{equation}
  \label{eq:VI.136}
  \tMee c^2=-\hbar c\int\frac{d\Omega}{4\pi}\,\left[\frac{g_R^2(\vartheta)}{\sin^3\vartheta}+\sin\vartheta\cdot g_S^2(\vartheta)\right]\cdot\int dr\,r^2\;\bAo(r)\cdot\frac{\tOp^2(r)+\tOm^2(r)}{r}\;,
\end{equation}
i.\,e. by reference to the angular normalization \myrf{VI.131a}
\begin{equation}
  \label{eq:VI.137}
  \tMee c^2=-\hbar c\int dr\,r^2\;\bAo(r)\cdot\frac{\tOp^2(r)+\tOm^2(r)}{r}\;.
\end{equation}
Thus the electric Poisson identity reads also in the ortho-configuration in the spherically symmetric approximation
\begin{equation}
  \label{eq:VI.138}
  \tNGee\doteqdot\ERee-\tMee c^2\equiv0\;,
\end{equation}
where $\ERee$ is given by \myrf{VI.134} and $\tMee c^2$ by \myrf{VI.137}.
The important thing here is that, apart from the different normalizations of the wave functions (cf. the para-normalization \myrf{VI.36} vs. the present ortho-normalization \myrf{VI.131b}), both mass equivalents \myrf{VI.35} and \myrf{VI.137} are formally identical as expected; and this similarity holds then also for the corresponding Poisson identities, cf. \myrf{VI.33} vs. \myrf{VI.138}. However, it must be stressed that these formal similarities between the ortho- and para-configurations are valid only for the spherically symmetric approximation because here the different angular dependencies become hidden behind the separate normalization to unity of just the angular parts of the wave amplitudes!

With all the individual contributions to the ortho-energy $\tEO$ \myrf{VI.126} being now specified, it is again very satisfying to find the former relativistic ortho-system \myrf{VI.113a}-\myrf{VI.116b} being identical to just the extremal equations of the present ortho-functional (i.\,e. $\delta\tEO=0$ $\leadsto$ {\em principle of minimal energy}\/). Especially, the solutions of the electric Poisson equation for the ortho-configurations
\begin{equation}
  \label{eq:VI.139}
  \frac{d^2\bbAo(r)}{dr^2}+\frac{2}{r}\frac{d\bbAo(r)}{dr}=-\as\,\frac{\tOp^2(r)+\tOm^2(r)}{r}
\end{equation}
extremalize the energy functional $\tEO$ \myrf{VI.126} with respect to $\bbAo(r)$ within the class of spherically symmetric trial functions $\bbAo(r)$. Therefore this equation \myrf{VI.139} is then to be considered as the spherically symmetric approximation to the original Poisson equation \myrf{VI.91} for the ortho-configurations, i.\,e. the ortho-counterpart of the para-equation \myrf{VI.38}.

Of course, the similarity of the present ortho-equation \myrf{VI.139} with its
para-counterpart \myrf{VI.38} raises now the general question of the extent to which the
former para-eigenvalue problem and the present ortho-eigenvalue problem are representing
essentially the {\em same}\/ mathematical problem? Indeed, this ortho/para equivalence
(within the framework of the spherically symmetric approximation) holds not only for the
above-mentioned electric Poisson equations but equally well for both eigenvalue systems
(cf. the para-case \myrf{VI.23a}-\myrf{VI.23b} to its ortho-analogy
\myrf{VI.115a}-\myrf{VI.115b}); and it holds also for both energy functionals $\tEPhi$
\myrf{VI.37} and $\tEO$ \myrf{VI.126}. Obviously, both eigenvalue problems are completely
equivalent from the mathematical viewpoint which in physical terms is nothing else than
the expected ortho/para degeneracy. And therefore it must be clear that both eigenvalue
problems do generate, albeit only {\em in the spherically symmetric approximation}\/, the
same arrangement of energy levels (fig.~2)! This is to be conceived as the RST counterpart
of the conventional energy degeneracy of ortho- and para-positronium when the magnetic
interactions are neglected, see subsection {\bf V.1} ({\em Conventional Multiplet
  Structure}\/).

But the present RST picture of this ortho/para-degeneracy is somewhat more intricate than
its conventional competitor, namely because it elucidates also the fact that the
ortho/para-degeneracy becomes broken not only if the magnetic interactions are taken into
account but also if the spherically symmetric approximation is dropped. In the latter case
the reduced wave amplitudes $\sRpm,\,\sSpm$ \myrf{VI.19a}-\myrf{VI.19b} for the
para-configurations and their ortho-counterparts $\ssRpm,\,\ssSpm$
\myrf{VI.111a}-\myrf{VI.111b} will depend also on the polar angles $\vartheta$ and $\phi$;
and additionally the electric potential $\bpAo(\vec{r})$ will become a function of $r$
{\em and}\/ $\vartheta$, even if the magnetic interactions are kept inactive. Here it
should be evident that such an angle-dependent charge density ${}^{(b)}k_0(\vec{r})$, as
it is observed for the ortho-configurations (fig.~5), will entail a rather different
interaction energy (of the electric type) in comparison to the ``almost'' spherically
symmetric charge densities ${}^{(p)}k_0(\vec{r})$ ($\sim\frac{1}{\sin\vartheta}$) of the
para-configuration, cf. \myrf{VI.17}. Thus there remains in RST the problem to clarify the
mutual numerical relationship of the anisotropic and the magnetic corrections which are to
be added to the predictions of the spherically symmetric approximation.

\vspace{3ex}
\begin{center}
\emph{\large Anisotropic and Magnetic Effects}
\end{center}

A first preliminary treatment of the anisotropic effects can be carried out most easily in
the non-relativistic approximation. Recall here the remarkable fact that both relativistic
eigenvalue systems \myrf{VI.113a}-\myrf{VI.113b} and \myrf{VI.115a}-\myrf{VI.115b} have
the same non-relativistic limit \myrf{VI.118}. However, observe also that this common
limit form of the eigenvalue equations does not imply the numerical equality of the
corresponding energy levels if the anisotropic or magnetic (or simultaneously both)
effects are taken into account. Actually, it is only in the electrostatic and spherically
symmetric approximation that both coinciding limit forms \myrf{VI.118} of the relativistic
eigenvalue equations do imply the same energy levels. The reason is here that both
relativistic versions \myrf{VI.113a}-\myrf{VI.113b} and \myrf{VI.115a}-\myrf{VI.115b} are
to be linked to different angular parts of the charge density ${}^{(b)}k_0(\vec{r})$, even
in the spherically symmetric approximation.

In order to present a simple example of this charge-density dichotomy of the
ortho-configurations, one may consider the smallest possible value for the ortho-number
$\lO$, i.\,e. $\lO=2$ (see the first line of the table given in appendix E). For this case
the angular functions $g_R(\vartheta)$ \myrf{VI.105a} and $g_S(\vartheta)$ \myrf{VI.105b} are
given by
\begin{subequations}
  \begin{align}
  \label{eq:VI.140a}
  g_R(\vartheta)&=\rho_3\cdot\sin^3\vartheta\\
  \label{eq:VI.140b}
  g_S(\vartheta)&=\sigma_1\cdot\cos\vartheta\;.
  \end{align}
\end{subequations}
Furthermore, the Dirac density ${}^{(b)}k_0(\vec{r})$ is seen from equation \myrf{VI.130} to factorize into an angular and a radial part according to
\begin{equation}
  \label{eq:VI.141}
  \bko(r,\vartheta)=\bbko(r)\cdot\bbko(\vartheta)
\end{equation}
where the radial part is given by
\begin{equation}
  \label{eq:VI.142}
  \bbko(r)=\frac{\tOp^2(r)+\tOm^2(r)}{r}\;,
\end{equation}
and similarly for the angular part
\begin{equation}
  \label{eq:VI.143}
  \bbko(\vartheta)=\frac{1}{4\pi}\left[\frac{g_R^2(\vartheta)}{\sin^3\vartheta}+\sin\vartheta\cdot g_S^2(\vartheta)\right]\,.
\end{equation}
This latter part is the marked difference between both types of ortho-configurations and becomes with reference to the chosen simplest situation \myrf{VI.140a}-\myrf{VI.140b}
\begin{equation}
  \label{eq:VI.144}
  \bbko(\vartheta)\Rightarrow \frac{\rho_3^2}{4\pi}\left[\sin^3\vartheta+\frac{\sigma_1^2}{\rho_3^2}\,\sin\vartheta\cos^2\vartheta\right]\,.
\end{equation}

The claimed density ambiguity of the ortho-configurations arises now through the circumstance that the ratio $\displaystyle\frac{\sigma_1}{\rho_3}$ can adopt two different values, namely for $\lO=2$ either
\begin{equation}
  \label{eq:VI.145}
  \frac{\sigma_1}{\rho_3}=-1
\end{equation}
for the first system \myrf{VI.113a}-\myrf{VI.113b}; or alternatively for the second system \myrf{VI.115a}-\myrf{VI.115b}:
\begin{equation}
  \label{eq:VI.146}
  \frac{\sigma_1}{\rho_3}=3\;,
\end{equation}
see equations \myrf{E15}-\myrf{E16} and the first line of the table in appendix E. Accordingly, one
obtains two different angular factors $\bbko(\vartheta)$ \myrf{VI.143} which both can be
required to obey the separate normalization condition \myrf{VI.131a} and do then appear as
follows:
\begin{subnumcases}{\label{eq:VI.147}
  \bbko(\vartheta)=\frac{\sin\vartheta}{\pi^2}\cdot}
  1 & \label{eq:VI.147a}\\
  \frac{1}{3}\,(1+8\cos^2\vartheta)\;, & \label{eq:VI.147b}
\end{subnumcases}
see fig.~5. Of course, such an angle-dependent density would generate also an
angle-dependent potential $\bAo(r,\vartheta)$ but this becomes truncated in the {\em
  spherically symmetric approximation}\/ to a symmetric potential
(i.\,e. $\bAo(r,\vartheta)\Rightarrow\bbAo(r)$) so that the angular discrimination between
both types of ortho-configurations becomes ineffectual.

But, strictly speaking, both states (due to the same value of $\lO$ but different ratio $\displaystyle\frac{\sigma_1}{\rho_3}$) must be expected to carry different energy $\ET$, even under the neglection of magnetism! The reason is that this quantity $\ET$ reacts sensitively to any redistribution of the {\em electric}\/ charge density $\bko(\vec{r})$, namely via the electrostatic interaction energy $\ERe$ \myrf{VI.10} which may be rewritten in terms of the Dirac density $\bpko(\vec{r})$ as
\begin{equation}
  \label{eq:VI.148}
  \ERe=-\iint d^3\vec{r}\,d^3\vec{r}\,'\;\frac{\bpko(\vec{r})\cdot\bpko(\vec{r}\,')}{\|\vec{r}-\vec{r}\,'\|}\;.
\end{equation}
Nevertheless, both charge distributions for given quantum number $\lO$ do generate the
same electrostatic interaction energy $\ERee$ \myrf{VI.134} {\em provided that one resorts
  to the spherically symmetric approximation}\/! The reason is that the approximate
Poisson equation \myrf{VI.139} implies the validity of the associated Poisson identity
\myrf{VI.138} with the electrostatic interaction energy $\ERee$ being specified by
equation \myrf{VI.134} and its mass equivalent $\tMee c^2$ by \myrf{VI.137}. However,
neither of the latter two quantities $\ERee,\,\tMee c^2$ is influenced by the angular
parts $g_R(\vartheta)$ and $g_S(\vartheta)$ of the wave amplitudes as long as one is
satisfied with the spherically symmetric approximation! From this reason one finally
arrives at a multiple degeneracy phenomenon: namely {\bf (i)} the mathematical equivalence
of the para-eigenvalue problem \{\myrf{VI.23a}-\myrf{VI.23b},\myrf{VI.37},\myrf{VI.38}\}
and the ortho-eigenvalue problem
\{\myrf{VI.115a}-\myrf{VI.115b},\myrf{VI.126},\myrf{VI.139}\} says that the solutions of
both eigenvalue problems must be the same and therefore must carry also the same energy
$\ET$; and {\bf (ii)} the solutions of both {\em ortho-subsystems}\/
\myrf{VI.113a}-\myrf{VI.113b} and \myrf{VI.115a}-\myrf{VI.115b} do also carry the same
energy (albeit only in the spherically symmetric and non-relativistic approximation).

Furthermore, it should be obvious that one has to expect also a certain interplay between these anisotropic and the magnetic effects; for a more distinct anisotropy of the wave amplitudes will imply also a more marked anisotropy of the azimuthal current $\bkphi$ \myrf{V.57c}. For instance, the latter quantity becomes quite generally by means of our ortho-ansatz \myrf{VI.111a}-\myrf{VI.111b}
\begin{equation}
  \label{eq:VI.149}
  \bkphi=\frac{\tOp(r)\cdot\tOm(r)}{2\pi r}\left\{\frac{g_R^2(\vartheta)}{\sin^2\vartheta}-\sin^2\vartheta\,g_S^2(\vartheta)-\frac{2\cos\vartheta\,g_R(\vartheta)\cdot g_S(\vartheta)}{\sin\vartheta}\right\}\,,
\end{equation}
and if we are satisfied for the moment with the simplest possible ortho-configuration
(i.\,e. the first line of the table in appendix E with $\lO=2$) one is led to the
following result (see also fig.~6)
\begin{subnumcases}{\label{eq:VI.150}
  \bkphi\Rightarrow\frac{\tOp(r)\cdot\tOm(r)}{2r}\left(\frac{2}{\pi}\,\sin\vartheta\right)^2\cdot}
  1\;, & $\displaystyle\frac{\sigma_1}{\rho_3}=-1\qquad$ \label{eq:VI.150a}\\
  \frac{1}{3}\,(1-16\cos^2\vartheta)\;, & $\displaystyle\frac{\sigma_1}{\rho_3}=3\;.$ \label{eq:VI.150b}
\end{subnumcases}

The key features of this ortho-current become obvious through a comparison to its
para-counterpart $\pkphi$ \myrf{VI.30}, see also fig.~1. The first observation is here
that the para-current $\pkphi$ is finite (and therefore singular) on the $z$-axis
($\vartheta=0,\pi$) whereas the ortho-current $\bkphi$ \myrf{VI.150a}-\myrf{VI.150b}
vanishes for $\vartheta=0,\pi$. This is physically plausible because the angular momentum
$J_z$ of the para-current is zero, cf. equation \myrf{V.46}, which may be associated with
a very close revolution of the quantum fluid around the $z$-axis (fig.~1). On the other
hand, the ortho-current $\bkphi$ \myrf{VI.150a}-\myrf{VI.150b} is zero on the $z$-axis and
is maximal off the $z$-axis which intuitively is to be associated with a larger angular
momentum; in fact, the eigenvalue of the present ortho-configuration $\psi_\mathcal{O}$
corresponds to one quantum of angular momentum, cf. equation \myrf{VI.119}. Clearly, such
a difference of the angular distribution of the currents will entail also a different
magnetic interaction energy of the para- and of the ortho-(sub)configurations, beyond the
different relative sign of the individual magnetic fields $\vec{H}_1$ and $\vec{H}_2$ due
to $\vec{H}_p$ and $\vec{H}_b$, cf. the discussion of the magnetic energy difference
$\Delta\!^\textrm{(m)}E_\textrm{T}$ \myrf{V.30}-\myrf{V.31}.

But clearly it remains to be clarified whether those density ambiguities as for $\bbko$ \myrf{VI.147} or for $\bkphi$ \myrf{VI.150} are artefacts due to the product ansatz \myrf{VI.111a}-\myrf{VI.111b} in connection with the spherically symmetric approximation, or whether that ambiguity is a real physical effect which would persist also for the exact solutions of the eigenvalue problem?

\vspace{3ex}
\begin{center}
\emph{\large Non-Relativistic Energy Functional}
\end{center}

For a first rough estimate of the energy content of the ortho-configurations it may again
be sufficient to consider merely the non-relativistic approximation. Its deduction follows
essentially those paths having been described already for the para-configurations,
cf. \myrf{VI.50}; however, there arise here some few points which are somewhat more subtle
and therefore need a closer inspection.  The main difference is in the present context
that the range of the ortho-number $\lO$ is restricted to $\lO\geq2$, whereas the
para-number $\lP$ can adopt any non-negative integer; and furthermore the passage from
formally $\lO\Rightarrow-\lO$ is associated with a distinctly different ortho-density
$\bbko(\vartheta)$, cf. \myrf{VI.147a} vs. \myrf{VI.147b}, whereas the para-density $\pko$
\myrf{VI.17} does not react at all to the substitution $\lP\Rightarrow-\lP$, cf. the
angular dependence of the para-amplitudes $\ptRpm,\,\ptSpm$
\myrf{VI.25a}-\myrf{VI.27b}. These differences must now be elaborated in greater detail in
order to better understand the ortho/para degeneracy and its breaking by the magnetic
interactions and by the anisotropy of the electric interaction potential.

It is true, both relativistic versions \myrf{VI.113a}-\myrf{VI.113b} and
\myrf{VI.115a}-\myrf{VI.115b} of the eigenvalue equations for the ortho-configurations are
related to each other through the formal substitution $\lO\Rightarrow-\lO$ which then
would correspond to the analogous replacement ($\lP\Rightarrow-\lP$) for the para-system
\myrf{VI.23a}-\myrf{VI.23b}. The same change of sign would then also apply for both
kinetic energies
${}^{[p]}\tilde{T}_\textrm{kin}={}^{[p]}\tilde{T}_{(r)}+{}^{[p]}\tilde{T}_{(\vartheta)}$
\myrf{VI.32a}-\myrf{VI.32b} and $\btTkin$ \myrf{VI.133}, but the case with the charge
densities $\bpko(\vec{r})$ is somewhat different. Namely, while the replacement
$\lP\Rightarrow-\lP$ leaves invariant the charge density $\pko$ as the source of the
electric potential $\pAo$, cf. \myrf{VI.17}, the passage from the first ortho-case
\myrf{VI.113a}-\myrf{VI.113b} to the second ortho-case \myrf{VI.115a}-\myrf{VI.115b} is
accompanied by an essential change of the ortho-density $\bko(\vec{r})$: its angular part
$\bbko(\vartheta)$ \myrf{VI.143} receives a considerable change of the magnitude of both
angular functions $g_R(\vartheta)$ and $g_S(\vartheta)$. This relative magnitude may be
measured by the ratio of the coefficients $\displaystyle\frac{\sigma_1}{\rho_3}$,
cf. equations \myrf{E15}-\myrf{E16} of appendix E. A special example of this effect is
given by the equations \myrf{VI.145}-\myrf{VI.147b}. It is only on account of the
spherically symmetric approximation that this change of the charge density $\bko(\vec{r})$
leaves us without any effect since its angle-dependent part is integrated off for the
spherically symmetric form of the energy functional, see for example the deduction of the
mass renormalization factor $\tZ_\Omega^2$ \myrf{VI.132} or also the deduction
of the kinetic energies
${}^{[b]}\tilde{T}_{r},\,{}^{[b]}\tilde{T}_{\vartheta},\,{}^{[b]}\tilde{T}_{\phi}$ in
appendix F.

On the other hand, both relativistic ortho-systems \myrf{VI.113a}-\myrf{VI.113b} and
\myrf{VI.115a}-\myrf{VI.115b} have the same non-relativistic approximation \myrf{VI.118},
albeit only within the framework of the spherically symmetric approximation. This
non-relativistic limit form \myrf{VI.118} of both ortho-versions is the same as the
corresponding limit form of the para-versions, cf. \myrf{VI.42}; and also the {\em
  relativistic}\/ energy functionals $\tEPhi$ \myrf{VI.50} and $\tEO$ \myrf{VI.126} are
the immediate counterparts of each other. Finally, the electric Poisson equations are also
formally the same since the non-relativistic limit of \myrf{VI.139}
\begin{equation}
  \label{eq:VI.151}
  \frac{d^2\bbAo(r)}{dr^2}+\frac{2}{r}\frac{d\bbAo(r)}{dr}=-\as\,\frac{\tilde{\Omega}(r)^2}{r}
\end{equation}
is again the immediate ortho-counterpart of its para-predecessor \myrf{VI.44}.

Thus there remains the last problem, namely of setting up the non-relativistic version $\tEEO$ of the present relativistic ortho-functional $\tEO$ \myrf{VI.126}. Clearly, one expects here that this will turn out again as the immediate ortho-counterpart of its para-predecessor $\tEEPhi$ \myrf{VI.50}; i.\,e. the expectation is the following
\begin{equation}
  \label{eq:VI.152}
  \tEEO=2E_\textrm{kin}+\ERee+2\lambda_\textrm{S}\cdot\tNNO+\lambda_\textrm{G}^\textrm{(e)}\cdot\tNNGee
\end{equation}
so that we are left with the problem to validate this assertion \myrf{VI.152} by deducing the non-relativistic forms of all the relativistic contributions to the functional $\tEO$ \myrf{VI.126}. To this end, one will of course follow again the path traced out already by the analogous para-calculations, see equations \myrf{VI.45}-\myrf{VI.49}:
\begin{enumerate}[\bf (i)]
\item {\em mass renormalization $\tilde{\Z}_\Omega^2$ \myrf{VI.132}:}
\begin{equation}
  \label{eq:VI.153}
  2\,Mc^2\tilde{\Z}_\Omega^2\Rightarrow 2\,Mc^2-\frac{\hbar^2}{M}\int dr\,r\,\left\{\left(\frac{d\tilde{\Omega}}{dr}\right)^2+\frac{\lO^2}{r^2}\,\tilde{\Omega}^2\right\} \mp\frac{\hbar^2}{M}\,\lO\int dr\;\frac{d}{dr}\left(\tilde{\Omega}(r)^2\right)
\end{equation}
\item {\em kinetic energy $\bbtTkin$ \myrf{VI.133}:}
\begin{subequations}
  \begin{align}
  \label{eq:VI.154a}
  \bbtTkin&\Rightarrow E_\textrm{kin}\pm\frac{\hbar^2}{4M}\,\lO\int dr\;\frac{d}{dr}\left(\tilde{\Omega}(r)\right)^2\\
  \label{eq:VI.154b}
  \big(\,E_\textrm{kin}&\doteqdot\frac{\hbar^2}{2M}\int dr\,r\,\left\{\left(\frac{d\tilde{\Omega}}{dr}\right)^2+\frac{\lO^2}{r^2}\,\tilde{\Omega}^2\right\}\,\big)
  \end{align}
\end{subequations}
\item {\em electric Poisson identity \myrf{VI.138}}
\begin{equation}
  \label{eq:VI.155}
  \tNGee\Rightarrow\tNNGee\doteqdot\ERee-\tMMee c^2\equiv0\;.
\end{equation}
Here the interaction energy $\ERee$ is formally the same for both the relativistic and the
non-relativistic situation, cf. \myrf{VI.134}; but for the non-relativistic form of the
mass equivalent $\tMMee c^2$ one neglects again the ``negative'' wave amplitude $\tOm(r)$
against its ``positive'' counterpart $\tOp(r)$, i.\,e. one puts
\begin{equation}
  \label{eq:VI.156}
  \tMee c^2\Rightarrow\tMMee c^2=-\hbar c\int dr\,r\;\bbAo(r)\cdot\left(\tilde{\Omega}(r)\right)^2\;,
\end{equation}
\item {\em normalization condition \myrf{VI.131b}:}
\begin{equation}
  \label{eq:VI.157}
  \tilde{N}_\Omega\Rightarrow\tNNO\doteqdot\int dr\,r\;\tilde{\Omega}(r)^2-1\equiv0\;.
\end{equation}
\end{enumerate}
Thus the non-relativistic ortho-functional $\tEEO$ \myrf{VI.152} is uniquely fixed; and it
is again a rather satisfying aspect that the boundary term ($\sim\lO$) cancels through the
cooperation of mass renormalization and kinetic energy:
\begin{equation}
  \label{eq:VI.158}
  2Mc^2\,\tilde{\Z}_\Omega^2+4\,\bbtTkin\Rightarrow2E_\textrm{kin}\;,
\end{equation}
with the one-particle kinetic energy $E_\textrm{kin}$ being specified by equation
\myrf{VI.154b} and the rest mass $2Mc^2$ being omitted. This cancellation is an essential
fact for the non-relativistic limit because it just ensures its uniqueness ($\leadsto$
both relativistic ortho-versions \myrf{VI.126} with $\pm\lO$ in the kinetic energy
$\bbtTkin$ \myrf{VI.133} generate the one non-relativistic form \myrf{VI.152} with
quadratic dependence upon $\lO$, cf. \myrf{VI.154b}).

Finally, it is also a very satisfying exercise to identify {\em both}\/ the former
non-relativistic eigenvalue equation \myrf{VI.118} {\em and}\/ the present Poisson
equation \myrf{VI.151} as the extremal equations ($\delta\tEEO=0$) due to the claimed
energy functional $\tEEO$ \myrf{VI.152}. This proves the mathematical equivalence of the
{\em non-relativistic}\/ eigenvalue problems for the ortho- and para-configurations, {\em
  albeit only in the electrostatic and spherically symmetric approximation}\/. The
physical consequence hereof is the ortho/para degeneracy as the RST counterpart of the
conventional degeneracy \myrf{V.13}. For the breaking of this ortho/para-degeneracy one
first has to work out more exact solutions of the present RST eigenvalue problems,
especially with regard to the anisotropy of the electric interaction potential
$\bpAo(\vec{r})$.

\renewcommand{\theequation}{\Alph{section}.\arabic{equation}}

  \newpage
  \setcounter{section}{1}
  \setcounter{equation}{0}
  \begin{center}
  {\textbf{\Large Appendix \Alph{section}:}}\\
  \emph{ \textbf{\Large Manifest Gauge Invariance of the Energy Functional}} \textbf{\Large~$\mathbf\tETT$}
  \end{center}
  \vspace{2ex}
  
  As it stands, the proposed energy functional $\tETT$ \myrf{IV.20} with its physical part
  $\Eiv$ \myrf{IV.23} is surely not {\em manifestly}\/ invariant with respect to the
  magnetic gauge transformations \myrf{III.20a}-\myrf{III.20c}, because there are two
  kinds of contributions which lack that desired invariance: these are {\bf (i)} the
  kinetic energies $\Tkina$ ($a=1,2,3$), as they are defined by equations
  \myrf{IV.26}-\myrf{IV.27}, and {\bf (ii)} the magnetic constraint term $\NGm$
  \myrf{IV.5}. Perhaps a somewhat critical term (concerning the gauge invariance) is
  $\nGm$ \myrf{IV.6} which contains the integral of the scalar product of
  $\vec{\mathfrak{A}}$ and $\vec{l}$ and transforms as follows:
\begin{equation}
  \label{eq:a.1}
  \int d^3\vec{r}\,\big(\vec{\mathfrak{A}}\sdot \vec{l}\,\big)\,\rightarrow\,\int d^3\vec{r}\,\big(\vec{\mathfrak{A}}'\sdot \vec{l}\,\big)\,=\,\int d^3\vec{r}\,\big(\vec{\mathfrak{A}}\sdot \vec{l}\,\big)-\int d^3\vec{r}\;\vec{\nabla}(\alpha_2-\alpha_3)\sdot \vec{l}\;.
\end{equation}
Thus the gauge invariance demands the vanishing of the last integral
\begin{equation}
  \label{eq:a.2}
  \int d^3\vec{r}\;\vec{\nabla}(\alpha_2-\alpha_3)\sdot \vec{l}\,=\,-\int d^3\vec{r}\;(\alpha_2-\alpha_3)\vec{\nabla}\sdot\vec{l}\,\stackrel{!}{=}0\;,
\end{equation}
and this would ensure the gauge invariance of $\nGm$ \myrf{IV.6}. This requirement
\myrf{a.2}, however, is satisfied for the \emph{stationary} bound systems. Namely, these systems
are to be expected to have source-free currents, cf. \myrf{II.18b}-\myrf{II.18c}
\begin{equation}
  \label{eq:a.3}
  \nabla^\mu k_{a\mu}\equiv0\,\Rightarrow\,B^\mu h_\mu-\Bstar^\mu\hstar_\mu=0\;,
\end{equation}
and this assumption lets vanish the source of the entanglement current $l_\mu$ \myrf{III.47} for the stationary situation
\begin{equation}
  \label{eq:a.4}
  \nabla^\mu l_\mu=\vec{\nabla}\sdot\vec{l}\equiv0
\end{equation}
which obviously validates the invariance requirement \myrf{a.2}.

Now it should be self-evident that the claim of invariance of $\tETT$ can be true only if
the {\em sum}\/ of the two non-invariant terms {\bf (i)} and{\bf (ii)} is itself
manifestly invariant. This, however, can easily be proven by first decomposing that
magnetic constraint term $\NGm$ \myrf{IV.5} into a gauge-dependent part $\tNGm$ and a
manifestly gauge-independent part $\hNGm$, i.\,e. we put
\begin{equation}
  \label{eq:a.5}
  \NGm=\tNGm+\hNGm\;,
\end{equation}
with
\begin{subequations}
\begin{align}
  \label{eq:a.6a}
  \tNGm&=-\frac{\hbar c}{2}\int d^3\vec{r}\,\left(\vAI\sdot\vec{j}_1+\vAII\sdot\vec{j}_2+\vAIII\sdot\vec{j}_3\right)\\
  \label{eq:a.6b}
  \hNGm&=\frac{\hbar c}{4\pi\as}\int d^3\vec{r}\,
     \left[
      \left( \vec{\nabla}\times\vec{A}_1\right) \sdot \left(\vec{\nabla}\times\vec{A}_2\right) +
      \left( \vec{\nabla}\times\vec{A}_2\right) \sdot \left(\vec{\nabla}\times\vec{A}_3\right)\right.\\
      & \left.+\left( \vec{\nabla}\times\vec{A}_3\right) \sdot \left(\vec{\nabla}\times\vec{A}_1\right) 
    \right]\,.\nonumber
\end{align}
\end{subequations}
But here the Maxwellian three-currents $\vec{j}_a\vr$ ($a=1,2,3$) in \myrf{a.6a} are identical (up to sign) to the Dirac currents $\vec{k}_a\vr$ \myrf{II.17a}-\myrf{II.17c} so that one has (e.\,g. for the second particle)
\begin{equation}
  \label{eq:a.7}
  -\int d^3\vec{r}\;\vAII\sdot\vec{j}_2=\int d^3\vec{r}\;\vAII\sdot\vec{k}_2=\int d^3\vec{r}\;\vAII\sdot\left[\zpp^\dagger\vec{\sigma}\zpm+\zpm^\dagger\vec{\sigma}\zpp\right]\,.
\end{equation}
On the other hand, the kinetic energy $\Tkinz$ \myrf{IV.27} reads in terms of the Pauli spinors $\zppm\vr$
\begin{equation}
  \label{eq:a.8}
  \Tkinz=-i\,\frac{\hbar c}{2}\int d^3\vec{r}\,\left[\zpm^\dagger\vec{\sigma}\sdot\vec{\nabla}\zpp+\zpp^\dagger\vec{\sigma}\sdot\vec{\nabla}\zpm\right]
\end{equation}
so that the sum of both non-invariant terms for the second particle ($a=2$) actually appears now in a manifestly gauge invariant form $\mathcal{T}_{\textrm{kin}(2)}$:
\begin{equation}
  \begin{split}
  \label{eq:a.9}
  \mathcal{T}_{\textrm{kin}(2)}&\doteqdot\Tkinz-\frac{\hbar c}{2}\int d^3\vec{r}\;\vAII\sdot\vec{j}_2\\
  &=-i\,\frac{\hbar c}{2}\int d^3\vec{r}\,\left[\zpm^\dagger\vec{\sigma}\sdot\big(\vec{\nabla}+i\vAII\big)\zpp+\zpp^\dagger\vec{\sigma}\sdot\big(\vec{\nabla}+i\vAII\big)\zpm\right]\,.
  \end{split}
\end{equation}

When the first ($a=1$) and third ($a=3$) particle are treated in the same way, the energy functional emerges finally in its manifestly gauge-invariant form:
\begin{equation}
\begin{split}
  \label{eq:a.10}
  \tETT&=\Z_{(1)}^2\cdot M_\textrm{p}c^2+\Z_{(2)}^2\cdot M_\textrm{e}c^2+\Z_{(3)}^2\cdot M_\textrm{e}c^2+ 2\mathcal{T}_\textrm{kin}\\
  &\hspace{-2.5em}+\sum_{a=1}^3\lDa\cdot\NDa+\lGe\cdot(\NGe-\nGe)+\lGm\cdot(\hNGm-\nGm)+\lGh\cdot(\NGh-\nGh)\\
  &\hspace{-2.5em}+\lGg\cdot(\NGg-\nGg)\;.
\end{split}
\end{equation}
Here, the first line constitutes now again the physical part of the RST energy $E_T$
(which, however, is now manifestly gauge invariant) whereas the second and third line
represent the constraints with $\NGm$ being replaced by its invariant part $\hNGm$
\myrf{a.6b}. But as a consequence of this rearrangement, the residual magnetic constraint
term $\hNGm-\nGm$ contributes now non-trivially to $\tETT$.

  \newpage
  \stepcounter{section}
  \setcounter{equation}{0}
  \begin{center}
  {\textbf{\Large Appendix \Alph{section}:}}\\
  \emph{\textbf{\Large Variational Deduction\\ of the Non-Relativistic Eigenvalue Equations}}
  \end{center}
  \vspace{2ex}

  The detailed variational deduction of the RST field equations is an instructive exercise
  since it mediates a better feeling of how to handle the magnetic gauge structure. In
  order to present a brief example of this, consider the deduction of the Pauli eigenvalue
  equations \myrf{IV.40a}-\myrf{IV.40c}. First recall here the fact that all the gauge
  field energies \myrf{IV.28}-\myrf{IV.31} do not depend at all upon the matter fields
  $\ap$; and therefore the corresponding variations ($\dkmat$, say) of the energy
  functional $\tEET$ \myrf{IV.48} with respect to the conjugate Pauli spinors
  $\aphip^\dagger$ becomes simply
\begin{equation}
  \label{eq:A.1}
  \begin{split}
  \dkmat\tEET &= \dkmat\,\Bigg[\Ekin+\hbar c\int d^3\vec{r}\left\{\vec{H}_I\sdot\vmSe+\vec{H}_{II}\sdot\vmSz+\vec{H}_{III}\sdot\vmSd\right\}\\ &+\sum_{a=1}^3\lPa\cdot\NNDa+\lGe\cdot\NNGe\Bigg]\,+\dkmat\,\Big[\lGh\cdot\NNGh+\lGg\cdot\NNGg\Big]\;.
  \end{split}
\end{equation}
Here, carrying out the variational procedure for the first part proceeds along standard rules and yields
\begin{equation}
  \label{eq:A.2}
  \begin{split}
  \dkmat\left[\Ekin+\ldots+\lGe\cdot\NNGe\right]&=
  \int d^3\vec{r}\left\{\big(\delta\epp\big)^\dagger\left[\hat{\H}_{(I)}+\lPe\right]\epp\right.\\
  &{}\hspace{-7em}\left.+\big(\delta\zpp\big)^\dagger\left[\hat{\H}_{(II)}+\lPz\right]\zpp
  +\big(\delta\dpp\big)^\dagger\left[\hat{\H}_{(III)}+\lPd\right]\dpp\right\}
  \end{split}
\end{equation}
where the one-particle Hamiltonians ($\hat{\H}$) have already been defined through equations \myrf{IV.42a}-\myrf{IV.42c}, and the previous value \myrf{IV.22b} for the multiplier $\lGe$ has also been used.

However, the variational formalism for the second part (i.\,e. exchange part in
\ref{eq:A.1}) requires a more thorough discussion. It is true, the ``electric'' part
($\NNGh$) hereof presents no problem because the matter fields $\appm$ enter the exchange
density $h_0(\vr)$ in the form \cite{MaSo2}
\begin{equation}
  \label{eg:A.3}
  h_0\doteqdot\zpp^\dagger\,\dpp+\zpm^\dagger\,\dpm
\end{equation}
and thus its non-relativistic approximation $\Ih_0(\vr)$ is obtained by simply neglecting the ``negative'' Pauli spinors $\aphim$:
\begin{equation}
  \label{eg:A.4}
  h_0\Rightarrow\Ih_0\doteqdot\zpp^\dagger\,\dpp\;.
\end{equation}
Therefore the variation ($\dkmat$) of the non-relativistic constraint term $\NNGh$
\myrf{IV.12} with respect to the spinors $\aphip^\dagger$ becomes
\begin{equation}
  \label{eq:A.5}
  \begin{split}
  \dkmat\NNGh=&-\frac{\hbar c}{2}\,\dkmat\int d^3\vec{r}\left\{B_0h_0+\Bstar_0\hstar_0\right\}\\
  =&-\frac{\hbar c}{2}\int d^3\vec{r}\left\{B_0\cdot\big(\delta\zpp\big)^\dagger\dpp+\Bstar_0\cdot\big(\delta\dpp\big)^\dagger\zpp\right\}\,.
  \end{split}
\end{equation}

But the variation of the magnetic constraint term $\NNGg$ (in the second line of (\ref{eq:A.1})) represents a more subtle problem. Indeed, here it is necessary to first split up the non-relativistic approximation $\vec{\Ih}$ of the exchange current $\vec{h}$ into a convection part $\vec{\bb}$ and a polarization part $\vec{\zz}$, i.\,e.
\begin{equation}
  \label{eg:A.6}
  \vec{h}\Rightarrow\vec{\Ih}\doteqdot\vbb+\vzz\;,
\end{equation}
see the preceding paper \cite{MaSo2} for a more detailed discussion of this splitting. The
convection current $\vbb$ is given by
\begin{equation}
  \label{eg:A.7}
  \vbb=\frac{i\hbar}{2M_ec}\left\{\left[\big(\vec{\nabla}+i\vAII\big)\zpp\right]^\dagger\dpp-(\zpp)^\dagger\big(\vec{\nabla}+i\vAIII\big)\dpp\right\}
\end{equation}
and the polarization current $\vzz$ emerges as the curl of the magnetic polarization density $\vZZ$
\begin{subequations}
  \begin{align}
  \label{eq:A.8a}
  \vzz&=\vec{\nabla}\times\vZZ\\
  \label{eq:A.8b}
  \vZZ&\doteqdot\frac{\hbar}{2M_ec}\zpp^\dagger\vec{\sigma}\dpp\;.
  \end{align}
\end{subequations}
Clearly as a consequence of this splitting, the {\em exchange mass equivalent}\/ $\MMg
c^2$ of the magnetic type (see ref.~\cite{MaSo2})
\begin{equation}
  \label{eq:A.9}
  \MMg c^2=\frac{\hbar c}{2}\int d^3\vec{r}\left\{\vec{B}\sdot\vec{\Ih}+\vBstar\sdot\vIhstar\right\}\,,
\end{equation}
which for the present intention is the relevant part of the constraint term $\NNGg$ \myrf{IV.15}, must also be split up into two components
\begin{equation}
  \label{eq:A.10}
  \MMg c^2=\MMgconv c^2+\MMgpol c^2\;.
\end{equation}
Evidently, the convection part $\MMgconv c^2$ appears here as
\begin{equation}
  \label{eq:A.11}
  \MMgconv c^2\doteqdot\frac{\hbar c}{2}\int d^3\vec{r}\left\{\vec{B}\sdot\vec{\bb}+\vBstar\sdot\vbbstar\right\}\,,
\end{equation}
and similarly for the polarization part $\MMgpol c^2$
\begin{equation}
  \label{eq:A.12}
  \MMgpol c^2\doteqdot\frac{\hbar c}{2}\int d^3\vec{r}\left\{\vec{B}\sdot\vec{\zz}+\vBstar\sdot\vzzstar\right\}\,.
\end{equation}

The reason for such a close inspection of the mass equivalent $\MMg c^2$ is of course that the matter variation of the constraint term $\NNGg$ (\ref{eq:IV.15}) is identical with that of the mass equivalent:
\begin{equation}
  \label{eq:A.13}
  \dkmat\NNGg=-\dkmat\big(\MMg c^2\big)=-\dkmat\big(\MMgconv c^2\big)-\dkmat\big(\MMgpol c^2\big)\;.
\end{equation}
Turning here first to the convection part $\MMgconv c^2$, one recasts this term through integration by parts into the following gauge-invariant form:
\begin{equation}
  \begin{split}
  \label{eq:A.14}
  \dkmat\big(\MMgconv c^2\big)&=-\frac{i\hbar^2}{4M_e}\int d^3\vec{r}\,\big(\delta\zpp\big)^\dagger\left[\covnab\sdot\vec{B}+2\vec{B}\sdot(\vec{\nabla}+i\vAIII)\right]\dpp\\
  & -\frac{i\hbar^2}{4M_e}\int d^3\vec{r}\,\big(\delta\dpp\big)^\dagger\left[\covnab\sdot\vBstar+2\vBstar\sdot(\vec{\nabla}+i\vAII)\right]\zpp\;.
  \end{split}
\end{equation}
Next, consider the non-relativistic polarization part $\MMgpol c^2$ (\ref{eq:A.12}). Since the exchange polarization current $\vec{\zz}$ (\ref{eq:A.8a}) is the curl of the exchange polarization density $\vec{\ZZ}$, one can convert the polarization part (through integration by parts) to the following gauge-invariant form
\begin{equation}
  \label{eq:A.15}
  \dkmat\big(\MMgpol c^2\big)=\frac{\hbar^2}{4M_e}\int d^3\vec{r}\left\{\big(\delta\zpp\big)^\dagger\vec{\sigma}\sdot\vec{Y}\,\dpp +\big(\delta\dpp\big)^\dagger\vec{\sigma}\sdot\vYstar\,\zpp\right\}\,.
\end{equation}

Finally, one puts together the variation of the second part of $\dkmat\tEET$, in the second line on the right-hand side of equation (\ref{eq:A.1}), by means of the partial results (\ref{eq:A.5}), (\ref{eq:A.14}) and (\ref{eq:A.15}) and thus one finds
\begin{equation}
  \label{eq:A.16}
  \begin{split}
  \dkmat\left[\lGh\cdot\NNGh+\lGg\cdot\NNGg\right]&=2\,\dkmat\left[\NNGh-\NNGg\right]\\
  &\hspace{-2em}=\int d^3\vec{r}\left\{\big(\delta\zpp\big)^\dagger\,\hat{\mathfrak{h}}_{(II)}\dpp+\big(\delta\dpp\big)^\dagger\,\hat{\mathfrak{h}}_{(III)}\zpp\right\}\,.
  \end{split}
\end{equation}
Here the exchange Hamiltonians ($\hat{\mathfrak{h}}$) have already been defined through equations \myrf{IV.43a}-\myrf{IV.43b} and the Lagrangean multipliers $\lGh,\,\lGg$ are taken from equation \myrf{IV.22c}. But now that both essential parts (\ref{eq:A.2}) and (\ref{eq:A.16}) are explicitly known, one arrives at
\begin{equation}
  \label{eq:A.17}
  \begin{split}
  \dkmat\tEET=&\int d^3\vec{r}\,\Bigg\{\big(\delta\epp\big)^\dagger\Big[\hat{\H}_{(I)}+\lPe\Big]\epp+ \big(\delta\zpp\big)^\dagger\Big[\big(\hat{\H}_{(II)}+\lPz\big)\zpp\\ &\qquad\;+\hat{\mathfrak{h}}_{(II)}\dpp\Big]+\big(\delta\dpp\big)^\dagger\Big[\big(\hat{\H}_{(III)}+\lPd\big)\dpp+\hat{\mathfrak{h}}_{(III)}\zpp\Big]\Bigg\}\;.
  \end{split}
\end{equation}
If this variation is now put to zero according to the RST principle of minimal energy
\begin{equation}
  \label{eq:A.18}
  \dkmat\tEET=0\;,
\end{equation}
one just ends up with the claimed eigenvalue equations \myrf{IV.40a}-\myrf{IV.40c}, provided one identifies the non-relativistic Lagrangean multipliers $\lPa$ with the Pauli energy eigenvalues $\EPa$ \myrf{IV.41a}-\myrf{IV.41c}
\begin{equation}
  \label{eq:A.19}
  \lPa=-\EPa\;.
\end{equation}
In this way, the logical consistency requirement of commutativity on p.\,\pageref{commutative} is actually realized.

  \newpage
  \stepcounter{section}
  \setcounter{equation}{0}
  \begin{center}
  {\textbf{\Large Appendix \Alph{section}:}}\bigskip\\
  \emph{\textbf{\Large Variational Deduction of the Gauge Field Equations}}
  \end{center}
  \vspace{2ex}

  The set of {\em non-relativistic}\/ gauge field equations is chosen to be formally the
  same as in the {\em relativistic}\/ case, with the exception that for the charge and
  current densities one has to substitute their non-relativistic forms \cite{MaSo2}. Therefore
  one can deduce the desired gauge field equations from the {\em relativistic}\/ energy
  functional $\tETT$ \myrf{IV.20} rather than from its {\em non-relativistic}\/
  approximation $\tEET$ \myrf{IV.48}. Here it is advantageous to consider the electric,
  magnetic and exchange subsystems separately because any subsystem does require for its
  treatment only a few terms of the whole functional $\tETT$.

Thus turning first to the deduction of the electric source equations (\ref{eq:III.29a})-(\ref{eq:III.29c}), one denotes the variation of $\tETT$ with respect to $\aAn$ by $\adn$ ($a=1,2,3$) and then finds for the first gauge field mode $\iAn$
\begin{equation}
  \label{eq:B.1}
  \idn\big(\tETT\big)=\idn\big(\ERe\big)+\lGe\cdot\idn\big(\NGe\big)\stackrel{!}{=}0\;,
\end{equation}
whereas for the second and third modes $\aAn$ ($a=2,3$)one has
\begin{equation}
  \label{eq:B.2}
  \adn\big(\tETT\big)=\adn\big(\ERe\big)-\adn\big(\ECh\big)+\lGe\cdot\adn\big(\NGe\big)+\lGh\cdot\adn\big(\NGh\big)+\lGg\cdot\adn\big(\NGg\big)\stackrel{!}{=}0\;.
\end{equation}
Obviously the first mode ($a=1$) contributes to the variation of the functional $\tETT$ only two terms, cf. (\ref{eq:B.1}), whereas any of the identical particles ($a=2,3$) contributes five terms, cf. (\ref{eq:B.2}). The resaon for this is that the second and third potentials $\iiAn$ and $\iiiAn$ (but not the first potential $\iAn$) do influence the electric exchange energy $\ECh$ \myrf{IV.30} via the exchange field strength $\vec{X}$ (\ref{eq:III.25a}); and furthermore these potentials enter also both constraint terms $\NGh$ \myrf{IV.12} and $\NGg$ \myrf{IV.15} via $\mathfrak{A}_0$ and $\AM$. As a consequence, the (generalized) Poisson equation for the modified mode $\IA_0$ (\ref{eq:III.8a}) is found from (\ref{eq:B.1}) to look relatively simple, i.\,e.
\begin{equation}
  \begin{split}
  \label{eq:B.3}
  \Delta\IA_0&=-4\pi\as\Ij_0\\
  (\Ij_0&\doteqdot\iij_0+\iiij_0)\;,
  \end{split}
\end{equation}
whereas the occurence of the exchange effect in the variational equations (\ref{eq:B.2}) lets appear the corresponding Poisson equations for the other two potentials $\IIA_0$ and $\IIIA_0$ somewhat more intricate:
\begin{subequations}
  \begin{align}
  \label{eq:B.4a}
  \Delta\IIA_0+2\vec{\nabla}\sdot\vec{W}&=-4\pi\as(\IIj_0-l_0)\\
  \label{eq:B.4b}
  \Delta\IIIA_0-2\vec{\nabla}\sdot\vec{W}&=-4\pi\as(\IIIj_0+l_0)\\
  (\IIj_0\,\doteqdot\,\ij_0+\iiij_0,&\quad\IIIj_0\,\doteqdot\,\ij_0+\iij_0)\;.\nonumber
  \end{align}
\end{subequations}
It is true, at first glance these electric Poisson equations may appear somewhat unaccustomed, but by means of the equations for $\vec{W}$ (\ref{eq:III.52a}) and $l_0$ (\ref{eq:III.57}) it is easy to show that the present Poisson equations for $\IA_0,\,\IIA_0,\,\IIIA_0$ (\ref{eq:B.3})-(\ref{eq:B.4b}) are actually equivalent to the original source equations (\ref{eq:III.29a})-(\ref{eq:III.29c}) for the electric field strengths $\vec{E}_a$ ($a=1,2,3$).

A similar situation occurs for the case of the magnetic Poisson equations. Denoting here the variations of $\tETT$ with respect to the three-vector potentials $\vec{A}_a$ by $\adup$, one finds again that the variation with respect to the first mode $\vec{A}_1$ yields only two terms, i.\,e.
\begin{equation}
  \label{eq:B.5}
  \idup\big(\tETT\big)=-\idup\big(\ERm\big)+\lGm\cdot\idup\big(\NGm\big)\;,
\end{equation}
whereas the variation with respect to the other two potentials $\vec{A}_a$ ($a=2,3$) produces six contributions
\begin{equation}
  \begin{split}
  \label{eq:B.6}
  \adup\big(\tETT\big)&=-\adup\big(\ERm\big)-\adup\big(\ECh\big)+\adup\big(\ECg\big)+\lGm\cdot\adup\big(\NGm\big)\\ &+\lGh\cdot\adup\big(\NGh\big)+\lGg\cdot\adup\big(\NGg\big)\;.
  \end{split}
\end{equation}
This corresponds to the circumstance that the energy functional $\tETT$ is gauge-invariant with respect to the magnetic transformations, namely through the use of the covariant derivative $\covnab$ (\ref{eq:III.26}) which relies on both vector potentials $\vec{A}_2$ and $\vec{A}_3$, but not on $\vec{A}_1$. As a consequence, the first magnetic Poisson equation emerging from (\ref{eq:B.5}) looks again rather simple
\begin{equation}
  \begin{split}
  \label{eq:B.7}
  \Delta\vec{A}_I&=-4\pi\as\vec{j}_I\\
  (\vec{j}_I&\doteqdot\vec{j}_2+\vec{j}_3)\;,
  \end{split}
\end{equation}
whereas its second and third counterpart ($a=2,3$) necessarily must appear somewhat more complicated
\begin{subequations}
  \begin{align}
  \label{eq:B.8a}
  \Delta\vec{A}_{II}-2(\vec{\nabla}\times\vec{V})&=-4\pi\as(\vec{j}_{II}-\vec{l})\\
  \label{eq:B.8b}
  \Delta\vec{A}_{III}+2(\vec{\nabla}\times\vec{V})&=-4\pi\as(\vec{j}_{III}+\vec{l})\\
  (\vec{j}_{II}\,\doteqdot\,\vec{j}_1+\vec{j}_3,&\quad\vec{j}_{III}\,\doteqdot\,\vec{j}_1+\vec{j}_2)\;.\nonumber
  \end{align}
\end{subequations}
Here it is again a simple exercise to show that the present (generalized) Poisson equations (\ref{eq:B.7})-(\ref{eq:B.8b}) are actually equivalent to the original source equations (\ref{eq:III.30a})-(\ref{eq:III.30c}) (hint: use the equations (\ref{eq:III.52b}) and (\ref{eq:III.58})).

The situation with the exchange counterparts of the electric and magnetic Poisson equations needs also sufficient attention because here the electric and magnetic field strengths $\vec{E}_a$ and $\vec{H}_a$ are again influenced by both the electromagnetic and exchange potentials; see equations (\ref{eq:III.23b})-(\ref{eq:III.23c}) and (\ref{eq:III.24b})-(\ref{eq:III.24c}). Keeping this in mind, one finds the following five contributions to the variation ($\Bdo$, say) of $\tETT$ with respect to the electric exchange potential $B_0$:
\begin{equation}
  \label{eq:B.9}
  \Bdo\big(\tETT\big)=\Bdo\big(\ERe\big)-\Bdo\big(\ECh\big)+\lGe\cdot\Bdo\big(\NGe\big)+\lGh\cdot\Bdo\big(\NGh\big)+\lGg\cdot\Bdo\big(\NGg\big)\;.
\end{equation}
Carrying here through all five variations yields the {\em electric exchange equation}\/ in the following form:
\begin{equation}
  \label{eq:B.10}
  \covlap B_0+\AM^2B_0-2i\vec{B}\sdot(\vfE-\vec{W})=-4\pi\as\hstar_0
\end{equation}
where the gauge-covariant Laplacean $\covlap$ is nothing else than the square of the
corresponding gradient operator $\covnab$ (\ref{eq:III.26}),
i.\,e. $\covlap=\covnab\sdot\covnab$. Clearly, the present exchange equation
(\ref{eq:B.10}) is again equivalent to the source equation (\ref{eq:III.31a}) for the
exchange field strength $\vec{X}$. This equation is also well-suited in order to
demonstrate that oversimplified linearization procedure of ref. \cite{MaSo2}. Indeed,
truncating the exchange equation (\ref{eq:B.10}) by means of the replacements
\begin{subequations}
  \begin{align}
  \label{eq:B.11a}
  \covlap&\Rightarrow\Delta\\
  \label{eq:B.11b}
  \AM^2&\Rightarrow\frac{1}{\aM^2}\\
  \label{eq:B.11c}
  \vec{B}\sdot(\vfE-\vec{W})&\Rightarrow0
  \end{align}
\end{subequations}
lets appear that equation (\ref{eq:B.10}) as
\begin{equation}
  \label{eq:B.12}
  \Delta B_0+\frac{1}{\aM^2}B_0=-4\pi\as\hstar_0\;,
\end{equation}
and this is just the equation (III.10a) of ref. \cite{MaSo2}.

But clearly, it should be self-evident that the most intricate case refers to the variational deduction of the curl equation (\ref{eq:III.31b}) for the magnetic exchange field strength $\vec{Y}$. Indeed, one finds that the variation ($\Bdup$, say) of $\tETT$ is built up by seven contributions:
\begin{equation}
  \begin{split}
  \label{eq:B.13}
  \Bdup\big(\tET\big)&=\Bdup\big(\ERe\big)-\Bdup\big(\ERm\big)-\Bdup\big(\ECh\big)+\Bdup\big(\ECg\big)\\ &+\lGe\cdot\Bdup\big(\NGe\big)+\lGh\cdot\Bdup\big(\NGh\big)+\lGg\cdot\Bdup\big(\NGg\big)\;.
  \end{split}
\end{equation}
However, carrying properly through all the variations yields finally the expected curl
equation (\ref{eq:III.31b}), provided the Lagrangean multipliers adopt those values being
specified by equations (\ref{eq:IV.22b})-(\ref{eq:IV.22c}). Moreover, the somewhat
oversimplified equation (III.10b) of ref. \cite{MaSo2} for the magnetic exchange field
$\vec{B}$ can of course be deduced from the present correct equation (\ref{eq:III.31b}) by
destroying the magnetic gauge invariance in the following way:
\begin{subequations}
  \begin{align}
  \label{eq:B.14a}
  &\AM\vec{X}\Rightarrow0\,,\quad B_0\vfE+i\vec{B}\times\vfH\Rightarrow0\\
  \label{eq:B.14b}
  \covnab\times\vec{Y}&\Rightarrow\vec{\nabla}\times\vec{Y} \Rightarrow\vec{\nabla}\times(\vec{\nabla}\times\vec{B}) =-\Delta\vec{B}+\vec{\nabla}(\vec{\nabla}\sdot\vec{B})\nonumber\\
  &\Rightarrow-\Delta\vec{B}+\frac{i}{\aM}\vec{\nabla}B_0 \Rightarrow-\Delta\vec{B}-\frac{1}{\aM^2}\vec{B}\;.
  \end{align}
\end{subequations}
Of course, this chain (\ref{eq:B.14b}) of neglections (``$\Rightarrow$'') demonstrates clearly in what way the magnetic gauge invariance becomes spoiled by the various steps of ``linearization''.

\renewcommand{\theequation}{\Alph{section}.\arabic{equation}}

\setcounter{section}{4}
\setcounter{equation}{0}

\begin{center}
{\textbf{\Large Appendix D:}}\\[2ex]
\emph{\textbf{\Large Electric Poisson Identity}}
\end{center}

The computation of the gauge field energy~$\ERee$(\ref{eq:VI.68}) or its mass
equivalent~$\tMMee$ (\ref{eq:VI.69}) is much more intricate than the analogous case with the
kinetic energy~$\Ekin$ (\ref{eq:VI.65}) and therefore the numerical
equality (\ref{eq:VI.70}), as the immediate consequence of the electric Poisson identity, may
be used as a welcome check for the correct calculation of the interaction energy.

Considering here first the part of the gauge field energy~$\ERee$ (\ref{eq:VI.68}), the
associated potential function~$\epot(\nu)$ (\ref{eq:VI.70}) arises in the form
\begin{equation}
  \label{eq:D1}
 \epot(\nu) = \int_0^\infty dy\, y^2\left(\frac{d\tilde{a}_\nu(y)}{dy}\right)^2 
\end{equation}
which requires to first determine the field strength in dimensionless form. Trying for
this the following ansatz
\begin{equation}
  \label{eq:D2}
  \frac{d\tilde{a}_\nu(y)}{dy} = -\frac{1}{y^2}\left\{1-\left(1+\tilde{f}_\nu(y)\right)e^{-y}\right\}\ ,
\end{equation}
the equation for the ansatz function~$\tilde{f}_\nu(y)$ is to be deduced from the Poisson
equation in the dimensionsless form (\ref{eq:VI.71}) and thus is found as
\begin{equation}
  \label{eq:D3}
  \frac{d\tilde{f}_\nu(y)}{dy}-\tilde{f}_\nu(y)-1=-\frac{y^{2\nu+1}}{\Gamma(2\nu+2)}\ .
\end{equation}
In order that the field strength (\ref{eq:D2}), i.e.\ the derivative of the
potential~$\tilde{a}_\nu(y)$, remains finite at the origin~$(y=0)$  one demands as an
initial condition for~$\tilde{f}_\nu(y)$
\begin{equation}
  \label{eq:D4}
  \tilde{f}_\nu(0)=0\ ,
\end{equation}
and then the solution for~$\tilde{f}_\nu(y)$ contains the variational parameter~$\nu$ in the
following form:
\begin{equation}
  \label{eq:D5}
  \tilde{f}_\nu(y)=e^y-1-y^{2\nu+2}\cdot\sum_{n=0}^\infty \frac{y^n}{\Gamma(2\nu+3+n)}\ ,
\end{equation}
which simplifies for integer values of~$2\nu$ to
\begin{equation}
  \label{eq:D6}
  \tilde{f}_\nu(y) = \sum_{n=1}^{2\nu+1}\frac{y^n}{n!}\ .
\end{equation}

This solution may now be substituted back into the field strength (\ref{eq:D2}) which in
the general case (\ref{eq:D5}) then reappears as
\begin{equation}
  \label{eq:D7}
  \frac{d\tilde{a}_\nu(y)}{dy}=-e^{-y}\sum_{n=0}^\infty\frac{y^{2\nu+n}}{\Gamma(2\nu+3+n)}\ .
\end{equation}
Obviously the electric field strength is finite at the origin~$(y=0)$ for~$\nu=0$ but is
zero for~$\nu>0$, see fig.~1 of ref.~\cite{BMS}. Furthermore, by use of the result
(\ref{eq:D7}) the potential function~$\epot(\nu)$ (\ref{eq:D1}) becomes now
\begin{equation}
  \label{eq:D8}
  \epot(\nu) = \frac{1}{2^{4\nu+3}}\sum_{m,n=0}^\infty\frac{1}{2^{m+n}}\,
  \frac{\Gamma(4\nu+3+m+n)}{\Gamma(2\nu+3+m)\cdot\Gamma(2\nu+3+n)}\ .
\end{equation}
For integer values of~$2\nu$ (i.e.~$2\nu=0,1,2,3,\ldots$) this potential function becomes
simplified again considerably: first, one recasts the original definition (\ref{eq:D1}) by
means of some simple manipulations into the following form
\begin{equation}
  \label{eq:D9}
  \int dy\,y^2 \left(\frac{d\tilde{a}_\nu(y)}{dy}\right)^2 =  \int dy\,e^{-2y}
  \left(\frac{\tilde{f}_\nu(y)}{y} \right)^2 - 2\int dy\,\left(e^{-y}-e^{-2y} \right)
  \cdot\frac{\tilde{f}_\nu(y)-y}{y^2}\ ,
\end{equation}
and if the special function~$\tilde{f}_\nu(y)$ (\ref{eq:D6}) is substituted herein, one
arrives at~\cite{MaSo2}
\begin{equation}
  \label{eq:D10}
  \epot(\nu) = 2\cdot\sum_{m,n=1}^{2\nu+1}\frac{1}{2^{m+n}}\cdot\frac{(m+n-2)!}{m!n!} -
  2\cdot\sum_{n=2}^{2\nu+1}\frac{1}{n(n-1)}\cdot\left(1-\frac{1}{2^{n-1}}\right)\ .
\end{equation}
Thus for the lowest values of~$\nu$ one gets the following table (for~$\nu=0$ the second
sum in (\ref{eq:D10}) is to be omitted):

\begin{flushleft}
\begin{tabular}{|c||c|c|c|c|c|c|c|c|c|}
\hline
$\nu$ & 0 & $\frac{1}{2}$ & 1 &$\frac{3}{2}$ & 2 & $\frac{5}{2}$ &3 &$\frac{7}{2}$ \\ \hline\hline
$\epot(\nu)$ & $\frac{1}{2}$ & $\frac{5}{16}$ & $\frac{11}{48}$ & $\frac{93}{512}$
&$\frac{193}{1280}$ & $\frac{793}{6144}$ & $\frac{1619}{14336}$ & $\frac{26333}{262144}$ \\ 
&  {\scs=0,500} & {\scs =0,3125} & {\scs =0,2291\ldots} & {\scs =0,1816\ldots}
&{\scs  =0,1507\ldots} &{\scs = 0.1290\ldots} & {\scs =0.1129\ldots} & {\scs
  = 0.1004\ldots}  \\ \hline
\end{tabular}
\end{flushleft}
\vspace{3ex}

But as the dimensionsless Poisson identity (\ref{eq:VI.70}) says, the positron-electron
interaction energy may also be viewn to be represented by the mass equivalent~$\tMMee c^2$
(\ref{eq:VI.69}) and then the potential function~$\epot(\nu)$ appears in the following
alternative form
\begin{equation}
  \label{eq:D11}
  \epot(\nu)=\frac{1}{\Gamma(2\nu+2)}\int dy\, y\,\tilde{a}_\nu(y)\cdot\tilde{\Phi}(y)^2
  =\frac{1}{\Gamma(2\nu+2)}\int dy\, y^{2\nu+1} e^{-y}\cdot\tilde{a}_\nu(y)\ .
\end{equation}
If one wishes to calculate the potential function~$\epot(\nu)$ in this alternative way (in
contrast to (\ref{eq:D1})), one obviously has first to calculate the electric
potential~$\tilde{a}_\nu(y)$. The differential equation for~$\tilde{a}_\nu(y)$ is given by
(\ref{eq:D2}), with~$\tilde{f}_\nu(y)$ given by (\ref{eq:D5}) in the general case, or by
(\ref{eq:D6}) for integer values of~$2\nu$, resp. However, for elaborating the
solution~$\tilde{a}_\nu(y)$ of (\ref{eq:D2}) for given~$\tilde{f}_\nu(y)$ one has to fix
the value of the potential~$\tilde{a}_\nu(y)$ at the origin~$(y=0)$ which works as an
initial condition for that first-order equation (\ref{eq:D2}). But how to fix that initial
condition~$\tilde{a}_\nu(0)$?

The solution of this problem comes from the standard potential~$\ppAn(r)$
(\ref{eq:VI.39}) which reads in the non-relativistic approximation
\begin{equation}
  \label{eq:D12}
  \ppAn(r) = \frac{\as}{8}\int\frac{d^3\vec{r}\,'}{r'}\frac{\tilde{\Phi}(r')^2}{||\vec{r}-\vec{r}\,' ||}
\end{equation}
or, resp., in dimensionless form for the selected trial function~$\tilde{\Phi}(y)$
(\ref{eq:VI.58})
\begin{equation}
  \label{eq:D13}
  \tilde{a}_\nu(y)=\frac{1}{4\pi\cdot\Gamma(2\nu+2)}\int d^3\vec{y}\,'\;
  \frac{e^{-y'}\cdot y'^{\,2\nu-1}}{||\vec{y}-\vec{y}\,'||}\ .
\end{equation}
Here it is obvious that this solution has the right limit form (i.e.\ Coulomb potential)
in the asymptotic region~$(y\to\infty)$, i.e.
\begin{equation}
  \label{eq:D14}
  \lim_{y\to\infty}\tilde{a}_\nu(y) = \frac{1}{y}
\end{equation}
cf.\ (\ref{eq:III.60}); and furthermore its value at the origin~$(y=0)$ is found by
explicit calculation of the integral in (\ref{eq:D13}) as
\begin{equation}
  \label{eq:D15}
  \tilde{a}_\nu(0) = \frac{1}{2\nu+1}\ .
\end{equation}

But with both boundary conditions (\ref{eq:D14})-(\ref{eq:D15}) being thus fixed, the
solution of equation (\ref{eq:D7}) for~$\tilde{a}_\nu(y)$ is unique and looks as follows:
\begin{equation}
  \label{eq:D16}
  \tilde{a}_\nu(y)=\frac{1}{2\nu+1}\left\{1-e^{-y}\cdot\sum_{n=0}^{\infty}\frac{n}{\Gamma(2\nu+2+n)}
  \,y^{2\nu+n}\right\}\ .
\end{equation}
For integer values of~$2\nu$ i.e.\ ($2\nu=0,1,2,3\ldots$) this result is converted to
\begin{equation}
  \label{eq:D17}
  \tilde{a}_\nu(y)=\frac{e^{-y}}{2\nu+1}\cdot\sum_{n=0}^{2\nu-1}\frac{y^n}{n!} +
  \frac{1}{y}\left(1-e^{-y}\cdot\sum_{n=0}^{2\nu}\frac{y^n}{n!}\right)
\end{equation}
(for~$\nu=0$ the first term is to be omitted). And finally, the potential
function~$\epot(\nu)$ (\ref{eq:D11}) emerges by means of the present potential~$
\tilde{a}_\nu(y)$ (\ref{eq:D16}) as
\begin{equation}
  \label{eq:D18}
  \begin{split}
  \epot(\nu) &=\frac{1}{\Gamma(2\nu+2)}\int dy\,y^{2\nu+1}e^{-y}\cdot\tilde{a}_\nu(y)\\*
  &= \frac{1}{2\nu+1}
  \left(1-\frac{1}{2^{4\nu+2}}\cdot\sum_{n=0}^\infty\frac{n}{2^n}
    \frac{\Gamma(4\nu+2+n)}{\Gamma(2\nu+2)\cdot\Gamma(2\nu+2+n)}\right)\ ,  
  \end{split}
\end{equation}
or for integer values of~$2\nu$, resp.
\begin{equation}
  \label{eq:D19}
\epot(\nu)=\frac{1}{2\nu+1} + \frac{1}{(2\nu+1)!} \left(\frac{1}{2\nu+1}\sum_{n=0}^{2\nu-1}
\frac{(2\nu+1+n)!}{n!\,2^{2\nu+2+n}} - \sum_{n=0}^{2\nu}\frac{(2\nu+n)!}{n!\,2^{2\nu+1+n}}
\right)\ .
\end{equation}
Clearly, this latter result can equally well be used in order to generate those numbers
which are displayed in the table below equation (\ref{eq:D10}); for a plot of~$\epot(\nu)$
see fig.~3.

\setcounter{section}{5}
\setcounter{equation}{0}

  \begin{center}
  {\textbf{\Large Appendix \Alph{section}:}}\\
  \emph{\textbf{\Large Angular Part of the Ortho-Wave Amplitudes \\ $\bMRpm,\bMSpm$}}
  \end{center}
  \vspace{2ex}

The angular dependency of the wave amplitudes must necessarily transcribe to all the physical densities which originate from these amplitudes. Therefore it is worthwile to inspect somewhat more closely the angular functions $g_R(\vartheta)$, $g_S(\vartheta)$ of the product ansatz \myrf{VI.111a}-\myrf{VI.111b}. Here it should be clear that the physical difference between the ortho- and para-configurations will originate from just the different angular structure of the corresponding densities. For instance, the ortho-analogy $\bkphi$ of the para-current $\pkphi$ \myrf{VI.30} will surely appear in a rather different way because the angular dependency of $g_R(\vartheta)$ and $g_S(\vartheta)$, as the solutions of the coupled system \myrf{VI.106a}-\myrf{VI.106b}, cannot be of that simple form as their para-counterparts $f_R(\vartheta)$ and $f_S(\vartheta)$ \myrf{VI.26a}-\myrf{VI.26b}. Quite generally speaking, both ortho-functions $g_R(\vartheta)$ and $g_S(\vartheta)$ do enter the ortho-current $\bkphi$ in the following way:
\begin{eqnarray}
  \label{eq:E1}
  \bkphi&=& \frac{1}{2\pi}\left\{\left(\frac{g_R(\vartheta)}{\sin\vartheta}\right)^2\cdot\frac{\ssRp(r)\,\ssRm(r)}{r} -\big(\sin\vartheta\cdot g_S(\vartheta)\big)^2\cdot\frac{\ssSp(r)\,\ssSm(r)}{r}\right\}\nonumber\\ 
  && \qquad-\frac{\cot\vartheta}{2\pi}\,g_R(\vartheta)g_S(\vartheta)\cdot\frac{\ssRp(r)\,\ssSm(r)+\ssRm(r)\,\ssSp(r)}{r}\;.
\end{eqnarray}
Indeed, this result is obtained by simply substituting the transformations $\big\{\bMRpm,\bMSpm\big\}\Rightarrow\big\{\ssRpm,\ssSpm\big\}$ \myrf{VI.111a}-\myrf{VI.111b} with general angular functions $g_R(\vartheta),\,g_S(\vartheta)$ into the general expression \myrf{V.57c} for the ortho-current $\bkphi$. Consequently, the wanted angular dependency of the ortho-current $\bkphi$ is determined essentially by the (yet unknown) functions $g_R(\vartheta)$ and $g_S(\vartheta)$. This becomes more explicit by resorting to the identifications \myrf{VI.112} which yields the following product structure of the ortho-current $\bkphi$ \myrf{E1}:
\begin{equation}
  \label{eq:E1'}
  \bkphi=\frac{\tOp\cdot\tOm}{2\pi r}\cdot\sin^2\vartheta\left\{\left(\frac{g_R(\vartheta)}{\sin^2\vartheta}\right)^2-g_S^2(\vartheta)-2\cos\vartheta\,g_S(\vartheta)\,\frac{g_R(\vartheta)}{\sin^3\vartheta}\right\}\,.
\end{equation}
Obviously, this ortho-current $\bkphi$ displays an angular characteristic quite different from its para-counterpart $\pkphi$ \myrf{VI.30}!

The determination of the angular functions can be performed most conveniently by resorting to the decoupled second-order equations \myrf{VI.108a}-\myrf{VI.108b}. Here it is advanageous to introduce the auxiliary variables $x$ and $z$ by
\begin{subequations}
  \begin{align}
  \label{eq:E2a}
  x&\doteqdot\sin\vartheta\\
  \label{eq:E2b}
  z&\doteqdot\cos\vartheta\;.
  \end{align}
\end{subequations}
This is to be combined with the hypothesis that $g_R(\vartheta)$ depends exclusively upon $x$ and $g_S(\vartheta)$ exclusively upon $z$, i.\,e. we put
\begin{subequations}
  \begin{align}
  \label{eq:E3a}
  g_R(\vartheta)&\doteqdot G_R(x)\\
  \label{eq:E3b}
  g_S(\vartheta)&\doteqdot G_S(z)\;.
  \end{align}
\end{subequations}
This arrangement recasts the second-order system \myrf{VI.108a}-\myrf{VI.108b} to the corresponding system for the new functions $G_R(x)$ and $G_S(z)$:
\begin{subequations}
  \begin{align}
  \label{eq:E4a}
  (1-x^2)\,\frac{d^2G_R(x)}{dx^2}-\frac{2-x^2}{x}\cdot\frac{dG_R(x)}{dx}&=(\dlO\ddlO)\cdot G_R(x)\\
  \label{eq:E4b}
  (1-z^2)\,\frac{d^2G_S(z)}{dz^2}-3z\,\frac{dG_S(z)}{dz}&=(\dlO\ddlO)\cdot G_S(z)\;.
  \end{align}
\end{subequations}

The general shape of the latter equations suggests now to try some ansatz in terms of a power series expansion, i.\,e. we put
\begin{subequations}
  \begin{align}
  \label{eq:E5a}
  G_R(x)&=\sum_{\rho=3}^{n_R}\rho_nx^n\\
  \label{eq:E5b}
  G_S(z)&=\sum_{n=1}^{n_S}\sigma_nz^n
  \end{align}
\end{subequations}
with constant coefficients $\rho_n$ and $\sigma_n$. For these coefficients one deduces by means of standard techniques the following recurrence formulae:
\begin{subequations}
  \begin{align}
  \label{eq:E6a}
  \rho_{n+2}&=\frac{(\dlO\cdot\ddlO)+n(n-2)}{(n+2)(n-1)}\cdot\rho_n\\
  \label{eq:E6b}
  \sigma_{n+2}&=\frac{(\dlO\cdot\ddlO)+n(n+2)}{(n+2)(n+1)}\cdot\sigma_n\;.
  \end{align}
\end{subequations}
Since the proposed power series expansions \myrf{E5a}-\myrf{E5b} are adopted to stop at certain maximal integers $n_R$ or $n_S$, resp., there arises a halt condition for each series, namely 
\begin{subequations}
  \begin{align}
  \label{eq:E7a}
  (\dlO\cdot\ddlO)&=-n_R(n_R-2)\\
  \label{eq:E7b}
  (\dlO\cdot\ddlO)&=-n_S(n_S+2)\;.
  \end{align}
\end{subequations}
Thus the maximal powers $n_R$ and $n_S$ are linked to each other by
\begin{eqnarray}
  \label{eq:E8}
  n_R&=&n_S+2\\
  \nonumber
  (n_S&=&1,3,5,7,\,...)
\end{eqnarray}
and the allowed values for the product of the constants $\dlO$ and $\ddlO$ turn out as
\begin{equation}
  \label{eq:E9}
  (\dlO\cdot\ddlO)=-n_R\cdot n_S\Rightarrow-3,-15,-35,\,...
\end{equation}
A possible physical meaning of this product of the constants is seen more clearly by
building the {\em quantum number of orbital angular momentum}\/ $\lO$:
\begin{equation}
  \label{eq:E10}
  \lO^2\doteqdot1-(\dlO\cdot\ddlO)=(n_S+1)^2\Rightarrow4,16,36,\,...
\end{equation}

Naturally, it is very temptive to interprete this in the way that ortho-positronium owns an {\em orbital}\/ angular momentum of magnitude $\lO\hbar$ ($\Rightarrow2\hbar,\,4\hbar,\,6\hbar,\,...$). But this would just meet with the intuitive expectation that each positronium constituent owns the same integer number of units ($\hbar$) of (orbital) angular momentum. However, one should not forget here that, in RST (as a fluid-dynamic theory), the physical observables are obtained by integration of the corresponding densities, not as the eigenvalues of certain Hermitian operators. Therefore such quantities as the above-mentioned discrete numbers $\lO$ \myrf{E10} and their para-counterparts $\lP$ \myrf{VI.20a}-\myrf{VI.20b}, or similarly the energy eigenvalue $E_*$ in the Schr\"odinger-like eigenvalue equations \myrf{VI.42} and \myrf{VI.118}, are to be conceived as auxiliary objects which help to build up the true physical observables (e.\,g. the value $\ET$ of the energy functional $\tETT$ upon some solution of the RST eigenvalue equations).

At first glance, the spacing of the ortho-numbers $\lO$ ($\Rightarrow2,\,4,\,6,\,...$) is
twice the spacing of their para-analogies $\lP=0,1,2,3,\,...$ (fig.~2); but a more
thorough scrutiny reveals a certain dichotomic character of any value of $\lO$ so that the
total number of physical states is (roughly) the same for the ortho- and
para-configurations. In order to realize this dichotomy of the ortho-configurations more
clearly, reconsider the definition of the ortho-number $\lO$ \myrf{E10} and combine this
with the equation \myrf{VI.117b}
\begin{equation}
  \label{eq:E11}
  \dlO+\ddlO=-2
\end{equation}
so that the two constants $\dlO$ and $\ddlO$, being originally introduced by equations \myrf{VI.106a}-\myrf{VI.106b}, become fixed now in terms of $\lO$ in the following dichotomic way, namely either
\begin{subequations}
  \begin{align}
  \label{eq:E12a}
  \dlO&=\lO-1\\
  \label{eq:E12b}
  \ddlO&=-(\lO+1)
  \end{align}
\end{subequations}
or the other way round (substituting $\lO\Rightarrow-\lO$)
\begin{subequations}
  \begin{align}
  \label{eq:E13a}
  \dlO&=-(\lO+1)\\
  \label{eq:E13b}
  \ddlO&=\lO-1\;.
  \end{align}
\end{subequations}
Indeed, both equations \myrf{E10} and \myrf{E11} are obeyed by each of these two possibilities.

But once the origin of the claimed dichotomy has become clear, one can look now for the physical consequences. From the mathematical viewpoint, one first wishes to see the integration constants $\sigma_1$ and $\rho_3$ of the solutions \myrf{E5a}-\myrf{E5b} being completely fixed. But this is not possible because one integration constant ($\rho_3$, say) must remain free for the sake of normalization of the angular part of the wave functions (see equation \myrf{VI.131a}). Therefore it is possible to nail down only the ratio $\frac{\sigma_1}{\rho_3}$. This, however, can easily be done by substituting the solutions $g_R(\vartheta)$ and $g_S(\vartheta)$, being associated with the auxiliary functions $G_R(x)$ and $G_S(z)$ \myrf{E5a}-\myrf{E5b}, back into the original first-order system \myrf{VI.106a}-\myrf{VI.106b} which then yields
\begin{equation}
  \label{eq:E14}
\frac{\sigma_1}{\rho_3}=\left\{\begin{array}{ll}\ds -3\,\frac{\dlO}{\lO^2-1}=\frac{3}{\ddlO} & ,\,{\rm for}\;n_S=1,5,9,\,...\\ \ds +3\,\frac{\dlO}{\lO^2-1}=-\frac{3}{\ddlO} & ,\,{\rm for}\;n_S=3,7,11,\,...\end{array}\right.
\end{equation}
Furthermore, the pair of auxiliary constants $\dlO,\,\ddlO$ has already been shown to occur in two versions, cf. \myrf{E12a}-\myrf{E13b}, and this of course transcribes then also to the ratio of integration constants \myrf{E14}, i.\,e.
\begin{equation}
  \label{eq:E15}
\frac{\sigma_1}{\rho_3}=\left\{\begin{array}{l}\ds \,-\frac{3}{\lO+1}\\[2ex]
\ds +\,\frac{3}{\lO-1}\end{array}\right.\;\mbox{\bf for\;\boldmath$\lO=2,6,10,\,...$}
\end{equation}
and
\begin{equation}
  \label{eq:E16}
  \frac{\sigma_1}{\rho_3}=\left\{\begin{array}{l}\ds +\,\frac{3}{\lO+1}\\[2ex] 
   \ds -\,\frac{3}{\lO-1}\end{array}\right.\;\mbox{\bf for\;\boldmath$\lO=4,8,12,\,...$}
\end{equation}
The upper case of each of both equations (\ref{eq:E15})-(\ref{eq:E16}) refers here to the
first eigenvalue system \myrf{VI.113a}-\myrf{VI.113b}; and the lower case to the second
system \myrf{VI.115a}-\myrf{VI.115b}.  The subsequent table displays a collection for the
three lowest-order cases $\lO=2,\,4,\,6$. Observe here that for the pairs $\dlO,\,\ddlO$
(third column) and for the ratios $\frac{\sigma_1}{\rho_3}$ (last column) one obtains
always two different possibilities for a fixed value of $\lO$ which entails the existence
of two different physical states for any admitted $\lO$.  In the non-relativistic limit,
\emph{both} relativistic systems (\ref{eq:VI.113a})-(\ref{eq:VI.113b}) \emph{and}
(\ref{eq:VI.115a})-(\ref{eq:VI.115b}) collapse to the \emph{same} non-relativistic
equation (\ref{eq:VI.118}).

\begin{center}
\begin{tabular}[t]{|c||c|c|c|c|c|c|}
\hline
$n_S$ & $n_R$ & $\begin{array}{rcr} \dlO & ; & \ddlO \end{array}$ & $\lO$ & $G_S(z)$ & $G_R(x)$ & $\frac{\sigma_1}{\rho_3}$ \\\hline\hline
1	&	3	&	$\begin{array}{rcr} 1 & ; & -3 \\\hline -3 & ; & 1 \end{array}$ & 2 & $\sigma_1\cdot z$ & $\rho_3\cdot x^3$ & $\begin{array}{r} -1 \\\hline 3 \end{array}$\\\hline
3 & 5 & $\begin{array}{rcr} 3 & ; & -5 \\\hline -5 & ; & 3 \end{array}$ & 4 & $\sigma_1\cdot(z-2z^3)$ & $\rho_3\cdot(x^3-\frac{6}{5}\,x^5)$ & $\begin{array}{r} \frac{3}{5} \\\hline -1 \end{array}$\\\hline
5 & 7 & $\begin{array}{rcr} 5 & ; & -7 \\\hline -7 & ; & 5 \end{array}$ & 6 & $\sigma_1\cdot(z-\frac{16}{3}\,z^3+\frac{16}{3}\,z^5)$ & $\rho_3\cdot(x^3-\frac{16}{5}\,x^5+\frac{16}{7}\,x^7)$ & $\begin{array}{r} -\frac{3}{7} \\\hline \frac{3}{5} \end{array}$ \\\hline
\end{tabular}
\end{center}


\setcounter{section}{6}
\setcounter{equation}{0}

  \begin{center}
  {\textbf{\Large Appendix \Alph{section}:}}\\
  \emph{\textbf{\Large Kinetic Energy $\btTkin$ of the Ortho-Configurations\\}}
  \end{center}
  \vspace{2ex}

The most intricate contribution to the energy functional $\tEO$ \myrf{VI.126} is surely the kinetic energy $\btTkin$. First, the kinetic energy \myrf{IV.26}-\myrf{IV.27} reads in terms of the Pauli spinors $\bppm$ for the ortho-configurations
\begin{equation}
  \label{eq:F1}
  \bTkin=i\,\frac{\hbar c}{2}\int d^3\vec{r}\,\left[\bpmk\,\vec{\sigma}\sdot\vec{\nabla}\,\bpp+\bppk\,\vec{\sigma}\sdot\vec{\nabla}\,\bpm\right]\,,
\end{equation}
and when the Pauli spinors are decomposed here as shown by equations \myrf{V.50a}-\myrf{V.50b} the kinetic energy appears as a sum of three contributions
\begin{equation}
  \label{eq:F2}
  \bTkin=\bTr+\bTth+\bTph\;.
\end{equation}
The radial part $\bTr$ itself consists of two contributions where the first of them refers to the wave amplitudes $\bMRpm(\vec{r})$ and the second to the amplitudes $\bMSpm(\vec{r})$:
\begin{equation}
  \label{eq:F3}
  \bTr=\bTr[\MR]+\bTr[\MS]
\end{equation}
with
\begin{eqnarray}
  \nonumber
  \bTr[\MR]&=&-\frac{\hbar c}{16\pi}\int d^3\vec{r}\,\left\{\bMRpS\!\cdot\!\frac{\partial\bMRm}{\partial r}+\bMRp\!\cdot\!\frac{\partial\bMRmS}{\partial r}-\bMRm\!\cdot\!\frac{\partial\bMRpS}{\partial r}-\bMRmS\!\cdot\!\frac{\partial\bMRp}{\partial r}\right.\\
  \label{eq:F4}
  &&\hspace{8em}\left.+3\,\frac{\bMRpS\cdot\bMRm+\bMRmS\cdot\bMRp}{r}\right\}\,,
\end{eqnarray}
and analogously for $\bTr[\MS]$ where merely the replacement $\MR\Rightarrow\MS$ must be applied. Just as the radial part $\bTr$ is a collection of all derivative terms with respect to the radial variable $r$, the longitudinal term $\bTth$ collects the derivative terms with respect to the angle $\vartheta$:
\begin{eqnarray}
  \nonumber
  \bTth&=&\frac{\hbar c}{16\pi}\int\frac{d^3\vec{r}}{r}\,\left\{{\rm e}^{2i\phi}\!\left[ \bMRm\!\cdot\!\frac{\partial\bMSpS}{\partial\vartheta}+\bMRp\!\cdot\!\frac{\partial\bMSmS}{\partial\vartheta}-\bMSpS\!\cdot\!\frac{\partial\bMRm}{\partial\vartheta}-\bMSmS\!\cdot\!\frac{\partial\bMRp}{\partial\vartheta}\right]\right.\\
  \label{eq:F5}
  &&\hspace{.5em}\left.+{\rm e}^{-2i\phi}\!\left[ \bMRpS\!\cdot\!\frac{\partial\bMSm}{\partial\vartheta}+\bMRmS\!\cdot\!\frac{\partial\bMSp}{\partial\vartheta}-\bMSm\!\cdot\!\frac{\partial\bMRpS}{\partial\vartheta}-\bMSp\!\cdot\!\frac{\partial\bMRmS}{\partial\vartheta}\right]\right\}\,.
\end{eqnarray}

Finally, the azimuthal part $\bTph$ represents a collection of all the derivative terms with respect to the azimuthal angle $\phi$ and looks as follows:
\begin{eqnarray}
  \nonumber
  \bTph&=&\frac{i\hbar c}{8\pi}\int\frac{d^3\vec{r}}{r}\,\left\{ \bMRmS\!\cdot\!\frac{\partial\bMRp}{\partial\phi}-\bMRm\!\cdot\!\frac{\partial\bMRpS}{\partial\phi}-\bMSmS\!\cdot\!\frac{\partial\bMSp}{\partial\phi}+\bMSm\!\cdot\!\frac{\partial\bMSpS}{\partial\phi}\right\}\\
  \nonumber
  &&\hspace{-3em}+\frac{i\hbar c}{16\pi}\int\frac{d^3\vec{r}}{r}\,\cot\vartheta\!\left\{{\rm e}^{2i\phi}\!\left[ \bMRm\!\cdot\!\frac{\partial\bMSpS}{\partial\phi}-\bMSpS\!\cdot\!\frac{\partial\bMRm}{\partial\phi}+\bMRp\!\cdot\!\frac{\partial\bMSmS}{\partial\phi}-\bMSmS\!\cdot\!\frac{\partial\bMRp}{\partial\phi}\right]\right.\\
  \label{eq:F5'}
  &&\left.+{\rm e}^{-2i\phi}\!\left[ \bMSm\!\cdot\!\frac{\partial\bMRpS}{\partial\phi}-\bMRpS\!\cdot\!\frac{\partial\bMSm}{\partial\phi}+\bMSp\!\cdot\!\frac{\partial\bMRmS}{\partial\phi}-\bMRmS\!\cdot\!\frac{\partial\bMSp}{\partial\phi} \right]\right\}\,.
\end{eqnarray}

In the next step, one substitutes herein the spherically symmetric approximations \myrf{VI.105a}-\myrf{VI.105b} for the wave amplitudes $\bMRpm,\,\bMSpm$ and thus obtains for both radial parts $\bTr[\MR]$ and $\bTr[\MS]$:
\begin{subequations}
  \begin{align}
  \nonumber
  \bTr[\MR]\Rightarrow\bTr[R'']&=-\frac{\hbar c}{8\pi}\int\! d^3\vec{r}\,\left(\frac{g_R(\vartheta)}{\sqrt{\sin^3\vartheta}}\right)^2\cdot\left\{ \frac{\ssRp(r)}{r}\cdot\frac{d\ssRm(r)}{dr}-\frac{\ssRm(r)}{r}\cdot\frac{d\ssRp(r)}{dr}\right.\\
  \label{eq:F6a}
  &\hspace{15em}\left.+3\,\frac{\ssRp(r)\cdot\ssRm(r)}{r^2}\right\}\\
  \nonumber\\
  \nonumber
  \bTr[\MS]\Rightarrow\bTr[S'']&=-\frac{\hbar c}{8\pi}\int\! d^3\vec{r}\,\left(\sqrt{\sin\vartheta}\!\cdot\! g_S(\vartheta)\right)^2\!\!\cdot\!\left\{\frac{\ssSp(r)}{r}\!\cdot\!\frac{d\ssSm(r)}{dr}-\frac{\ssSm(r)}{r}\!\cdot\!\frac{d\ssSp(r)}{dr}\right.\\
  \label{eq:F6b}
  &\hspace{15em}\left.+3\,\frac{\ssSp(r)\cdot\ssSm(r)}{r^2}\right\}\,.
  \end{align}
\end{subequations}
Now one resorts to the ortho-identification of the wave amplitudes \myrf{VI.112} and thus finds the radial kinetic energy $\bTr$ \myrf{F3} in the following product form:
\begin{eqnarray}
  \label{eq:F7}
  \bTr\Rightarrow\btTr&=&-\frac{\hbar c}{2}\int\frac{d\Omega}{4\pi}\,\left\{\left(\frac{g_R(\vartheta)}{\sqrt{\sin^3\vartheta}}\right)^2+\left(\sqrt{\sin\vartheta}\cdot g_S(\vartheta)\right)^2\right\}\\
  \nonumber
  &&\hspace{1.5em}\cdot\int dr\,r^2\left\{\frac{\tOp(r)}{r}\cdot\frac{d\tOm(r)}{dr}-\frac{\tOm(r)}{r}\cdot\frac{d\tOp(r)}{dr}+3\,\frac{\tOp(r)\cdot\tOm(r)}{r^2}\right\}\,.
\end{eqnarray}
But this is indeed now a very pleasant result because the product structure allows us to apply here the separate normalization condition \myrf{VI.131a} for the angular factor which then leaves the radial kinetic energy in the final form
\begin{eqnarray}
  \nonumber
  \btTr\Rightarrow\bbtTr&=&-\frac{\hbar c}{2}\int dr\,r^2\left\{\frac{\tOp(r)}{r}\cdot\frac{d\tOm(r)}{dr}-\frac{\tOm(r)}{r}\cdot\frac{d\tOp(r)}{dr}+3\,\frac{\tOp(r)\cdot\tOm(r)}{r^2}\right\}\,.\\
  \label{eq:F8}
\end{eqnarray}

In quite a similar way one may calculate the longitudinal kinetic energy $\bTth$
\myrf{F5}, and one finds the result again in form of a product structure, i.\,e.
\begin{eqnarray}
  \nonumber
  \bTth\Rightarrow\bbtTth\hspace{-.5em}&=&\hspace{-.5em}\hbar c\int\frac{d\Omega}{4\pi}\left[\frac{g_R(\vartheta)}{\sqrt{\sin^3\vartheta}}\!\cdot\!\frac{\partial}{\partial\vartheta}\left(\sqrt{\sin\vartheta}\,g_S(\vartheta)\right)-\sqrt{\sin\vartheta}\,g_S(\vartheta)\!\cdot\!\frac{\partial}{\partial\vartheta}\left(\frac{g_R(\vartheta)}{\sqrt{\sin^3\vartheta}}\right)\right]\\
  \label{eq:F9}
  \vphantom{\rule{0pt}{25pt}}
  &&\cdot\int dr\,r^2\,\frac{\tOp(r)\cdot\tOm(r)}{r^2}\;.
\end{eqnarray}
For the angular part one obtains here by use of the first-order angular system \myrf{VI.106a}-\myrf{VI.106b}
\begin{eqnarray}
  \label{eq:F10}
  \int\hspace{-.75em}&\displaystyle\frac{d\Omega}{4\pi}&\hspace{-.75em} \left[\frac{g_R(\vartheta)}{\sqrt{\sin^3\vartheta}}\cdot\frac{\partial}{\partial\vartheta}\left(\sqrt{\sin\vartheta}\,g_S(\vartheta)\right)-\sqrt{\sin\vartheta}\,g_S(\vartheta)\cdot\frac{\partial}{\partial\vartheta}\left(\frac{g_R(\vartheta)}{\sqrt{\sin^3\vartheta}}\right)\right]\\
  \nonumber
  \vphantom{\rule{0pt}{30pt}}
  &=&\hspace{-1em}\int\frac{d\Omega}{4\pi}\,\left\{2\cos\vartheta\,\frac{g_R(\vartheta)\!\cdot\! g_S(\vartheta)}{\sin^2\vartheta}-\frac{g_R^2(\vartheta)}{\sin^3\vartheta}+\sin\vartheta\!\cdot\! g_S^2(\vartheta)\pm\lO\cdot\left[\frac{g_R^2(\vartheta)}{\sin^3\vartheta}+\sin\vartheta\!\cdot\! g_S^2(\vartheta)\right]\right\}\,.
\end{eqnarray}
The term containing the quantum number $\lO$ (of orbital angular momentum) would present
no problem because one could apply again the normalization condition \myrf{VI.131a} for
the angular part of the charge density $\bko(\vec{r})$ \myrf{VI.130}. (The upper sign in
front of $\lO$ refers to the first possibility \myrf{E12a}-\myrf{E12b} and the lower sign
to the second possibility \myrf{E13a}-\myrf{E13b}). But the other three terms on the
right-hand side of \myrf{F10} would not fit into the right picture of the
ortho-configurations.

But fortunately, there is a third contribution to the kinetic energy $\bTkin$ \myrf{F2}, i.\,e. the azimuthal part $\bTph$. And indeed, this term is found to be again of the following product form:
\begin{eqnarray}
  \nonumber
  \bTph\Rightarrow\bbtTph &=&\hbar c\int\frac{d\Omega}{4\pi}\, \left\{\frac{2g_R^2(\vartheta)}{\sin^3\vartheta}-2\cos\vartheta\,\frac{g_R(\vartheta)\cdot g_S(\vartheta)}{\sin^2\vartheta}\right\}\cdot\int dr\,r^2\,\frac{\tOp(r)\cdot\tOm(r)}{r^2}\;.\\
  \label{eq:F11}
\end{eqnarray}
Obviously, it is just this form which is needed in order to bring the transversal kinetic energy ($\bbtTth+\bbtTph$) into the expected shape:
\begin{eqnarray}
  \nonumber
  \bbtTth + \bbtTph  &=&(1\pm\lO)\,\hbar
  c\int\frac{d\Omega}{4\pi}\,\left[\frac{g_R^2(\vartheta)}{\sin^3\vartheta}+\sin\vartheta\cdot
    g_S^2(\vartheta)\right]\cdot\int dr\,r^2\,\frac{\tOp(r)\cdot\tOm(r)}{r^2}\ ,\\
  \label{eq:F12}
\end{eqnarray}
or if the angular normalization \myrf{VI.131a} is used again
\begin{eqnarray}
  \label{eq:F13}
  \bbtTth+\bbtTph&=&(1\pm\lO)\,\hbar c\int dr\,r^2\,\frac{\tOp(r)\cdot\tOm(r)}{r^2}\;.
\end{eqnarray}
Observe the difference of this ortho-result with respect to its para-counterpart \myrf{VI.32b}!

Summarizing, one puts together all three partial results \myrf{F8} and \myrf{F13} and then ultimately finds for the desired kinetic energy $\bTkin$ \myrf{F2}
\begin{eqnarray}
  \nonumber
  \bTkin\hspace{-.25em}\Rightarrow\hspace{-.25em}\bbtTkin\hspace{-.75em}&=&\hspace{-.75em}-\frac{\hbar c}{2}\hspace{-.25em}\int \hspace{-.5em}dr\,r^2\!\left\{\hspace{-.25em}\frac{\tOp(r)}{r}\!\cdot\!\frac{d\tOm(r)}{dr}-\frac{\tOm(r)}{r}\!\cdot\!\frac{d\tOp(r)}{dr}+(1\mp2\lO)\frac{\tOp(r)\!\cdot\!\tOm(r)}{r^2}\hspace{-.25em}\right\},\\
  \label{eq:F14}
\end{eqnarray}
where the upper/lower sign in front of $\lO$ refers to the first/second possibility \myrf{VI.113a}-\myrf{VI.113b}/\myrf{VI.115a}-\myrf{VI.115b}.

\begin{center}
\epsfig{file=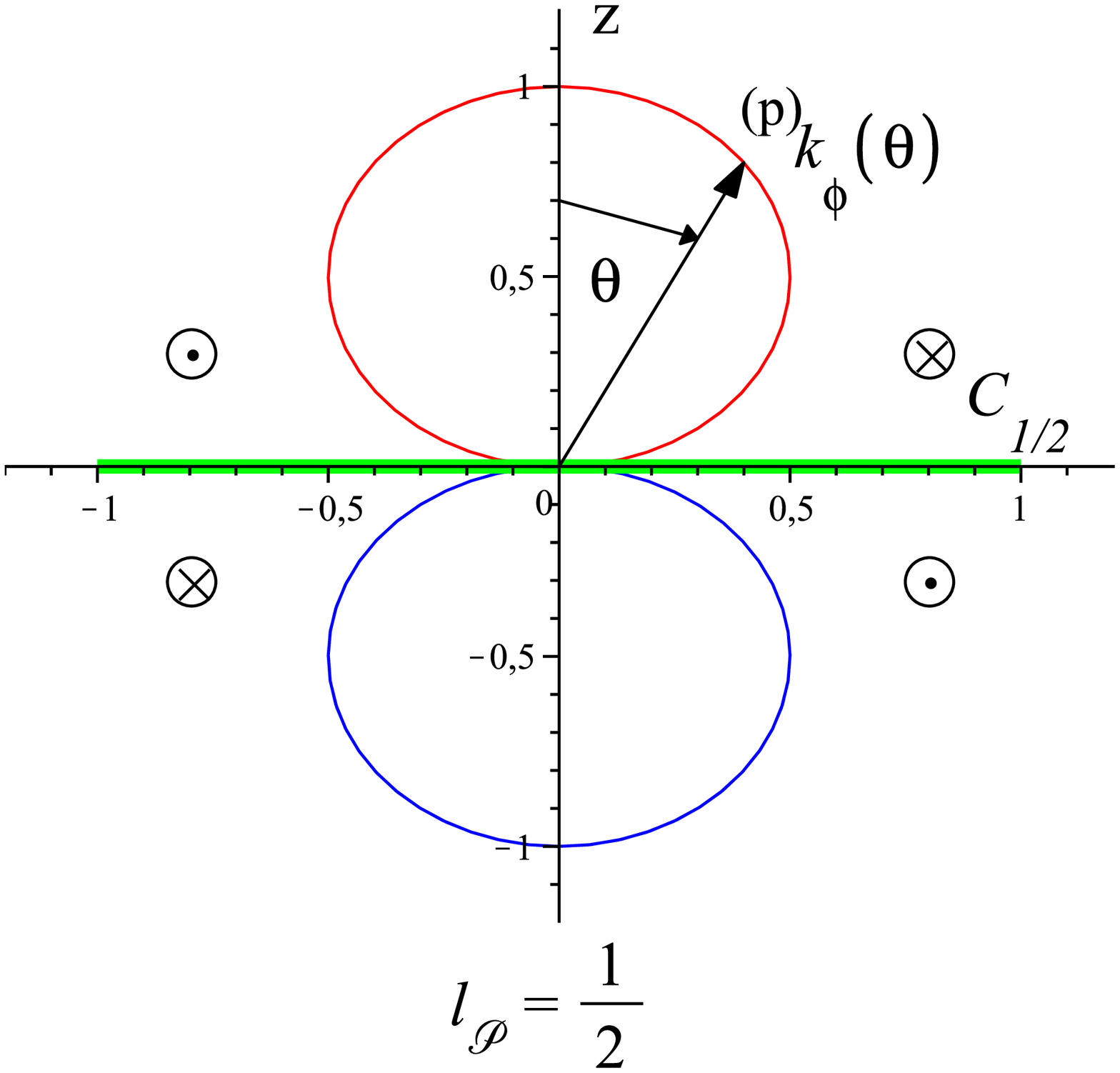,height=6cm}
\end{center}
\begin{center}
\epsfig{file=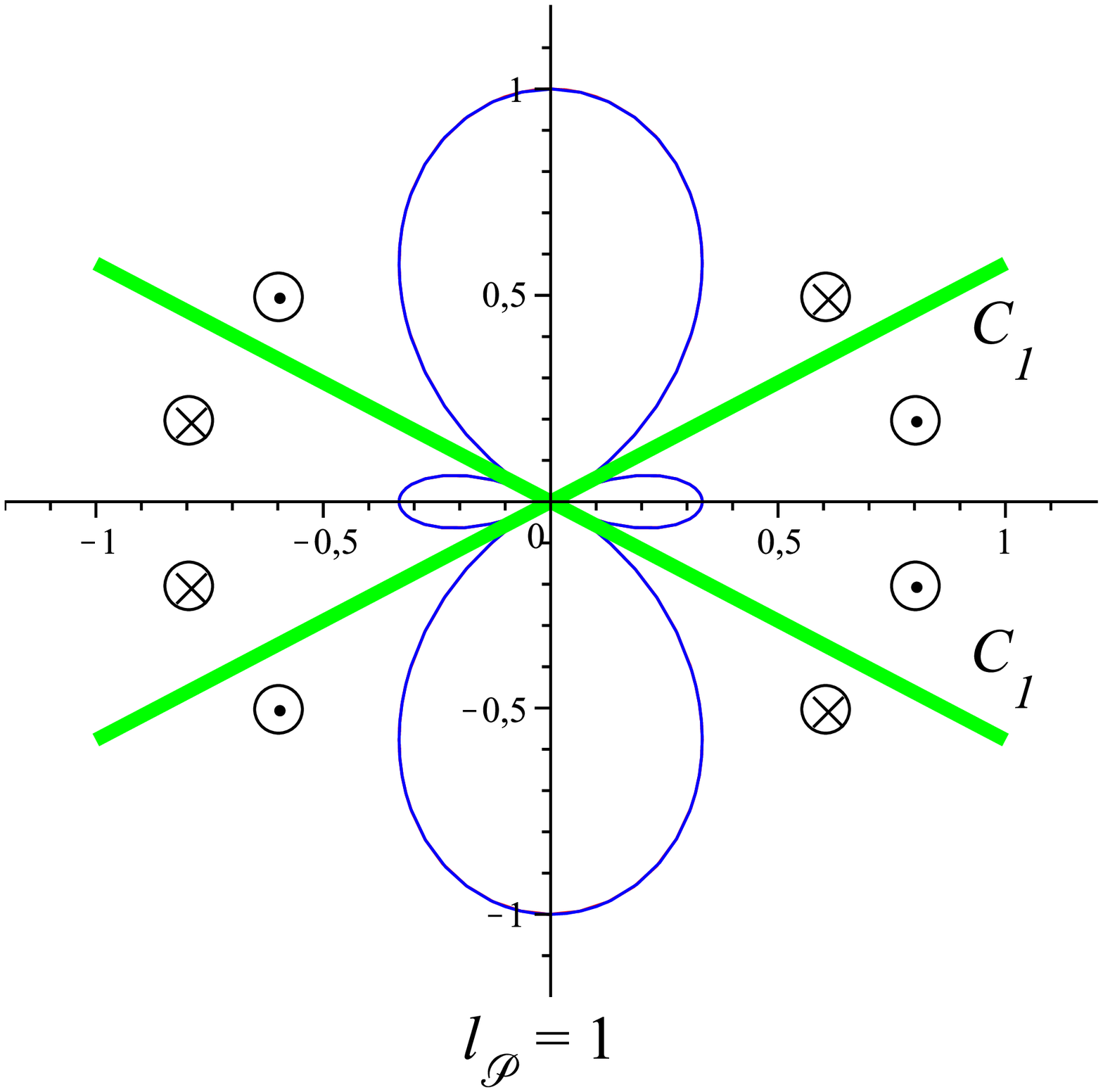,height=6cm}
\end{center}
\begin{center}
\epsfig{file=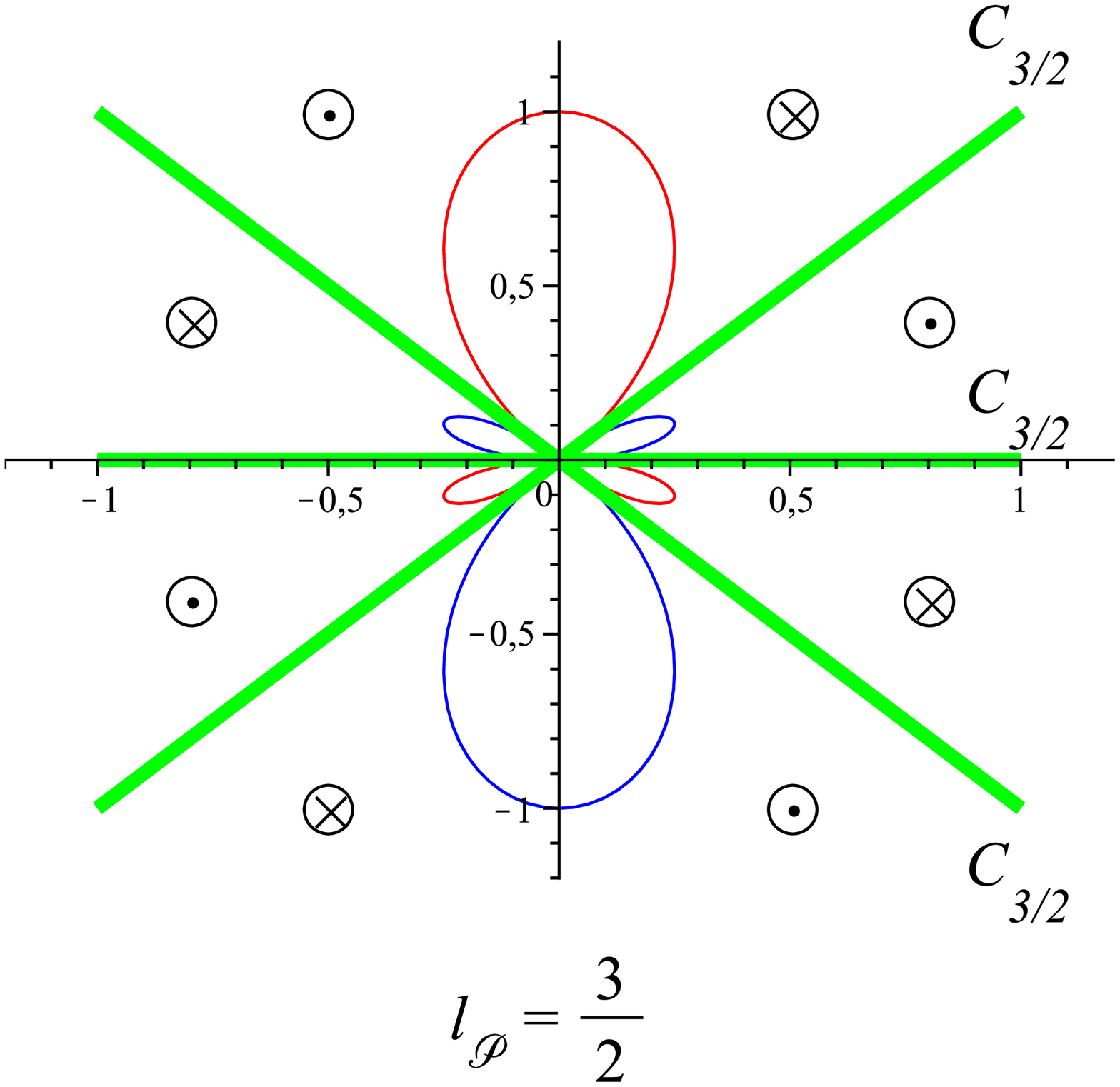,height=6cm}
\end{center}
{\textbf{Fig.\ 1:}\hspace{0.5cm} \emph{\large\textbf{Azimuthal Para-Current}~$\mathbf{
      {}^{[p]}k_\phi(\vartheta)}$}}

\pagebreak
{\textbf{Fig.\ 1:}\hspace{0.5cm} \emph{\large\textbf{Azimuthal Para-Current}~$\mathbf{ {}^{[p]}k_\phi(\vartheta)}$}}
\indent

The azimuthal current~${}^{(p)}k_\phi$ (VI.30) is sketched for fixed radial
variable~$r$. The current strength vanishes on the cones~$C_{1/2},C_1,C_{3/2},\ldots$
where it changes sign, and it becomes maximal between the cones, especially on the
z-axis~$(\vartheta=0,\pi)$. Therefore the magnetic field strength~$\vec{H}_p$ is singular on
the z-axis. For~$l_{\cal P}=0$ (i.e.\ the para-groundstate) the polar curve is a circle
centered in the origin. The flow direction of the para-current~$\vec{k}_p=
{}^{(p)}k_\phi\cdot \vec{e}_\phi$ becomes inverted across the cones (~$\bigotimes$
off-going;~$\bigodot$ approaching).

\begin{center}
\epsfig{file=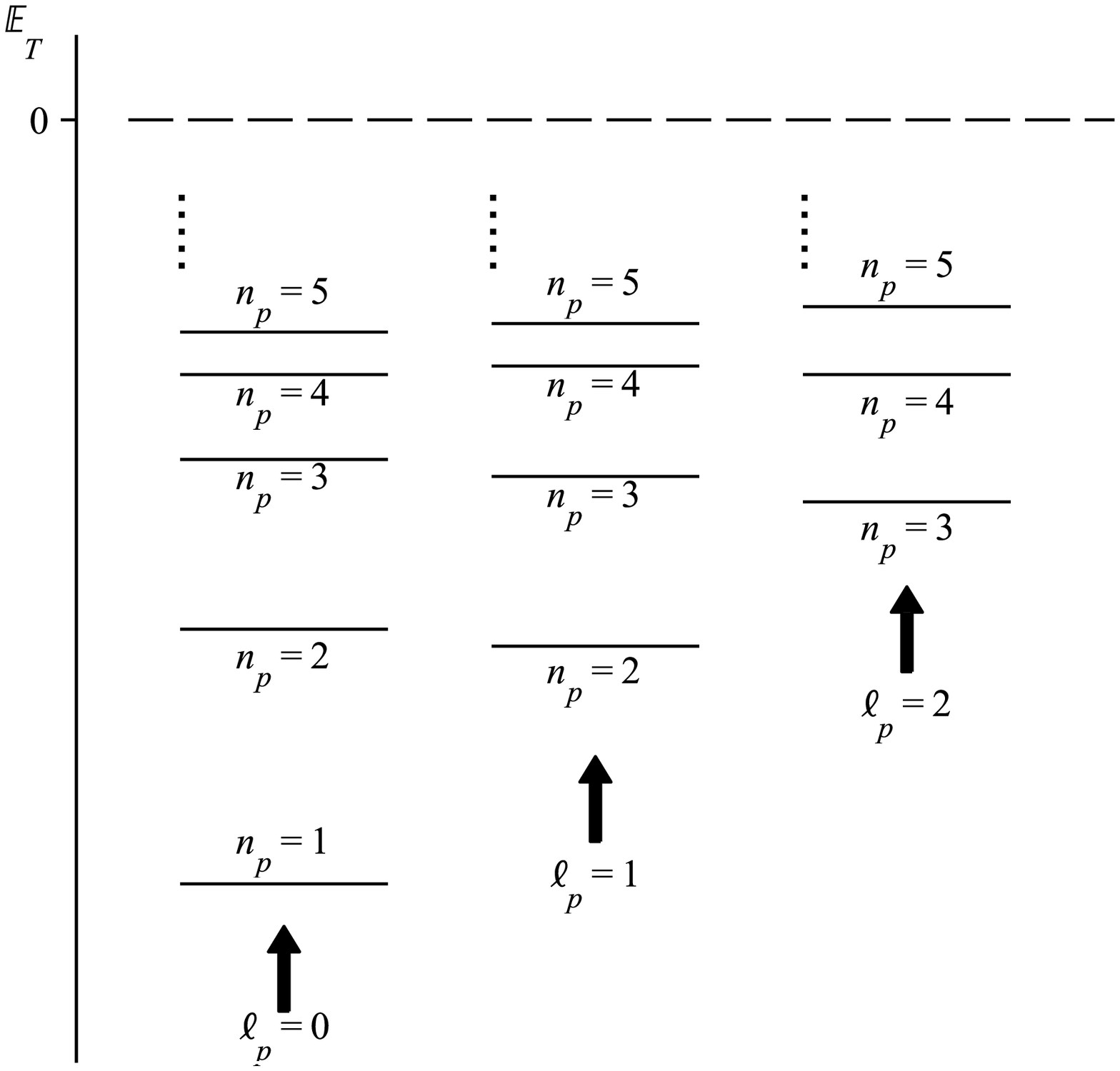,height=15cm}
\end{center}
{\textbf{Fig.\ 2:}\hspace{0.5cm} \emph{\large\textbf{Non-Relativistic Para-Spectrum}}}
\indent

The non-relativistic energy levels of para-positronium can be classified by the quantum
number~$\lP$ of orbital angular momentum~$(\lP=0,1,2,3,\ldots)$. For any (integer) value
of~$\lP$ there exists a subspectrum of excited states
which may be enumerated by the ``principal quantum number''~$\nP$. Degeneracy does occur
if states with the same~$\nP$ (but different~$\lP$) carry the same energy~$\EET$; whether
or not this is possible in the electrostatic, spherically symmetric and non-relativistic
approximation can be clarified not until more exact solutions of the para-eigenvalue
problem (\ref{eq:VI.42})-(\ref{eq:VI.44}) are at hand. For the use of \emph{approximate}
variational solutions, see fig.s~4a-4b. Observe also that the discreteness of the RST
spectrum arises here without any reference to the (anti)-commutation of operators.

\begin{center}
\epsfig{file=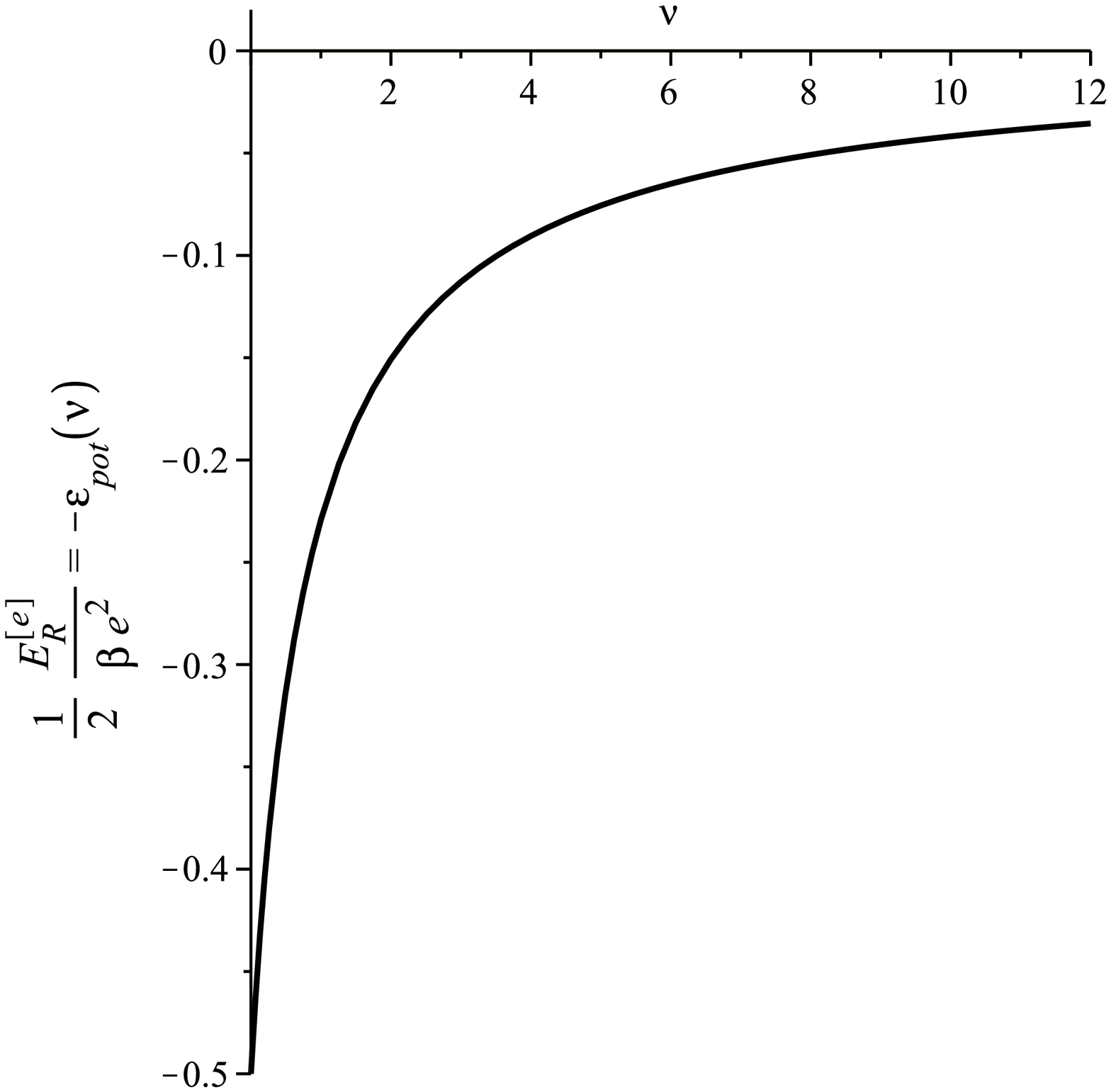,height=15cm}
\end{center}\vspace{1cm}
{\textbf{Fig.\ 3:}\hspace{0.5cm} \emph{\large\textbf{Interaction Energy}} {\large \ $\mathbf{\ERee=\tMMee} c^2$ }
\emph{\large\textbf{(\ref{eq:VI.68})-(\ref{eq:VI.70})}}}\\[-2ex]
\indent

The interaction energy (\ref{eq:VI.68})-(\ref{eq:VI.70}) for para-positronium is
essentially determined by the potential function~$\epot(\nu)$ (\ref{eq:VI.70}) and is
continuously increasing (up to zero) with increasing trial parameter~$\nu$
(\ref{eq:VI.55}). Concerning the values of~$\epot(\nu)$ for the half-integer values
of~$\nu$ see the table in Appendix~D. Since for~$\lP>0$ the kinetic energy~$\Ekin$
(\ref{eq:VI.65})-(\ref{eq:VI.66}) becomes infinite for~$\nu\to 0$, the total
energy~$\EET(\nu)$ (\ref{eq:VI.84}) develops a minimum (fig.~4) which represents the
binding energy of para-positronium.

\begin{center}
\epsfig{file=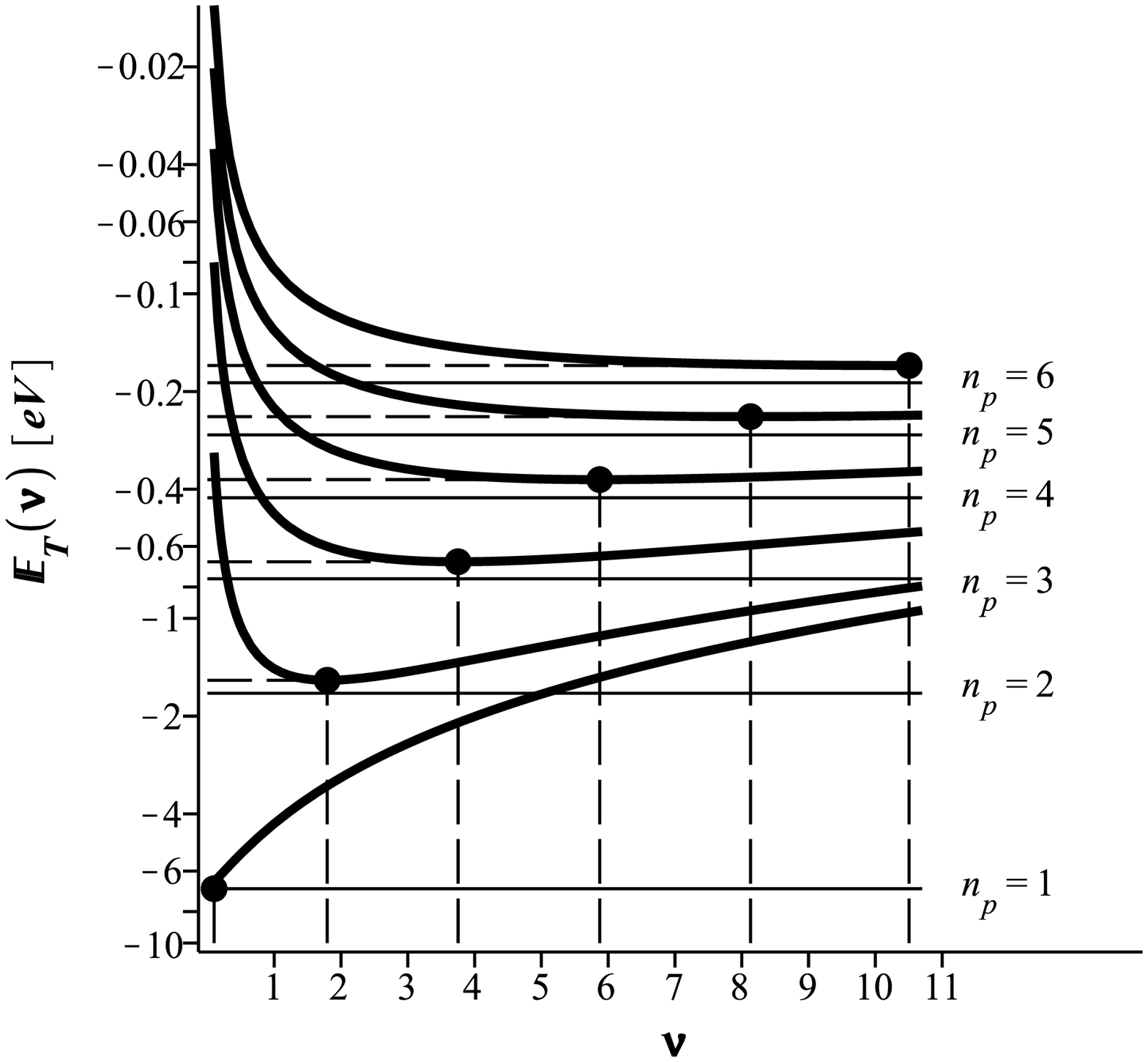,height=15cm}
\end{center}
{\textbf{Fig.\ 4a:}\hspace{0.5cm} \emph{\large\textbf{Energy Curves}} \large\textbf{$\EET(\nu)$}
  \emph{\large\textbf{ (\ref{eq:VI.84})}}}
\indent

The energy function~$\EET(\nu)$ (\ref{eq:VI.84}) has for any quantum number~$\lP$
\emph{one} minimum which represents the positronium binding energy. For~$\lP=0\ (\nP=1)$
there occurs a boundary minimum at~$\nu=0$ which then yields \emph{exact} coincidence of
the corresponding RST prediction and its conventional counterpart~$\Ec$
(\ref{eq:VI.85}). If one is willing to accept also weakly singular wave functions
(i.e.~$\nu<0$) there occurs a \emph{local} minimum also for~$\lP=0$, see
fig.~4b. For~$\lP>0$ the local minima of the energy curves~$\EET(\nu)$ predict the excited
positronium energies not better than up to a deviation of some 10\% because the chosen
trial function~$\tPhi(r)$ (\ref{eq:VI.55}) is too simple for the excited states, see the
table on p.~\pageref{tabE}.

\begin{center}
\epsfig{file=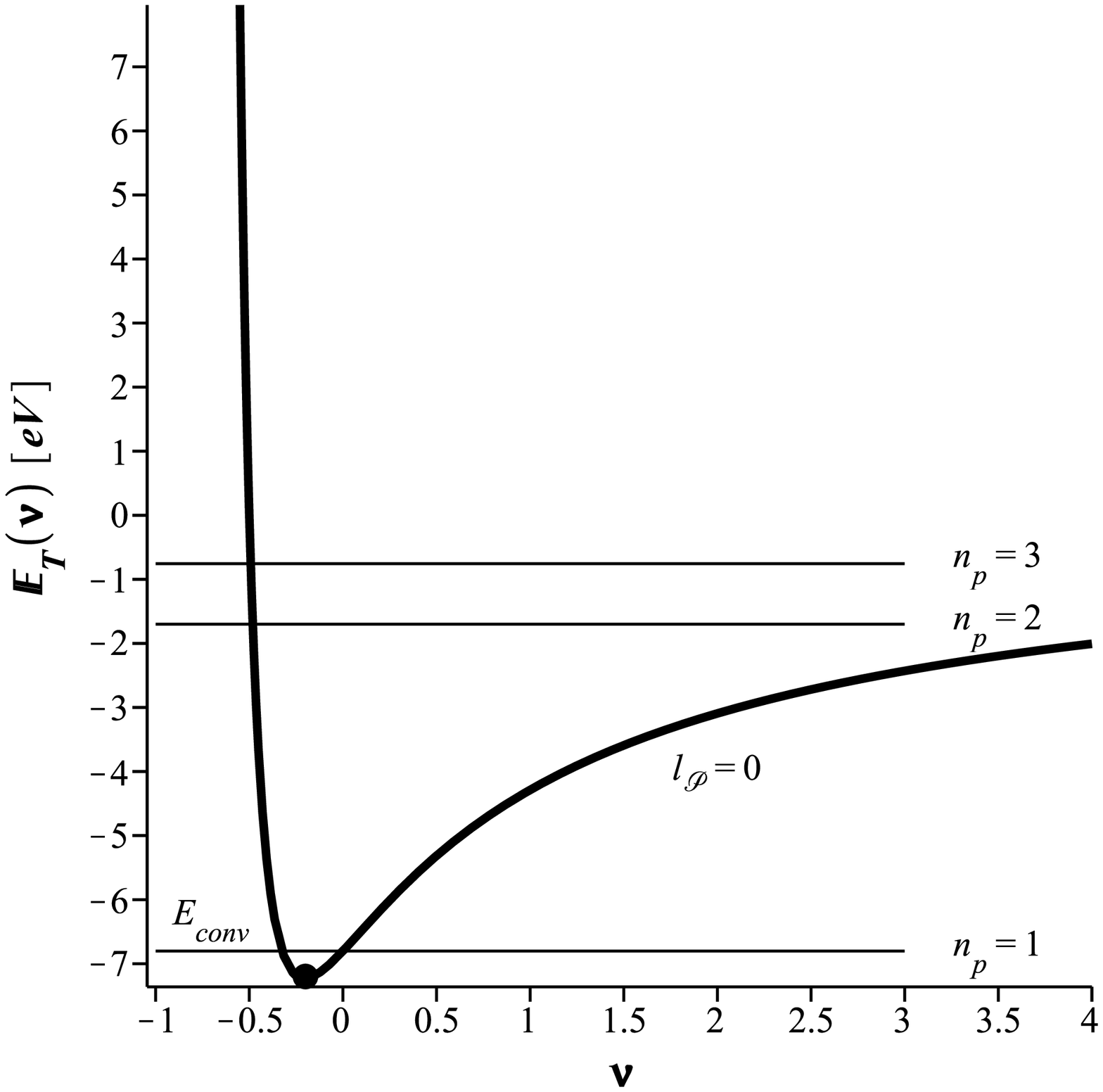,height=15cm}
\end{center}
{\textbf{Fig.\ 4b:}\hspace{0.5cm} \emph{\large\textbf{Non-Relativistic Groundstate Energy}} }
\indent

The energy curve~$\EET(\nu)$ (\ref{eq:VI.84}) for the groundstate~$(\lP=0)$ hits the
conventional prediction~$\Ec$ (\ref{eq:VI.85}) for vanishing variational
parameter~$(\nu=0)$. This then yields a boundary minimum if one wishes to exclude singular
solutions which have~$\nu<0$ (fig.~4a). However if one admits a weak
singularity~$(\nu\lesssim 0)$ one finds the true (local) minimum of the groundstate
curve~$(\lP=0)$ at $-7,23\ldots[eV]$ for~$\nu=-0,2049\ldots$. This supports the previous
prediction of~$-7,68\ldots[eV]$ which has been obtained by the use of hydrogen-like wave
functions, see ref.~\cite{MaSo}. This latter method must be considered to yield a more
exact estimation because it is based upon trial functions with more than the two
variational parameters~$(\beta,\nu)$ due to the present choice~$\tilde{\Phi}(r)$
(\ref{eq:VI.55}).

\begin{center}
\epsfig{file=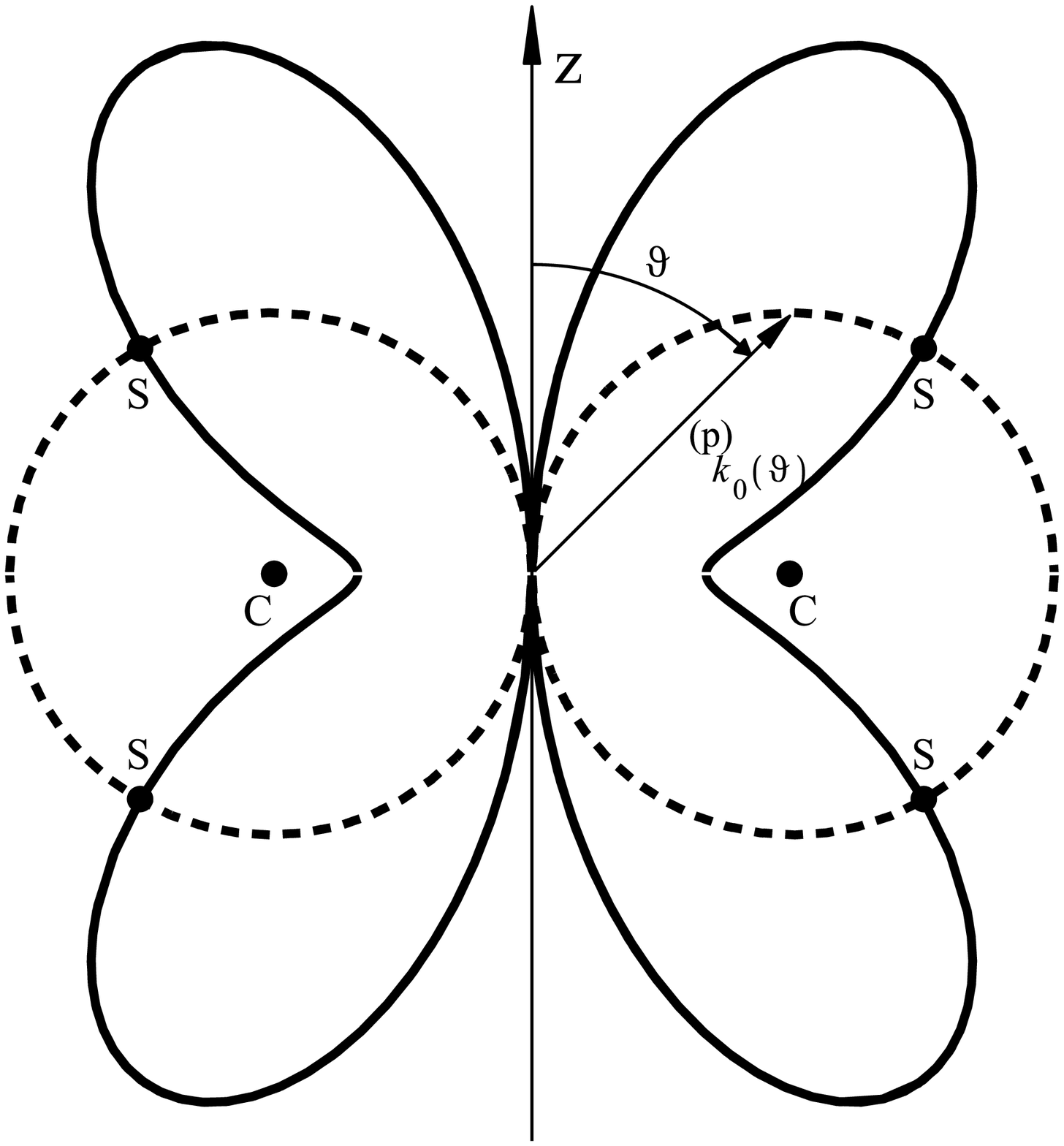,height=15cm}
\end{center}
{\textbf{Fig.\ 5:}\hspace{0.5cm} \emph{\large\textbf{Angular Factors}}\ \ \ \large\textbf{$\mathbf{\bbko(\vartheta)}$} 
  \emph{\large\textbf{ (\ref{eq:VI.147}) for Ortho-Positronium}}}
\indent

The ortho-configurations do occur in two forms which are characterized by their different
angular distributions~$\bbko(\vartheta)$ (\ref{eq:VI.147a})-(\ref{eq:VI.147b}). The first
distribution (\ref{eq:VI.147a}) (broken lines) is characterized by circles with center~C
at a distance~$1/2\pi^2$ from the z-axis. The second distribution (\ref{eq:VI.147b})
(solid curves) equals the first one (\ref{eq:VI.147a}) on the intersections~$S$ with the
three-cones of vertex angle~$\vartheta(s)=\pi/3,2\pi/3$.

\emph{For the spherically symmetric approximation} both types of charge
distributions~$\bbko(r,\vartheta)$ are replaced by the symmetric
density~$\tilde{\Omega}(r)/4\pi r$ which then generates the spherically symmetric
potential~$\bAn(r)$ according to the Poisson equation (\ref{eq:VI.151}). By this
neglection of the anisotropy, there arises the desired \emph{ortho/para degeneracy} which
thus appears as a very crude approximation within the framework of RST.

\begin{center}
\epsfig{file=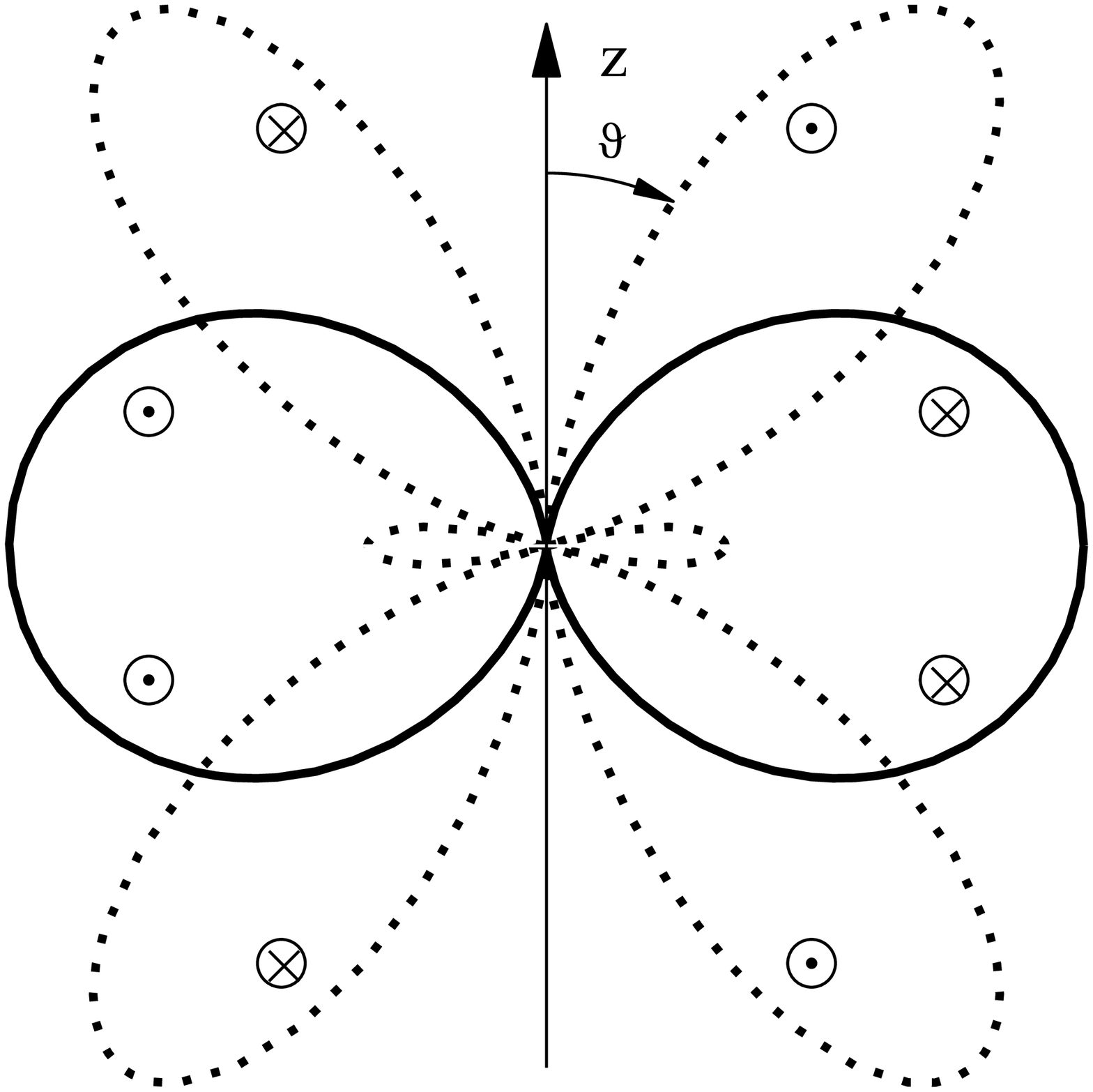,height=15cm}
\end{center}
{\textbf{Fig.\ 6:}\hspace{0.5cm} \emph{\large\textbf{Azimuthal Ortho-Current}}\ \ \ \large\textbf{$\mathbf{\bbkphi(\vartheta)}$} 
  \emph{\large\textbf{ (\ref{eq:VI.150})}}}
\indent

The ortho-current~$\bkphi(r,\vartheta)$ (\ref{eq:VI.150}) factorizes into a radial and a longitudinal part 
\begin{equation*}
  \bkphi(r,\vartheta)=\bbkphi(r)\cdot\bbkphi(\vartheta)\ ,
\end{equation*}
quite analogously to its para-counterpart~$\pkphi(r,\vartheta)$ (\ref{eq:VI.30}), see
fig.~1. However, whereas the angular part~$\ppkphi(\vartheta)$ of the para-current is
unique for any value of~$\lP$, the present ortho-case~$\ppkphi(\vartheta)$
(\ref{eq:VI.150}) for~$\lO=2$ splits up into the two subcases
(\ref{eq:VI.150a})-(\ref{eq:VI.150b}) with
\begin{equation*}
\bbkphi(\vartheta) = \left(\frac{2}{\pi}\sin\vartheta\right)^2\cdot
\begin{cases}
  1\ , & \text{solid line}\\
  \frac{1}{3}\left(1-16\cos^2\vartheta \right)\ , &\text{broken line\ .}
\end{cases}
\end{equation*}
Observe here that in the second case (broken line) the ortho-current~$\bkphi(r,\vartheta)$
changes sign on the three-cones determined by~$\cos^2\vartheta=1/16$.

\renewcommand{\refname}{{


\section*{Complete List of RST Papers}
\begin{itemize}
\item \textit{Relativistic Schr\"odinger Equations}\\
M. Sorg, Preprint (1992)
\* 
\* 
\item 
\textit{Fermions and Expanding Universe}\\
U. Ochs and M. Sorg, Int. Journ. Theor. Phys. \textbf{32}, 1531 (1993)
\* 
\* 
\item 
\textit{Kinematics of Dirac's Spinor Field}\\
M. Mattes and M. Sorg, J. Phys. A \textbf{26}, 3013 (1993)
\* 
\* 
\item 
\textit{Relativistic Schr\"odinger Equations and the Bohm-Aharonov Effect}\\
M. Mattes and M. Sorg, J. Phys. Soc. Jap. \textbf{63}, 2532 (1994)
\* 
\* 
\item 
\textit{Cosmic Spinor Fields and the Early Universe}\\
U. Ochs and M. Sorg, Int. J. Theor. Phys. \textbf{33}, 2157 (1994)
\* 
\* 
\item 
\textit{Integrability Conditions, Wave Functions, and Conservation Laws for
the Relativistic Schr\"odinger Equations}\\
M. Mattes and M. Sorg, Nuovo Cimento \textbf{109B}, 1097 (1994)
\* 
\* 
\item 
\textit{Spin Precession and Expansion of the Universe}\\
M. Sorg, Nuovo Cimento \textbf{109B}, 465 (1994)
\* 
\* 
\item 
\textit{Cosmological Principle and Primordial Spinor Fields}\\
M. Sorg, Lett. Math. Phys. \textbf{33}, 113 (1995)
\* 
\* 
\item 
\textit{Relativistic Schr\"odinger Equations and the Particle-Wave Duality}\\
U. Ochs and M. Sorg, J. Phys. Soc. Jap. \textbf{64}, 1120 (1995)
\* 
\* 
\item 
\textit{Relativistic Generalization of Bohm's Quantum Potential}\\
M. Mattes and M. Sorg, Nuovo Cimento \textbf{110}, 1323 (1995)
\* 
\* 
\item 
\textit{Non-linear Generalization of the Relativistic Schr\"odinger Equations}\\
U. Ochs and M. Sorg, Z. Naturforsch. \textbf{51a}, 965 (1995)
\* 
\* 
\item 
\textit{Non-singular Cosmological Solutions for the Coupled Dirac-Einstein Equations}\\
U. Ochs and M. Sorg, J. Phys. A \textbf{28}, 7263 (1995)
\* 
\* 
\item 
\textit{Cosmological Solutions for the Coupled Einstein-Yang-Mills-Higgs Equations}\\
U. Ochs and M. Sorg, Gen. Rel. Grav. \textbf{28}, 1177 (1996)
\* 
\* 
\item 
\textit{Matter Production in the Early Universe}\\
M. Mattes, U. Ochs and M. Sorg, Int. J. Theor. Phys. \textbf{35}, 155
(1996)
\* 
\* 
\item 
\textit{Zur Theorie der\*  Relativistischen Schr\"odingergleichung}\\
U. Ochs, PhD thesis, Stuttgart (1996)
\* 
\* 
\item 
\textit{Identities and Conservation Laws in Relativistic Schr\"odinger Theory}\\
M. Sorg, Nuovo Cimento \textbf{112B}, 23 (1997)
\* 
\* 
\item 
\textit{Quantum Mixtures in the Early Universe}\\
T. Sigg and M. Sorg, Gen. Rel. Grav. \textbf{29}, 1557 (1997)
\* 
\* 
\item 
\textit{Topological Currents in Relativistic Schr\"odinger Theory}\\
M. Sorg, J. Phys. \textbf{A 30}, 5517 (1997)
\* 
\* 
\item 
\textit{Mixtures and Pure States in Relativistic Schr\"odinger Theory}\\
M. Mattes and M. Sorg, Int. Journ. Theor. Phys. \textbf{30}, 395 (1997)
\* 
\* 
\item 
\textit{Exchange Effects in Relativistic Schr\"odinger Theory}\\
T. Sigg and M. Sorg, Nuov. Cim. \textbf{B113}, 1261 (1998)
\* 
\* 
\item 
\textit{Second-Order Mixtures in Relativistic Schr\"odinger Theory}\\
M. Mattes and M. Sorg, Journ. Math. Phys. \textbf{40}, 71 (1999)
\* 
\* 
\item 
\textit{Two-Particle Systems in Relativistic Schr\"odinger Theory}\\
M. Mattes and M. Sorg, J. Phys \textbf{A32}, 4761 (1999)
\* 
\* 
\item 
\textit{External and Internal Motion in Relativistic Schr\"odinger Theory}\\
M. Mattes and M. Sorg, Nuov. Cim. \textbf{B114}, 815 (1999)
\* 
\* 
\item 
\textit{Zitterbewegung and Quantum Jumps in Relativistic Schr\"odinger Theory}\\
S. Rupp, T. Sigg, and M. Sorg, Int. Journ. Theor. Phys. \textbf{39,}
1543 (2000)
\* 
\* 
\item 
\textit{Fermion-Boson Dichotomy of Matter and RelativisticSchr\"odinger Theory}\\
S. Rupp and M. Sorg, Preprint (2000)
\* 
\* 
\item 
\textit{Mixing Group for Relativistic 2-Particle Quantum States}\\
S. Rupp and M. Sorg, Phys. Rev. \textbf{A 63}, 022112 (2001)
\* 
\* 
\item 
\textit{Electromagnetic (Self-)Interactions in Relativistic Schr\"odinger
Theory}\\
M. Mattes, S. Rupp, and M. Sorg, Can. J. Phys. \textbf{79,} 879 (2001)
\* 
\* 
\item 
\textit{Mixture Degeneracy in\*  Relativistic} \textit{Schr\"odinger Theory}\\
S. Rupp and M. Sorg, Nuov. Cim. B \textbf{116}, 739 (2001)
\* 
\* 
\item 
\textit{Positive and Negative Mixtures in Relativistic Schr\"odinger Theory}\\
S. Rupp and M. Sorg, Int. Journ. Theor. Phys.\*  \textbf{40,} 1817
(2001)
\* 
\* 
\item 
\textit{Geometry and Topology of Relativistic Two-Particle Quantum Mixtures}\\
S. Hunzinger, M. Mattes and M. Sorg, http://xxx.lanl.gov/abs/hep-th/0301205
\* 
\* 
\item 
\textit{2-Teilchen
Systeme in der Relativistischen Schr\"odingertheorie}\\
S.Rupp, PhD thesis, Stuttgart (2001)\\ http://elib.uni-stuttgart.de/opus/volltexte/2002/996
\* 
\* 
\item 
\textit{Exchange Degeneracy of Relativistic Two-Particle Quantum States}\\
S. Rupp, S. Hunzinger and M. Sorg, Found. Phys.\*  \textbf{32,\* }
705 (2002)
\* 
\* 
\item 
\textit{Exchange Interactions and Mass-Eigenvalues of Relativistic Two-Particle
Quantum Mixtures}\\
S. Rupp and M. Sorg, Nuov. Cim. B \textbf{117}, 549 (2002)
\* 
\* 
\item 
\textit{Relativistic Wave Equations for Many-Particle Quantum Systems}\\
S. Rupp, Phys. Rev. A \textbf{67}, 034101 (2003)
\* 
\* 
\item 
\textit{Topological Quantum Numbers of Relativistic Two-Particle Mixtures}\\
S. Pruss-Hunzinger, S. Rupp and M. Sorg, http://arxiv.org/abs/hep-th/0305227 
\* 
\* 
\item 
\textit{Relativistic Schr\"odinger Theory and the Hartree-Fock Approach}\\
M. Verschl and M. Sorg, Found. Phys. \textbf{33}, 913 (2003)
\* 
\* 
\item 
\textit{Quantum Entanglement in Relativistic Three-Particle Systems}\\
P. Schust, M. Mattes and M. Sorg, Found. Phys. \textbf{34}, 99 (2004)
\* 
\* 
\item 
\textit{Energy Functional in Relativistic Schr\"odinger Theory}\\
S. Rupp and M. Sorg, Nuov. Cim. \textbf{B} 118, 259 (2003)
\* 
\* 
\item 
\textit{Relativistic Effects in Bound Two-Particle Systems}\\
M. Verschl, M. Mattes and M. Sorg, Eur. Phys. J. \textbf{ A 20}, 211 (2004)
\* 
\* 
\item 
\textit{Fermions in Relativistic Schr\"odinger Theory}\\
S. Pruss-Hunzinger and M. Sorg, Nuov. Cim. \textbf{B} 118, 903 (2003)
\* 
\* 
\item 
\textit{U(N)-Eichtheorien mit abel'scher Symmetrie-Brechung f\"ur relativistische N-Teilchen Systeme}\\
M. Verschl, Diplomarbeit, Universit\"at Stuttgart (2003)
\* 
\* 
\item 
\textit{Relativistic Schr\"odinger Theory as a Lorentz Covariant Generalization of the Hartree-Fock Approach}\\
M. Verschl, preprint (2003)
\* 
\* 
\item 
\textit{Magnetic Interactions in Relativistic Two-Particle Systems}\\
P. Schust and M. Sorg, http://arxiv.org/abs/hep-th/0410023 
 \* 
\* 
\item 
\textit{Relativistic Ground-State of the Heavy Helium-like Ions}\\
S. Pruss-Hunzinger, M. Mattes and M. Sorg, Nuov. Cim. \textbf{B} 119, 277 (2004)
\* 
\* 
\item 
\textit{Self-Energy and Action Principle in Relativistic Schr\"odinger Theory}\\
P. Schust, F. Stary, M. Mattes and M. Sorg, Found. Phys. \textbf{35}, 
1043 (2005) 
\* 
\* 
\item 
\textit{Relativistic Ionization Energies of the Helium-like Ions}\\
S. Pruss-Hunzinger,  F. Stary, M. Mattes and M. Sorg, Nuov. Cim. \textbf{B} 120, 467 (2005)
 \* 
\* 
\item 
\textit{Ortho- and Para-helium in Relativistic Schr\"odinger Theory}\\
F. Stary and M. Sorg, Found. Phys. \textbf{36}, 1325 
(2006) 
\* 
\* 
\item 
\textit{Variationsprinzip f\"ur die Relativistische Schr\"odinger Theorie mit Anwendung auf Zwei-Elektronen-Atome}\\
F. Stary, Diplomarbeit, Universit\"at Stuttgart (2005) 
\* 
\* 
\item 
\textit{Helium Multiplet Structure in Relativistic Schr\"odinger Theory}\\
R. Gr\"abeldinger, T. Beck, M. Mattes and M. Sorg, preprint (2005),\\
http://arxiv.org/abs/physics/0602087
\* 
\* 
\item 
\textit{Relativistic Energy Levels of Para-helium}\\
R. Gr\"abeldinger, P. Schust, M. Mattes and M. Sorg, preprint (2006),\\
http://arxiv.org/abs/physics/0609081
\* 
\* 
\item 
\textit{Selbst-Energie des Dirac Elektrons in der Relativistischen Schr\"odinger Theorie}\\
R. Gr\"abeldinger, Diplomarbeit, Universit\"at Suttgart (2006) 
\* 
\* 
\item 
\textit{Das Dreiteilchen-Problem in der Relativistischen Schr\"odinger Theorie}\\
T. Beck, Diplomarbeit, Universit\"at Suttgart (2006) 
\* 
\* 
\item 
\textit{Positive and Negative Charges in Relativistic Schr\"odinger Theory}\\
T. Beck and M. Sorg, preprint (2006), http://arxiv.org/abs/hep-th/0609164
\* 
\* 
\item 
\textit{Two- and Three-Particle Systems in Relativistic Schr\"odinger Theory}\\
T. Beck and M. Sorg, Found. Phys. \textbf{37}, 1093 (2007)
\* 
\* 
\item 
\textit{Positronium Groundstate in Relativistic Schr\"odinger Theory}\\
T. Beck, M. Mattes and M. Sorg, preprint (2007),\\
http://arxiv.org/abs/0704.3810
\* 
\* 
\item  
\textit{Principle of Minimal Energy in Relativistic Schr\"odinger Theory}\\
M. Mattes and M. Sorg, preprint (2007), http://arxiv.org/abs/0708.1489
\* 
\* 
\item 
\textit{Non-Relativistic Positronium Spectrum in Relativistic Schr\"odinger Theory}\\
M. Mattes and M. Sorg, preprint (2008), http://arxiv.org/abs/0803.2289
\* 
\* 
\item 
\textit{Exchange Interactions and Principle of Minimal Energy in Relativistic Schr\"odinger Theory}\\
M. Mattes and M. Sorg, preprint (2008), http://arxiv.org/abs/0809.4692
\* 
\* 
\end{itemize}

\end{document}